%% file: thesis.tex
\documentclass{ucbthesis}
\usepackage{amsmath}
\usepackage{amssymb} 
\usepackage{graphicx}
\usepackage{empheq} 
\newcommand*\widefbox[1]{\fbox{\hspace{2em}#1\hspace{2em}}}

\usepackage[utf8]{inputenc}
\DeclareUnicodeCharacter{0301}{\'\i}
\DeclareUnicodeCharacter{1E84}{\"W}

\newtheorem{exercise}{Exercise}  
\def\>{\rangle}
\def\<{\langle}
\def\t{\dagger}
\def\T{\intercal}
\def\->{\rightarrow}
\def\=>{\implies}
\def\6{\partial}
\def\Tr{\text{Tr}}
\def\real{\text{Re}}
\def\imag{\text{Im}}

\usepackage{color} 

\newcommand{\subheading}[1]{\textbf{#1}}

\newcommand{\major}[1]{\textit{#1}} 
\newcommand{\ped}[1]{\textbf{#1}} 
\newcommand{\definition}[1]{\textit{#1}} 
\newcommand{\erf}[1]{Eq.~(\ref{#1})} 
\newcommand{\erfs}[1]{Eqs.~(\ref{#1})} 

\newcommand{\nn}{\nonumber}

\begin{document}


\title{\normalfont Quantum feedback for measurement and control}
\author{Leigh Martin}
\degreesemester{Summer}
\degreeyear{2019}
\degree{Doctor of Philosophy}
\cochairs{Professor Irfan Siddiqi}{Professor Birgitta Whaley}
\othermembers{Professor Hartmut H\"affner\\
  Professor Eric Neuscamann}
\numberofmembers{4}
\field{Physics}
\campus{Berkeley}


\maketitle
\copyrightpage

\include{abstract}

\begin{frontmatter}

\begin{dedication}
\null\vfil
\begin{center}
To my parents, my step parents, and my amazing sister Willow.
\end{center}
\vfil\null
\end{dedication}


\tableofcontents
\clearpage
\listoffigures
\clearpage
\listoftables

\begin{acknowledgements}
The fact that research is not an isolated endeavor is one of the most important lessons I learned from my time at Berkeley. 
The quality of one's research, and the joy of doing it are a direct result of one's collaborators, colleagues and friends. I feel incredibly fortunate to have worked with such a talented, curious and kind group of people. I first wish to thank my advisors, Irfan Siddiqi and Birgitta Whaley for taking a risk on me by supporting my joint work in theory and experiment. The chance to work with both research groups has been an amazing and irreplaceable opportunity. I wish to thank both of them for providing the perfect balance of encouragement and critique, and guidance and freedom. In my experimental work, I am deeply grateful for the guidance of and collaboration with Shay Hacohen-Gourgy and Emmanuel Flurin, who showed me the ropes of experimental work, helped me discover my flaws and strengths, and never turned down an chance to discuss a crazy idea (no matter how sure they were that it was wrong!). I am also indebted to Mollie Schwartz, who helped give me the opportunity to work in Irfan's group and introduced the field to me. Her warmth and encouragement made an enormous difference in embarking on a new path in research. In my theory work, I am indebted to the guidance and sharp intuition of Felix Motzoi, whose initial suggestion for a project carried me through a PhD's worth of theory research, as well as Mohan Sarovar, whose mentorship gave me confidence and stability. I also wish to thank Mahrud Sayrafi, Sissi Wang, Yitian Chen, Song Zhang and Yuxiao Jiang for the meetings all over Berkeley while we carried out our joint projects. I looked forward and enjoyed each and every one of these discussions, which consistently took us in exciting and unexpected directions. In my experimental work, I especially wish to thank Vinay Ramasesh, who taught me the importance of cordial competition and open communication, and William Livingston, who always helped me see the light in the darkness of challenge or my own stubbornness. Many of the general insights that I attempt to convey in this thesis are of their making. I greatly appreciate Machiel Blok's support, and all of the late afternoons spent tossing around ideas (also my apologies to Esther Blok for the countless times that I made Machiel late). I also wish to thank Sydney Schreppler, Kevin O'Brien, John Mark Kreikebaum, Andrew Eddins, David Toyli and Norman Yao for their support, collaborations and friendship. Finally, an enormous thank you to my parents, step parents, sister and friends in Berkeley, whose support was everything.
\end{acknowledgements}

\end{frontmatter}

\pagestyle{headings}


\include{IntroQuantum}
\include{IntroCQED}
\include{OpenSystemsMeasurement}
\include{MeasControl}
\include{NonCommutingObservables}

\include{AdaptivePhase}

\include{OptimalMeas}
\include{Outlook}


\bibliographystyle{plain}
\bibliography{LeighBibliography}

\end{document}

%% file: abstract.tex

\begin{abstract}



The standard quantum formalism introduced at the undergraduate level treats measurement as an instantaneous collapse. In reality however, no physical process can occur over a truly infinitesimal time interval. A more subtle investigation of open quantum systems lead to the theory of continuous measurement and quantum trajectories, in which wave function collapse occurs over a finite time scale associated with an interaction. Within this formalism, it becomes possible to ask many new questions that would be trivial or even ill-defined in the context of the more basic measurement model. In this thesis, we investigate both theoretically and experimentally what fundamentally new capabilities arise when an experimental apparatus can resolve the continuous dynamics of a measurement. Theoretically, we show that when one can perform feedback operations on the timescale of the measurement process, the resulting tools provide significantly more control over entanglement generation, and in some settings can generate it optimally. We derive these results using a novel formalism which encompasses most known quantum feedback protocols. Experimentally, we show that continuous measurement allows one to observe the dynamics of a system undergoing simultaneous non-commuting measurements, which provides a reinterpretation of the Heisenberg uncertainty principle. Finally, we combine the theoretical focus on quantum feedback with the experimental capabilities of superconducting circuits to implement a feedback controlled quantum amplifier. The resulting system is capable of adaptive measurement, which we use to perform the first canonical phase measurement.

\end{abstract}

%% file: IntroQuantum.tex
\chapter{The Building Blocks of Quantum Devices: Qubits, Harmonic Oscillators and Quantum Optics}
\label{ch:IntroQuantum}


\section{Introduction}

The ability to trap and control individual electrons, atoms and molecules turned the most bizarre thought experiments of quantum mechanics into tangible, thoroughly verified demonstrations. Once-contentious interpretational issues have been explored at the level of single, well-controlled degrees of freedoms. The double-slit experiment has been performed on single electrons sent and detected one-by-one, confirming wave-particle duality \cite{bach2013controlled,back2013youtube}. Quantum superposition of increasingly macroscopic objects is becoming routine. In the last few years, an experiment spanning an entire university campus succeeded in violating Bell's inequality and many loopholes that could have preserved a more classical way of thinking\cite{Hensen2015}. In short, modern atomic, molecular and optical experiments have demanded a very different way of thinking about physics. 

One of the strangest and most subtle differences between classical and quantum mechanics is measurement. Measurement bridges the conceptual and operational gap between wave functions and classical probabilities via the Born rule. At the most basic level, the Born rule is stated as an instantaneous collapse of the wave function into an eigenstate determined by the detection apparatus. In reality however, no process in nature is truly instantaneous. Atoms spontaneously emit via a continuous interaction with the electromagnetic vacuum, for example. Furthermore, no measurements are perfectly sharp, meaning that they do not determine precisely one measurement outcome with perfect certainty. Position measurements are limited by the de Broglie wavelength of the probe, and even discrete-outcome measurements like the readout of a digital memory are corrupted by some amount of noise. On the surface, neither of these inevitable constraints are captured by the Born rule's most basic statement.

When applied and interpreted carefully, the Born rule leads to a natural generalization of instantaneous projective measurements that meshes consistently with the above limitations. It predicts that wave function collapse is actually a continuous process, and hence wave function coherence is perturbed rather than destroyed over sufficiently short timescales. This modification forces the notion of a measurement rate dictated by the interaction strength between the system and its environment. The resulting dynamics are called quantum trajectories, which represent the finest-grained picture of a measurement. In most atomic, molecular and optical (AMO) systems, trajectories are exquisitely difficult to observe. The fundamental challenge is one of efficiency; the coupling strength of a quantum system to its environment is likely greater than that to a detector in our possession. For example, if we monitor the spontaneous emission from an atom with a localized detector, there is a good chance that we will miss the majority of emitted photons. Consequently, by the time we have enough information about the system to to say anything definitive, the environment at large has already collapsed the wave function for us.

\begin{figure}
\centering
{\includegraphics[width = 160mm]{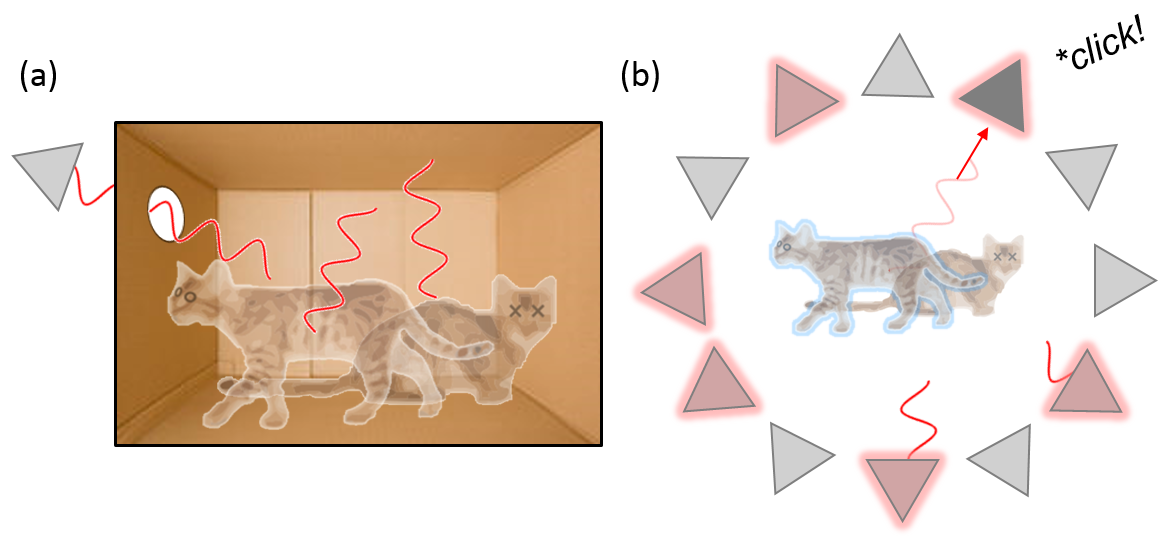}}
\caption{(a) An imperfectly isolated quantum system quickly shares information with its environment. An external detector typically acquires only a small fraction of the total available information, and hence sees no residual coherence during the measurement process. (b) In highly controlled systems such as superconducting circuits, one can arrange that the system interacts most strongly with our detector, allowing one to resolve wave function collapse as a continuous-time process.} 
\label{fig:BelavkinsCat}
\end{figure}

In recent years, many foundational quantum phenomena have also been observed in artificial quantum systems like circuits. In much the same way that atomic energy levels consist of orbiting electrons, oscillating and circulating currents in electric circuits can also exhibit quantized energy levels. Such artificial atoms offer a greater range of control, as both their energy levels and their environment can be tailored. For superconducting circuits, they can be made to interact most strongly with a 1-dimensional waveguide, so that all information emitted by the system can be collected by a single detector. Combined with innovations in quantum-limited amplifiers, these systems recently enabled the first experimental observation of quantum trajectories\cite{Murch:2013ur}, though important experimental advances along these lines had already been achieved with Rydberg atoms\cite{gleyzes2007quantum}.

With the ability to observe quantum trajectories comes many new experimental capabilities. Indeed, the full scope of these new capabilities is still as much a subject of theoretical discovery as experimental realization. The scope of opportunities is vast. Despite the apparent abstractness of quantum mechanics, the proposed applications of coherent quantum technology ranges from computation to astrophysical observation, and many derive from the ability to enhance measurement devices with uniquely quantum effects. 

Quantum trajectories not only resolve measurement at its finest scales, but also provide the opportunity to act on a system in unique ways. This thesis asks what fundamentally new capabilities arise from continuous measurement, and experimentally implements two foundational examples. One basic example is the ability to measure non-commuting observables simultaneously. In the most basic statement of the Born rule, such a task is ill-defined. However realistic systems interact with their environments via multiple degrees of freedom, and hence this process is occurring constantly in nature. 

The other new operation afforded by quantum trajectories is \definition{quantum feedback}, whereby one acts upon a system during the measurement process. From a theoretical vantage point, we consider tasks that are commonly achieved probabilistically with measurement, such as entanglement generation, and show that they can be made to operate deterministically. In a few cases, we are able to prove that the resulting protocols are optimal. In the final experimental work, we apply feedback not to a system that we wish to control but rather to the measurement apparatus itself. By changing the measurement basis during wave function collapse, we implement the first canonical phase measurement.


\subheading{Outline of this thesis} Theses are a primary route of knowledge transfer within fields and outline the necessary background for research. To this end, we include somewhat more introductory material than average, in hopes that the majority of content becomes accessible to an undergraduate who has taken upper-division quantum mechanics. Chapters and sections begin with basic, general calculations and end with a qualitative overview of important material outside the scope of our presentation. To balance familiar and specialized material, we often give a slightly unconventional presentation when reviewing standard topics. Chapter 1 starts with basic quantum mechanics to introduce some concepts that are often skipped in an undergraduate course, like density matrices, frame transformation and the rotating wave approximation. Taking a slightly more information-based approach, we walk through the basics of qubit manipulation and control at a level that keeps the material relevant to all AMO-like systems. Exercises are included in the first three chapters as a way to flag concepts that are easily proven but divert from the flow of presentation. The interesting results are typically given in the problem statement, so the reader can view them as extensions of the text without working through them if desired.

Chapter 2 turns to the experimental system of this work, superconducting circuits. We overview at a high level the essential elements of superconductivity that arise in the field, and then jump into the quantization of lumped-element and beyond-lumped-element circuits. After explaining the basic classes of qubits, we take our first foray into open quantum systems in our introduction to parametric amplifiers. Chapter 3 builds a foundation for the theoretical contribution of this thesis, describing weak measurement, quantum trajectories and decoherence. Chapter 4 describes our work on quantum feedback, outlining methods for designing quantum feedback protocols. We apply the formalism to the task of remote entanglement generation and show how one can make ordinarily stochastic processes deterministic in realistic systems. 

Chapter 5 presents the first experimental work of this thesis. We explain the details of achieving high quantum efficiency measurements in circuit QED, an essential ingredient for quantum feedback. We then describe a novel circuit QED measurement technique, which we use to observe the dynamics of a system under simultaneous non-commuting measurements. As non-commuting observables are necessary to observe non-classical effects, the experiment brings a qualitatively new platform to the study of quantum foundations and their consequences, one that closely resembles the interactions found in nature. As a first step in this direction, we show theoretically and experimentally how the continuous disturbance induced by measurement is bounded by a modern reformulation of the Heisenberg uncertainty principle.

Chapter 6 fully merges the experimental and theoretical works, applying quantum feedback to a parametric amplifier to implement a canonical, or `ideal' phase measurement. The broader significance of the result is that quantum feedback can turn a standard measurement device into something much more general, and hence able to observe a larger class of physical properties. Finally, in chapter 7 we return to the feedback protocols of chapter 4, proving optimality of a number of feedback protocol and discussing the implications.

\section{Qubits and Harmonic Oscillators}

There are two common approaches to quantum physics. The standard approach transitions from classical mechanics to the Schr\"odinger equation, emphasizing wave mechanics and only ending with atomic physics and other discrete systems. For review, we briefly detour through a less common but potentially more modern approach, which begins with quantum information in order to build up the framework, and then connects to atoms (and circuits) later. In order to avoid excessive abstraction, we focus on the dynamics that are commonly involved in experimental qubit characterization: Rabi oscillations, Ramsey oscillations and phenomena associated with the Jaynes-Cummings Hamiltonian. The advantage of the abstraction of this chapter is that the material is relevant to most physical implementations of a qubit, including trapped ions, cold neutral atoms, nitrogen vacancy centers, nuclear spins, quantum dots and of course superconducting circuits.

\subheading{Qubits in abstraction} Qubits are the smallest and most fundamental objects in quantum information. A qubit is any system that may be described by a two-dimensional Hilbert space, either because it only has two available states (such as an electron spin) or because it is somehow prevented from leaving a two-dimensional subspace (for instance by only driving transitions between a single pair of states in a multi-level system).
\begin{equation}
    |\psi_q\> = \alpha |0\> + \beta |1\>
\end{equation}
is a general qubits state, where $\alpha$ and $\beta$ are complex and normalized such that $|\alpha|^2 + |\beta|^2 = 1$. According to the Born rule, if one measures an operator $\mathcal{O}$ (which for a qubit is a $2\times2$ Hermitian matrix), then the possible states after measurement are the eigenvectors of $\mathcal{O}$ $|\lambda\>$, which are typically labeled by their corresponding eigenvalues $\lambda$. In general, the probability for a given measurement outcome $\lambda$ to occur is then given by
\begin{equation} \label{eq:WaveFuncBornRule}
    P(\lambda|\psi_q) = |\<\lambda|\psi_q\>|^2.
\end{equation}
Note that as $P$ is unchanged if we replace $|\psi_q\>$ with $e^{i\phi}|\psi_q\>$, states that differ only by a global phase are physically indistinguishable.\footnote{One also needs the fact that $|\psi_q\>$ and $e^{i\phi}|\psi_q\>$ evolve identically under the Schr\"odinger equation, which follows from its linearity.} A primary focus of this thesis is the physical process underlying measurement, which we describe beginning in chapter \ref{ch:OpenSystems}.

The direct relationship between wave functions and probabilities has lead many to suggest that at its heart, quantum mechanics is as much a generalization of mathematical probability theory as a physical theory. As we will see shortly, the inherent richness of complex numbers allow for interference between distinct possibilities via time (Hamiltonian) evolution. However the wave function by itself does not incorporate classical uncertainty, so strictly speaking, it is more a distinct case than a generalization of a probability. The most general case must include quantum systems in classically uncertain states. For instance, one might have randomly prepared states $|\psi_i\>$ with probabilities $P_i$ such that $0\leq P_i\leq 1$ and $\sum_i P_i = 1$. It is convenient to define the \definition{density matrix} for this scenario as
\begin{equation} \label{eq:DefRho}
    \rho \equiv \sum_i P_i |\psi_i\>\<\psi_i|,
\end{equation}
which again applies for any quantum system, not just a qubit. At first glance, one might think that the density matrix does not suffice to completely characterize a quantum state; given a density matrix, there may be many possible ways to decompose it in terms of pure states $|\psi_i\rangle$ and probabilities $P_i$. Surprisingly, if two probabilistic ensembles of pure states yield the same density matrix, then they are physically indistinguishable! To show this, we observe how the density matrix allows one to restate the Born rule so as to take into account the randomness of state preparation. Noting that $|\lambda\>\<\lambda|$ is the density matrix associated with an eigenstate of $\mathcal{O}$, we define the measurement outcome probability as
\begin{equation} \label{eq:BornRuleRho}
    P(\lambda|\rho) \equiv \Tr[|\lambda\>\<\lambda|\rho] = \sum_i P_i \<\lambda|\psi_i\>\<\psi_i|\lambda\> = \sum_i P_i P(\lambda|\psi_i)
\end{equation}
where we have used the cyclic property of the trace ($\Tr[AB]=\Tr[BA]$) and the fact that $\Tr[a]=a$ if $a$ is a scalar. \erf{eq:BornRuleRho} is just a weighted average of the measurement outcome probabilities for each possible state preparation $|\psi_i\>$, so the right hand side reproduces the expected probability based on \erf{eq:WaveFuncBornRule}. A similar argument shows that $\<\mathcal{O}\> = \Tr[\mathcal{O}\rho]$, where the expectation value of an operator is defined over its eigenvalues \textit{i.e.} as $\<\mathcal{O}\> \equiv \sum_\lambda P(\lambda)\lambda$. As the measurement probabilities are uniquely determined by $\rho$, $\rho$ completely characterizes the state of a system.

\begin{exercise} \label{ex:RhoPositive}
Show that $\rho=\rho^\t$, that its trace always equals 1, and that the eigenvalues of the density matrix are greater than or equal to $0$ (\textit{i.e.} that it is positive semidefinite). Any matrix satisfying these three properties is a valid density matrix.  
\end{exercise}

\begin{exercise} \label{ex:PureStates}
Pure states: show that if $\Tr[\rho^2]=1$, then $\rho$ is rank 1 and thus can be written in the form $|\psi\>\<\psi|$. If $\Tr[\rho^2]<1$, $\rho$ is referred to as a mixed state.
\end{exercise}

Returning to the specific example of a qubit, the most general qubit state is conveniently expressed in terms of the Pauli matrices as
\begin{align} \label{eq:DefBlochVec}
    &\rho_q = \frac{\sigma_0 + x \sigma_x + y \sigma_y + z \sigma_z}{2} \\ \nonumber
    &\sigma_x = 
    \begin{pmatrix}
    0 & 1 \\
    1 & 0
    \end{pmatrix} ~~~
    \sigma_y = 
    \begin{pmatrix}
    0 & -i \\
    i & 0
    \end{pmatrix} ~~~
    \sigma_z = 
    \begin{pmatrix}
    1 & 0 \\
    0 & -1
    \end{pmatrix}
\end{align}
where $\sigma_0$ is the $2\times 2$ identity matrix and $x$, $y$ and $z$ must be real to keep $\rho$ Hermitian. Using the identity $\Tr[\sigma_i \sigma_j] = 2\delta_{i,j}$, one can show the following useful facts:
\begin{enumerate}
    \item $x=\<\sigma_x\>\equiv \Tr[\sigma_x \rho]$, and likewise for $y$ and $z$.
    \item The purity of $\rho_q$ as per exercise \ref{ex:PureStates} is $(1+r^2)/2$ where $r=\sqrt{x^2+y^2+z^2}$. Thus $\rho$ is pure if and only if $r=1$.
    \item From 2,  it follows that $\rho_q$ is a valid density matrix as per exercise \ref{ex:RhoPositive} if and only if $0\leq r \leq 1$.
\end{enumerate}
As the Pauli matrices together with $\sigma_0$ span the vector space of $2\times 2$ Hermitian matrices, every possible qubit state can be represented uniquely by specifying $x$, $y$ and $z$. These facts justify the representation of a qubit as a 3-vector, called the \definition{Bloch vector}; $x$, $y$ and $z$ are measurable and fully specify $\rho_q$. In many physical systems, the Bloch vector corresponds to a physical orientation in space, such as the magnetic dipole moment of an electron. Although the Pauli matrices generalize nicely to the Gell-Mann matrices for larger systems, the statements analogous to 2 and 3, even for a qutrit (3-level system), are somewhat more complicated (see Eqs (13-18) of \cite{goyal2016geometry}).

\subheading{Schr\"odinger equation for wave functions and density matrices} The Bloch vector offers a convenient visual representation for intuiting the dynamics of a qubit. A physical qubit has an associated Hamiltonian, which governs its dynamics. For an isolated qubit, the only term in the Hamiltonian is an intrinsic energy splitting between the ground and excited states, conventionally taken to be along the $\sigma_z$ axis \textit{i.e.} $H_0/\hbar = \omega_q \sigma_z/2$. The Schr\"odinger equation is
\begin{align}
    \frac{\6 |\psi\>}{\6 t} = -\frac{i}{\hbar} H_0|\psi\> \implies \frac{\6\rho}{\6 t} = -\frac{i}{\hbar}[H_0,\rho].
\end{align}
The version for density matrices may be derived by taking the time derivative of \erf{eq:DefRho}. As the Hamiltonian is already diagonalized, the Schr\"odinger equation decouples into two first order differential equations, one for $\alpha$ and one for $\beta$. The general solution for an initially pure state is
\begin{equation} \label{eq:QubitLabFrameSolution}
    |\psi_q(t)\> = \alpha_0 e^{-i\omega_q t/2}|0\> + \beta_0 e^{i\omega_q t/2} |1\> \implies \rho(t) =
    \begin{pmatrix}
        |\alpha_0|^2 & \alpha_0 \beta_0^* e^{-i\omega_q t}\\
        \alpha_0^* \beta_0 e^{i\omega_q t} & |\beta_0|^2 
    \end{pmatrix}
\end{equation}
where we have taken slightly unfortunate but ubiquitous convention that $|1\rangle$ designates the ground state. Thus under the evolution of $H$, the relative phase between $|0\>$ and $|1\>$ varies, but the populations in each state do not. In the density matrix, these observations are apparent in the off-diagonal and diagonal elements, respectively.

\begin{figure}
\centering
{\includegraphics[width = 160mm]{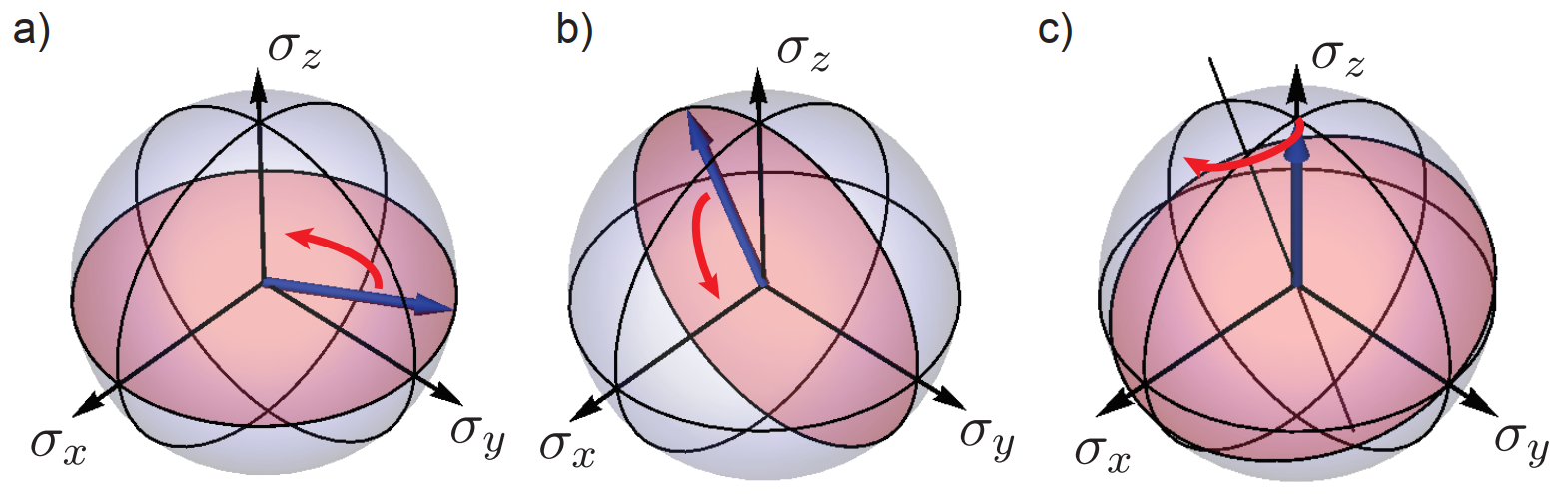}}
\caption{(a). Precession of a qubit in the lab frame. The direction of rotation may be obtained by the right-hand rule, and the angular frequency is $\omega_q$. (b). Rabi oscillations about the $\sigma_x$ axis ($\delta=0$) in the rotating frame. The angular frequency is $\Omega_R = g\alpha_0/2$. (c). Rabi oscillations for a detuned drive ($\Delta > \Omega_R$), shown in the frame of the frame of the classical drive field $\omega_d$.} 
\label{fig:BlochVectorDynamics}
\end{figure}

The dynamics of the Bloch vector are shown in Fig. \ref{fig:BlochVectorDynamics}a. The Bloch vector precesses around the $\sigma_z$ axis at a frequency $\omega_q$. The Bloch vector of a mixed state is the average Bloch vector of each pure state in the underlying statistical ensemble, so the result applies to mixed states as well. The alignment between the rotation axis and the Hamiltonian illustrates another useful property of the Bloch vector. As the Hamiltonian of an isolated qubit may always be expressed in terms of the Pauli matrices, we can also associate to it a vector. As is evident from the solution \erf{eq:QubitLabFrameSolution}, the vector associated with $H_0$ is simply the rotation axis. Furthermore, the vector associated with $\rho$ and $H_0$ transform identically under a change of (Hilbert space) basis, so a solution for $H\propto \sigma_z$ is in fact a general solution for all time-independent Hamiltonians. For example, $H\propto \sigma_y$ would generate rotations about the $y$ axis instead.

\subheading{Rabi oscillations and rotating frames} To control the state of a qubit, and in particular to change the populations in $|0\>$ and $|1\>$, we need to generate rotations about an axis other than $\sigma_z$. To do so, we must somehow couple our qubit to the outside world. The simplest control method is to drive oscillations with a classical control field. Most physical qubits couple to external fields via some kind of dipole moment. Electron spins couple to magnetic fields, atomic energy levels to electric fields etc. The resulting Hamiltonian is called the Rabi model, and the dynamics are called Rabi oscillations.

It is convenient to express the dipole operator $\sigma_x$ in terms of the ladder operator $\sigma = (\sigma_x - i \sigma_y)/2$, which satisfies $\sigma^2=0$, $\sigma|0\> = |1\>$, and $\sigma^\dagger|1\>=|0\>$.\footnote{We will justify $\sigma_x$ as a dipole operator in chapter \ref{ch:IntrocQED}. Why we single out $\sigma_x$ and not $\sigma_y$ is a matter of convention, but nevertheless carries some physical meaning. We will see how this works when we consider physical qubits in the following chapter.} Parameterizing the unitless classical control field as $\alpha(t)$, the qubit Hamiltonian with a dipolar coupling is
\begin{equation}
    H/\hbar = \omega_q \frac{\sigma_z}{2} + g \alpha(t)(\sigma+\sigma^\t)
\end{equation}
where $g$ parameterizes the coupling strength. 

Typically $g \alpha \ll \omega_q$, so the first term dominates the dynamics. However we understand the dynamics induced by the first term already and would like to focus on the second term. The first term can be eliminated by going into a \definition{rotating frame}, which is similar to a rotating coordinate system in classical mechanics. To do so, we define a transformed state $|\tilde{\psi}(t)\> = U(t)|\psi(t)\>$, where $U(t)$ is a time-dependent unitary matrix that acts as a basis transformation. Substitution of $|\psi(t)\>=U^\t(t) |\tilde{\psi}(t)\>$ into the Schr\"odinger equation yields
\begin{empheq}[box=\widefbox]{align} \label{eq:RotatingFrame}
    \frac{\6 |\tilde{\psi}(t)\>}{\6 t} &= -\frac{i}{\hbar} \tilde{H}|\tilde{\psi}(t)\> \\ \nonumber
    \tilde{H} &= U(t) H U^\t(t) - i\hbar U(t) \dot{U}^\t(t) \\ \nonumber
    &= e^{-i X t} H e^{i X t} + \hbar X
\end{empheq}
where in the last line we have defined $U(t)=\exp(-i X t)$, which is guaranteed to be unitary if $X$ is Hermitian. By substitution into the original Schr\"odinger equation, one can show that $U(t)|\psi(0)\>$ is a solution to the Schr\"oodinger equation when $H/\hbar=X$, so that $U(t)$ acts as a time evolution operator. The above transformation is a key method with many applications, and we will use it frequently. The first term of $\tilde{H}$ acts as a basis transformation of the Hamiltonian, while the second term is somewhat analogous to a Coriolis potential. 

\begin{figure}
\centering
{\includegraphics[width = 100mm]{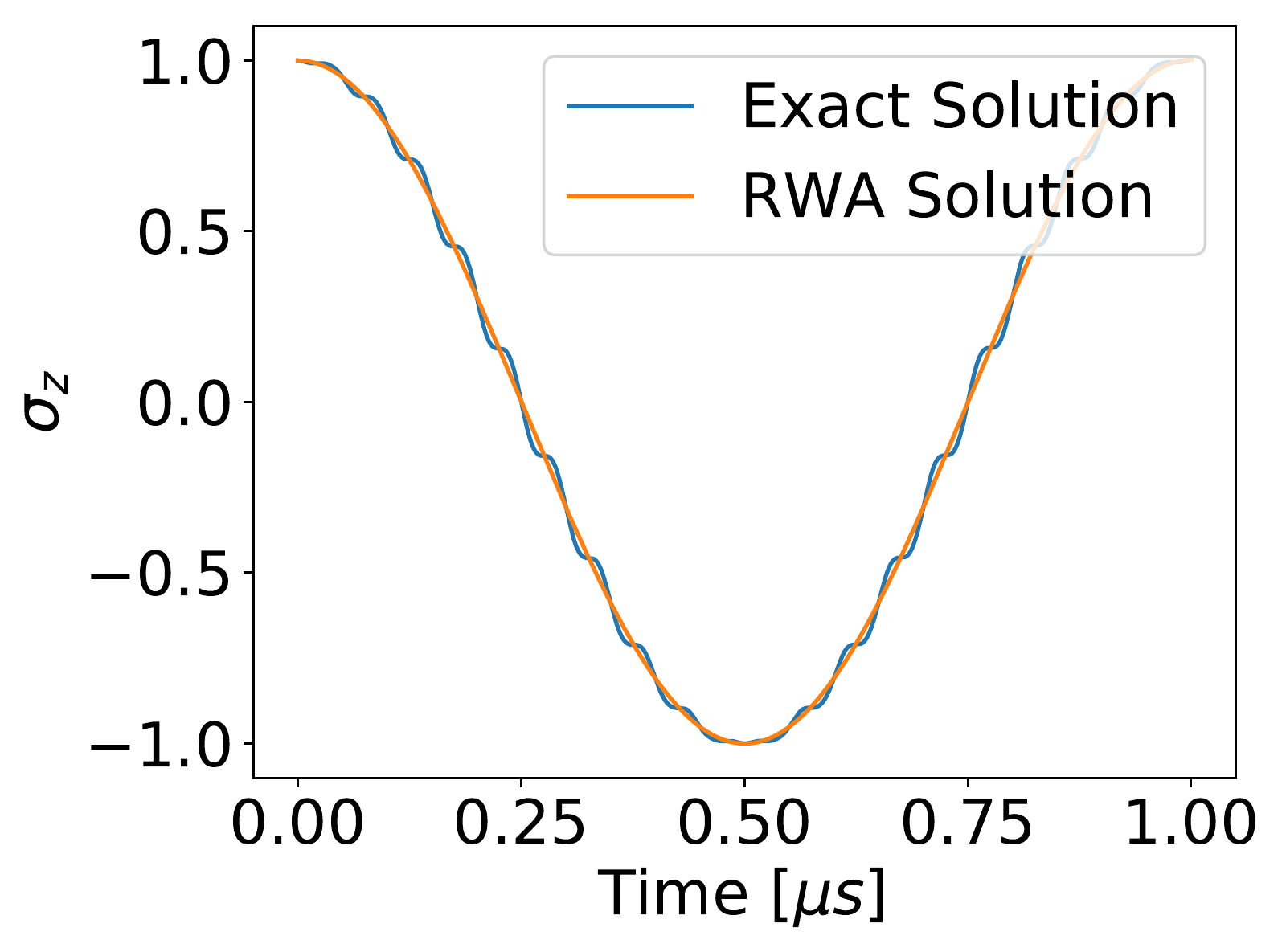}}
\caption{A plot of the z component of the Bloch vector vs. time with $\tilde{H}$ and $\tilde{H}_\text{RWA}$. The exact solution oscillates rapidly around the approximate solution but never deviates far from it. Parameters for the graphic are $\Omega_R/2\pi = 0.5$ MHz and $\omega_q/2\pi = 20$ MHz. In most realistic systems, the separation in scales between $\Omega_R$ and $\omega_q$ is even greater.} 
\label{fig:RWA}
\end{figure}

Intuitively, we wish to transform into a basis that corotates with the dynamics induced by $H_0$, which is accomplished by setting $X=-H_0/\hbar$. As $H_0$ is diagonal, it is straightforward to exponentiate $X$ (for example via a Taylor series) and calculate $\tilde{H}$
\begin{equation} \label{eq:HQubitRotatingFrame}
    \tilde{H}/\hbar = g \alpha(t)(\sigma e^{-i\omega_q t} + \sigma^\dagger e^{i \omega_q t})
\end{equation}
To control the qubit, one can create a resonant interaction by letting $\alpha(t) = \alpha_0 \cos(\omega_d t + \delta)$ with the drive frequency $\omega_d = \omega_q$
\begin{equation}
    \tilde{H}/\hbar = \frac{g \alpha_0}{2}(\underbrace{\sigma e^{i\delta}+\sigma^\t e^{-i\delta}}_\text{co-rotating terms} + \underbrace{\sigma e^{-2i \omega_q t - i\delta} + \sigma^\t e^{2i\omega_q t+i\delta}}_\text{counter-rotating terms}).
\end{equation}
The Schr\"odinger equation is not simple to solve if we consider all four terms. Fortunately, solution is straightforward when only the co-rotating terms are considered, which is called the \definition{rotating wave approximation}. In Fig. \ref{fig:RWA}, we plot numerical solutions to the Schr\"odinger equation with and without counter-rotating terms, which agree very well.

To see why the counter-rotating terms contribute negligibly to the dynamics, first consider the solution in the absence of the counter-rotating terms, which we call $|\tilde{\psi}_\text{RWA}(t)\>$. $|\tilde{\psi}_\text{RWA}(t)\>$ changes on the time scale of $1/\alpha g$, which is slow compared to $\omega_q$. To see if it is a good approximate solution, we compute the first order change in the wave function due to the counter-rotating terms by integrating the Schr\"odinger equation with only those terms present (\textit{i.e.} using time-dependent perturbation theory). The result is 
\begin{align} \label{eq:RWAIntegral}
|\tilde{\psi}(t)\> - |\tilde{\psi}_\text{RWA}(t)\> 
 \approx \int_0^t \frac{g \alpha_0}{2} \left[\sigma e^{-2i \omega_q t' - i\delta} + \sigma^\t e^{2i\omega_q t'+i\delta} \right] |\tilde{\psi}_\text{RWA}(t')\>dt'.
\end{align}
As $|\tilde{\psi}(t)\>$ is very nearly constant over the timescale $1/\omega_q$, the integrand oscillates rapidly and therefore the integral vanishes to good approximation. Thus adding in the counter-rotating terms has little effect on the solution, and we may replace $\tilde{H}$ with
\begin{align} \label{eq:HQubitRWA}
    \tilde{H}_\text{RWA}/\hbar &= \frac{\Omega_R}{2}\sigma_\delta \\ \nonumber
     \sigma_\delta &\equiv \sigma e^{i\delta}+\sigma^\t e^{-i\delta} = \sigma_x \cos(\delta) + \sigma_y \sin(\delta)
\end{align}
where $\Omega_R = g \alpha_0$ is called the Rabi frequency. 
As $\tilde{H}_\text{RWA}$ is equivalent to $H_0$ up to a basis change, it still induces oscillations, except this time about the $\sigma_\delta$ axis. The dynamics are shown in Fig. \ref{fig:BlochVectorDynamics}b.

\begin{exercise} \label{ex:DetunedRabi}
In the case that the control field is detuned from the qubit resonance frequency, one can make $\tilde{H}_\text{RWA}$ time independent by instead choosing $X=-i\omega_d\sigma_z/2$. Show that now $\tilde{H}_\text{RWA}/\hbar = \Delta \sigma_z/2 + \Omega_R \sigma_\delta/2$, where $\Delta = \omega_q - \omega_d$. Diagonalize this Hamiltonian to show that the Rabi frequency is $\Omega_R = \sqrt{\Omega_R^2 + \Delta^2}$. See Fig. \ref{fig:BlochVectorDynamics}c.
\end{exercise}

\begin{exercise}
Using the solution to exercise \ref{ex:DetunedRabi}, let $|\tilde{\psi}(0)\> = |1\>$ and suppose that $\Delta$ sweeps slowly from a large negative value to a large positive value. What is the final state? Recall (or read about) adiabatic theorem.
\end{exercise}

It is interesting to pause and contrast these dynamics to those of a classical bit. If a classical bit is prepared in an equal mixture of $0$ and $1$, no reversible transformation can take it to a definite state. By comparison, a qubit in the state $(|1\>+|0\>)/\sqrt{2}$ is indistinguishable from a classical mixture if one performs $\sigma_z$ measurements, but can be converted to $|1\>$ (or $|0\>$) alone under the action of $\tilde{H}_\text{RWA}$! Such a transformation would map $|1\> \rightarrow (|1\>+|0\>)/\sqrt{2}$ and $|0\> \rightarrow (|1\>-|0\>)/\sqrt{2}$, which interfere to yield $|1\>$. At the level of quantum information, this interference is no different than the interference of a particle travelling through a pair of slits to yield a diffraction pattern. The Hilbert space is just smaller.

\begin{figure}
\centering
{\includegraphics[width = 120mm]{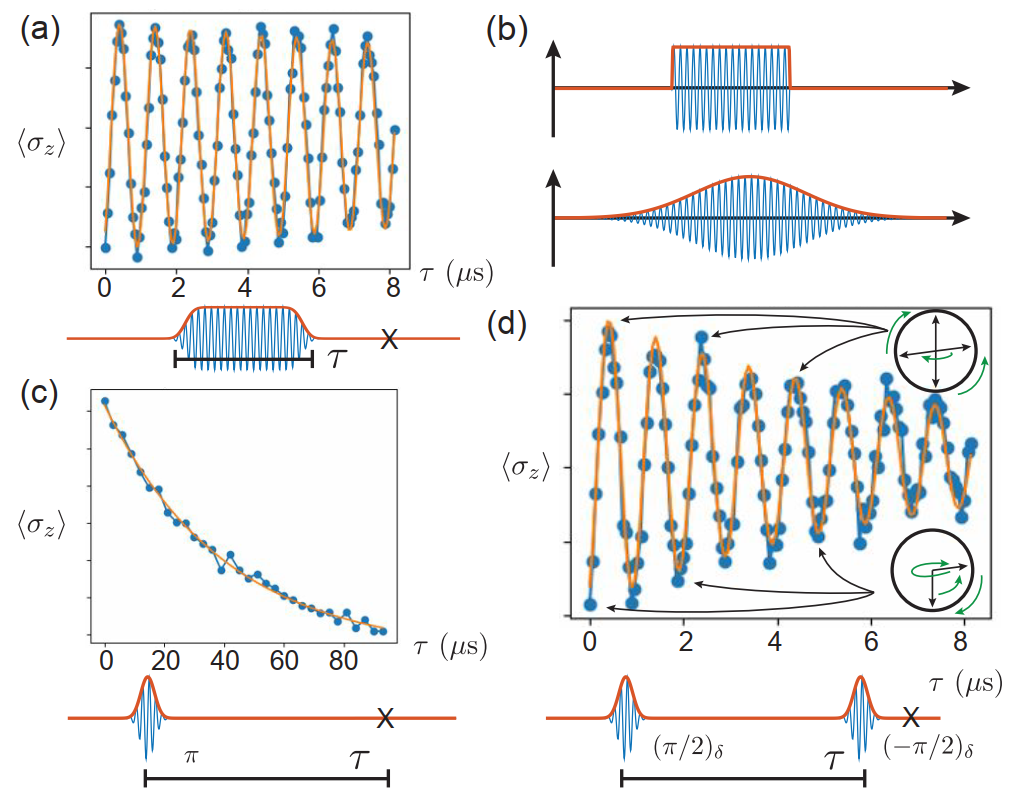}}
\caption{(a) Rabi oscillations. A best-fit line and the corresponding experimental pulse sequence implemented to measure them. To measure a single point on this graph, one drives the qubit for a duration $\tau$ and then measures in the $\sigma_z$ basis. Each point is an average $\<\sigma_z\>$ over many runs. Measurement occurs at the `X'. (b) The electric field profile and envelope of a square and a Gaussian pulse. (c) Similar to (a) but a $T_1$ measurement. (d) Ramsey oscillation and the associated pulse sequence. Inset Bloch spheres visualize the qubit dynamics at various points on the trace. The oscillation frequency is set by the qubit-pulse detuning, and the exponential decay is the coherence time $T_2^*$. Best fit indicates an exponential decay time of $7.3~\mu$s.} 
\label{fig:ExperimentalRabiT1Ramsey}
\end{figure}

Driving Rabi oscillations is a primary technique in experimental quantum information. Under the evolution of $\tilde{H}_\text{RWA}$, the qubit rotates endlessly about the $\sigma_\delta$ axis, but fortunately, the dynamics are still exactly solvable when the drive amplitude $\alpha_0$ is time-dependent. The exact same rotation occurs, except at a time-dependent rate. This is most easily seen in basis that diagonalizes the Hamiltonian $H\rightarrow \Omega_R(t)/2 \sigma_z$.  The solution to the Schr\"odinger equation becomes 
\begin{equation} \label{eq:TimeDependentOmegaSolution}
|\tilde{\psi}_\text{RWA}(t)\> = \alpha_0 \exp\left(-i\int_0^t \Omega_R(t')dt'/2\right) + \beta_0 \exp\left(i \int_0^t \Omega_R(t')dt'/2\right),
\end{equation}
which can be verified by direct substitution into the Schr\"odinger equation. Therefore the qubit's rotation angle accumulates proportionally to the drive strength. \erf{eq:TimeDependentOmegaSolution} is still of course only an approximate solution for $\tilde{H}$, and the rotating wave approximation can fail if $\Omega(t)$ changes too rapidly. If $\alpha_0(t)$ changes significantly over a period of $2\omega_q$, we can no longer guarantee that the integral \erf{eq:RWAIntegral} will vanish. Although a closed-form solution is not be available for arbitrary time dependence of $\delta$ and $\omega_d$, the intuition that $\tilde{H}$ acts as a rotation always holds, offering a way to induce arbitrary qubit dynamics. Rabi oscillations measured on a real superconducting qubit are shown in Fig. \ref{fig:ExperimentalRabiT1Ramsey}a.

\subheading{Qubit pulses and characterization} The above result gives us the tool to induce arbitrary time dependence on the state. However, just as classical digital computation is simpler to work with than analog, quantum information tasks also simplify if we impose discreteness. One typically uses Rabi oscillations to implement only $\pi/2$ and $\pi$ rotations. The obvious way to implement a $\pi$ pulse is to set $\Omega_R$ to some maximum value and then turn it off again after the correct time interval. The resulting boxcar pulse is shown in Fig. \ref{fig:ExperimentalRabiT1Ramsey}b. However, due to the presence of sharp edges, the Fourier transform of this pulse has support (is nonzero) out to very high frequencies, which leads to several difficulties in practice. In the present model, the only issue would be a breakdown of the rotating wave approximation. Additional issues arise in realistic settings. Firstly, physical qubits often contain other unused energy levels, and a broadband pulse can easily drive undesired transitions. Furthermore, if $\alpha(t)$ represents an electromagnetic signal travelling through a medium, a wide-bandwidth pulse is more susceptible to distortion. A better choice would be the Gaussian pulse envelope also shown in Fig. \ref{fig:ExperimentalRabiT1Ramsey}b. Although it also has support out to high frequencies, the spectral power decays faster for frequencies away from $\omega_d$ (exponential decay as opposed to $\mathcal{O}(1/\omega)$). Furthermore, the pulse bandwidth may be reduced arbitrarily by increasing the pulse duration.

$\pi$ and $\pi/2$ pulses suffice for most qubit characterization, which is an integral part of experimental quantum information science. The two most basic characterization tools aside from Rabi oscillations are $T_1$ and Ramsey oscillation measurements. $T_1$, the excited state lifetime, is measured by applying a $\pi$ pulse to a qubit and waiting for it to decay back to its ground state. Just as in radioactive processes, the excited state probability typically follows an exponential decay. An experimental measurement of $T_1$ is shown in Fig. \ref{fig:ExperimentalRabiT1Ramsey}c. A general and widely applicable model of qubit decay is presented in chapter \ref{ch:OpenSystems}.

Ramsey measurements are a somewhat more subtle but also more useful tool, as they allow measurement of three quantities: $T_2$ and $T_2^*$, which characterize how well the qubit maintains a superposition, and the precise qubit frequency. As we use Ramsey measurements to determine the precise qubit frequency, we should analyze the case where $\omega_q$ is not already known to the experimenter, and thus the drive frequency $\omega_d$ may not exactly match it. We saw in exercise \ref{ex:DetunedRabi} that the rotating frame Hamiltonian in this case is\footnote{\erf{eq:HQubitRWADetunedDrive} is written in the frame of $\omega_d$. One can also write it in the frame of $\omega_q$, in which case the $\Delta \sigma_z/2$ term does not appear but $\delta$ becomes time dependent. This picture of a time-varying rotation axis is equally valid and intuitive for understanding Ramsey oscillations.}
\begin{equation} \label{eq:HQubitRWADetunedDrive}
\tilde{H}_\text{RWA}/\hbar = \Delta \sigma_z/2 + \Omega_R(t) \sigma_\delta/2.
\end{equation}
The sequence for a Ramsey experiment consists of two $\pi/2$ pulses with a variables waiting time $\tau$ in between and a measurement at the end, as shown in Fig. \ref{fig:ExperimentalRabiT1Ramsey}d. We assume that $\Omega_R(t)$ is large compared to $\Delta \equiv \omega_q - \omega_d$ for most of the duration of the pulse. Thus during the pulses, \erf{eq:HQubitRWADetunedDrive} is well approximated by \erf{eq:HQubitRWA} and the pulses implement the same rotations that they would on resonance. 

During the waiting interval $\tau$, the qubit precesses around the $\sigma_z$ axis at the mismatch frequency $\Delta$. When $\tau=0$, the $\pi/2$ pulses occur one after the other and qubit ends in the excited state. The same is true if $\tau \Delta$ is an integer multiple of $2\pi$, as illustrated in Fig. \ref{fig:ExperimentalRabiT1Ramsey}d. On the other hand, if $\tau \Delta$ is an odd multiple of $\pi$, then the second pulse rotates the qubit from the equator to the ground state instead. In general, the excited state population oscillates at a rate $\Delta$, allowing us to measure $\Delta$ with high precision. Note that the oscillations only occur because during the waiting interval, the cosine term in $\alpha(t) = \alpha_0(t) \cos(\omega_d t + \delta)$ continues to oscillate at the same frequency $\omega_d$. Experimentally, this means that we need a classical oscillator, called a \definition{local oscillator} to keep time throughout each run of the experiment, even when we are not applying pulses.

Again we have only considered an isolated qubit, but like $T_1$ measurements, much of their utility applies primarily to imperfect qubits. The primary imperfection in most qubits manifests as an apparent instability in $\omega_q$. For example, as both electron spins and loops of current couple to magnetic fields, many types of qubits, from superconducting qubits to trapped ions have their frequencies changed by an external magnetic field. A fluctuating external field will cause the qubit to rotate randomly about the $\sigma_z$ axis, so that the average Bloch vector gets shorter for long $\tau$ (see exercise \ref{ex:RamseyDecay}). $T_2^*$ is the exponential decay time of the Bloch vector length, which manifests as decay of the Ramsey oscillations. Such a decay is visible in Fig. \ref{fig:ExperimentalRabiT1Ramsey}d. As measurement destroys coherence, Ramsey oscillations also decay if an external system acquires information about the qubit's energy.




\begin{exercise} \label{ex:RamseyDecay}
Semiclassical model of Ramsey oscillations: Show that if $\Delta$ is constant but randomly drawn from a zero mean Gaussian distribution with variance $\sigma$ for each measurement, then the off-diagonal elements of $\rho$ and hence the Ramsey profile decay as $e^{-\sigma^2 t^2}$. If instead $\Delta$ changes during each measurement, then the relative phase between $|0\>$ and $|1\>$ undergoes a random walk. In this case, show that the Ramsey decay is now exponential. What happens in each of these cases if you add an echo pulse? For both parts, use the fact that $\< e^{i\phi}\> = e^{-\<\phi^2\>}$ if $\phi$ is a zero-mean Gaussian distributed random variable.
\end{exercise}

\subheading{Harmonic oscillators} Having introduced the basics of qubits, we now turn to the other extreme of infinite dimensional systems, specifically the harmonic oscillator. The harmonic oscillator often serves as a bridge between classical and quantum mechanics. Although one initially develops it to treat a particle in a parabolic potential, it is ubiquitous in quantum mechanics, also modelling LC circuits, phonons, and even photons\ped{\cite{gerry2005introductory}}, as we will see in the following chapters. 

Focusing on a particle for concreteness, the wave function now forms a complex probability distribution over the position variable $x$. States of definite position are Dirac delta functions $|\psi_{x_0}(x)\> \propto \delta(x-x_0)$, which are eigenstates of the position operator $\hat{x}=x$ (loosely speaking). The de Broglie relation $p = \hbar k$ motivates identifying plane waves $|\psi_k(x)\> \propto e^{i k x}$ as states of definite momentum, which in turn motivates the definition of the momentum operator $\hat{p} = -i\hbar \partial/\partial x$ (as $\hat{p}|\psi_k(x)\> = p|\psi_k(x)\>$). With these operators in hand, it is natural to take the Hamiltonian of the harmonic oscillator to be $H_0 = p^2/2m + k x^2/2$, where $k=m \omega^2$ is the spring constant defining the parabolic potential and $\omega$ is the classical oscillation frequency (we drop the hats to avoid clutter).


The time-independent Schr\"odinger equation may be solved with the help of Hermite polynomials, and the ground state of $H_0$ is a Gaussian with variance $\sqrt{\hbar/m\omega}$. However, Dirac introduced a much simpler algebraic method for finding the energy spectrum. Introducing the latter operator
\begin{equation} \label{eq:DefLadderOp}
    a \equiv \frac{m \omega}{2\hbar}\left(x+\frac{i}{m\omega}p\right)
\end{equation}
one can express the Hamiltonian in terms of $a^\t a$
\begin{align}
    a^\t a &= \frac{m\omega}{2\hbar}\left(x^2 + \frac{p^2}{(m\omega)^2} + \frac{i}{m\omega}[x,p]\right) = \frac{1}{\hbar\omega}H_0 - \frac{1}{2} \\ \nonumber
    &\implies H_0 = \hbar \omega \left(a^\t a + \frac{1}{2}\right)
\end{align}
where we have used the fact that $[x,p]=i\hbar$. Using this commutation relation, one can also show that
\begin{align} \label{eq:aCommutationRelations}
    &[a,a^\t] = 1 \\ \nonumber
    &[a,a^\t a] = a \\ \nonumber
    &a^\t |n\> = \sqrt{n+1} |n+1\> \\ \nonumber
    &a|n\> = \sqrt{n} |n-1\>
\end{align}
where $|n\>$ is the eigenvector of $a^\t a$ with non-negative integer eigenvalue $n$. These eigenvectors are also eigenvectors of the Hamiltonian, and are in one-to-one correspondence with the Hermite polynomial solutions. The associated energies are $\hbar\omega (n+1/2)$. Notice that the ladder operators $a$ are just like the ladder operators $\sigma$ for a qubit. Both raise and lower a state among different energy levels.

\begin{exercise}
If you need the review, derive \erfs{eq:aCommutationRelations} and use them to argue that $n$ is an integer greater than zero. Recall that $\<\psi|a^\t a|\psi\> \geq 0$ for all $|\psi\>$ as it is the square of the state $a|\psi\>$.
\end{exercise}

The harmonic oscillator eigenstates have progressively larger variance in $x$ and $p$ (which correspond to $\mathbf{E}$ and $\mathbf{B}$ for photons, $Q$ and $I$ for an LC oscillator or lattice displacement for phonons) as we go up in energy. However, these states are also time-independent and have zero mean in $x$ and $p$, which does not line up with our classical intuition of a vibrating mass. To recover these classical states, we study the effect of a time-dependent force by adding $F(t) x = \epsilon_d(t)(a+a^\t)$ to the Hamiltonian. The problem is exactly analogous to the driven qubit problem that we have already encountered, and is solved by similar methods. We will make use extensive use of the Baker-Campbell-Hausdorff (BCH) formula and its corollaries, which we state upfront for future reference
\begin{align} \label{eq:BCH}
    &e^X e^Y = \exp\left(X + Y + \frac{1}{2}[X,Y] + \frac{1}{12}([X,[X,Y]] + [Y,[Y,X]]) + ...\right) \\ \nonumber 
    &e^{X+Y} = e^X e^Y \exp\left(-\frac{1}{2}[X,Y]\right) \exp\left(\frac{1}{6}(2[Y,[X,Y]]+[X,[X,Y]])\right)... \\ \nonumber 
    & e^X Y e^{-X} = Y + [X,Y] + \frac{1}{2!}[X,[X,Y]]+\frac{1}{3!}[X,[X,[X,Y]]]+...
\end{align}
Subsequent terms involve fourth and higher order commutators of X and Y. The higher order terms are only easily guessed for the last equation. The first identity is the standard BCH formula. The second is called the Zassenhaus formula, which forms the basis for a quantum simulation algorithm called the Trotter expansion. Although they look formidable, they often truncate or lead to simple closed-form expressions, particularly for the harmonic oscillator.

To treat the driving term, we use \erf{eq:RotatingFrame} just as before, setting $X = -H_0/\hbar$ in that formula. The rotating frame Hamiltonian is
\begin{align}
    \tilde{H} &= e^{i H_0 t/\hbar} H e^{-i H_0 t/\hbar} + \hbar X = H_0 - H_0 + \epsilon_d(t)(a+a^\t) \\ \nonumber
    &= \epsilon_d(t)(a e^{-i \omega t} + a^\t e^{i \omega t})
\end{align}
where we have used the third BCH formula to evaluate 
\begin{align}
\exp(i\omega a^\t a t) a \exp(-i\omega a^\t a t) &= \sum_{n=0}^\infty \frac{1}{n!} \underbrace{[i \omega a^\t a, [i \omega a^\t a, [...[i\omega a^\dagger a}_n, a]...]  \\ \nonumber
 &= \sum_n \frac{(i\omega t)^n}{n!} (-1)^n a = a e^{-i \omega t}.
\end{align}
A simple solution arises because the nested commutators like $[a^\t a, [a^\t a, a]] = -[a^\t a, a] = a$ always simplify to $\pm a$. (Note that we could have also used this formula to calculate $e^{i\omega_q \sigma_z t/2} \sigma e^{-i\omega_q \sigma_z t/2}$ in \erf{eq:HQubitRotatingFrame}). 

Setting $\epsilon_d(t) = 2\epsilon_0 \cos(\omega t + \delta)$ again yields a resonant interaction
\begin{equation}
\tilde{H}_\text{RWA} = \epsilon_0 (a e^{i\delta} + a^\t e^{-i\delta})
\end{equation}
where we have dropped counter-rotating terms for the same reason that applied in the qubit case. Solving for the driven dynamics is only slightly more involved than the analogous qubit Hamiltonian \erf{eq:HQubitRWA} with the help of the BCH formulas. If the system starts in its ground state $|0\>$, then the state after time $t$ is formally
\begin{align} \label{eq:DisplacementOp}
|\psi(t)\> &= U(t)|0\> = \exp(-i H_0 t/\hbar)|0\> \\ \nonumber
&= \exp(\alpha a^\t - \alpha^* a)|0\> \\ \nonumber
\alpha &\equiv -i\epsilon_0 e^{-i\delta} t /\hbar
\end{align}
We have dropped tildes, since we will only work in the rotating frame through the rest of this section. We recognize a potential instance to apply the second BCH formula (the Zassenhaus formula), letting $X=\alpha a^\t$ and $Y=\alpha^* a$. As $[a^\t,a] = -1$, all higher-order commutators vanish, leaving us with only the first three terms
\begin{align} \label{eq:CoherentState}
|\psi(t)\> &= e^{\alpha a^\t} e^{-\alpha^* a} e^{-|\alpha|^2/2}|0\> = e^{-|\alpha|^2/2} e^{\alpha a^\t}|0\> \\ \nonumber
&= e^{-|\alpha|^2/2} \sum_n \frac{\alpha^n}{n!} {a^\t}^n |0\> = e^{-|\alpha|^2/2} \sum_n \frac{\alpha^n}{\sqrt{n!}} |n\> \\ \nonumber
&\equiv |\alpha\>
\end{align}
The fact that $a|0\> = 0$ leads to substantial simplification and is responsible for the second equality. $|\alpha\>$ is known as a \definition{coherent state}, a family of non-orthogonal states parameterized by the complex displacement scalar $\alpha$. Unlike the qubit case, where $|0\>$ periodically returns to itself under driving, $\tilde{H}$ continually couples $|\psi\>$ to higher and higher energy levels, spreading the state out over the ladder of states as it climbs. The coherent states are fundamental to quantum optics and have numerous important properties:
\begin{itemize}
\item Coherent states are eigenstates of $a$. Specifically $a|\alpha\> = \alpha |\alpha\>$. It follows that $\<a+a^\t\>/2 = \real[\alpha]$ and $\<a-a^\t\>/2i = \imag[\alpha]$, so that $\alpha$ represents the mean position of the particle in phase space (note that from \erf{eq:DefLadderOp}, $a+a^\t$ and $(a-a^\t)/i$ are proportional to $x$ and $p$ respectively).

\item In the lab frame, coherent states rotate in phase space just like classically displaced states in classical phase space. $|\alpha(t)\> = |\alpha(0)e^{-i\omega t}\>$.

\item Like the ground state $|0\>$, $|\alpha\>$ is a minimum uncertainty state, meaning that $\Delta x\Delta p$ saturates the Heisenberg uncertainty principle. This relatively low quantum uncertainty partially explains why these states are semiclassical.

\item $|\<n|\alpha\>|^2$ is a Poisson distribution in $n$ with mean $|\alpha|^2$.

\item As is the case for $|0\>$, $|\<\psi_x|\alpha\>|^2$ and $\<\psi_k|\alpha\>$ are both Gaussian distributions with the same variances as those of $|0\>$.

\item Coherent states are not orthogonal, but satisfy $\<\alpha|\beta\> = \exp(\alpha^*\beta - |\alpha|^2/2-|\beta|^2/2)$ (which yields $|\<\alpha|\beta\>|^2 = e^{-|\alpha-\beta|^2}$). While this means that they cannot be used as an orthonormal basis, they nevertheless form an overcomplete basis satisfying $\int d^2 \alpha |\alpha\>\<\alpha| = \pi$.
\end{itemize}

We will make use of these properties enough that they will hopefully become intuitive and easy to remember.

In deriving the expression for $\alpha$, we also introduced a very important operator called the \definition{displacement operator}
\begin{equation} \label{eq:DisplacementOperator}
D[\alpha] \equiv \exp(\alpha a^\t - \alpha^* a)
\end{equation}
which also has properties worth noting:
\begin{itemize}
\item $D^\t[\alpha] = D[-\alpha] = D^{-1}[\alpha]$

\item The composition of two displacements is also a displacement. Specifically, $D[\alpha]D[\beta] = e^{(\alpha\beta^* - \alpha^*\beta)/2} D[\alpha + \beta]$ (this follows from the first BCH formula, the one that we haven't used so far). The global phase is physically irrelevant except in the case where the displacement depends on the state of another quantum system. In this case, it can lead to entanglement between the oscillator and that external system. 

\item A sequence of displacements that maps a closed path in the complex plane may nevertheless impart a global phase. This phase is $e^{2i A}$, where $A$ is the area in phase space enclosed by the path\cite{vutha2018displacement}.

\item Just as the Pauli matrices form a complete orthonormal basis under the Hilbert-Schmidt norm, so do the displacement operators. 

\item $D^\t[\alpha] a D[\alpha] = a+\alpha$. Just as we have made heavy use of rotating frames by exponentiating $H_0$, one can use this identity to work in displaced frames. It is often useful to choose $\alpha$ to displace the oscillator state to its ground state. We will see several examples of this chapter \ref{ch:OpenSystems}.
\end{itemize}

Note that just as a general qubit drive always implemented a rotation of some kind, $D[\alpha]$ always implements a displacement.

\begin{figure}
\centering
{\includegraphics[width = 160mm]{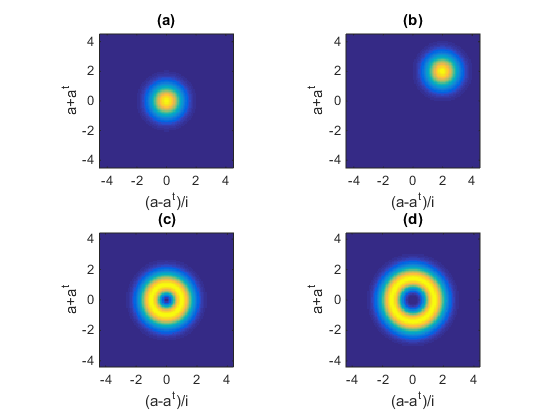}}
\caption{Q functions for (a) the ground/vacuum state (b) a coherent state (c) the $n=1$ Fock state (d) the $n=2$ Fock state.} 
\label{fig:QFuncs}
\end{figure}

So far, we have covered almost all of the same material for both qubits and harmonic oscillators. The main thing we lack on the oscillator is a simple state visualization like the Bloch vector. We have saved this for last because the most convenient visualizations all rely on coherent states. The first is called the \definition{Husimi Q representation}
\begin{equation} \label{eq:QFunction}
Q(\alpha) \equiv \frac{1}{\pi} \<\alpha|\rho|\alpha\>.
\end{equation}
$Q(\alpha)$ looks somewhat like a probability distribution, and indeed it does represent the probability of outcome $\alpha$ if one were to measure in the coherent state basis. The only catch is that $|\alpha\>$ are not orthogonal states, so orthogonal states do not necessarily have orthogonal $Q$ representations. $Q$ is nevertheless quite a useful quantity, in particular because it depicts the state in phase space as a 2D image, similar to classical phase space. It is also invertible, meaning that one can recover $\rho$ uniquely from $Q(\alpha)$ (though see exercise \ref{ex:QFuncCatStates} for an important caveat). Finally, the Q function can be measured using heterodyne detection or phase-preserving amplification, as we will see in the next chapter. The $Q$ functions for several quantum states are plotted in Fig. \ref{fig:QFuncs}.

As we will only use the $Q$ representation for the purpose of visualization, we refer the reader to \cite{gerry2005introductory} chapter 3 for a more extensive treatment. Also covered are the equally important Glauber-Sudarshan $P$ and Wigner representations.

\begin{exercise} \label{ex:QFuncCatStates}
Cat states: Compute the $Q$ function for $|\psi\> \propto |\beta\> + |{-\beta}\>$. Don't worry about overall constant factors like normalization. How would the calculation differ if we replaced $|\psi\>$ with classical mixture of the same two coherent states ($\rho \propto |\beta\>\<\beta| + |{-\beta}\>\<{-\beta}|$)? Show that the difference between these two $Q$ functions scales as $e^{-|\beta|^2}$, indicating that they are difficult to distinguish visually. The Wigner function is much better for this purpose.
\end{exercise}


\section{Quantum Optics: The Jaynes-Cummings Hamiltonian}

Qubits and harmonic oscillators form the basis for most of quantum information. To use them in interesting ways, we must combine them and allow them to interact. Of the three possible combinations (qubit-qubit, qubit-oscillator and oscillator-oscillator), the qubit-oscillator system offers the richest dynamics from a quantum optics standpoint. The oscillator expands the number of possible states, while the qubit provides a nonlinearity that is necessary for universal control of the system. To see the power of this combination, note that while the qubit Hamiltonian considered above allowed us to prepare any qubit state, the cavity Hamiltonian was only capable of producing coherent states. This limitation is a fundamental impediment to quantum information processing in harmonic oscillators arising from their linearity; it is not possible to drive the $|i\>\leftrightarrow |j\>$ transition without also driving the $|i+1\>\leftrightarrow |j+1\>$ transition. In addition to providing the required nonlinearity for full control, qubit-oscillator interactions also provide a medium through which qubits interact in most physical systems, and we will see examples of how to generate a two-qubit gate using a harmonic oscillator as an intermediary in future chapters.



So far we have only considered isolated quantum systems. To treat interactions, we need a way to model multiple objects with a single wave function, which is accomplished using the \definition{tensor product}. This generalization is perhaps most clear for continuous variable states like harmonic oscillators. Given the states $\psi_1(x)$ and $\psi_2(x)$ of two different particles, the wave function of the joint state is simply $\psi_1(x_1)\psi_2(x_2)$. Once again, the connection to classical probabilities is evident; the joint probability distribution of two independent probability distributions is also given by the product. Furthermore, the measurement outcome probabilities satisfy $P(x_1,x_2) = |\psi_1(x_1)|^2|\psi_2(x_2)|^2=P_1(x_1)P_2(x_2)$. Finally, just as correlated probability distributions $P(x_1,x_2)$ cannot be written in separable form, there also exist quantum states $\psi(x_1,x_2)$ that cannot be written as a single tensor product. These are precisely the entangled states, which play a fundamental role in quantum information and computation.

For discrete systems, the tensor product construction is essentially the same as the above, except treating the continuous variables $x_i$ as discrete indices ($x_i=0,1$ for a qubit, for example). This notation would suggest treating the composite system $\psi(x_1,x_2)$ as a matrix, or in the case of more than two subsystems, as a tensor. Although such notation has its uses, one typically prefers to treat the wave function as column vector for compatibility with the Schr\"odinger equation. To do so, we simply stack the columns of $\psi(x_i,x_j)$ on top of one another. The general construction may be concisely defined for any operators $A$ and $B$ (column vectors included) as
\begin{align}
A\otimes B = \begin{bmatrix}
A_{1,1} B & A_{1,2} B & \hdots & A_{1,n} B \\
A_{2,1} B & \ddots & & A_{2,n} B \\
\vdots & 	& & \vdots \\
A_{m,1} B & A_{m,2} B	& \hdots & A_{m,n} B
\end{bmatrix}.
\end{align}
Note that each entry in the above table is a block of the matrix $A\otimes B$, and the vertical and horizontal dimensions of the whole are therefore the product of the dimensions of A and B. This particular arrangement is useful because it satisfies $(A\otimes B)\times (C\otimes D) = (A\otimes C)\times (B\otimes D)$. This relation allows us to evolve composite systems with matrix multiplication just as we did with isolated systems (\textit{i.e.} two isolated systems evolve as $|\psi_{1,2}(t)\> = (U_1(t)|\psi_1(0)\>)\otimes (U_2(t)|\psi_2(0)\>) = (U_1(t)\otimes U_2(t))(|\psi_1(0)\>\otimes |\psi_2(0)\>$, meaning that the time evolution operator for a composite state is simply the tensor product of the original time evolution operators).

\begin{exercise} \label{ex:Schmidt decomposition}
As an example of when it's useful to think of composite quantum states as tensors instead of column vectors, the singular value decomposition ($A=U\Sigma V^\t$ where $\Sigma$ is a diagonal matrix and $U$ and $V$ are unitary) of the matrix $\psi_{i,j}$ yields a useful construction called the \definition{Schmidt decomposition}: $|\psi\> = \sum_i \Sigma_{i,i}|i_1\>\otimes|i_2\>$ (the first $\sum$ is a summation symbol, while the second is a diagonal component of the matrix $\Sigma$). Look up the Schmidt decomposition in Nielsen and Chuang\cite{Nielsen2010}. This decomposition is a simple example of a \definition{tensor network}, which includes useful state representations like \definition{matrix product states}\cite{orus2014practical}.
\end{exercise}

\textbf{The Jaynes Cumming Hamiltonian} is the most basic form of a qubit coupled to a harmonic oscillator. From an applied standpoint, the coupled qubit-oscillator models countless physical systems, from light-matter interactions to lattice defect states coupled to phonons
%
\begin{equation} \label{eq:JaynesCummingsHamiltonian}
H/\hbar = \omega_q \frac{\sigma_z}{2} + \omega_o a^\t a + g(a+a^\t)(\sigma + \sigma^\t).
\end{equation}
Each operator acts on only one tensor factor of the composite system, so implicitly $a\rightarrow a\otimes I_q$, $\sigma\rightarrow I_o \otimes \sigma$ \textit{etc.} Like the single qubit and oscillator models seen already, the essence of the coupling term is a dipole approximation. The new ingredient here is that we treat the dipole driving term as another quantum system. 

\begin{figure}
\centering
{\includegraphics[width = 0.5\textwidth]{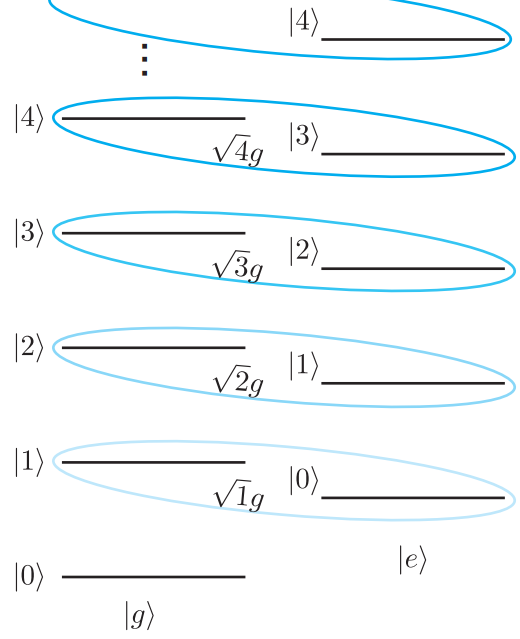}}
\caption{The level structure for the Jaynes Cummings Hamiltonian. In the rotating wave approximation, only pairs of levels with the same total excitation number interact, as illustrated with the blue ellipses. Due to the increasing second moment of states as we move up the harmonic oscillator ladder, the coupling strength increases with $n$.} 
\label{fig:JaynesCummingsLadder}
\end{figure}

We can perform calculations with $H$ most easily if we choose a frame in which it is still time independent. This works only if we rotate the qubit and oscillator frames at the same frequency \textit{i.e.} $U(t) = \exp(i(\omega \sigma_z/2 + \omega a^\t a))$. Both $\omega=\omega_q$ and $\omega=\omega_o$ are useful choices, so we go with the latter. The Jaynes Cummings Hamiltonian in the rotating wave approximation is
\begin{equation}
\tilde{H}/\hbar = \Delta \frac{\sigma_z}{2} + g(a^\t \sigma + a \sigma^\t)
\end{equation}
where $\Delta = \omega_q-\omega_o$. Many sources justify dropping counter-rotating terms $a \sigma$ and $a^\t \sigma^\t$ because they appear to violate energy conservation, exciting or deexciting the oscillator and the qubit simultaneously\footnote{Arguments about energy conservation were irrelevant to previous sections, as $H$ was time-dependent.}. However the Hamiltonian \textit{defines} the energy of a system, so this argument does not hold. What can be said is that if $g \ll \omega_o,\omega_q$, then the interaction acts like a perturbation, and we expect that the energies not to change much when we add it in. $a \sigma$ and $a^\dagger \sigma^\dagger$ change the bare energies by a lot, so we might already expect their contributions to be small. Loosely speaking, time-energy uncertainty allows counter-rotating terms to create or destroy two excitations, as long as these excitations don't last long. This suggests that the contribution of these terms is non-zero, but cancels out on average, as we saw when justifying the rotating wave approximation with time-dependent perturbation theory. This gives us a physical argument for why these terms probably do not contribute. While these arguments hold for many physically reasonable systems, we will see an example in chapter \ref{ch:OpenSystems} (single- and multi-quadrature measurements) in which these terms play a significant role in the dynamics.

\begin{exercise} \label{ex:CoupledHarmonicOscillators}
The solution for two coupled harmonic oscillators resembles that of the Jaynes-Cummings Hamiltonian. Let $H/\hbar=\omega_a a^\t a + \omega_b b^\t b + g(a^\t b + a b^\t)$. The coupling term can be derived from the energy of two masses sharing a common spring: $k(x_a - x_b)^2/2$. Show that $H$ can be diagonalized by introducing the normal mode operators $\tilde{a} = \alpha a + \beta b$ and $\tilde{b} = \alpha b-\beta a$ with $H/\hbar = \tilde{\omega}_a \tilde{a}^\t \tilde{a} + \tilde{\omega}_b \tilde{b}^\t \tilde{b}$ and $\alpha^2+\beta^2=1$ (chosen so that $[\tilde{a},\tilde{a}^\t] = 1$, $[\tilde{a}, \tilde{b}^\t]=0$ \textit{etc}.). You don't have to solve for these parameters unless you want to do exercise \ref{ex:CoupledHarmonicOscillators2}.
\end{exercise}

\begin{exercise} \label{ex:CoupledHarmonicOscillators2}
Show that $\alpha = \cos(\theta/2)$ and $\beta = \sin(\theta/2)$ with $\theta = \arctan(2g/\Delta)$ and $\Delta = \omega_a-\omega_b$. Show that the normal mode frequencies satisfy $\tilde{\omega}_a + \tilde{\omega}_b = \omega_a + \omega_b$ and $\tilde{\omega}_a - \tilde{\omega}_b = \sqrt{(2g)^2 + \Delta^2}$.
\end{exercise}

The Jaynes-Cummings Hamiltonian can be solved exactly in the rotating wave approximation. The elimination of $a \sigma$ and $a^\t \sigma^\t$ terms reduce the problem to that of an effective qubit. To see how, note that the action of $H$ takes the state $|n,g\>$ to $|n-1,e\>$, but then can only map the latter back to $|n,g\>$. Thus each component of the wave function evolves in a two-dimensional subspace of fixed total excitation number, as depicted in Fig. \ref{fig:JaynesCummingsLadder}. We can then break $H$ up into many of these qubit subspaces and solve each individually. Computing the action of $H$ in a subspace amounts to computing a few matrix elements
\begin{align} \label{eq:JaynesCummingsQubitModel}
H_n &\equiv \begin{pmatrix}
\<e,n-1|H|e,n-1\> & \<e,n-1|H|g,n\> \\
\<g,n|H|e,n-1\> & \<g,n|H|g,n\>
\end{pmatrix} \\ \nonumber
&= \hbar\begin{pmatrix}
\Delta/2 & \sqrt{n} g \\
\sqrt{n}g & -\Delta/2
\end{pmatrix}
\end{align}
This effective Hamiltonian is just like \erf{eq:HQubitRWADetunedDrive} with $\delta=0$. We have not yet written down the fully general solution, so let's do that now by computing the eigenvectors and eigenvalues
\begin{align} \label{eq:JaynesCummingsQubitSol}
\frac{E_{\pm,n}}{\hbar} &= \pm \frac{1}{2}\sqrt{\Delta^2 + n(2g)^2} \\ \nonumber
|+,n\> &= \cos(\theta_n/2) |e,n-1\> + \sin(\theta_n/2) |g,n\> \\ \nonumber
|-,n\> &= \cos(\theta_n/2) |g,n\> - \sin(\theta_n/2) |e,n-1\> \\ \nonumber
\tan(\theta_n) &\equiv -\frac{2\sqrt{n} g}{\Delta}.
\end{align}
%
%
Note the similarity to the results of exercises \ref{ex:CoupledHarmonicOscillators} and \ref{ex:CoupledHarmonicOscillators2}.

There are two interesting limits to consider in the solution to Jaynes-Cummings. The simplest limit is $\Delta \rightarrow 0$, in which case the eigenvectors become equally weighted between $|n,g\>$ and $|n-1,e\>$ and the effective qubit Hamiltonian \erf{eq:JaynesCummingsQubitModel} reduces to the resonant Rabi Hamiltonian \erf{eq:HQubitRWA}. If we start the oscillator in its ground state and the qubit in its excited state (the $n=1$ manifold), then the excitation swaps coherently between the qubit in the cavity at an angular frequency of 2g. These oscillations are called vacuum Rabi oscillations.

The other regime, and the last piece of physics that we introduce quantitatively in this chapter is $\Delta \gg g$, called the dispersive regime. In this case, we can treat the off-diagonal elements of \erf{eq:JaynesCummingsQubitModel} as a perturbation to the detuning terms $\pm \Delta/2$. We could use a Taylor expansion of \erf{eq:JaynesCummingsQubitSol}, but instead let's use second-order perturbation theory. Breaking $H_n$ into its diagonal and off-diagonal parts $H_0$ and $H'$ respectively, the new eigenvalues and eigenvectors are
\begin{align} \label{eq:DispersiveHPert}
E_{e,n-1}' &\approx \frac{\Delta}{2} + \frac{|\<g,n|H'|e,n-1\>|^2}{\Delta/2 - (-\Delta/2)} \\ \nonumber
&= \frac{\Delta}{2} + \frac{n g^2}{\Delta}  \\ \nonumber
E_{g,n}' &\approx 
 -\frac{\Delta}{2} - \frac{n g^2}{\Delta} \\ \nonumber
|e,n-1\>' &\approx |e,n-1\> + \frac{\<g,n|H'|e,n-1\>}{\Delta/2 - (-\Delta/2)} \\ \nonumber
&= |e,n-1\> + \frac{g}{\Delta} |n,g\> \\ \nonumber
|g,n\>' &\approx |g,n\> - \frac{g}{\Delta}|e,n-1\>.
\end{align}
The qubit and cavity states become `dressed' by one another and pick up a small energy shift called the dispersive shift. The energy shift $\pm n g^2/\Delta$ takes a form that can be easily reexpressed as a Hamiltonian, called the dispersive Hamiltonian
\begin{align} \label{eq:DispersiveH}
H_\text{Dispersive} &= \chi a^\t a \sigma_z \\ \nonumber
\chi &\equiv \frac{g^2}{\Delta},
\end{align}
which is technically in the dressed basis $|g,n\>'$, $|e,n-1\>'$. Written in this way, we see that the dispersive Jaynes-Cummings Hamiltonian leads to a qubit-state-dependent frequency shift of the cavity, or a cavity-state-dependent shift of the qubit frequency. This form of coupling is quite useful in practice. For instance, one can measure measure the state of the qubit by measuring the frequency of a harmonic oscillator to which it is coupled. This is one of many reasons why the dispersive Hamiltonian arises extensively in circuit QED.


\begin{exercise} \label{ex:SchriefferWolff}
An alternative and powerful way to derive \erf{eq:DispersiveH} is called the Schrieffer-Wolff transformation. Using the Baker-Campbell-Hausdorff formula at second order $e^{X}He^{-X} = H + [X,H]+[X,[X,H]]/2$ and $X= (a \sigma^\t - \sigma a^\t) g/\Delta$, derive \erf{eq:DispersiveH} from \erf{eq:JaynesCummingsHamiltonian}. In general, one can use this transformation to diagonalize a Hamiltonian $H_0 + H'$ to first order in $H'$, provided we can find an $X$ such that $[X,H_0] = -H'$.
\end{exercise}

\section{Applications of Quantum Systems}

It is no coincidence that the vast scientific progress of the 20th century coincided with rapid technological development. Without quantum theory, the transistor, LED, laser, MRI and many forms of atomic scale microscopy would have remained ideas of science fiction. Despite the magnitude of recent developments, there is reason to believe that some of the most significant quantum technologies are yet to come.
Just as a motor is an electronic device that performs an essentially mechanical task, most 20th century technologies based on quantum mechanics still perform fundamentally classical tasks, such as a transistor switching a current or a laser generating an oscillating electric field. Devices that operate on truly quantum degrees of freedom may offer comparably novel capabilities, analogous to classical computers, CCDs and the internet. The full capacity of these technologies, and in many cases even proofs that they supersede the best possible classical implementation are unknown. The field is still wide open.

The experimental and theoretical tools used in this thesis both have a vast range of potential applications. The field of quantum information is still in its infancy, and we only touch on a small range of its potential scope. Although this thesis does not focus specifically on quantum computation, the basic concepts of that field are essential for our work and for quantum information science in general. Below we outline the main anticipated categories of technology that are expected to be impacted by quantum information, and then give a short, high-level overview of quantum computation, referring the reader to reference \ped{\cite{Nielsen2010}} and other references of this section for details.

\subheading{Quantum simulation}: The classical representation of a quantum wave function grows exponentially as degrees of freedom are added to it. While methods like density functional theory (DFT) and the density matrix renormalization group (DMRG) render many systems tractable, a classical computer's ability to model a general quantum system is fundamentally limited. Richard Feynman pointed out that this exponential growth should make quantum systems powerful simulators of other quantum systems. Long-term ambitions for this field include understanding and developing high-temperature superconductors, understanding and designing chemicals and simulating high-energy physics processes, for example with lattice gauge theories.

\subheading{Quantum Computation}: Inspired by Feynman's insight, researchers developed models of computation that involve quantum mechanics at their core, allowing for superposition and entanglement. In the early 90s, evidence rapidly accumulated that a computer operating on quantum degrees of freedom could be fundamentally more powerful than a Turing machine, in violation of the extended Church-Turing Thesis. Perhaps the most striking example is the realization by Peter Shor that a quantum computer can factor numbers about as efficiently as any computer can multiply them, which is exponentially faster than the fastest known classical algorithm. For review of the potential near and far-term applications, see \cite{preskill2018nisq}, though the field is moving so quickly that there were significant updates within a year of its publication\cite{tang2018recommendation}.\footnote{See also  https://www.scottaaronson.com/blog/?p=3880}


\subheading{Novel quantum devices} Quantum devices have many potential applications beyond computation. One important example is quantum cryptography, which uses quantum correlations to guarantee security and secrecy. Along these lines, researchers have also proposed unforgeable quantum money and certifiable random number generators. In a different vein, a general class of advantages comes from quantum metrology. Whereas classical statistical uncertainties on $N$ repeated measurements scale as $1/\sqrt{N}$, quantum measurements can be made to scale as $1/N$\major{\cite{giovannetti2006quantum}}. This discovery pointed out the importance of a quantity called the quantum Fisher information, which predicts the ultimate capabilities of a measurement device via the quantum Cram\'er-Rao bound. There is hope, and in some cases existing demonstrations showing that molecular imaging, magnetometry, interferometry (such as the use of squeezing at LIGO), clocks and very-long-baseline interferometry for telescopes could all be improved with quantum effects.


\subheading{Classical technologies at their quantum limits} Finally, as classical technologies approach their quantum limits, quantum physics will be required to design them. This is already happening in classical computing, where transistors are approaching the atomic scale. Photonic computation and communication will also become more quantum as signals approach the single-photon level.

While many of the above capabilities do not superficially relate to quantum computation, quantum information nevertheless forms a foundation and common language for understanding them. All of the above examples (though in particular the first three) are naturally described in terms of quantum circuits. For example, the fundamental limits of interferometry may be understood in terms of an algorithmic primitive called the quantum phase estimation algorithm\cite{wiseman2009adaptive}. In subsequent chapters, we will understand open systems, quantum trajectories and quantum feedback as quantum circuits. Therefore an understanding of quantum computation is beneficial for understanding the ultimate limits of quantum technology in general.

\subheading{A brief overview of quantum computation} In the current understanding, quantum algorithms gain their advantage over classical computers through two essential primitives. The first is \definition{Grover's search algorithm}\major{\cite{grover1996fast}}, which provides an intrinsic advantage in searching an unsorted database. This database can be thought of as a function $f(x)$ which maps integer inputs to integer outputs in some fashion ($x$ might be encoded by its binary representation over a register of qubits). To find an entry $x$ for which $f(x)$ produces some desired output $y$, a classical computer must resort to querying each value of $f$ individually until the right $x$ is found. Thus if $f$ is defined over $N$ values of $x$, then the average number of queries before success is $N/2$ (denoted $\mathcal{O}(N)$). Surprisingly, a quantum computer can find the right value in $\mathcal{O}(\sqrt{N})$ by passing multiple values of $x$ in superposition on each query! While this necessitates implementing $f(x)$ in the quantum computer (so we cannot use this trick to find a book at a disorganized bookstore), it offers a quadratic speedup for an enormous range of problems.

It is often said that a quantum computer achieves its speedup by computing on all possible inputs in superposition. While it is true that a quantum computer can evaluate $f$ on all values of $x$ by preparing each qubit of $x$ in a superposition, the output state is also a superposition over all possible answers. As measurement collapses the wave function, we have no way to access all of these answers in a single run. If we could, or if we could guarantee that the measurement outcome would correspond to $y$, then such an algorithm would achieve an enormous speedup on the unstructured search problem instead of quadratic. The actual intuition behind Grover's search algorithm is somewhat more subtle, though it is satisfyingly straightforward to understand quantitatively and visually. We refer the reader to reference \ped{\cite{Nielsen2010}} for details.

If only quadratic speedups were available, quantum computation might not be worth the trouble. However quantum systems appear to be exponentially more difficult to simulate than classical, hinting at something more powerful that Grover's search. The second algorithmic primitive for computation is the \definition{quantum Fourier transform}\cite{bernstein1997quantum}, which provides the exponential speedup found in Shor's factoring algorithm. The quantum Fourier transform exploits the fact that the discrete Fourier transform is itself a unitary operation, and therefore can be applied directly to a wave function. It allows a quantum computer to find the period of a function in only a few function calls, and forms the basis for another important routine called quantum phase estimation mentioned above. Again we refer the reader to reference \ped{\cite{Nielsen2010}} for more.

The above algorithms may be implemented efficiently using single and two-qubit operations. Which gates will be available and the connectivity between qubits will ultimately be determined by the physical implementation of the computer. A surprising range of physical systems provide viable, equally powerful models of computation. For example, given photon counters single photon states as inputs, linear optics (beam splitters and phase shifters) suffice to perform any quantum algorithm\major{\cite{knill2001scheme}}. This construction is particularly remarkable because many of the basic operations upon which it relies fail with high probability. Another significant example is measurement-based computation, in which no single-or two-qubit gates are allowed\ped{\cite{nielsen2006cluster}}. Provided a particular initial state, called a cluster state, universal computation is implemented by sequentially and adaptively measuring different parts of the computer. Finally, it is possible to encode the output of any quantum circuit into the ground state of a local Hamiltonian $H_P$ (meaning that only nearby qubits are coupled). By smoothly varying $H$ from a simple, non-interacting Hamiltonian to $H_P$, one can efficiently implement any algorithm\major{\cite{aharonov2008adiabatic}}. Adiabatic theorem ensures that the system remains in its ground state at all times, and the discreteness of the original quantum circuit appears to be lost.

When we consider quantum computation in realistic systems, the continuous nature of the wave function poses a challenge. Classical models of computation are digital, and dissipative processing like latching and copying ensure resilience to noise. Neither of these tricks work in quantum systems, where dissipative interactions destroy coherence and copying is forbidden via the no-cloning theorem\ped{\cite{Nielsen2010}}. In the early days of quantum computing, researchers worried that even the smallest errors in computation would compound exponentially and eventually ruin a realistic device. 

An important conceptual shift in noisy quantum computation was the insight that although errors are continuous, there always exists an equivalent error model that is discrete (we show this result in chapter \ref{ch:OpenSystems} exercise \ref{ex:NonUniqueKraus}). For example, if a qubit undergoes a random $\sigma_z$ rotation of arbitrary angle, we can equally well model this process as random rotations by only $0$ or $\pi$. With a discrete set of errors to worry about, quantum error correcting codes were devised to correct them one-by-one. A single `logical' qubit can be encoded in multiple physical qubits using entanglement, and errors can be corrected with nonlocal measurements. While encoding introduces overhead, an important result called the threshold theorem showed that this scaling is favorable, and that errors can be suppressed to arbitrary precision without so much overhead that they null the benefits of quantum computation.

\begin{figure}
    \centering
    \includegraphics[width=\textwidth]{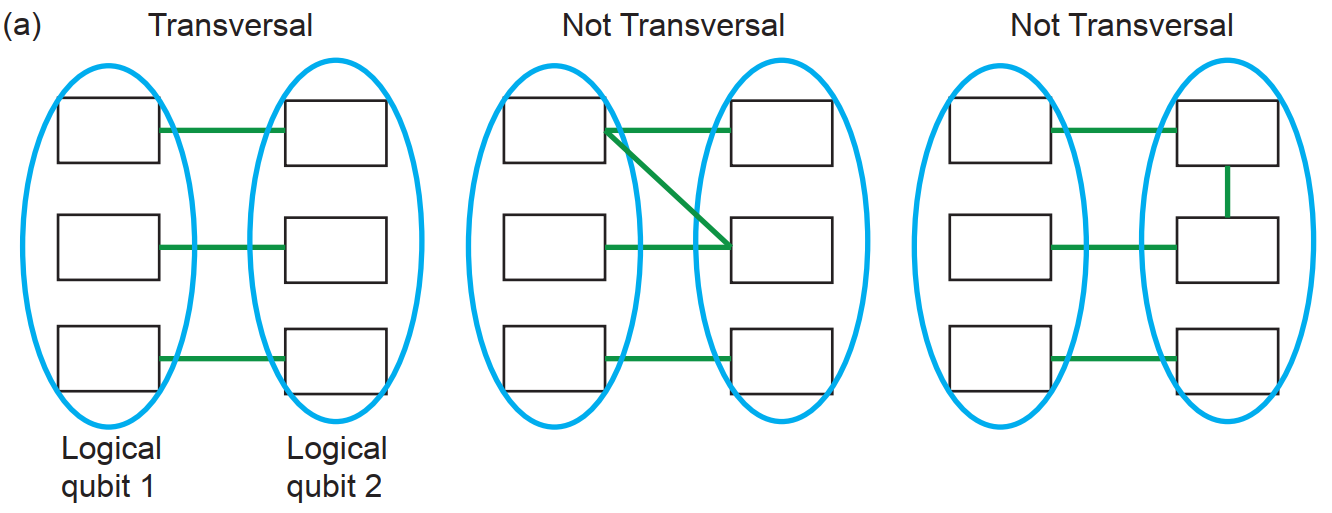}
    \caption{(a) Graphical illustration of two logical qubits each encoded into 3 physical qubits (black boxes). A transversal gate operates as shown. (b) A non-transversal gate. (c) Another non-transversal gate.}
    \label{fig:TransversalGates}
\end{figure}

There is error correction and then there is fault tolerance. Error correction means that we can hold a quantum state in an encoded memory for longer than if we were to store it in a single physical qubit. Fault tolerance means that we can perform computations on these qubits while preserving their coherence, and in particular ensuring that a single error does not spread and pollute the rest of the system. A standard method to ensure fault tolerance is to implement gates transversely. In the context of an error correcting code, a transversal gate between two encoded logical qubits only allows for pairwise gates between the physical qubits. This is best illustrated with pictures, as in Fig. \ref{fig:TransversalGates}. 

Unfortunately, restricting to transversal gates limits the range of operations that may be performed on a logical qubit. Eastin-Knill theorem\cite{eastin2009restrictions} shows that transversal gates cannot be universal\cite{dawson2005solovay}, meaning that not all quantum algorithms can be implemented with the available gate set.  There are many proposed routes to circumvent this impediment to universal, fault-tolerant quantum computation. One proposed route is magic state distillation, in which one imperfectly prepares a state that cannot be prepared fault-tolerantly and then uses fault tolerant gates and measurements to purify it to a near-perfect state\cite{bravyi2005universal}. Magic states also have the remarkable property that one can be `consumed' to perform gates that otherwise could not be implemented fault tolerantly. Another route to fault tolerant quantum computation are topological quantum states like the surface code. Most, but not all proposals for topological computing also require magic state distillation.




\section{Recommended Reading} \label{sec:RecommendedReading}

Perhaps the most daunting aspect of starting work in a new field of research is the vast body of potentially relevant literature. Graduate courses provide overviews and calculational tools for many fields, but rarely cover even slightly narrower topics like quantum electrodynamics, let alone superconducting circuits. While the main goal in writing this thesis is to bridge this gap, no two research projects require the same knowledge base and no two students come in with the same background. To provide a more flexible introduction, below is a list of particularly useful books and papers along with recommended sections, roughly sorted with the most accessible texts at the top. The usefulness of these texts depends on one's interest of course, but these are chosen for their breadth and clarity.

\begin{itemize}
\item \textbf{Quantum Computation and Quantum Information} \textit{Nielsen and Chuang}\ped{\cite{Nielsen2010}}: Conversational and perfect for self-learning, this is the single resource that I recommend to everyone entering the field. Part I chapter 1 presents a broad overview of quantum information science, and chapter 2 contains a nice summary of quantum theory from a more informational perspective. Part II, quantum computing, is a great place to learn about algorithms, circuits and the abstract details of experimental implementation. Part III chapters 8-10 is a must-read for anyone interested in understanding decoherence, open quantum systems and quantum error correction.

\item \textbf{Introductory Quantum Optics} \textit{Gerry and Knight}\ped{\cite{gerry2005introductory}}: When we deal with light quantum mechanically, we are studying quantum electrodynamics (QED). While quantum field theory is a class well-worth taking, not everyone has the time or desire to trudge through Peskin and Schroeder. Fortunately, QED is a particularly accessible subset of QFT that doesn't require most of its machinery. \textit{Gerry and Knight} covers the full quantum theory of light at an undergraduate level. Chapter 1 alone is well worth reading. 

\item \textbf{Quantum Optics} \textit{Scully and Zubairy}\ped{\cite{scully1999quantum}}. There are many great resources available on quantum optics, but this one strikes a particularly nice balance between pedagogy and completeness. While the Jaynes Cummings Hamiltonian is never referenced by name, it is developed in the first few pages of chapter 6 and studied extensively. Another comparable resource with a very different approach is Haroche's \textit{Exploring the Quantum: Atoms, Photons and Cavities}, which is perhaps a bit more accessible\cite{haroche2006exploring}.

\item \textbf{Microwave Engineering} \textit{Pozar}\ped{\cite{pozar2009microwave}}: A standard reference, you may find this to be a piece of cake after \textit{Jackson}! It provides a practically useful introduction to the theory of standard microwave components like resonators (chapter 6), splitters, circulators, and directional couplers (chapter 7) while emphasizing essential concepts like impedance matching (chapter 2), scattering matrices (4.3) and reciprocity. Chapter 3 is useful for design of superconducting devices. See \textbf{\cite{montgomery1987principles}} for a more advanced and complete text on the subject.


\item \textbf{Introduction to Superconductivity} \textit{Tinkham}\ped{\cite{tinkham2004introduction}}: Provides a concise but fairly broad introduction to the major concepts of superconductivity. Most of the book is dedicated to phenomenological models, with one chapter dedicated to an accessible first-principles derivation of the Bardeen-Cooper-Schrieffer (BCS) model of superconductivity. It is difficult to outline what knowledge is necessary for research on superconducting circuits as the field grows. A few terms to keep an eye out for on more specialized topics include quasiparticles and their dissipation, vortices, Andreev bound states and the concept of conduction channels. Not all of these areas are covered in detail in \textit{Tinkham}, but are commonly discussed in the literature.

\item \textbf{Introduction to Quantum Noise, Measurement and Amplification} \textit{Clerk et al.}\ped{\cite{clerk2010noise}}: A Review of Modern Physics article that covers many experimentally relevant calculations for quantum amplifiers and general measurement theory. The appendices are perhaps the most useful, explaining many standard calculations tools and their application. 
You'll see this referenced all over the place.

\item \textbf{Quantum Noise} \textit{Gardiner and Zoller}\ped{\cite{gardiner2004noise}}: A dense but great book on quantum optics with a strong emphasis on open quantum systems. It's a standard reference if you're embarking on a serious quantum optics calculation. Chapter 3 derives the quantum Langevin equation, chapter 5 derives the master equation and another handy trick called quantum regression theorem. Some topics of this book have simpler, more modern descriptions elsewhere, such as the SLH formalism for modeling very general networks of quantum systems\ped{\cite{combes2017slh}}. 
\end{itemize}

The reader is advised to recognize the above reference numbers to see their applicability in the following introduction. We have also put in \major{italic} citations of particular significance for the field. It is hoped that the reader will get a big-picture overview of the field by reading enough of these papers to understand the main results, without working through the details or proofs unless desired. For learning concepts in detail, we highlight particularly useful or pedagogical references in \ped{bold} to aid self-study.

Finally, there is no substitute for reading current research papers to get a sense of the open questions in the field and how people are trying to solve them. With multiple potentially relevant papers being published every day, it is easy to become overwhelmed. The following tips may be useful for managing the deluge:
\begin{itemize}
    \item You do not have time to read everything, and you don't have to. Choose which papers to read based on their importance to your work or how interesting they sound to you. If a paper piques your interest but is not directly related to your project, it's not a waste of time to go through it. Reading papers that interest you is far more likely to inspire your future work than diligently reading only the papers that are relevant to your current project.
    \item If you decide to open a paper, spend as much or as little time as is useful. For a crucial or foundational papers that are directly related to your research, it can be worth it to work through derivations and detailed results. If you only want to spend half an hour (or five minutes) on a paper, try to extract the essence of their results rather than get bogged down in details and derivations. What are the main results of the experiment? Why was it challenging? What are the most important pieces of intuition behind a theory paper?
    \item Sometimes the most important part of a paper are the introduction and references therein. This section is a great place to get a sense of the most important results in a field.
    \item Set time aside every week to read papers. Make it a part of your routine. Short term demands can make it difficult to fit in, but the long-term payoff is substantial.
    \item Papers aren't the only way to learn about research. More and more seminar and conference talks are recorded on YouTube, and explanations are often better than those in the corresponding manuscripts.
\end{itemize}

%% file: IntroCQED.tex
\chapter{Introduction to circuit QED}
\label{ch:IntrocQED}

\section{Introduction}

It is somewhat remarkable that we can even consider encoding a qubit in a superconducting current. Most systems that have been used as qubits encode information in a single degree of freedom, like the orbital state of an electron in an atom or the spin of a nucleus. In contrast, superconducting circuits encode information in the collective motion of a billions of billions of electrons. Coherently encoding information in such a system is analogous to the famous Schr\"odinger's Cat thought experiment, in which a system is put in a superposition of two macroscopically distinct states. In fact, the original proposal for observing coherence in a superconducting circuit was motivated not by quantum information, but rather by testing if quantum mechanics remained valid for macroscopic systems\major{\cite{leggett1980macroscopic}}. This work suggested such a system ``would probably be as near as we are likely to get to a laboratory version of Schr\"odinger's Cat,'' and that preserving its coherence long enough to measure it would ``likely to present serious, though possibly not insuperable, difficulties.'' These difficulties indeed yielded to experimentalists several years later when the macroscopic coherence of the phase difference $\varphi$ across a junction was demonstrated in reference \major{\cite{martinis1987experimental}}. Since then, steady progress has lead to qubits with near-millisecond coherence times and large-scale efforts to do quantum computation and simulation. However even in the most applied work, the original goal of demonstrating macroscopic coherence is never far away. Efforts to improve qubit coherence and implement progressively larger error correcting codes will push quantum theory to untested limits.

In this chapter, we outline the essential features of circuit quantization and superconducting qubits. We begin with a broad, qualitative overview of superconductivity, highlighting the features of the theory that are relevant for qubit design. This section includes a quantitative derivation of the Josephson effect, the circuit element that makes quantum information processing possible. Section \ref{sec:CircuitQuantization} describes the process of circuit quantization and illustrates the connections between Josephson junctions and inductors. We do not cover the general quantization of a circuit with multiple degrees of freedom in much depth, but instead provide overview of the available methods for treating them (including a high-level overview of black-box quantization). Section \ref{sec:ArtificialAtoms} summarizes the known sources of decoherence in superconducting qubits and common qubit designs invented to circumvent them. In section \ref{sec:IOTheory}, we take our first excursion into open quantum systems, introducing input-output theory in the context of microwave amplification. 



\section{Superconductivity: Broad Brush Strokes}



In general, electrons in a solid are strongly interacting due to Coulomb forces. However we typically study solids using a band structure theory, which maps out electron energy as a function of its momentum while completely ignoring electron-electron interactions. The surprising success of the band-structure way of thinking derives from the Fermi liquid, or Landau liquid model. In this model, it is shown that electron-electron interactions dress the single-electron states to give them a different effective mass (think of a particle moving through molasses). Most importantly, the result is again a non-interacting model which the structure of the Fermi sea. The Landau liquid model is correct to all orders in perturbation theory, but this does not guarantee that it never fails. In particular, Cooper showed that the Fermi sea is unstable if there is an attractive interaction between electrons; an arbitrarily small binding energy between electrons of equal and opposite momentum destabilizes the Fermi sea and leads to bound pairs of electrons that are not captured at any order in perturbation theory\ped{\cite{tinkham2004introduction}}. A net attractive force can arise between electrons via interactions with the underlying lattice. These bound pairs, composed of two fermions, behave like a boson and condense in a manner similar (though not identical) to a Bose Einstein condensate. The resulting state is the Bardeen-Cooper-Schrieffer (BCS) state of superconductivity, named for its discoverers.

Superconductivity plays several crucial roles in circuit QED. The obvious role is the elimination of resistance, which allows oscillations in an LC circuit to persist for a long time, and DC currents to persist almost indefinitely in macroscopic samples.\footnote{AC currents dissipate because superconducting currents have finite inductance. Therefore an AC current induces a non-zero voltage gradient inside and around the material, which causes the flow of currents that are not in the superconducting phase or are exterior to the superconductor, and thus have finite resistance.} Another useful consequence of superconductivity is that the electron condensate behaves like a rigid fluid of particles that move as one collective degree of freedom. It costs a finite amount of energy, superconducting gap $E_g$, to `break' electrons out of this state. Thus we are able to treat the \textit{quantum state} of a chunk of metal as if its only degree of freedom were the net current flowing through it, which allows us to encode and safely store quantum states in a macroscopic material. Finally, when a superconducting wire is interrupted with a thin insulating barrier, a surprising phenomenon called the Josephson effect imbues this simple circuit element with the properties of a dissipationless non-linear inductor. This non-linearity turns out to be necessary to make qubits and amplifiers. For the sake of completeness, we give a qualitative description of superconductivity to introduce the essential concepts that often arise in circuit QED. We then cover the Josephson effect in more detail, as it is the essential ingredient for making a qubit. 

Conventional superconductors, those that may be modeled by a phonon-mediated electron-electron interaction and the standard BCS theory, may be characterized by three physical parameters: the gap $E_g$, the London coherence length $\lambda_p$ and the Pippard coherence length $\xi$, all of which are temperature dependent. The gap $E_g(T)$ is the minimum energy required to break a single Cooper pair. The superconductor will not absorb energy below $E_g$, which provides robustness to thermal fluctuations and noise. The gap also sets the temperature at which the metal begins to superconduct by the relation $T_c = E_g(0)/3.528 k_b$, where $T_c$ is called the critical temperature. $E_g(T_c)=0$ and increases to its maximal value at $T=0$. When energetic photons, phonons or other radiation sources deposit energy above the gap break a Cooper pair, the result is a quasiparticle. Quasiparticles move throughout the material and lead to dissipation. In particular, if a quasiparticle tunnels across a Josephson junction in a qubit, it can add or remove energy from the qubit state, leading to spurious relaxation and excitation\cite{serniak2018hot}. At typical operating temperatures, a thermal distribution should leave no quasiparticles in a block of metal the size of the Earth!\footnote{$T_c$ for aluminum is $1$ K, so $E_g=3.528 k_b$. Operating at $30$ mK, the Boltzmann suppression factor is $e^{-E_g/k_b T}\approx 10^{-51}$. This wins out against the charge carrier density of aluminum multiplied by the volume of the Earth ($(2.1\times 10^{29}\text{m}^{-3})(1.1\times 10^{21} \text{m}^3)$). This calculation is somewhat optimistic, since we are neglecting that there is a continuum of quasiparticle states above the ground state, but you get the gist.} However, much higher quasiparticle concentration is observed in practice, and they are a major source of decoherence\cite{serniak2018hot,serniak2019direct}, though it may be possible to eliminate them through improved shielding. Quasiparticles have been observed in circuit QED both through their affect on qubits, and by trapping them in a superconducting nanowires\cite{levenson2014single}.

While the BCS model ultimately provided a physical explanation of superconductivity, phenomenological models still play an important role in its understanding. The London and Pippard coherence lengths are best understood using \definition{Landau-Ginzburg theory}, a phenomenological model that was discovered before BCS. The essential ingredient of Landau-Ginzburg theory is the replacement of the macroscopic many-electron wave function with an unnormalized wave function $\psi(x)$, and the Schr\"odinger equation with the non-linear Landau-Ginzburg equation. The local density of superconducting electrons is given by $n_s=|\psi(x)|^2/n$ where $n$ is the total electron density. The Landau-Ginsberg equations predict that superconducting currents are rigid, in the sense that they can only support curl-free flow \textit{i.e.} $\nabla \times J(\mathbf{r}) = 0$. This leads to the \definition{Meisner effect}, in which magnetic fields are completely expelled from a superconductor. However the magnetic field cannot go from a finite value to zero perfectly at the interface between the vacuum and the metal, as this would require an infinite current density at the surface. The London penetration depth gives the length scale over which an externally imposed magnetic field decays near an interface. In general, $\lambda_L=\sqrt{m^*/\mu_0 n_s e^{*2}}$, where $n_s$ is the bulk superconducting current density and $m^*$ and $e^*$ are respectively the effective mass and charge of the current carriers (so $e^*=2e$ for Cooper pairs).

The phase of $\psi$ is the physical phase in the BCS wave function, and relates to the Pippard coherence length.\footnote{For those who have worked through BCS before, the phase of $\psi$ is simply the phase factor that appears in the BCS wave function $\psi_\text{BCS} = \prod_k (u_k + e^i\varphi v_k c^\t_{k\uparrow} c^\t_{-k\downarrow}|0\>$. It is often ignored in derivations, but the important feature is that it is forced to have the same value in every term of the product, so that it becomes a macroscopic parameter.} This phase has a physical importance similar to that of an ordinary wave function. For instance, spatial variation of the phase is associated with momentum and leads to current flow. However a superconductor can only support oscillations up to a certain spatial frequency before the associated energy exceeds the gap. This enforces a distance scale over which the superconducting phase remains coherent, which is the Pippard coherence length $\xi$. One can think of $\xi$ as the effective size of the Cooper pairs. $\xi$ scales inversely with $T_c$, so that superconductors with a large gap have a shorter coherence length. It is possible for thermal fluctuations to change the phase coherence between two regions of superconductor by $2\pi$, particularly in constrictions that are narrow compared to the coherence length. This process is called an incoherent phase slip, and can lead to dissipation of a supercurrent. Coherent phase slips were recently observed\cite{astafiev2012coherent}, which are phase slips arising from tunneling instead of thermal activation.



$\xi$ and $\lambda_L$ are independent parameters, and one can have $\xi>\lambda_L$ or $\lambda_L>\xi$. This comparison turns out to be enormously consequential for the properties of a superconductor. The qualitative difference is so significant that superconductors with $\xi/\lambda_L > \sqrt{2}$ are called \definition{type I} while those with $\xi/\lambda_L < \sqrt{2}$ are called type II. This ratio determines how the material responds to external magnetic fields. For both type I and type II, the Meissner effect breaks down at sufficiently strong fields, and the material returns to its normal metal state. In a type I material, there is an energy penalty to having a normal metal region next to a superconducting region. Loosely speaking, this arises because the Cooper pairs are larger than $\lambda_L$, and thus are exposed to magnetic field. Consequently, if the material goes normal, it tends to go normal in large regions, so as to minimize the surface area between normal and superconducting. In contrast, type II have a negative surface energy, so that any penetrating magnetic field will fragment into many tiny normal regions called vortices. Each vortex contains a quantum unit of flux $\phi_0 = h/2e$ and is surrounded by a circulating current that screens this flux from the bulk.\footnote{One can imagine that the condensate picks up an Aharonov-Bohm phase as it travels around a loop enclosing a magnetic field. For $\psi$ to be single-valued, the minimum phase Aharonov-Bohm phase is $2\pi$, which limits the amount of flux that can penetrate to $\phi_0$.} As type II superconductors can support some magnetic field penetration without going normal, they have higher critical fields and are more practical for creating superconducting magnets. In principle wandering vortices pose a threat to superconducting qubit lifetimes, though they can be pinned in place by a number of mechanisms.


\subheading{Josephson junctions} We now turn to a quantitative description of the most important superconductivity effect for circuit QED. At first glance, a superconductor that is interrupted by an insulating barrier looks like an unlikely candidate for an inductor, which we normally associate with coils of wire and a significant amount of energy stored in a magnetic field. Indeed, when this problem was considered in the early days of superconductivity, it was assumed that single-electron tunneling across the barrier would dominate and nothing too unusual would happen. This expectation changed drastically when Josephson showed that coherent tunneling of Cooper pairs were not only significant, but would also generate surprising electrical properties resembling those of an inductor.

To model a tunnel junction, we begin by considering two superconducting regions separated by a thin insulating barrier. In superconducting circuits, this is often a layer of aluminum oxide that is grown on aluminum. Anticipating that current will end up flowing between them, we parameterize the quantum state of the system according to the number of Cooper pairs on each island as $|m\> = |N_L-m,N_R+m\>$. The tunneling of Cooper pairs is described by the Hamiltonian\cite{girvin2011superconducting,schrieffer2018theory}
\begin{equation} \label{eq:HJosephsonDiscrete}
H_T = -\frac{1}{2} E_J \sum_{m=-\infty}^{\infty} |m\>\<m+1| + |m+1\>\<m|.
\end{equation}
$E_J$ may be calculated by modelling electron tunnelling microscopically. It may also be estimated from a physical device by measuring the junctions room temperature resistance $R_T$ and using the  \definition{Ambegaokar-Baratoff relation} $E_J = h\Delta/8e^2 R_T$, where $\Delta$ is the superconducting gap.

$H_T$ translates Cooper pairs across the junction. Its eigenstates must be a uniform superposition of all values of $m$, otherwise acting $H_T$ on the state would change it. The eigenstates are
\begin{align} \label{eq:HJosephsonEigs}
    |\varphi\> = \frac{1}{\sqrt{2\pi}}\sum_{m=-\infty}^\infty e^{im \varphi} |m\>
\end{align}
The above states are physically unreasonable, as they contain states with an arbitrarily large charge imbalance. We could remedy this by adding Coulomb attraction to $H_T$, which would energetically suppress states of large $m$. However our aim is eventually to do lumped element circuit analysis, so we defer treatment of the Coulomb energy until we consider capacitance.

%
%
%
%

The above basis diagonalizes $H_T$, allowing us to write it in terms of a phase operator
\begin{equation} \label{eq:HJosephsonPhi}
H_T = -E_J \cos(\hat{\varphi}), ~~~ \hat{\varphi} = \int_{0}^{2\pi} \varphi |\varphi\>\<\varphi| d\varphi.
\end{equation}
The phase $\varphi$ turns out to be the difference in superconducting phases across the junction, the one that we discussed in reference to Landau-Ginzburg theory. Note that $\hat{\varphi}$ is a somewhat unusual operator, as its definition is ambiguous. Taking the integration bounds to be from $-\pi$ to $\pi$ would yield the same Hamiltonian but a different operator $\hat{\varphi}$. This ambiguity will be a bit of a nuisance, but poses no fundamental problem if we are careful.

\begin{exercise} \label{ex:HJosephsonEigs}
Verify that $|\varphi\>$ is an eigenstate of $H_T$ with eigenvalue $-E_J \cos(\varphi)$.
\end{exercise}

\begin{exercise} \label{ex:ExpPhiOp}
Show that $e^{i\hat{\varphi}} = \sum_{m=-\infty}^\infty |m\>\<m+1|$ and $e^{-i\hat{\varphi}} = \sum_m |m+1\>\<m|$. This phase operator does not suffer from ambiguity in its definition. Note the similarity to the results of exercise \ref{ex:HJosephsonEigs}.
\end{exercise}

$H_T$ allows us to compute the current flowing through the junction using the Heisenberg equations of motion. Defining the Cooper pair number operator $\hat{n} = \sum_{m=-\infty}^\infty m |m\>\<m|$ lets us derive the current as
\begin{align} \label{eq:FirstJosephson}
I &\equiv 2 e \frac{d \hat{n}}{d t} = \frac{2 i e}{\hbar} [H_T, \hat{n}] = \frac{E_J e}{i\hbar} \sum_m (|m\>\<m+1| - |m+1\>\< m|)  \\ \nonumber
&= \frac{2 e E_J}{\hbar} \sin(\hat{\varphi}).
\end{align}
Just as a phase gradient in a wave function leads to momentum (recall that momentum eigenstates are $e^{ixp/\hbar}$), a finite change in phase across the junction leads to a flow of current. Indeed, the Josephson junction can be thought of as a form of nonlinear kinetic inductance.\footnote{Kinetic inductance is inductance arising from the kinetic energy of the electrons themselves, as opposed to energy arising from the magnetic field generated by current flow. It is only non-negligible in high-mobility conductors like superconductors, and becomes more significant in narrow constrictions such as nanowires (imagine fluid flow forced through a narrow constriction, so that kinetic energy increases).}
%
\erf{eq:FirstJosephson} is important enough to merit the title of the \definition{first Josephson relation}. Note that as $\sin$ is bounded between $-1$ and $1$, the amount of current that may flow through a Josephson junction is also bounded by the \definition{critical current} $I_c = 2 e E_J/\hbar$. If one manages to force more current than $I_c$ through the junction, the superconducting phase breaks down and a finite resistance develops.

To understand the electrical properties of the Josephson junction, we need to understand the dynamics of $\hat{\varphi}$. In isolation, $\hat{\varphi}$ commutes with the Hamiltonian and is therefore a constant of motion. To see any time dependence, we need to see how the junction responds to an externally applied voltage. An applied voltage across the junction adds a term $\hat{Q} V = - 2 e V \hat{n}$ to $H_T$. V could be a classical scalar or an operator that commutes with $\hat{n}$. 

We could compute the Heisenberg equation of motion for $\varphi$ directly, but the result is less clean due to the artificial choice of boundaries in its definition \erf{eq:HJosephsonPhi}. It is cleaner to first work with the periodic operator $e^{i\hat{\varphi}}$ from exercise \ref{ex:ExpPhiOp}. Both $\cos(\hat{\varphi})$ and $e^{i\hat{\varphi}}$ are diagonal in the $|\varphi\>$ basis and therefore commute, which leaves
\begin{align} \label{eq:SecondJosephson}
\frac{d e^{i\hat{\varphi}}}{dt} &= \frac{i}{\hbar} [-2 e V \hat{n}, e^{i\hat{\varphi}}] \\ \nonumber
&= \frac{2i eV e^{i\hat{\varphi}}}{\hbar}.
\end{align}
using the result of exercise \ref{ex:ExpPhiOp}. $e^{i\hat{\varphi}}$ is not Hermitian and therefore not observable, but we can use the chain rule to recover an equation of motion for $\hat{\varphi}$
\begin{align} \label{eq:SecondJosephsonRelation}
\frac{d e^{i\hat{\varphi}}}{dt} &= i e^{i\hat{\varphi}} \frac{d \hat{\varphi}}{dt} = \frac{2 i e V e^{i\hat{\varphi}}}{\hbar} \\ \nonumber
\frac{d \hat{\varphi}}{dt} &= \frac{2 e V}{\hbar}
\end{align}
%
which is known as the \definition{second Josephson relation}. Combining these relations, we immediately see some bizarre effects. For constant $V$, $\hat{\varphi}$ grows linearly in time and the current through the junction oscillates at a frequency of $2 e V/h$ ($2e/h = 1/\phi_0 = 483.597.011$ MHz/$\mu$V). Thus the current spends half of its time flowing against the externally applied voltage! This phenomenon, called the \definition{AC Josephson effect} has several important applications in physics. As frequency may be measured with extremely high precision by counting oscillations over a long time interval, it provides one of the most precise methods to measure voltage. The Josephson relations will break down if the AC Josephson frequency exceeds the superconducting gap divided by $\hbar$, at which point the oscillation are so high frequency that they generate quasiparticles.

\begin{exercise} \label{ex:NPhiCommutator}
It is often stated that $[\hat{\varphi},\hat{n}] = i$, even when $\hat{n}$ is discrete as in \erf{eq:HJosephsonDiscrete}. Show that if $A$ has countably many eigenvalues (so that it may be written $\sum_n a_n |n\>\<n|$), then $[A,B]$ is hollow in the eigenbasis of A i.e. all the diagonal elements are zero. Argue therefore that $[\hat{\varphi},\hat{n}] = i$ is impossible
\end{exercise}

\begin{exercise} \label{ex:NPhiCommutatorDeriveRelations}
As we will see in the next section, it is often useful or even necessary to treat charge as continuous rather than discrete, in which case $[\hat{\varphi},\hat{n}] = i$ is perfectly valid. Use this commutation relation to derive the first and second Josephson relations. A useful lemma is that if two operators $A$ and $B$ commute with their commutator $[A,B]$, then $[A,f(B)] = [A,B]f'(B)$.
\end{exercise}

For our purposes, \erf{eq:HJosephsonPhi} will provide the main starting point for quantum circuits. However the Josephson junction's most important application for our purposes, its properties as a nonlinear inductor, are easily shown from the Josephson relations. For a linear inductor, $V = L \dot{I}$, so to cast the relations in this form, we take the time derivative of \erf{eq:FirstJosephson}
\begin{align}
\dot{I} &= \frac{d \hat{\varphi}}{dt} I_c \cos\hat{\varphi} = \pm \frac{2 e V}{\hbar}\sqrt{I_c^2 - I^2} \\ \nonumber
V&=\pm\frac{\hbar}{2e \sqrt{I_c^2 - I^2}} \dot{I}.
\end{align}
As written, the sign is ambiguous and must be determined from the sign of $\cos\hat{\varphi}$. The possibility for a negative inductance arises for the same reason that current flows against $V$, and is a manifestation of hysteresis. The voltage across the junction is proportional to the derivative of the current, but the proportionality constant itself depends on the current. To zeroth order in $I$, the Josephson inductance is 
\begin{align} \label{eq:LinearJosephsonInductance}
L_J = \frac{\hbar}{2e I_c} = \frac{\hbar^2}{4e^2 E_J}
\end{align}
and behaves as a linear inductor when $I$ is small relative to $I_c$. For a reasonable value like $E_J/\hbar = 12~\text{GHz}$, $\hbar/2e I_c \approx 100 \text{nH}$, which is quite large for a device that might only be $50\times 50$ nm!

As the Josephson junction comes up so much in circuit QED, it is worthwhile to spend a little more time developing intuition for its dynamics. For this purpose, there are two physical analogs of $H_T$. The first comes from noting that \erf{eq:HJosephsonDiscrete} looks like the kinetic energy energy, or hopping term for a 1 dimensional tight binding model Hamiltonian, where $m$ labels lattice positions. \erf{eq:HJosephsonEigs} shows that $\varphi$ plays the role of momentum. The elegant feature of this analogy comes in when we add an external voltage. The additional energy term $-2e V \hat{n}$ is equivalent to the application of a constant force $F x$. Under this force, $\varphi$ increases until it hits the edge of the Brillouin zone $\pm \pi$, at which point it Bragg reflects, entering from the opposite side of the Brillouin zone. The resulting Bloch oscillations offer one visualization of the AC Josephson effect.

\begin{figure}
\centering
{\includegraphics[width = \textwidth]{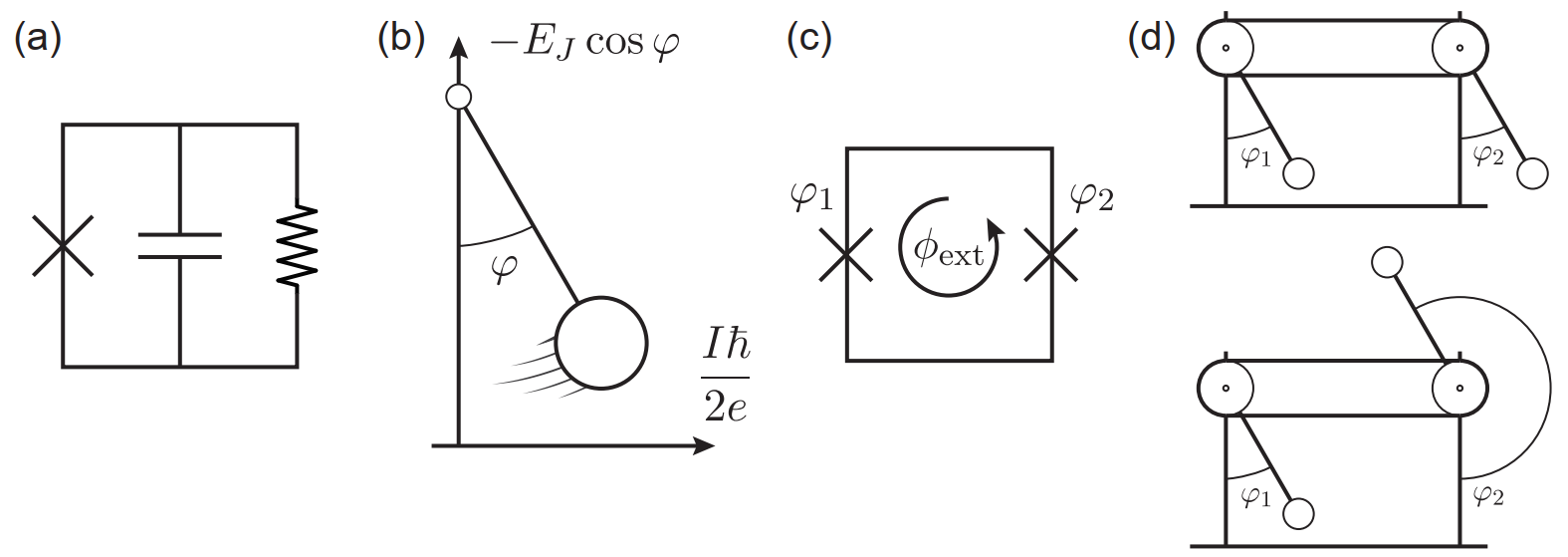}}
\caption{(a) A Josephson junction (typically denoted as an $X$ on a circuit diagram) in parallel with a capacitor and a resistor. (b) A mass coupled to a pendulum, which has the same Hamiltonian as (a). A SQUID threaded with an external magnetic flux $\phi_\text{ext}$. (d) A physical analog of the SQUID for $\phi_\text{ext} = 0$ (top) and $\phi_\text{ext} = \phi_0/2$ (bottom). In the latter case, the gravitational potential of the two masses cancel, effectively removing the junction from the circuit.
}
\label{fig:Ch2_JunctionOscillator}
\end{figure}

The second analogy requires us to think of a Josephson junction embedded in a parallel LCR circuit, as in Fig. \ref{fig:Ch2_JunctionOscillator}a. This simple circuit will form the basis for our two simplest qubits and the most common microwave amplifier. It is also necessarily the effective circuit for any realistic Josephson junction. As a Josephson junction consists of two superconductors separated by a dielectric, it is as much a capacitor as it is a junction. Thus as we noted earlier, we must include the Coulomb energy in any realistic model of a junction. Given the standard formula for a capacitor's energy, the resulting Hamiltonian is
\begin{align}
H = \frac{(2e)^2}{2C} \hat{n}^2 - E_J \cos(\varphi).
\end{align}
We defer the mathematical treatment of a resistor for later, but for now note qualitatively that it should tend to damp the system to a state of zero current ($\varphi = 0$ or $\varphi = \pi$). Fig. \ref{fig:Ch2_JunctionOscillator} suggests interpreting $\varphi$ as the angular coordinate of a pendulum, and $-E_J \cos(\varphi)$ as its gravitational potential energy. In contrast to the first analogy, the capacitance term now plays the role of kinetic energy. Kinetic energy is quantized due to the periodic nature of $\varphi$, just as in a quantum rigid rotor. Thus we can think of this JCR circuit as a nonlinear resonator, just like a pendulum.

If the capacitance in the above circuit is the intrinsic capacitance of the Josephson junction, then the resulting resonance frequency is called the \definition{Josephson plasma frequency}. This resonance can interfere with normal circuit operation if it is close enough to primary operating frequencies. In practice, our microwave circuits will have an operating frequency much lower than the plasma frequency.

The Josephson oscillator is an extremely rich physical system, and there are many important phenomena that can be derived from the above equations that we do not consider here, such as Shapiro steps and RSFQ logic. The pendulum analogy carries through to many standard examples. The last that we consider in detail is the superconducting quantum interference device, known as a SQUID. A SQUID is simply two Josephson junctions in parallel, as depicted in Fig. \ref{fig:Ch2_JunctionOscillator}c. If the junctions are identical, the effective Hamiltonian is simply $2H_T$. Things get interesting when we apply a magnetic field through the loop, which sets up a circulating current in the loop via the Meissner effect. We can use Maxwell's equations to find the effect on each of the junctions. Labelling the two junctions as 1 and 2 and using Stoke's theorem, we find
\begin{align}
    \nabla \times E &= -\frac{\partial B}{\partial t} \\ \nonumber
    \oint E\cdot dl &= - \frac{\partial \phi_\text{ext}}{\partial t} = V_1 + V_2 = \frac{\hbar}{2e}(\dot{\varphi}_1 + \dot{\varphi}_2)
\end{align}
where $\phi_\text{ext}$ is the magnetic flux passing through the loop. The above relation forces a boundary condition between $\varphi_1$ and $\varphi_2$
\begin{align}
    & \varphi_1 + \varphi_2 + \frac{2\pi}{\phi_0} \phi_\text{ext} = 0
\end{align}
where $\phi_0 = h/2e$ is the flux quanta. Using a basic trigonometric identity, we can reduce the Hamiltonian of two junctions in parallel to that of a single junction with effective $E_J$
\begin{align} \label{eq:SquidEJ}
    E_J(\phi_\text{ext}) = \cos\left(\frac{\pi \phi_\text{ext}}{\phi_0}\right)
\end{align}
for the case that $E_{J,1}=E_{J,2}$. 

There are many ways to reason about the SQUID intuitively. \erf{eq:SquidEJ} arises from the interference between two phase-shifted cosine terms in the Hamiltonian. We can interpret this interference as literal interference between Cooper pairs tunneling simultaneously across both junctions in parallel. When $\phi_\text{ext} = \phi_0/2$, the two channels interfere destructively, completely inhibiting the flow of current. A slightly more abstract model is shown in Fig. \ref{fig:Ch2_JunctionOscillator}d. The fixed phase relationship between $\varphi_1$ and $\varphi_2$ amounts to locking the relative phase of two pendula with a pulley\cite{altshuler2003josephson}. $\phi_\text{ext} = \phi_0/2$ corresponds to putting the two masses exactly out of phase, so that the gravitational (Josephson) energies cancel.

The SQUID has is a ubiquitous device in superconducting electronics. Its magnetic field sensitivity makes it the most sensitive magnetometer ever developed. It also provides a way to tune the resonant frequency of a JC circuit, at the expense of making it sensitive to ambient magnetic field noise. One can play a trade-off between tunability and noise susceptibility by unbalancing $E_{J,1}$ and $E_{J,2}$, so that the cancellation becomes incomplete. The SQUID also shows up unintentionally in large-area junctions. If the flux threading a subregion of the junction becomes comparable to $\phi_0$, then we must treat that single junction as many junctions in parallel, each with their own threaded flux. This makes even single Josephson junctions potentially sensitive to flux noise.






\section{Quantization of a circuit}
\label{sec:CircuitQuantization}

\begin{figure}
\centering
{\includegraphics[width = 130mm]{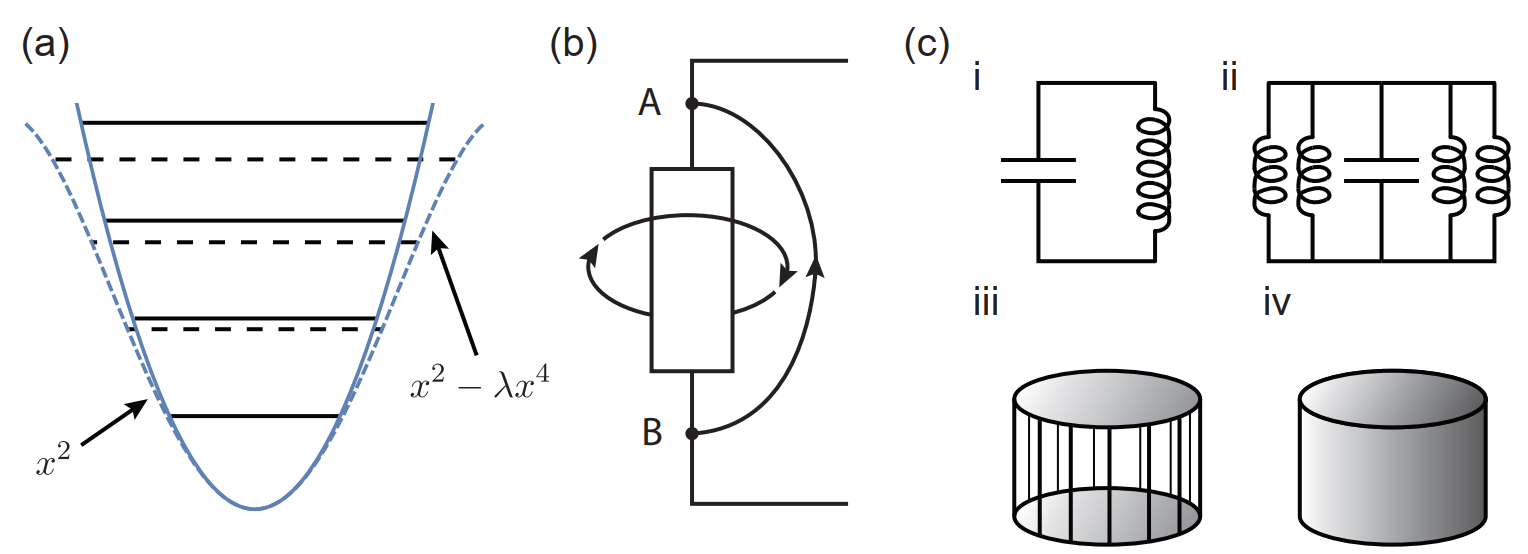}}
\caption{(a) A parabolic potential with equally spaced transitions (solid lines) and a quadratic-plus-quartic potential with unequal spacing. Transitions may be addressed individually in the anharmonic case, as they now differ in frequency. (b) Illustration of the integrals used to define branch flux (vertical curved path) and branch charge (loop) of a general circuit element. (c) progressively shrinking an LC circuit until it becomes a microwave cavity.
}
\label{fig:Ch2HarmonicOscillators}
\end{figure}

We have developed a Hamiltonian for the the Josephson junction, the essential ingredient for a superconducting qubit. This nonlinearity will allow us to turn a harmonic LC oscillator into a nonlinear JC oscillator, which in turn lets us selectively drive a single transition in the ladder of states as in Fig. \ref{fig:Ch2HarmonicOscillators}a. However before we describe superconducting qubits in further detail, it is important to understand how to quantize a general circuit in a bit more detail.

Classical LC circuits like the one shown in Fig. \ref{fig:Ch2HarmonicOscillators}c (i) behave like classical harmonic oscillators, and there are a number of ways to quantize them of varying levels of generality. We take a few shortcuts to get to the essential physics as quickly as possible, and refer to other references at the end of this section. We begin with the Hamiltonian for a classical LC circuit 
\begin{align}
    H_\text{LC} &= \frac{1}{2}L I^2 + \frac{Q^2}{2C}
\end{align}
where $I$ is the current through the inductor and $Q$ is the charge stored on the capacitor. If we were to begin with a Lagrangian for the circuit, then we could immediately identify canonically conjugate coordinates by the relation $p = \partial L /\partial \dot{q}$, and then quantize by promoting $p$ and $q$ to operators satisfying $[q,p]=i\hbar$. To get them from a Hamiltonian, we must find which choice of coordinates reproduces the correct classical equations of motion.\footnote{In general, both of these procedures can lead to ambiguities in quantization; classical operators commute, so there may be operator ordering ambiguities when we try to assign a quantum operator to a product of classical operators. However if there is one set of coordinates $p_i$ that only appear in the Lagrangian to second order, then partial derivatives of these terms lead to first-order canonical variables which cannot have ordering ambiguities. We will only deal with linear capacitors, which provides such a guarantee. Quantization of circuits with nonlinearities in both $Q$ and $I$ is an area of ongoing research, so it is surprisingly subtle.} We know the classical equation of motion that we seek to reproduce is $\ddot{Q} = \omega^2 Q$ with $\omega = 1/\sqrt{LC}$. In analogy with the harmonic oscillator, we expect the conjugate coordinates to be proportional to $I$ and $Q$. Let us name our canonically conjugate coordinates $q = \beta Q$ and (suggestively) $\phi = \alpha I$, where $\alpha$ and $\beta$ are constants to be determined. Hamilton's equations of motion give 
\begin{align}
\dot{q} &\equiv \frac{\partial H}{\partial \phi} = \frac{1}{\alpha} LI \\ \nonumber
\dot{\phi} &\equiv -\frac{\partial H}{\partial q} = -\frac{Q}{\beta C}.
\end{align}
Combining these by substitution, we find
\begin{align}
Q \frac{L}{C} \frac{1}{\alpha^2 \beta^2} = \ddot{Q}.
\end{align}
which shows that $\alpha \beta = L$. Alternatively, we could have simply observed that $\dot{Q} = I$, which also implies $\alpha \beta=L$. $\beta$ by itself is equivalent to choosing a coordinate system, so we are free to take $\beta=1$ so that $q=Q$. Thus to quantize the oscillator, we take $[\phi, Q] = i\hbar$ with $\phi=LI$. Representing these coordinates in terms of ladder operators, we have
%


%
\begin{align} \label{eq:PhiQZPFs}
    \phi &= \phi_\text{ZPF}(a+a^\t) \\ \nonumber
    Q &= -i q_\text{ZPF}(a-a^\t) \\ \nonumber
    \phi_\text{ZPF} &= \sqrt{\frac{\hbar Z_\text{eff}}{2}} \\ \nonumber
    Q_\text{ZPF} &= \sqrt{\frac{\hbar}{2 Z_\text{eff}}}
\end{align}
where $Z_\text{eff}=\sqrt{L/C}$ is the characteristic impedance of the oscillator.\footnote{We know that $\<0|H|0\> = \hbar \omega/2$ for any harmonic oscillator. The third and fourth equations of \erf{eq:PhiQZPFs} follow if we assume that this zero-point energy is divided evenly between $Q$ and $\phi$, as suggested by Virial theorem.} The prefactors $\phi_\text{ZPF}$ and $Q_\text{ZPF}$ quantify the amplitude of vacuum fluctuations in each coordinate of the circuit, as $\<0|\phi^2|0\>=\phi_\text{ZPF}^2$ and likewise for $Q$. These quantities play an important role the the ultimate behavior of a circuit, and show up in black box quantization.

We might hope that the linear inductor treated in this section can be related to the Josephson inductance of the previous section, and indeed our naming schemes of the variables $\phi$ and $\varphi$ are intentionally suggestive. The second Josephson relation connected $\varphi$ to $V$. The corresponding relation between $\phi$ and $V$ for an inductor is the standard relation
\begin{align} \label{eq:Inductor}
V = L \frac{d I}{dt} = \frac{d\phi}{dt}.
\end{align}
Thus \erf{eq:SecondJosephsonRelation} is also of the form of an inductor voltage relation if we take $\phi = \varphi \phi_0/2\pi$ for the junction, where $\phi_0 = h/2e$. With this definition of $\phi$ for a junction, $V=d\phi/dt$ becomes a restatement of the Josephson relations. If we combine the canonical commutation relation with this identification, then we can loosely write $[\hat{\varphi}, \hat{n}] = i$, as we used in exercise \ref{ex:NPhiCommutatorDeriveRelations}. 

\erf{eq:Inductor} is such a useful relation that it pays to turn things around and define $\phi$ not as $IL$ or $\varphi \phi_0/2\pi$, but instead as
\begin{align}
\phi \equiv \int_{-\infty}^t V(t')dt'
\end{align}
Defined as such, $\phi$ is called the \definition{branch flux}, and it is sufficiently general to be defined for virtually any circuit element. Equally useful is the \definition{branch charge}
\begin{align} \label{eq:BranchCharge}
Q = \int_{-\infty}^t I(t') dt'.
\end{align}
Though it might seem trivial, this definition is useful because it applies to any circuit element, not necessarily just a capacitor. Both the branch flux and branch charges may be defined from the fields generated by (and external to) a circuit element, rather than the detailed dynamics of the charges and currents
\begin{align} \label{eq:VoltageCurrentDefs}
V(t) &= \int_A^B \vec{E}\cdot d\vec{l} \\ \nonumber
I(t) &= \frac{1}{\mu_0} \oint \vec{B}\cdot d\vec{l}
\end{align}
where the contour integral encircles only the circuit element in question and $A$ and $B$ are its terminals as in Fig. \ref{fig:Ch2HarmonicOscillators}b. These relations, along with the associated branch variables allow one to quantize a very general circuit element\cite{vool2017introduction}.

Some resonators in superconducting circuits are true lumped element devices, but it is also common to use microwave cavities. To quantize these, there is a nice argument due to Feynman that directly relates an LC circuit to an electromagnetic resonator. Imagine that we wish to increase the resonant frequency of our LC circuit. $\omega = 1/\sqrt{LC}$, so we should shrink our capacitors and inductors as much as possible. Inductance can be decreased by placing multiple inductors in parallel, as in Fig. \ref{fig:Ch2HarmonicOscillators}c ii. Each inductance can be decreased by eliminating loops of wire until each is just a straight length of wire made as short as possible, as in Fig. \ref{fig:Ch2HarmonicOscillators}c iii. Once each wire is so short that it spans only the gap on the capacitor, then all that is left to do is to add more `inductors' until the interior space of the capacitor is completely enclosed, which results in a cavity! Thus the only distinction between a lumped element circuit and a cavity resonator is that an LC circuit stores electric and magnetic fields in spatially separate regions. Both may be quantized in the same way. This argument also works the other way around. \erf{eq:VoltageCurrentDefs} lets us define and quantize lumped element circuits via the electomagnetic fields that they generate, which is a common strategy.

So far, we have quantized LC and LJ oscillators, two of the simplest superconducting circuits. However there is a strange difference between the two. For the LC circuit, $Q$ and $\phi$ are analogous to $x$ and $p$, and therefore have their eigenvalues in $\mathbb{R}$. The treatment is not sensitive to the fundamental discreteness of charge. Capacitive energy depends on the spatial distribution of charge, which is continuous even through electrons come in integer amounts. In contrast, our treatment of the Josephson junction restricted the eigenvalues of $\varphi$ to be between $0$ and $2\pi$, and therefore $\hat{n}$ only had integer eigenvalues. The periodicity of $\varphi$ is also not easily circumvented, as it corresponds to the superconducting phase in the BCS wave function.

The resolution comes from the second Josephson relation \erf{eq:SecondJosephson} and the definition of branch flux. Different circuit elements can have different Hilbert spaces, discrete or continuous, but we can nevertheless couple them by computing how they generate and respond to potential differences. As a lumped element, the Josephson junction couples only to Cooper pairs located very close to the tunnel barrier, but coupling to the capacitor and capacitive energy favors Cooper pairs spread out over the entire superconducting island. The net effect is a device that behaves as if its charge operator is discrete, and the junction mediates tunneling of delocalized Cooper pairs from one side to the other.

We can also interpolate smoothly from a JC circuit to an LC circuit, which has the non-periodic current-phase relationship $I=\phi/L$. Suppose we replace a single Josephson junction with Josephson energy $E_J$ with $N$ junctions in series, each with an increased coupling energy $N E_J$. As $L_J \propto 1/E_J$, the effective linear inductance is the same as that of the original junction by \erf{eq:LinearJosephsonInductance}. For the branch flux of the entire network of junctions in series, we have $\phi_\text{tot} = \sum_i \phi_i$, so $\varphi_i = 2\pi \phi_\text{tot}/N \phi_0$ for each junction individually. Thus the $2\pi$ periodic symmetry is broken to a larger period of $2\pi N$. Furthermore, as we take $N\rightarrow \infty$, each $\varphi_i$ becomes very small, so that the nonlinear current-phase relationship for each junction individually becomes linear via a Taylor expansion
\begin{align}
I &= \frac{2 e N E_J}{\hbar} \sin\left(\frac{2\pi \phi_\text{tot}}{N\phi_0} \right) \\ \nonumber
&\approx \frac{\phi_\text{tot}}{L_J}.
\end{align}
Thus the chain of junctions in series becomes an inductor, and our JJ...JC circuit becomes an LC circuit. Qualitatively, it is as if charge becomes continuous because Cooper pairs can now delocalize over the chain.

We have reconciled how some circuits can have discrete charging energy while others have continuous, but what happens when we combine them? This difficulty arises if we put an inductor and a junction in parallel. The inductor breaks the periodic symmetry of the circuit, so that $\phi=0$ and $\phi=\phi_0$ are now physically distinct states with different amounts of energy. To extend the pendulum analogy that we used to understand the SQUID, it is as if the junction pendulum is connected by a pulley to a torque spring that winds as the pendulum rotates. When shunted by a capacitor, this circuit becomes that of a relatively modern qubit called the capacitively shunted flux qubit.

\subheading{Other methods of circuit quantization} 
The circuits considered in this section give relatively simple examples because the kinetic and potential energies are stored in a single capacitor and a single inductor respectively. Most circuits will not be so simple, and more general methods will be useful to quantize them. However the essential ingredient that we have used here, that a classical resonance may be replaced with a quantum harmonic oscillator, carries over to more general methods.
One is a standard procedures to go from a general lumped-element circuits to a quantum Hamiltonian via methods borrowed from classical circuit analysis\ped{\cite{vool2017introduction}}. This method does not require one to first diagonalize the circuit, and can handle a circuit with nonlinear inductors or nonlinear capacitors (but not both at the same time). Most of the subtlety lies in writing down a valid classical Hamiltonian for the lumped-element circuit and recognizing independent degrees of freedom. 

While this method is important for understanding circuit QED, it is not the most common way to analyze quantum circuits in practice. The current standard is called \definition{black-box quantization}\cite{minev2019catching}, which derives a quantum Hamiltonian from numerical finite-element simulations of a circuit. Standard numerical packages such as COMSOL and HFSS can diagonalize a linear electrodynamics problems into a set of resonant modes. To account for the presence of Josephson junctions, one takes the quadratic approximation of the Josephson potential \erf{eq:LinearJosephsonInductance} and places lumped element inductors at the locations of the junctions in simulation. The resulting eigenmodes with frequencies $\omega_i$ are then quantized simply by writing down
\begin{align}
    H_\text{linear} = \hbar \sum_i \omega_i a_i^\t a_i.
\end{align}
The full Hamiltonian reintroducing nonlinear terms of the Josephson potential is then
\begin{align}
    H &= H_\text{linear} - \sum_j E_J\left(\cos(\hat{\varphi}_j) - \frac{1}{2}\hat{\varphi}_j^2\right) \\ \nonumber
    \hat{\varphi}_j &= \frac{2\pi\hat{\phi}_j}{\phi_0}.
\end{align}
The $\hat{\phi}_j$ operators, one for each Josephson junction, may be expressed as linear combinations of the current quadrature of the finite element eigenmodes
\begin{align}
\hat{\phi}_j = \sum_i \phi_{ij}^\text{ZPF} (a_i^\t + a_i)
\end{align}
where $\phi_{ij}^\text{ZPF}$ are the zero point fluctuations from \erf{eq:PhiQZPFs}, except that now the current quadrature of mode $i$ is divided among the junctions and the electromagnetic energy of other linear parts of the circuit. The $\phi_{ij}^\text{ZPF}$ quantities are linearly related to the energy participation ratio, which may be extracted from the finite element simulations\cite{minev2019catching}. Black-box quantization is an ongoing area of research, and continual improvements are being made to its accuracy and computational efficiency. For further details, we refer the reader to the PyEPR github repository that implements black-box quantization.\footnote{https://github.com/zlatko-minev/pyEPR}

\section{Artificial Atoms}
\label{sec:ArtificialAtoms}

In the previous section, we wrote down some simple circuits and quantized them. We now look at the dynamics of these circuits and learn how they may be operated as qubits. Much of the subtlety of qubit design lies in understanding and minimizing decoherence, so we begin by overviewing the primary sources of it.

\begin{figure}
\centering
{\includegraphics[width = 0.8\textwidth]{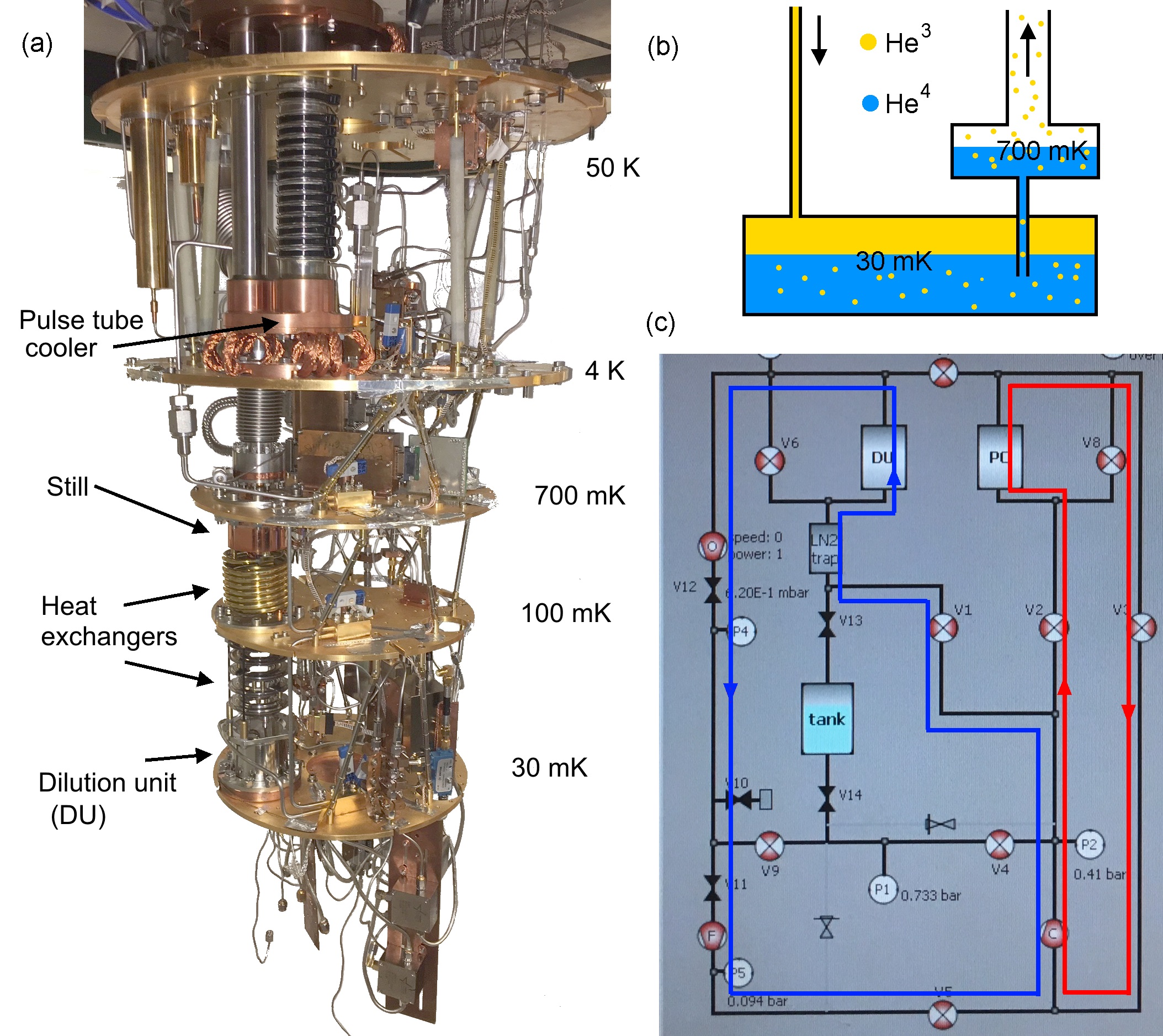}}
\caption{(a) An opened dilution refrigerator. Stages are cooled by 3 different mechanism. The 4 K plate is cooled by expansion cooling of Helium via a pulse tube cooler. A circuit of a He$^3$/He$^4$ mixture is pumped into the dilution unit and then pumped out of the still with a turbo pump (not pictured). The turbo pump cools the still via evaporative cooling of liquid He$^3$, which has a higher vapour pressure than He$^4$. (b) The principle of dilution refrigeration, which is analogous to evaporative cooling. He$^4$ is bosonic, and has a smaller de Broglie wavelength than He$^3$ due to its higher mass. Both of these properties allow a smaller interatomic spacing between He$^4$ atoms, which in turn leads to a lower energy via the Van der Waals force. Therefore at low temperatures, He$^3$ and He$^4$ phase-separate like oil and water, as depicted. However a He$^3$ atom can lower its energy by entering the He$^4$ phase. Due to the Pauli exclusion principle, He$^3$ atoms entering the He$^4$ phase must occupy increasingly higher energy states. Eventually the Fermi energy exceeds the energy gained by entering the He$^4$ phase. This balance occurs when the He$^4$ phase contains 6.4\% He$^3$, even at zero temperature! This allows finite cooling power all the way down to $T=0$ (unlike evaporative cooling, where the vapor pressure goes to zero below the boiling point). 
(c) Circulation of helium during the initial precooling of the fridge (red), used to go from room temperature to about 10 K using the pulse tube, and dilution cooling (blue).
}
\label{fig:DilutionFridge}
\end{figure}

\textbf{Decoherence mechanisms} 
First and foremost, superconducting circuits must remain cold to approach the idealized quantum behavior described above. Aside from working well below the superconducting critical temperature, low temperatures allow for thermal initialization of a qubit to its ground state. As typical superconducting circuits operate at frequencies of a few GHz, millikelvin temperatures are required to get low excited state population (1 GHz corresponds to a temperature of about 48 mK). These extremely low temperatures may be achieved with a commercially available device called a dilution refrigerator. The basic working principle of a dilution refrigerator is described in Fig. \ref{fig:DilutionFridge}. Low temperatures ensure that nearby degrees of freedom at the qubit frequency are also in their ground state, which mitigates many potential sources of decoherence.

As mentioned at the beginning of this chapter, resistive loss is finite at non-zero frequencies even in a superconductor. Finite inductance leads to potential differences across the superconductor, which in turn drives current through the fraction of electrons that are not superconducting. Fortunately at low temperatures, the residual non-superconducting electron density decreases exponentially with temperature ($1-n_s \propto e^{E_g/k_b T}$), and qubits tend to operate far below the superconducting critical temperature. However dissipation in other nearby materials such as the substrate onto which a qubit is fabricated and (in particular) surface interfaces pose major sources of loss. Dielectric loss is mitigated by using high-resistance substrates like silicon and sapphire, and by designing qubits that store more of their electromagnetic energy in the nearby vacuum\cite{minev2016planar,paik2011observation}. Some experiments have even etched away the substrate in the vicinity of small features where electric field is concentrated\cite{chu2016suspending}. A good deal of qubit fabrication development lies in minimizing these sources of loss by making and testing high-Q resonators.\footnote{$Q=\omega/\kappa$ is called the quality factor of a resonator. It measures the number of oscillations the resonator will undergo before the oscillations damp out due to loss.} 

Finally, microwave circuits can lose energy by radiating microwave photons. One must take care to place the qubit in an isolated environment with no resonances near the qubit frequency. Qubits are place in conducting enclosures that prevent dipole emission into free space. Finite element solvers like COMSOL and HFSS are used to identify electromagnetic resonances of a circuit and its enclosure to eliminate parasitic resonances. Even the substrate can support resonances that harm qubit coherence. These can be mitigated with through-silicon vias that gap these modes up to higher frequencies.

\begin{figure}
\centering
{\includegraphics[width = 130mm]{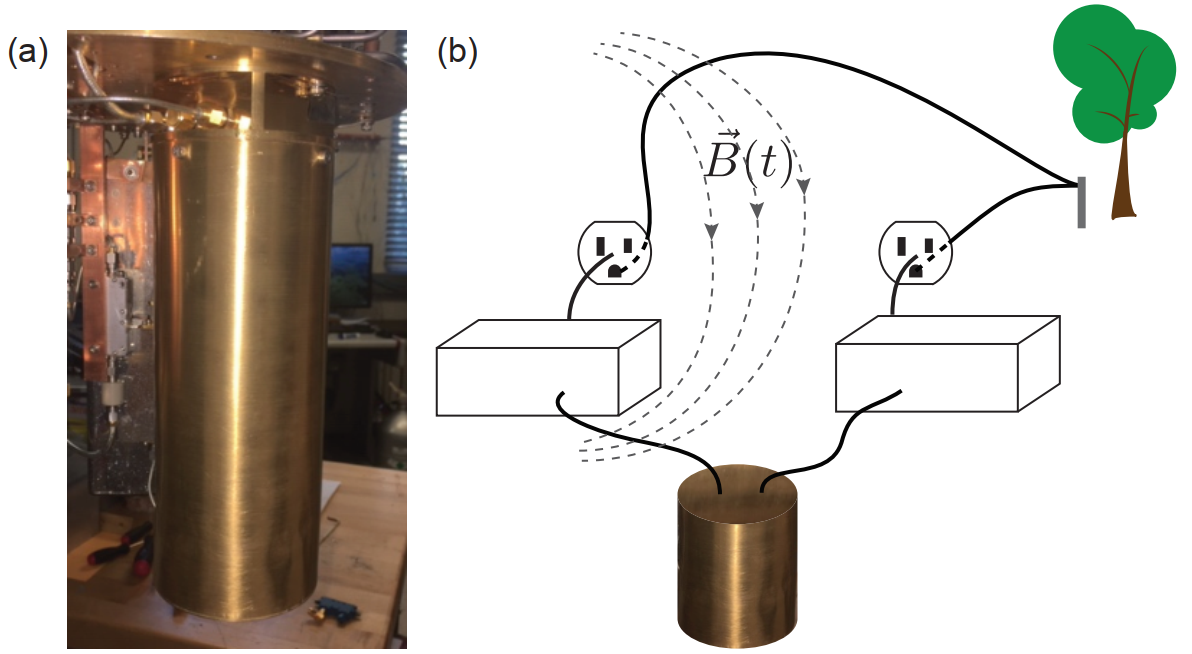}}
\caption{(a) A light-tight can for housing superconducting circuits. DC and microwave control lines enter through tightly sealed bulkheads. A venting hole is necessary to allow evacuation before cooling to mK temperatures. (b) A classic example of a ground loop. Ground loops can be shrunk  by running all equipment from the same breaker and ensuring electrical isolation.
}
\label{fig:LightTightCan}
\end{figure}

Quasiparticles constitute another major source of qubit decoherence. While in principle they can lead to finite resistance and hence could be characterized by making and testing resonators, their primary effect occurs when one tunnels across a Josephson junction. This is because the junction supports a large potential difference. When a quasiparticle tunnels across the junction, its kinetic energy can change by the qubit energy level spacing and changing the qubit state. Curiously, quasiparticles have been observed to excite the qubit more often than they lead to decay, indicating that they do not follow an equilibrium Boltzmann distribution. One way to mitigate quasiparticles is to enhance the rate at which they are dissipated by putting a normal metal in contact with the qubit\cite{riwar2016normal}. More commonly one attempts to limit how many are generated in the first place by placing infrared filters (such as Eccosorb) on microwave control lines 
and placing qubits in a sealed light-tight can, as in Fig. \ref{fig:LightTightCan}a (hermetic sealing is typically not possible, due to the need to evacuate air from the can during cool-down). It is also important to block radiation from hotter portions of a dilution refrigerator from entering the low temperature region. A commonly used method is to place aluminum tape over holes in various stages and check for light leaks using a flashlight, though it is unclear if this method is sufficient.

Loss due to resistance and quasiparticles has decreased steadily with improved fabrication and filtering. A decoherence source that has been more difficult to mitigate directly is charge fluctuations. The exact source is somewhat unclear, but it appears that the charges in the dielectric substrate or the junction fluctuate, so that the qubit experiences a gradually changing electric field.  
Charge fluctuations can be modelled by changing the capacitive energy to account for a randomly varying charge offset $n_g$
\begin{align}
H_\text{cap} = \frac{2 e^2}{C}(n-n_g)^2.
\end{align}
$n_g$ can fluctuate on both short and long timescales, and it is not at all uncommon for $n_g$ to shift by one or more units of charge over minutes to hours. It is commonly modelled to have a $1/f$ power spectrum. The outlook for eliminating charge fluctuations is somewhat unclear. Surface charges are a notoriously difficult problem in many quantum systems, and are one of the primary limiting factors for trapped ion qubits and near-surface NV qubits. Fortunately, superconducting qubits can be designed to be insensitive to $n_g$ as we will see. Much of the success of superconducting qubits has come from the transmon qubit, which is designed to be insensitive to charge noise.

In principle, any physical parameter of a qubit circuit can fluctuate, including $E_J$ or $E_C$, though these do not appear to change much in practice. The last source of decoherence we consider here is magnetic flux noise, which is in a sense dual to charge noise.\footnote{Note that in our analysis of the SQUID, $\varphi$ appeared in the argument of the Josephson energy, much as $n_g$ appears in the argument of the capacitive energy.} As we saw previously, a pair of Josephson junctions in a loop serves as an exquisitely sensitive magnetometer, which is great for sensing but terrible for a qubit! A qubit that contains a SQUID as its inductive element can be tuned by a magnetic field passing through the loop, which is sometimes worth the additional susceptibility to flux noise. Flux noise, like charge noise, often follows a $1/f$ power spectrum, and recent evidence suggests that it arises largely from diamagnetic impurities like oxygen. It can also come from wandering superconducting vortices in type II superconductors, which can be pinned in place by adding holes to any large-area metal regions (which is called `waffling').\footnote{Vortices can also lead to resistive dissipation, as they experience a Lorentz in the presence of current flow. As vortex motion can experience scattering, they offer a way to damp current flow in a superconductor. Type II superconductors would not have conductivities much better than ordinary metals without flux pinning.} An insidious source of flux noise is ground loops; if magnetic field fluctuations from the laboratory can drive current through various parts of the system at large, these currents can generate magnetic fields near the qubit. An example of a ground loop is shown in Fig. \ref{fig:LightTightCan}b. Since most qubits derive their inductance from Josephson junctions, the rest of the device tends to have very low mutual inductance with its surroundings. Therefore non-tunable qubits are insensitive to flux noise. However some qubit designs such as the flux qubit necessarily have loops containing Josephson junctions and are therefore sensitive to flux noise by design. Typically there exists some kind of sweet spot where the first-order sensitivity (\textit{i.e.} first derivative of the qubit energy vs. flux) goes to zero.

\subheading{Cooper pair box} Perhaps the simplest qubit is the \definition{Cooper pair box}, which is a Josephson junction and a capacitor in parallel as in Fig. \ref{fig:Ch2_JunctionOscillator}a. Including the possibility of a charge offset $n_g$, the Hamiltonian is 
\begin{align} \label{eq:CooperPairBox}
    H &= 4 E_C (n-n_g)^2 - E_J \cos(\varphi) \\ \nonumber
    E_C &= \frac{e^2}{2 C}
\end{align}
where $C$ is the capacitance, which includes any capacitance of the junction itself (which is in parallel with the lumped element capacitance of the circuit). $E_C$ is the capacitive energy associated with one electron's worth of charge displaced in the capacitor. The factor of four arises because $\hat{n}$ counts Cooper pairs, each of which contains two electrons. 

\begin{figure}
\centering
{\includegraphics[width = 130mm]{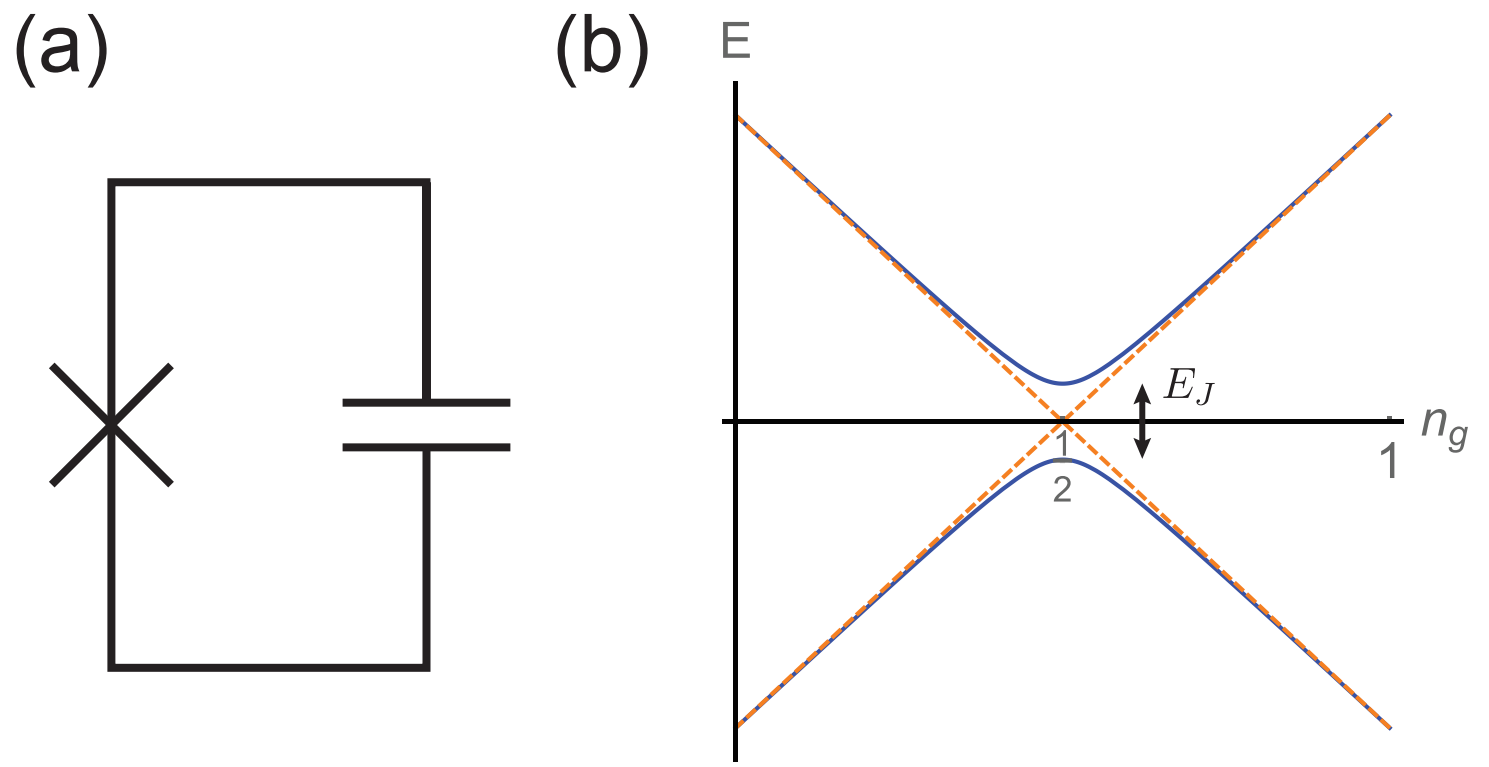}}
\caption{(a) The lumped-element circuit for a Cooper pair box (b) The energy levels of a Cooper pair box in the 2-level approximation. The sweet spot occurs at $n_g=1/2$.
}
\label{fig:CooperPairBox}
\end{figure}

In the Cooper pair box, $E_C>E_J$. A large $E_C$ energetically favors states with small amounts of charge. As $\hat{n}$ and $\hat{\varphi}$ do not commute, such states are highly delocalized in $\hat{\varphi}$ by the uncertainty principle. If we interpret the Cooper pair box circuit as a pendulum, as we did in Fig. \ref{fig:Ch2_JunctionOscillator}, then it is as if the mass is so small that it is completely delocalized, constantly spinning much like an electron bound to a nucleus. As $\hat{n}$ behaves like the (angular) momentum operator, the gate charge $n_g$ behaves like a magnetic field, so that the gate charge creates a Zeeman splitting between clockwise and counterclockwise motion.

\begin{exercise}
If the Hamiltonian for a charged particle in a magnetic field is unfamiliar to you, show that Hamilton's equations of motion applied to $H = (\vec{p}-e\vec{A})^2/2m + eV(x)$ reproduces the Lorentz force. Note the similarity between this Hamiltonian and \erf{eq:CooperPairBox}.
\end{exercise}

As the Cooper pair box eigenstates are localized in charge, they are more easily thought of in that basis
The energy levels of the qubit correspond to displacing different numbers of individual Cooper pair across the junction. A state with $\hat{n}$ close to $n_g$ are energetically favored. Supposing $0<n_g<1$, it becomes relatively straight-forward to think of the Cooper pair box as a qubit, with the qubit subspace given by states of 0 or 1 Cooper pairs displaced across the junction. The junction couples these two states, so that the actual eigenstates are superpositions of $n=0$ and $n=1$. Computing the Hamiltonian matrix elements on these states, we can reduce the Cooper pair box Hilbert space to that of a qubit, with an effective Hamiltonian\cite{schuster2007circuit}
\begin{align} \label{eq:CooperPairBoxQubit}
H = 2E_C \sigma_z (1-2n_g) - E_J\frac{\sigma_x}{2}
\end{align}
where we have dropped a scalar energy offset. \erf{eq:CooperPairBoxQubit} is the Rabi Hamiltonian of chapter \ref{ch:IntroQuantum}. The eigenstates of $H$ are the bare qubit states, analogous to the lab frame Hamiltonian from chapter \ref{ch:IntroQuantum}. The Cooper pair box can be controlled by applying an external electric field, which adds a time-dependent term proportional to $\sigma_z$. As per the Rabi model discussed in chapter \ref{ch:IntroQuantum}, oscillating this external field at the qubit frequency can drive transitions between eigenstates, allowing for qubit control.

To consider the coherence properties of the Cooper pair box, the eigenstates of $H$ are plotted in Fig. \ref{fig:CooperPairBox}c. There exists a `sweet spot' at $n_g=1/2$ where the transition frequency is first-order insensitive to fluctuations in the value of $n_g$, which is an avoided crossing between the $n=0$ and $n=1$ states split by the junction energy. The junction mediates rapid tunneling between the $|n=0\>$ and $|n=1\>$ states, and the new eigenstates are instead equal superpositions of $|0\>$ and $|1\>$ with different relative phases. Usefully, decreasing sensitivity to charge noise by operating at the sweet spot does not eliminate sensitivity to an oscillating control field.

From the standpoint of decoherence, this smearing out of the $|n=0\>$ and $|n=1\>$ states hides information about the qubit state from the environment. When the eigenstates are unequal superpositions of these states, they have different expectation values of charge and hence are easily distinguished by anything nearby with a dipole moment. Any form of measurement constitutes a source of decoherence, and thus these states are quickly dephased. At the sweet spot, the expectation value of charge is the same for both qubit states. To learn about the qubit state, the environment must be able to distinguish the $0$ and $1$ charge states faster than the period of oscillation $\hbar/E_J$. Otherwise the measurement outcome washes out over both states and the environment obtains no information of the qubit's state.

The Cooper pair box derives its stability from $E_J$ at the sweep spot, so fluctuations in this parameter set a fundamental limit on its coherence. In practice however, fluctuations in $n_g$ are severe, such that not only do second-order effects come into play, but also that it can be difficult to maintain the qubit near its sweet spot for long periods of time. However note that the larger the value of $E_J$, the greater the splitting and hence the wider the range over which the energy bands remain flat.

\subheading{The transmon qubit} The above reasoning suggests that we stand to lose little by further increasing $E_J$. 
as we increase $e_j$, states will be forced to localize near $\varphi=0$, and thus will delocalize beyond the $n=\{0,1\}$ subspace. this change of roles favors visualizing the eigenfunctions in the $\varphi$ basis, like the mass on a pendulum. the transmon regime $e_j \gg e_c$ corresponds to a region in which gravity is strong enough that the pendulum is well-localized around $\varphi=0$/major{\cite{koch2007transmon}}.

\begin{figure}
\centering
{\includegraphics[width = 130mm]{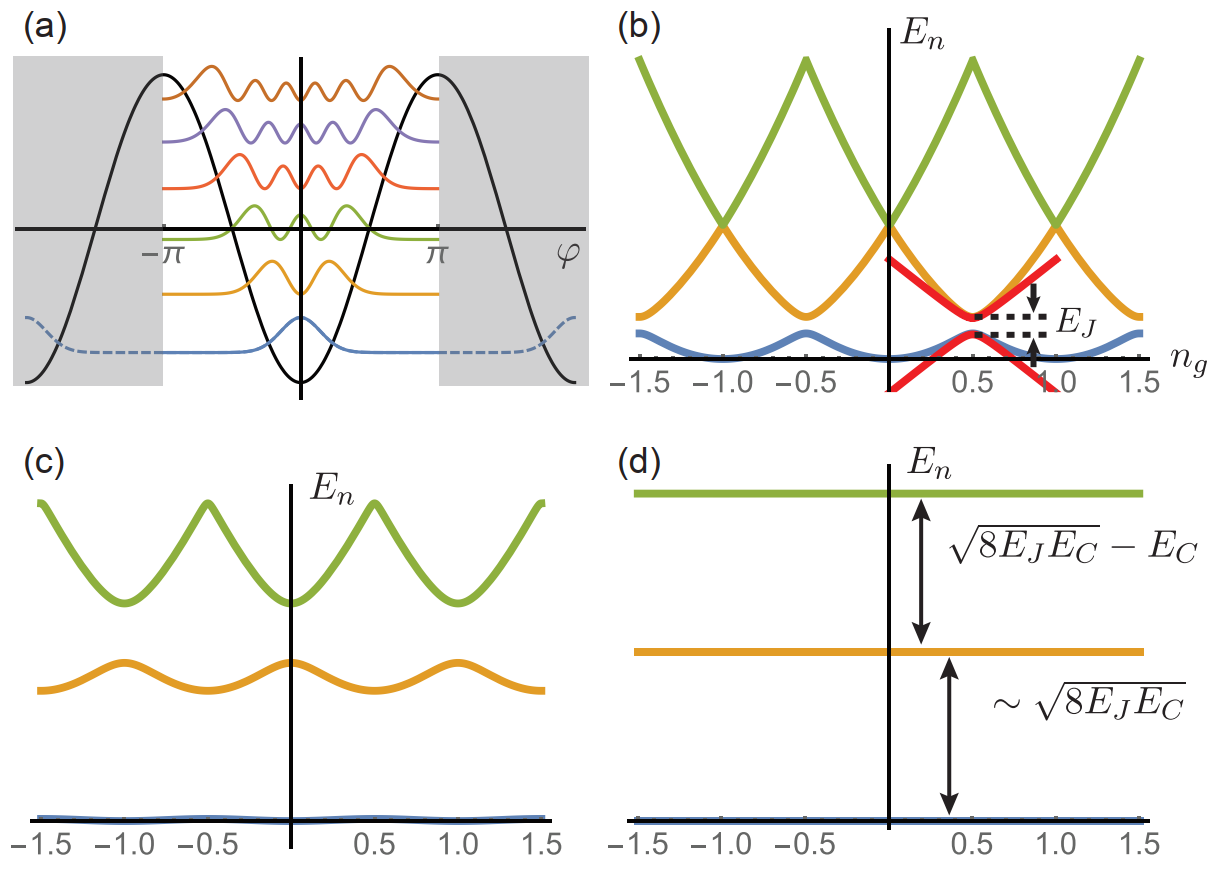}}
\caption{(a) Exact eigenstates of the transmon for $E_J/E_C = 50$, plotted in the $\varphi$ basis. The ground state is well-localized around $\varphi=0$, and is unlikely to tunnel into an adjacent well of the cosine potential. (b) Energy of the first three transmon/Cooper pair box eigenstates as a function of $n_g$ for $E_J/E_C = 1/2$. Approximate energies of the first two levels from Fig. \ref{fig:CooperPairBox} are replotted for comparison. (c) Same but with $E_J/E_C = 5$. (d) Same but with $E_J/E_C = 50$, at which point the energies depend very weakly on $n_g$. Figure based on \cite{koch2007transmon}.
}
\label{fig:Transmon}
\end{figure}

The eigenstates and eigenvalues of the transmon (and by extension the Cooper pair box) can be found exactly in terms of Mathieu functions. These eigenfunctions are plotted in Fig. \ref{fig:Transmon}a for $E_J/E_C=50$. As the state is localized near the bottom of the cosine potential, the lowest-energy eigenstates resemble those of a harmonic oscillator. We can use this observation to find a closed-form expression for the approximate eigenenergies, which is useful for building intuition. For states near $\varphi=0$, we can expand the cosine in a power series of $\varphi$ in the Hamiltonian
\begin{align} \label{eq:HarmonicOscillatorTransmon}
H \approx 4E_C(\hat{n}-n_g)^2 + E_J \frac{\hat{\varphi}^2}{2!} - E_J \frac{\hat{\varphi}^4}{4!}.
\end{align}
The Hamiltonian is no longer $2\pi$ periodic in $\phi$, so we should replace the charge and phase operators with their continuous counterparts \erf{eq:PhiQZPFs} (using $Q=2e \hat{n}$ and $\phi = 2\pi \varphi/\phi_0$). This approximation is equivalent to our approximation that $\varphi$ is localized. Eigenstates are so delocalized in $\hat{n}$ that the discreteness of charge can be neglected.

Expanding out the capacitive energy term, we recognize the gate charge as a drive term $8 E_C n_g \hat{n}$. From our discussion of coherent states and displacement operators in chapter \ref{ch:IntroQuantum}, we see that we can eliminate this offset charge by entering a displaced frame with $U=\exp(-i n_g \hat{\phi})$ (using $U (\hat{n}-n_g) U^\t = \hat{n}$). This is our first hint of reduced sensitivity to charge noise. We can also see that reducing charge noise sensitivity does not prevent us from driving the qubit. $U$ transforms out the static offset $\bar{n}_g$. If $n_g(t)$ oscillates about the mean, then $H$ contains a residual term $8 E_C (n_g(t) - \bar(n)_g) \hat{n}$. Fortunately the $1/f$ power spectrum of $n_g(t)$ becomes negligible near the qubit frequency where the rotating wave approximation would allow $n_g$-driven transitions, but we can generate such a term at will by applying a time-dependent electric field to the qubit.

Converting \erf{eq:HarmonicOscillatorTransmon} into ladder operators yields a harmonic oscillator with a non-linear correction
\begin{align}
    \sqrt{8 E_J E_C} (a^\t a + 1/2) - \frac{E_C}{12}(a+a^\t)^4
\end{align}
Computing the approximate eigenenergies is then a straightforward application of first-order perturbation theory 
\begin{align}
    E_n &\approx \sqrt{8 E_J E_C}(m+1/2) - \frac{E_C}{12}\<m|(a+a^\t)^4|m\> \\ \nonumber
    &= \sqrt{8 E_J E_C}(m+1/2) - \frac{E_C}{4}(2m^2 + 2m + 1)
\end{align}
where we have labeled the eigenstates with $m$ to avoid confusion with charge ($\hat{n}$). Crucially, $n_g$ does not appear in the spectrum, indicating insensitivity to charge noise. To use our system as a qubit, we require the $0 \leftrightarrow 1$ and $1 \leftrightarrow 2$ transitions are well-separated in frequency, so that we can preferentially address the lowest one. The relevant figure of merit is the relative anharmonicity
\begin{align} \label{eq:TransmonAnharmonicity}
    \alpha_r &\equiv \frac{E_{12}-E_{01}}{E_{01}} = \frac{E_C}{\sqrt{8 E_J E_C} - E_C} \\ \nonumber
    &\approx \sqrt{8 E_C/E_J}.
\end{align}
Thus as we enter the transmon regime by increasing $E_J/E_C$, it becomes more difficult to address the lowest-level transition without inadvertently driving higher transitions. In practice we require longer, spectrally narrower control pulses as the anharmonicity decreases.

\erf{eq:TransmonAnharmonicity} indicates a trade-off between anharmonicity and charge noise-induced decoherence. We also saw that the above treatment cannot estimate the effect of charge noise. To understand how charge noise comes in, it pays to return to the pendulum analogy. The charge offset acts as a magnetic field applied parallel to the axis of rotation. Classically, the applied field generates a radial Lorentz force, which has no effect on the dynamics. However quantization of angular momentum enforces boundary conditions on the allowed energy values. This boundary condition is effected by the applied magnetic field, and leads to the Zeeman splitting (via the Aharonov-Bohm phase). Thus a charge offset can only affect the energy spectrum if the probability for a $2\pi$ excursion in $\varphi$ is non-negligible; if $\psi(\pm \pi)=0$, then the periodic boundary condition is trivially satisfied and the magnetic field has no effect.

There are numerous ways to compute the charge noise sensitivity of the energy spectrum (called charge dispersion)\cite{koch2007transmon, girvin2011superconducting}. Here we merely aim to estimate how the effect scales with the ratio $E_J/E_C$. Given the above argument, we expect charge noise to scale with the probability for the state to tunnel by $2\pi$ into an adjacent well of the cosine potential. We can estimate the probability for this occurrence using the harmonic oscillator wave functions. Making the correspondence $x \leftrightarrow \hat{\varphi}$ and $p \leftrightarrow \hat{n}$, we have
\begin{align}
|\psi_0(x)|^2 &\propto \exp\left(-\frac{m \omega x^2}{\hbar} \right) \\ \nonumber
&\rightarrow \exp\left( -\sqrt{\frac{E_J}{8E_C}}\varphi^2 \right).
\end{align}
Thus the probability to tunnel by a distance $\varphi=2\pi$ decreases \textit{exponentially} with $\sqrt{E_J/8E_C}$. If we consider energy levels above $n=0$, the standard deviation of $|\psi(\varphi)|^2$ scales as $\sqrt{n}$,\footnote{$E \propto n^2$, and the classical turning point at which all energy is potential energy occurs when $E=m\omega^2 x_\text{max}^2/2$. Thus $x_\text{max} \propto \sqrt{n}$.} so that higher levels are more susceptible. These scaling factors match the results found using the WKB approximation and the analytic solution in terms of Mathieu functions. In Fig. \ref{fig:Transmon}b-d we plot the energies of the lowest few transmon eigenenergies as a function of $n_g$ for three values of the ratio $E_J/E_C$. Fig. \ref{fig:Transmon}d plots the bands for $E_J/E_C=50$, a common value that balances the trade-off well enough to make charge dispersion negligible. For a qubit with a resonant frequency of $6$ GHz, the absolute anharmonicity is $(E_{12}-E_{01})/h \approx 200$ MHz, allowing for control pulses lasting on the order of 10s of nanoseconds.





\subheading{Flux and phase qubits} In the following chapters, we only work with transmon qubits, which have been the workhorse of the field for some time. Coherence times have improved steadily since its introduction, and experiments have yet to hit any fundamental limits. Nevertheless one of the major appeals of the qubit is its simplicity as a circuit. As fabrication and qubit control has improved, researches continue to revisit old designs and explore new parameter regimes. The full range of circuits is relatively unexplored, and the coming years will likely bring many new forms of superconducting circuits with unanticipated properties. We now qualitatively outline the other major classes of qubits.

\begin{figure}
\centering
{\includegraphics[width = 120mm]{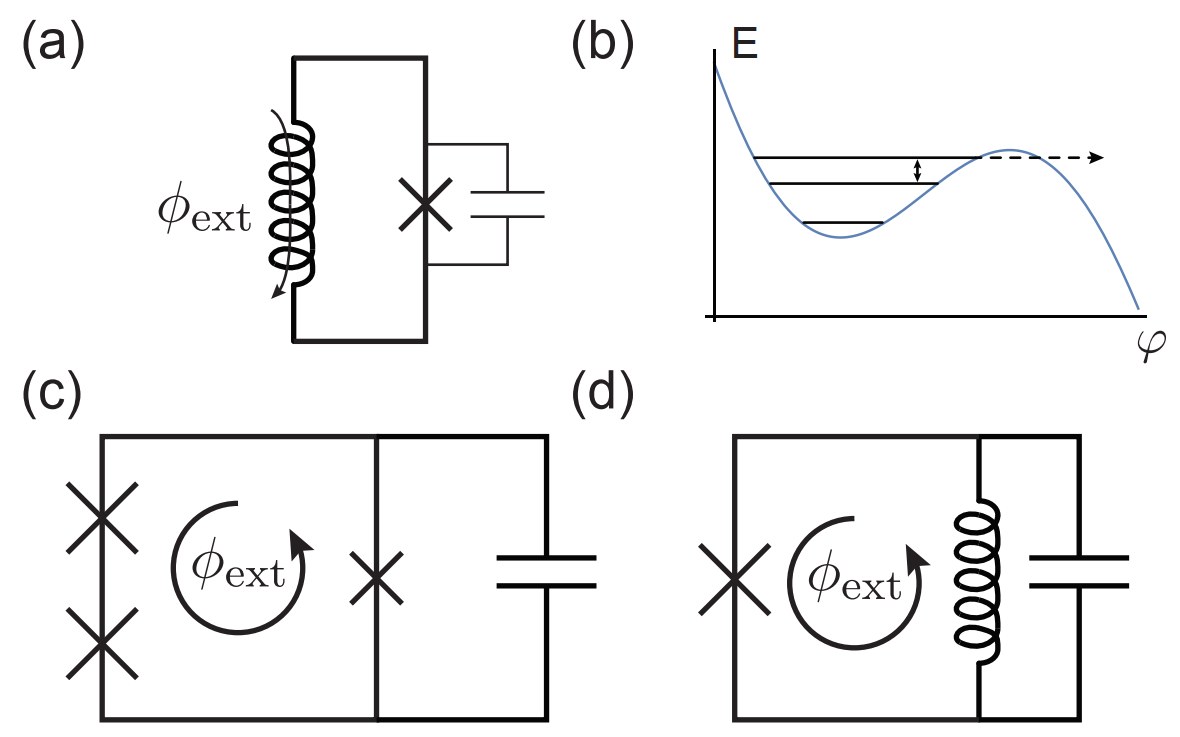}}
\caption{(a) Circuit for a phase qubit, which operates with a large flux bias $\phi_\text{ext}$. Circuit capacitance  is provided by the intrinsic capacitance of the junction. (b) Readout method of the phase qubit. $|e\>$ state is mapped to the third level, which tunnels out of a single well of the cosine potential and down the parabolic potential of the inductor, producing a macroscopic readout signal. (c) The capacitively shunted flux qubit (d) Fluxonium qubit. Inductance is provided by a chain of Josephson junctions in series.
}
\label{fig:Ch2_FluxQubits}
\end{figure}

The most well-explored class of qubits outside of the transmon are flux qubits, which modify the inductive part of the potential in some way, but are otherwise similar to the transmon in their topology. An early iteration is the phase qubit\cite{martinis2002rabi} shown in Fig. \ref{fig:Ch2_FluxQubits}. The flux potential consists of a parabola from the inductor modulated by the cosine potential of the Josephson junction. In these qubits, $E_J \gg E_C$ to the extent that the qubit is very nearly harmonic. The small capacitive energy greatly reduces sensitivity to charge noise. To operate as a qubit, the circuit's inductance is given a large current bias, which offsets the parabolic portion of the potential away from $\phi=0$ and thus introduces a large asymmetry in the potential as seen in Fig. \ref{fig:Ch2_FluxQubits}.

Phase qubits have a somewhat unusual readout mechanism shown in Fig. \ref{fig:Ch2_FluxQubits}b. The flux bias is chosen so that only a small number of states are bound. To read out the state, one pulses the $1\leftrightarrow 2$ transition. The $2$ state rapidly tunnels into the quasi-continuum of the parabolic potential, which creates a large change in current and hence a macroscopically large voltage across the junction via the second Josephson relation. 

The \definition{flux qubit}\cite{mooij1999josephson} is another major type of qubit that has garnered increasing attention recently after a period of relatively little use. The circuit for a modern incarnation, the capacitively shunted flux qubit\cite{yan2016flux} is shown in Fig. \ref{fig:Ch2_FluxQubits}c. In the simplest picture of this circuit, the $\varphi$ potential consists of a sum of $\cos(\varphi)$ and a $\cos(\varphi/2)$ terms. The relative phase between these terms can be modulated by applying an external flux, just as in the SQUID. In reality, Josephson junctions have intrinsic capacitances, so that the network of junctions has more than one degree of freedom and a 2D Josephson potential is useful for understanding the full picture.

\begin{figure}
\centering
{\includegraphics[width = \textwidth]{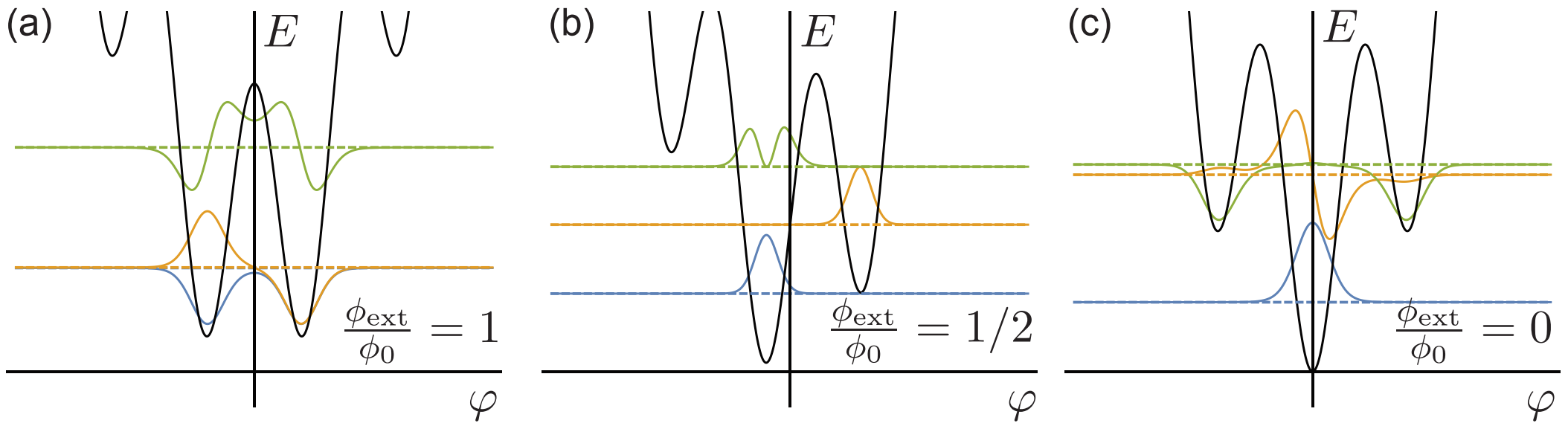}}
\caption{Potential for the fluxonium qubit, neglecting junction capacitances so that the $\varphi$ potential is 1D. Plots also include the lowest 3 eigenvectors placed vertically at their corresponding eigenenergies, derived numerically. Physical parameters match those from \cite{manucharyan2009fluxonium}, but with twice as large a value of $E_J$ to enhance the degeneracies. Flux bias shifts the cosine portion of the potential relative to the parabolic portion. (a) Flux bias which leads to a $\Lambda$-type system (b) Ladder system (c) V-type system.
}
\label{fig:QubitPotentials}
\end{figure}

A further incarnation is the fluxonium qubit\cite{manucharyan2009fluxonium} shown in Fig. \ref{fig:Ch2_FluxQubits}d, which contains a very large inductance generated by a chain of Josephson junctions in series. Both the flux and fluxonium qubits can have large inductive energies while preserving large nonlinearities. The effective potential of a fluxonium qubit is shown in Fig. \ref{fig:QubitPotentials}. Not only does the potential deviate substantially from a parabola, but the resulting level structure can be tuned by the external flux. Nearly degenerate states to not couple strongly to each other, so one can generate a fairly long-lived qubit in Fig. \ref{fig:QubitPotentials}a. The higher state can decay to either of the two lower states. This level configuration is called a $\Lambda$ system, which is the basis for many fundamental AMO effects like electromagnetically induced transparency (EIT) and stopped light. At another flux bias point like Fig. \ref{fig:QubitPotentials}c, the two higher levels become degenerate, which then both decay to a common lower state. This configuration is called a V system. An interesting effect that arises in this setting is quantum beats, in which spontaneous emission from the two levels can interfere, leading to coherent oscillations.





\section{Microwave amplification and detection}
\label{sec:IOTheory}

\subheading{Input-output theory} Inductors, capacitors and Josephson junctions form the essential ingredients of superconducting qubits. To understand readout and amplifiers quantum mechanically, we need a formalism to treat non-classical input and output fields. This section provides our first treatment of dissipation, in which a degree of freedom interacts irreversibly with our system. For amplifiers, it is convenient to work in the Heisenberg picture, and the resulting formalism is called input-output theory. We treat dissipation in the Schr\"odinger picture in chapter \ref{ch:OpenSystems}.

Several scenarios in circuit QED involve a system coupled to a continuum of harmonic oscillators. A qubit radiating into free space couples to an infinite number of normal modes of the vacuum. An infinitely long transmission line also has an infinite number of normal modes, so anything coupled to even a 1D waveguide can be described in roughly the same way. We can treat these systems very general with the following Hamiltonian
\begin{equation} \label{eq:IOCouplingH}
    H = H_\text{sys} + \sum_k \hbar \omega_k b_k^\t b_k - i \hbar (f_k a^\t b_k - f_k^* a b_k^\t).
\end{equation}
Here $k$ is an index that labels modes $b_k$ by their wave vector. Each mode has energy $\hbar \omega_k$. $H_\text{sys}$ is the system of interest, such as a qubit or amplifier, and $a$ is a system operator. The only assumption that we make about the system is that $[a,a^\t]=1$, so that \erf{eq:IOCouplingH} represents swap coupling to the modes $b_k$ like the Jaynes Cummings Hamiltonian. While this precludes $H_\text{sys}$ being an ideal qubit, all of the qubits that we will consider in circuit QED are actually nonlinear bosonic oscillators, so this commutation relation is be satisfied.

\begin{figure}
\centering
{\includegraphics[width = 120mm]{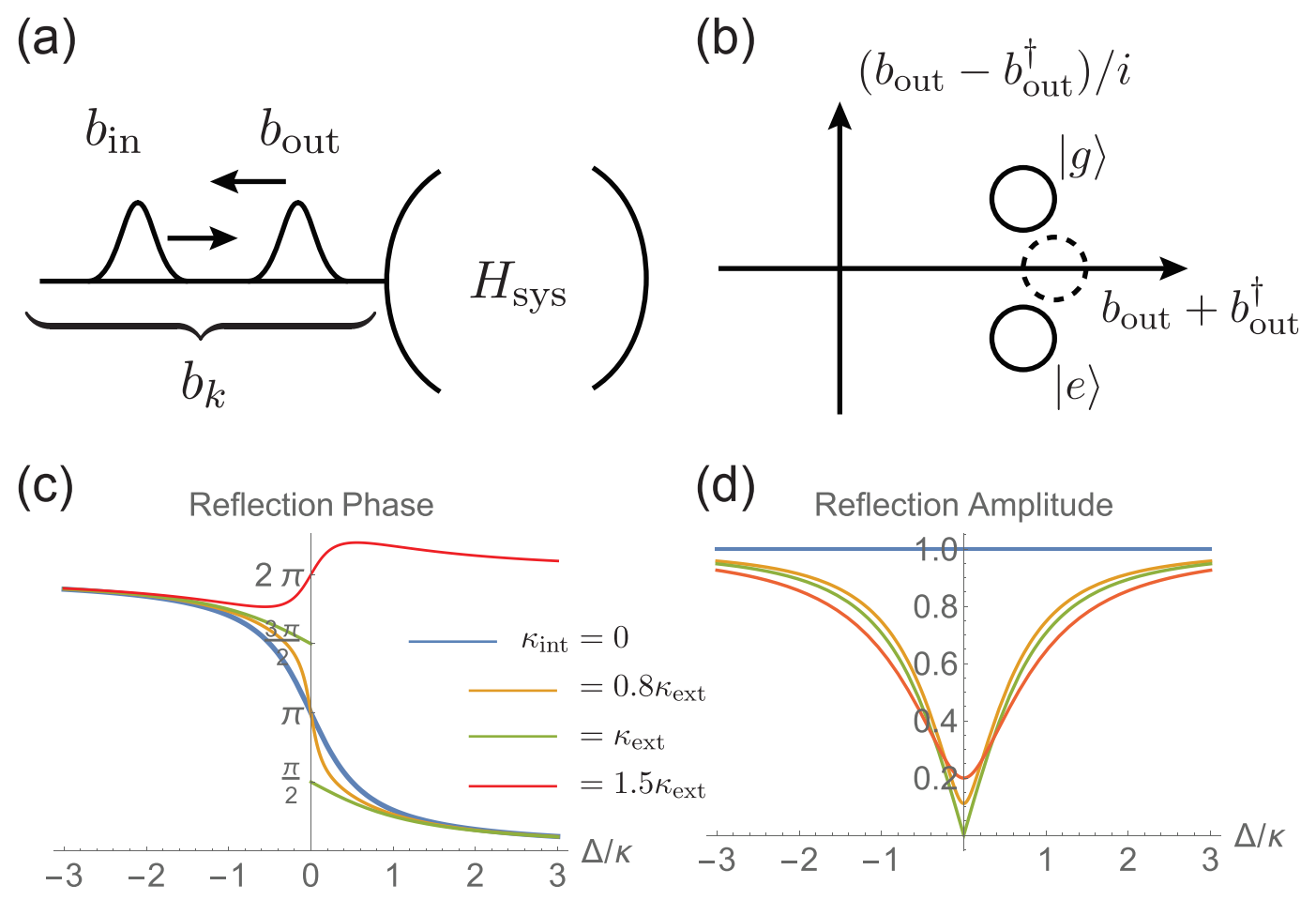}}
\caption{(a) A transmission line coupled to a cavity or other system. In input-output theory, we break the degrees of freedom on the transmission line into ingoing and outgoing modes. (b) Coherent state representation of qubit readout. Dashed circle represents in the input state, while the solid circles represent the reflected output states conditioned on the qubit state. (c) Reflected phase of a coherent state incident on a cavity as a function of detuning from the cavity resonance. Total phase shift across the resonance decreases as we increase the internal loss. (d) Amplitude of the reflected signal, which goes to zero at the critical coupling point $\Delta=0$, $\kappa_\text{int}=\kappa_\text{ext}$
}
\label{fig:Ch2_InputOutputTheory}
\end{figure}

We depict \erf{eq:IOCouplingH} loosely in Fig. \ref{fig:Ch2_InputOutputTheory}a. We wish to find the dynamics for the system operator $a$ in the Heisenberg picture, and ultimately to look at the relationship between incident and outgoing fields, which both consist of excitations of the $b_k$ modes. We know from basic wave analysis that a continuum of modes can be decomposed into travelling wave packets that are spatially localized. In a 1D waveguide in the absence of dispersion, we expect these temporal wave packets to be time independent, except at the instant when they overlap with our system. In what follows, we recover this intuition mathematically by solving for the dynamics induced by this instantaneous interaction, largely following \ped{\cite{clerk2010noise}}. We begin with the Heisenberg equations of motion for $b_k$
\begin{align} \label{eq:IOBequation}
    \dot{b}_k = \frac{i}{\hbar}[H, b_k] = -i \omega_k b_k + f_k^* a
\end{align}
This is an inhomogeneous first order differential equation in which the system acts as a drive on each $b_k$ mode. The standard solution is
\begin{align}
    b_k(t) = e^{-i\omega_k (t-t_0)} b_k(t_0) + \int_{t_0}^t dt' e^{-i\omega_k(t-t')}f_k^* a(t')
\end{align}
where the first term gives time evolution based on initial conditions and the second term is the $b_k$ mode Greens function convolved with the system operator and represents driving of the transmission line by the system. With the full solution of the $b_k$ mode in hand, we next find the equations of motion for $a$ and substitute our solution
\begin{align}
\dot{a} 	&= \frac{i}{\hbar}[H_\text{sys}, a] - \sum_k f_k b_k \\ \nonumber
		&= \frac{i}{\hbar}[H_\text{sys}, a] - \sum_k e^{-i \omega_k(t-t_0)} b_k(t_0) \\ \nonumber
		& ~~ - \sum_k |f_k|^2 \int_{t_0}^t e^{-i \omega_k (t-t')}a(t') dt'
\end{align}
The second term is the particular linear combination of the $b_k$ modes that drives the system at time $t$, which we recognize as the operator for a temporally localized wave packet. We label this operator $b_\text{in}[t]$, using brackets instead of parenthesis to emphasize that $t$ is a label like $k$; the $b_\text{in}[t]$ operators have no time dependence except at the moment that they interact with the system.\footnote{This is not to say that the wave packet associated with $b_\text{in}[t]$ does not change in time. It propogates at the group velocity and may also disperse as dictated by $\omega_k$. The point is rather that the excitation of the mode, described by a superposition of different numbers of photons, remains constant as it propogates.} The third term simplifies to a $\delta$ function if we assume make the \definition{Markov approximation} that $f_k$ is independent of $k$. We will have more to say about it in chapter \ref{ch:OpenSystems}. Applying these simplifications yields
\begin{align} \label{eq:DeriveIORelation}
&= \frac{i}{\hbar} [H_\text{sys}, a] - \kappa \int_{t_0}^t \delta(t-t') a(t')dt' - \sqrt{\kappa} b_\text{in} \\ \nonumber
b_\text{in}[t] &\equiv \frac{f}{\sqrt{\kappa}} \sum_k e^{-i \omega_k (t-t_0)} b_k(t_0)
\end{align}
where we have defined the rate $\kappa$ so that $\sum_k |f|^2 \exp(-i \omega_k(t-t')) = \kappa \delta(t-t')$, which naturally takes the density of states into account. While it may seem unusual that our $b_\text{in}$ operators have units of $\sqrt{\text{Hz}}$, the result appears natural when we compute the commutation relation
\begin{align}
[b_\text{in}[t], b^\t_\text{in}[t']] = \delta(t-t')
\end{align}
where we have used the fact that $[b_k, b_{k'}^\t] = \delta_{k,k'}$. The unusual units arise because the $\delta$ function has the inverse units of its argument. Physically, $b_\text{in}^\t b_\text{in}$ has units of Hz because it quantifies power, just as $a^\t a$ measures energy.

The second term of \erf{eq:DeriveIORelation} has an interesting interpretation. It arose from the system driving the $b_k$ modes, and then computing in turn how this contribution to the $b_k$ drives the system. The integral of half of a $\delta$ function is $1/2$, so the second term simplifies and we arrive at the important \definition{input-output relation}
%
\begin{empheq}[box=\widefbox]{align} \label{eq:IORelation}
\dot{a} = \frac{i}{\hbar} [H_\text{sys}, a] - \frac{\kappa}{2} a(t) - \sqrt{\kappa} b_\text{in}[t].
\end{empheq}
%
We see that this `back-reaction' term is actually damping, or radiative decay. The system's amplitude \textit{i.e.} its quadrature operators $a \pm a^\t$ damp at a rate of $\kappa/2$, just like a harmonic oscillator under the influence of friction.

Just as we derived a relation for the input field, one can derive a similar relation for the output field by solving \erf{eq:IOBequation} using the advanced Greens function instead of the retarded Greens function (the state of each $b_k$ at time $t$ is determined by the future evolution of $a(t')$ just as well as by the past). The end result is
\begin{align} \label{eq:IORelationBOut}
\dot{a} &= \frac{i}{\hbar}[H_\text{sys}, a] + \frac{\kappa}{2} a - \sqrt{\kappa} b_\text{out}[t] \\ \nonumber
b_\text{out}[t] &= \frac{f}{\sqrt{\kappa}} \sum_k e^{-i \omega_k (t-t_1)} b_k(t_1)
\end{align}
where $t_1$ is in the distant future just as $t_0$ was taken to be in the distance past. \erf{eq:IORelationBOut} is a sort of time-reversed version of \erf{eq:IORelation}. Subtracting \erf{eq:IORelation} from \erf{eq:IORelationBOut} yields a relation between the input and output fields
%
\begin{empheq}[box=\widefbox]{align}  \label{eq:IOBInOut}
b_\text{out}[t] = b_\text{in}[t] + \sqrt{\kappa} a(t).
\end{empheq}
%
The physical interpretation is that $b_\text{out}$ consists of input modes reflected off of the system and an additional field radiated by the system itself, as depicted in Fig. \ref{fig:Ch2_InputOutputTheory}a. In general, the $b_\text{out}$ modes could constitute a reflected or transmitted signal, depending on whether the $b_k$ operators describe a single-sided transmission line as depicted in the figure or a transmission line (or higher-dimensional region) that extends from the system in all directions. The mathematical description of both scenarios is essentially the same, though we will mostly encounter the single-sided case in circuit QED.

\erf{eq:IOBInOut} and \erf{eq:IORelation} are extremely useful, offering a simple way to treat many dissipative systems. In this chapter, we will use them to derive and understand parametric amplifiers, though they are ubiquitous in circuit QED and well worth committing to memory! Note that some conventions replace $b_\text{in}$ with $-b_\text{in}$.

Before turning to amplifiers, we first describe an important application that motivates microwave amplification in the first place: qubit readout. Qubit readout is also a simple application of the input-output relations, though some of its aspects are better treated with the formalism of the next chapter. Suppose we have a qubit dispersively coupled to a cavity as in chapter \ref{ch:IntroQuantum}. In the rotating frame, the system Hamiltonian is
\begin{align} \label{eq:IODispersiveH}
H_\text{sys}/\hbar = \chi a^\t a \sigma_z.
\end{align}
This equation applies to a transmon when we restrict it to lie in its lowest two energy levels. One can use this Hamiltonian to perform a \definition{quantum non-demolition} measurement of the qubit state. This term means that the measurement has no other action on the qubit but to project it into an eigenstate of the measurement operator, which in this case will be $\sigma_z$. To see how this works, let's write down the input-output relations for the cavity based on \erf{eq:IODispersiveH}
\begin{align} \label{eq:IOQubitReadout}
\dot{a} = -i \chi a \sigma_z - \frac{\kappa}{2}a - \sqrt{\kappa}b_\text{in}[t].
\end{align}
As in most circuit QED applications, we take the $b_\text{in/out}$ modes to describe excitations of a 1D transmission line, so that $k$ is a scalar rather than a vector.

Suppose the qubit is in either $|g\>$ or $|e\>$, and $b_\text{in}$ is in a coherent state. In this case, the system reduces to an off-resonantly driven cavity. We know from chapter \ref{ch:IntroQuantum} that the cavity should remain in a coherent state at all times.\footnote{Strictly speaking, we have not yet shown that this is the case once we add a coupling to an external system, however it turns out to be the case. It is a basic and essential fact of coherent states that they are mapped to coherent states under the action of Hamiltonians that are linear or quadratic ladder operators and contain no \definition{squeezing terms} $a_i^\t a_j^\t$ or $a_i a_j$.} Let us assume that the state of the system at time $t$ is $|\psi\> = |g/e, \alpha_{g/e}(t), \epsilon\>$, where $\alpha_{g/e}$ is the coherent state amplitudes of $a$ conditioned on the qubit state and $\epsilon$ is proportional to the coherent state amplitude of $b_\text{in}[t]$. Taking the expectation value of \erf{eq:IOQubitReadout} yields
\begin{align}
\frac{d}{dt} \<a\> = \dot{\alpha}_{g/e} = \pm i \chi \alpha_{g/e} - \frac{\kappa}{2} \alpha_{g/e} - \sqrt{\kappa}\epsilon.
\end{align}
Note that putting $b_\text{in}$ in a coherent state has the same effect as adding a term proportional to $i(a^\t-a)$ to $H_\text{sys}$, which explains how we generate this term in an experiment. We can easily solve for the steady-state cavity displacement by taking $\dot{\alpha}_{g/e}=0$, and then using this solution to compute the output field
\begin{align} \label{eq:ReadoutCoherentStates}
\alpha_{g/e} &= \frac{\sqrt{\kappa}\epsilon}{\pm i \chi - \kappa/2} \\ \nonumber
\<b_\text{out}\> &= \epsilon \left( 1+ \frac{\kappa}{\pm i \chi - \kappa/2} \right) \\ \nonumber
&= \epsilon \frac{\chi^2 - \kappa^2/4 \mp i\chi \kappa}{\chi^2 + \kappa^2/4}.
\end{align}
Thus the reflected phase of the output field depends on the qubit state. Fig. \ref{fig:Ch2_InputOutputTheory}b depicts the input and output fields in the IQ plane. The observed phase shift is a completely classical effect, analogous to an off-resonantly driven mass on a spring. When driven far below its resonance frequency, the mass moves in phase with the drive, whereas it responds $\pi$ out of phase when driven above resonance. Interference between the input and output fields leads to a full $2\pi$ phase shift as the drive tone is swept across the cavity resonance, as plotted in Fig. \ref{fig:Ch2_InputOutputTheory}d (the thick blue curve only). For qubit readout, the dispersive coupling shifts the cavity resonance conditioned on the qubit state, which generates a qubit-state-dependent phase shift of the drive.

Although $b_\text{out}$ is not Hermitian, and hence not observable, all one has to do is measure $(b_\text{out} - b_\text{out}^\t)/i$ (or actually $\int_0^t dt' (b_\text{out}[t'] - b_\text{out}^\t[t'])/i$ to get an appreciable signal-to-noise ratio; more on that in chapter \ref{ch:OpenSystems}), which picks out the imaginary component of the above expression. By performing many measurements of $(b_\text{out} - b_\text{out}^\t)/i$, we can infer $\<(b_\text{out} - b_\text{out}^\t)/i\>$, from which we can infer $\<\sigma_z\>$. 

If we can reach a signal-to-noise ratio much greater than 1 
then we can also infer the specific measurement outcome for each shot of the measurement. This more demanding task is called \definition{single-shot readout}, which is important because it lets us observe correlations between simultaneous measurements of different qubits, and the quantum trajectories of a single qubit (much more on that later). We can rarely measure anything noiselessly, particularly a microwave signal composed of a handful of photons. Any noise on the signal can mask the difference between the coherent state signals associated with $|e\>$ and $|g\>$, which prevents us from being sure of the measurement outcome.  We might hope that we can increase $\epsilon$, or else the duration of the measurement, but both have their limitations. Readout duration is limited by the excited state lifetime of the qubit, $T_1$. The intercavity photon number $|\alpha|^2$ can only go so high before the dispersive approximation on which \erf{eq:ReadoutCoherentStates} is based breaks down. Even before this critical photon number is reached, the readout power $b_\text{in}$ can start driving unwanted transitions in the system\cite{sank2016measurement}. To achieve a signal-to-noise ratio greater than one at our room temperature electronics, we need a low noise, low temperature microwave amplifier.


\begin{exercise} \label{eq:CoherentStateDecay}
Consider a cavity initially in a coherent state $\alpha$. If all of the $b_\text{in}[t]$ modes are in the vacuum, show that the coherent state amplitude decays as $\alpha(t) = \alpha(0) e^{-\kappa t/2}$. If you work in the rotating frame, then $H_\text{sys} = 0$. 
\end{exercise}

\begin{exercise} \label{ex:CriticalCoupling}
Lossy cavities and critical coupling: Realistic cavities have some internal loss in addition to their coupling to an external control line. We can model this loss with another decay rate $\kappa_\text{int}$ and an additional set of input and output modes. Show that when this additional loss channel is present, the reflected coherent state is
\begin{align}
\<b_\text{out}\> = \epsilon \frac{(\kappa_\text{int}-\kappa)/2 + i \Delta}{(\kappa_\text{int} + \kappa)/2 + i\Delta}
\end{align}
where $\Delta$ is the detuning between the cavity resonance and the drive. The amplitude and phase of this function are plotted in Fig. \ref{fig:Ch2_InputOutputTheory}c and d. Note that perfect transmission to the loss channel occurs when $\kappa=\kappa_\text{int}$. This scenario is called critical coupling, and is a simple example of impedance matching.
\end{exercise}

\textbf{Parametric amplifiers} Amplifiers necessarily have inputs and outputs, and thus are naturally described by input-output theory. While the most common classical amplifiers have seperate input and output connections, the simplest possible amplifier has only a single port that serves as both its input and output. We call these reflection-mode amplifiers, and they are often based on driven nonlinear cavities. In what follows, we first describe how to generate a \definition{squeezing Hamiltonian} using a weakly nonlinear JC circuit. We then show that our new favorite Hamiltonian ideally amplifies $b_\text{in}$. 

\begin{figure}
\centering
{\includegraphics[width = \textwidth]{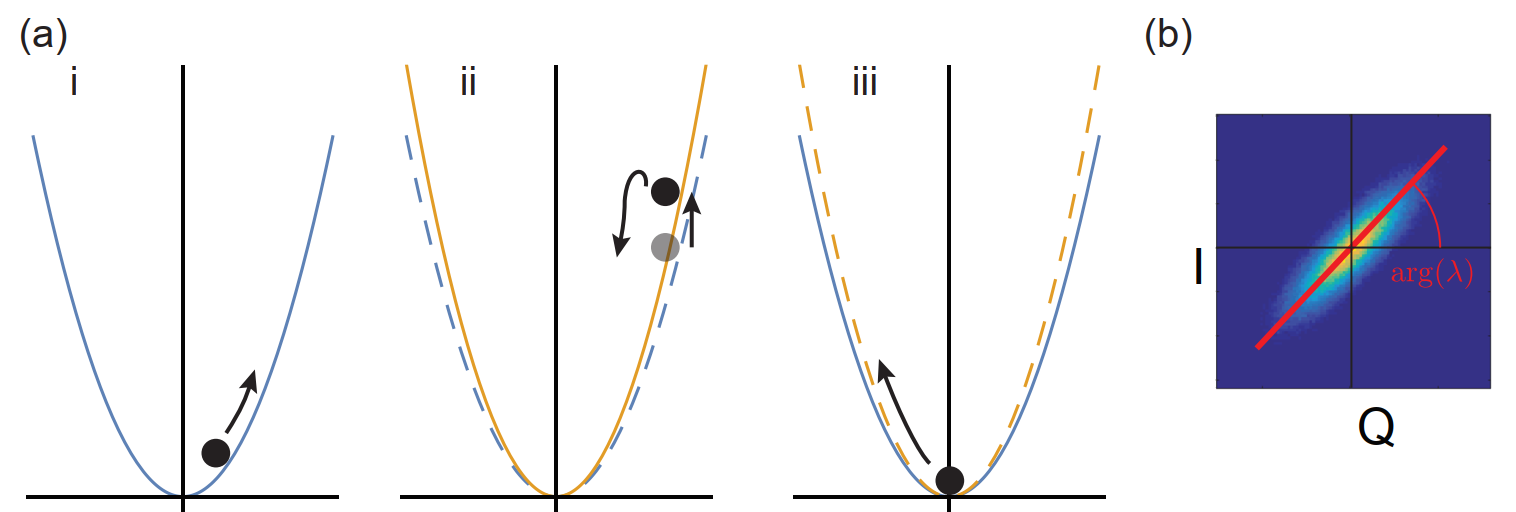}}
\caption{(a) Graphical illustration of parametric amplification. Ball gains potential energy at the top of the parabolic potential. (b) A squeezed state, which is the output of a quantum-limited amplifier.
}
\label{fig:Paramp}
\end{figure}

The simplest way to implement a squeezing Hamiltonian technically requires no non-linearity at all, though in practice we rely on the Josephson nonlinearity to generate it. The basic concept is explained in Fig. \ref{fig:Paramp}a, and can be visualized as a mechanical harmonic oscillator for all intents and purposes. We embue ourselves with the ability to change the spring constant, which amounts to changing the potential energy by a multiplicative factor. Normally the mass oscillates back and forth at a frequency $\omega$ in the potential. Suppose that at moment that the mass arrives at the classical turning point, we increase the spring constant slightly, so that the potential rises up as in Fig. \ref{fig:Paramp}b ii. Although at rest, the system has gained a small amount of energy. Furthermore, we can increase the overall amplitude of oscillation if we return the spring constant to its original value once the mass returns to $x=0$. If we continuously modulate the potential in this way, which is at twice the resonant frequency $\omega$, the amplitude of oscillation increases exponentially! This simple analogy, not so different from a kid on a swingset (the kind with rigid poles), describes parametric amplification.

\begin{exercise}
Using the same logic as above, argue that the motion will be deamplified if the mass oscillates $\pi/2$ out of phase relative to the scenario of Fig. \ref{fig:Paramp}.
\end{exercise}

The quantum version of a parametric amplifier is straight-forward to implement in a JC circuit. A SQUID allows us to periodically vary $E_J$, which will serve as our time-dependent spring constant. This method of driving a parametric amplifier is called flux pumping. As we saw in the previous section, we can ignore the discreteness of $\hat{n}$ in a JC circuit when $E_J \gg E_C$. We also saw that the non-linearity of the junction becomes negligible when $E_J$ is large enough to confine the wave function to the bottom of the cosine potential. While a small nonlinearity causes problems for a transmon qubit, it lines up nicely with our classical model of a parametric amplifier. Thus taking $E_J \gg E_C$ beyond the transmon limit lets us approximate
\begin{align}
H = 4 E_C n^2 - E_J(t) \cos \varphi \approx 4 E_C n^2 + E_J(t) \frac{\varphi^2}{2}
\end{align}
Motivated by our classical paramp, let $E_J(t) = E_{J,0} + \epsilon \cos(2\omega t + \delta)$ for some constant small $\epsilon$ that we won't worry about too much. A standard application of ladder operators converts the time-independent terms of $H$ to $\hbar \omega a^\t a$, leaving us with
\begin{align} \label{eq:FluxPump1}
H/\hbar = \omega a^\t a + 2 \lambda \cos(2\omega t + \delta) (a+a^\t)^2
\end{align}
where we have absorbed $\epsilon$ and $\phi_\text{ZPF}$ into a new constant $\lambda$. At this point, the reader may recognize another potential application of the rotating wave approximation, just like our treatment of the driven harmonic oscillator in chapter \ref{ch:IntroQuantum}. The main difference is that we now have terms like $a^2$. Entering a rotating frame applies a phase factor $e^{-i \omega t}$ to each $a$ individually, which transforms the whole Hamiltonian to
\begin{align} \label{eq:ParampSqueezingH}
H/\hbar &= \lambda (e^{2i \omega t + i \delta} + e^{-2i \omega t - i \delta}) (a^2 e^{-2i \omega t} + a^\t a + a a^\t + a^{\t 2} e^{2i \omega t}) \\ \nonumber
&\approx \lambda ^* a^2 + \lambda a^{\t 2}
\end{align}
where we have absorbed a constant $e^{i\delta}$ into lambda, making it complex. 

\erf{eq:ParampSqueezingH} is known as a squeezing Hamiltonian, because time evolution under $H$ leads to squeezed states like that depicted in Fig. \ref{fig:Paramp}b. In that figure, we start with the vacuum state $|\psi(t)\> = |0\>$, evolve for a fixed amount of time, and then plot the Q distribution of the final state. As $H$ contains only even powers of $a$ and $a^\t$, photons are only created in pairs. Physically, the nonlinearity of the SQUID is splitting single pump photons of energy $2\hbar \omega$ into two cavity photons of energy $\hbar \omega$. 
%
%

Although real parametric amplifiers have input and output ports, we can already see strong hints of amplification from \erf{eq:ParampSqueezingH}. Just as the displacement operator that we associated with $a+a^\t$ generated all coherent states, we can write down a squeezing operator by exponentiating $H$
\begin{align}
S(z) = \exp\left[ \frac{1}{2} \left(z^* a^2 - z a^{\t 2}\right) \right].
\end{align}
In analogy with the displacement operator identity $D^\t (\alpha) a D(\alpha) = a + \alpha$, the squeezing operator satisfies
\begin{align} \label{eq:SqueezingOpIdentity}
S^\t(z) a S(z) &= a \cosh|z| - e^{i \arg(z)} a^\t \sinh|z| \\ \nonumber
\end{align}
which is a standard application of the Baker-Campbell-Hausdorff formula \erf{eq:BCH}. \erf{eq:SqueezingOpIdentity} is useful because it lets us compute what happens to the the amplitude quadrature operators after evolution under $H$
\begin{align} \label{eq:SqueezingQuadratures}
S^\t(z) I_{\arg(z)} S(z) &= e^{-|z|}I_{\arg(z)} \\ \nonumber
S^\t(z) Q_{\arg(z)} S(z) &= e^{|z|}Q_{\arg(z)} \\ \nonumber
I_\phi &\equiv a e^{-i\phi/2} + a^\t e^{i \phi/2} \\ \nonumber
Q_\phi &\equiv (a e^{-i\phi/2} - a^\t e^{i\phi/2})/i
\end{align}
using $\cosh(r) \pm \sinh(r) = e^{\pm r}$. $I_\phi$ and $Q_\phi$ are quadrature operators rotated in phase space by $\phi$, which are exponentially amplified and deamplified respectively just like our classical model of Fig \ref{fig:Paramp}a. Curiously, the squeezed quadrature $I_\phi$ will have fluctuations less than that of the vacuum state. The Heisenberg uncertainty principle is saved by the fact that the conjugate $Q_\phi$ has larger fluctuations by the same factor, conserving the product of the uncertainties. Most importantly, \erf{eq:SqueezingQuadratures} clearly shows that our squeezing Hamiltonian acts as a linear amplifier of the $Q_\phi$ quadrature. The enhanced fluctuations of the output field correspond to amplified vacuum fluctuations.

\begin{exercise} \label{ex:SinglePumping}
One can also generate a squeezing Hamiltonian by driving a nonlinear cavity resonantly, which is called single pumping. Starting in the rotating frame with $H= \Delta a^\t a + \lambda (a^\t a)^2$, show that one can generate \erf{eq:ParampSqueezingH} by applying a strong drive \textit{i.e.} $a \rightarrow a+\alpha$ and setting $\Delta = -4 \lambda |\alpha|^2$. You will need to drop third and forth order terms in $a$, which amounts to assuming that quantum fluctuations are small. What is the physical interpretation for this kind of pumping?
\end{exercise}

Now that we recognize our pumped cavity as a linear amplifier, let's add an input/output mode to see how it amplifies an externally applied field\cite{eddins2017superconducting}. Just as we did with qubit readout, we will first solve for the internal system dynamics using \erf{eq:IORelation}, and then substitute the result into \erf{eq:IOBInOut} to find a relation between the input and output fields. Taking our squeezing Hamiltonian \erf{eq:ParampSqueezingH} for $H_\text{sys}$ in \erf{eq:IORelation} gives
\begin{align} \label{eq:ParampIORelation1}
\dot{a} = -i \lambda a^\t - \frac{\kappa}{2} a^\t - \sqrt{\kappa} b_\text{in}
\end{align}
We could solve for the steady state $a$ as before, but as the dynamics are more interesting, let us derive the full time-dependent solution this time around. As \erf{eq:ParampIORelation1} is linear, we can solve for $a(t)$  by Fourier transforming it
\begin{align} \label{eq:ParampIORelation2}
\int_{-\infty}^\infty \dot{a}(t) e^{i\Delta t} dt &= -i \lambda \int_{-\infty}^\infty a^\t(t) e^{i\Delta t} dt - \frac{\kappa}{2} \tilde{a}(\Delta) - \sqrt{\kappa} \tilde{b}_\text{in}(\Delta) \\ \nonumber
=-i \Delta \tilde{a}(\Delta) &= -\lambda \left[ \int_{-\infty}^\infty a(t) e^{-i \Delta t} dt \right]^\t - \frac{\kappa}{2} \tilde{a}(\Delta) - \sqrt{\kappa} \tilde{b}_\text{in}(\Delta) \\ \nonumber
&= -i \lambda \tilde{a}^\t (-\Delta) - \frac{\kappa}{2} \tilde{a}(\Delta) - \sqrt{\kappa} \tilde{b}_\text{in}(\Delta)
\end{align}
The Fourier transform converts the time derivative into multiplication by $-i \Delta$, which can be shown using integration by parts. We have used $\Delta$ as our frequency variable to remind ourselves that we work in a rotating frame, so $\Delta=0$ corresponds to fields oscillating at the paramp resonance frequency $\omega$.  The appearance of $\tilde{a}(-\Delta)$ is of fundamental importance; nonlinearity of the paramp symmetrically mixes modes of frequency $\pm \Delta$. If the signal of interest is at a frequency $\Delta$, then $-\Delta$ is called the idler mode.

Using \erf{eq:ParampIORelation2} and its complex conjugate evaluated at $-\Delta$, we can solve for $\tilde{a}(\Delta)$ in terms of $b_\text{in}(\Delta)$ and $b_\text{in}^\t(-\Delta)$. Some straightforward algebraic manipulations gives
\begin{align}
\tilde{a}(\Delta) = \sqrt{\kappa} \left( \frac{i \lambda \tilde{b}_\text{in}^\t(-\Delta) - (\kappa/2 - i\Delta) \tilde{b}_\text{in}(\Delta)}{(\kappa/2 - i\Delta)^2 - |\lambda|^2} \right).
\end{align}
With the cavity dynamics in hand, we can compute the input-output relation for a general parametric amplifier
%
%
\begin{empheq}[box=\widefbox]{align} \label{eq:ParampIORelation}
\tilde{b}_\text{out}(\Delta) &= -G_S \tilde{b}_\text{in}(\Delta) + \underbrace{i e^{i\arg(G_S)} \sqrt{|G_S|^2-1}}_{G_I} \tilde{b}_\text{in}^\t(-\Delta) \\ \nonumber
G_S &= \frac{\kappa^2/4 + \Delta^2 + |\lambda|^2}{(\kappa/2-i \Delta)^2 - |\lambda|^2}
\end{empheq}
%
where $G_S$ and $G_I$ are called the signal and idler gains. Note the close similarity between \erf{eq:ParampIORelation} and \erf{eq:SqueezingOpIdentity}. In particular, $G_S$ and $G_I$ satisfy $|G_S|^2 - |G_I|^2 = 1$ in analogy with the identity $\cosh^2(z) - \sinh^2(z) = 1$. Thus at least for $\Delta=0$, the relationship between $G_S$ and $G_I$ implies that we can recognize $\tilde{b}_\text{out}(0)$ as a squeezed copy of $\tilde{b}_\text{in}(0)$ (up to a minus sign, which we recognize as the $\pi$ phase shift caused by reflection). For $\Delta \neq 0$, $\tilde{b}_\text{in}(\Delta) \neq \tilde{b}_\text{in}(-\Delta)$, which breaks down the exact correspondence between \erf{eq:ParampIORelation} and \erf{eq:SqueezingOpIdentity}. In this case, $\tilde{b}_\text{out}$ is an example of a \definition{two-mode squeezed state}; vacuum fluctuations from the idler mode $\tilde{b}_\text{in}(-\Delta)$ are mixed into the amplified signal, which contributes noise. In the large gain limit $\lambda\rightarrow \kappa/2$, $G_S \approx G_I$, indicating that ideally half of the signal consists of amplified idler vacuum fluctuations.

The noise added by an amplifier is one of its most important figures of merit, and as hinted above, there are two important cases to consider. When $\Delta \neq 0$, the input quadratures are both amplified by $G_S$, but in the limit $G_S \gg 1$, substantial signal from the $-\Delta$ mode is mixed in. Even if the idler mode is in its ground state, vacuum fluctuations pollute the signal, so that the maximum possible signal-to-noise ratio is $1/2$. This mode of operation is called phase-preserving amplification, because the signal is amplified faithfully aside from the added noise. The situation differs when $\Delta=0$. In this case, the signal and idler modes are the same mode, and \erf{eq:ParampIORelation} matches up with the squeezing relation \erf{eq:SqueezingOpIdentity}. Just as in \erf{eq:SqueezingQuadratures}, there is one now one amplified quadrature (with gain of $2G_S$ instead of $G_S$) and one squeezed quadrature. This situation, called phase-sensitive gain, does not add any noise.

What determines if an amplifier operates in phase-sensitive or phase-preserving mode? Realistic signals have finite bandwidth, and so we cannot assign an exact frequency to them, but only a center frequency. In reality, the distinction between phase-sensitive and phase-preserving is largely a matter of perspective. If the Fourier component of the signal that is $2\Delta$ away from its center frequency is far out of the bandwidth of interest, then we would call the operation phase-sensitive, and the input idler mode to contribute noise. Conversely, if the signal of interest has time dependence on the time scale of $1/2\Delta$, then we resolve that time dependence as a (potentially rotating) single-mode squeezing of the input field.

One more important property of ideal parametric amplifiers is an intrinsic trade-off between gain and bandwidth, which we now derive. In line with \erf{eq:SqueezingQuadratures}, the phase-sensitive gain of an amplifier may be computed as
\begin{align}
G_\text{PS}(\Delta) &= |G_S| + |G_I| = \frac{\kappa/2 + \lambda + i \Delta}{\kappa/2 - \lambda - i \Delta} \\ \nonumber
|G_\text{PS}(\Delta)|^2 &= \frac{\kappa/2+\lambda)^2 + \Delta^2}{(\kappa/2-\lambda)^2 + \Delta}
\end{align}
The second line is power gain, which is also often labelled $G$ by convention. In the large gain limit $\lambda \approx \kappa/2$ and for in-band frequencies ($\Delta < \kappa$), we can drop the $\Delta^2$ term in the numerator. This simplification gives
\begin{align} \label{eq:GainBandwidthProduct}
|G_\text{PS}|^2 \approx \frac{|G_\text{PS}(0)|}{1+\dfrac{\Delta^2}{\Delta_\text{BW}^2}} \\ \nonumber
\Delta_\text{BW} = \kappa/2-\lambda \approx \frac{\kappa}{G_\text{PS}(0)}.
\end{align}
\erf{eq:GainBandwidthProduct} is a Lorenzian distribution with bandwidth $\Delta_\text{BW}$. As $\Delta_\text{BW}$ scales inversely with the peak of the gain profile, the gain-bandwidth product is constant $|G_\text{PS}|\Delta_\text{BW} = \kappa$ is constant as a function of gain. 

The appearance of a Lorentzian profile and the gain bandwidth trade-off are essential and ubiquitous predictions of input-output theory and driven resonators in general (note another appearance of a Lorentzian in the first line of \erf{eq:ReadoutCoherentStates} when we considered qubit resonance). Both can be circumvented if we violate the Markov approximation made at the beginning of this section, allowing the cavity-transmission line coupling to be frequency dependent. Indeed, parametric amplifiers without a gain-bandwidth trade-off have been made by introducing frequency dependence to the transmission line impedance\cite{roy2015broadband}.

\subheading{Realistic parametric amplifiers} Just like qubits, experimental parametric amplifiers do not behave exactly like their idealized theoretical counterparts. Here we treat the nonidealities that matter most for tuning up a real paramp, which are higher-order nonlinearities and internal loss. As we saw in exercise \ref{ex:SinglePumping}, another way to generate a squeezing Hamiltonian is to apply a strong resonant drive to a weakly nonlinear cavity. This mode of operation is called single-pumping. The essence of the single-pumping (and double-pumping, with the help of the rotating wave approximation) is that under a large displacement $a\rightarrow a+\alpha$, the nonlinear term $(a^\t a)^2$ expands out to $\alpha^2 a^{\t 2} + \alpha^{*2} a^2 + 4|\alpha|^2a^\t a$ plus higher-order terms that may be neglected at first order.\footnote{Yet another common method is double-pumping, in which two off-resonant tones detuned by $\pm \Delta_\text{sideband}$ create the same effective Hamiltonian.} Without these corrections (and the detuning term $a^\t a$ that may be cancelled by detuning the pump), the above treatment applies exactly, leading to parametric amplification. Higher-order corrections become non-negligible at high gain or high input power, when the fluctuations of $a$ about its mean $\alpha$ become large. As \erf{eq:ParampIORelation} predicts infinite gain at finite pump powers, high-order corrections inevitably affect realistic devices.

It is easiest to understand nonlinear effects in the case of single pumping, and there are few essential differences between that and flux pumping. We parameterize the nonlinear cavity Hamiltonian as
\begin{align}
H = \Delta a^\t a + \frac{\zeta}{2}a^{\t 2} a^2
\end{align}
where $\Delta$ is the detuning between the pump and the cavity resonance and $\zeta$ parameterizes the nonlinearity, which is called the Kerr nonlinearity in this context. Including finite internal loss of the cavity, the resulting input-output relation for the cavity using \erf{eq:IORelation} is
\begin{align}
\dot{a} = -i \Delta a - i \frac{\zeta}{2} a^\t a^2 - \frac{\kappa}{2} a - \sqrt{\kappa_\text{ext}} b_\text{in} - \sqrt{\kappa_\text{int}} b_\text{in,int}
\end{align}
where $\kappa = \kappa_\text{ext} + \kappa_\text{int}$. To include the effects of loss within the cavity, we have included another mode $b_\text{in,int}$, which provides an alternative decay channel. Assuming that this additional mode is in its ground state, the expectation value of the above expression leads to an equation of motion for the coherent displacement of the cavity
\begin{align}
\dot{\alpha} &\approx -i \frac{\zeta}{2} \alpha^* \alpha^2 - \left(\frac{\kappa}{2} + i \Delta \right)\alpha - \epsilon \\ \nonumber
\epsilon &\equiv \sqrt{\kappa_\text{ext}}\<b_\text{in}\>.
\end{align}
We are approximating the intercavity state as a coherent state, so the above represents a semiclassical approximation that neglects quantum fluctuations. A linear response calculation about the classical solution can recover more details than what we work out below, such as gain in the nonlinear regime\cite{yurke2006performance}. We simplify this calculation further by only considering the steady-state behavior, applying the condition $\dot{\alpha} = 0$. Writing $\alpha = e^{-i\phi}|\alpha|$, we have
\begin{align} \label{eq:ParampComplexPolynomial}
-i \frac{\zeta}{2} |\alpha|^3 - \left(\frac{\kappa}{2} + i \Delta \right)|\alpha| = \epsilon e^{i\phi}.
\end{align}
As the left-hand side contains only odd powers of $|\alpha|^2$, we can convert it to a third degree polynomial in $|\alpha|^2$ if we multiply both sides by their complex conjugates
\begin{align}
\frac{\zeta^2}{4}|\alpha|^6 + \zeta \Delta |\alpha|^4 + \left(\frac{\kappa^2}{4} + \Delta^2\right) |\alpha|^2 = |\epsilon|^2.
\end{align}
This equation has three roots, two of which may be complex. 

\begin{figure}
\centering
{\includegraphics[width=\textwidth]{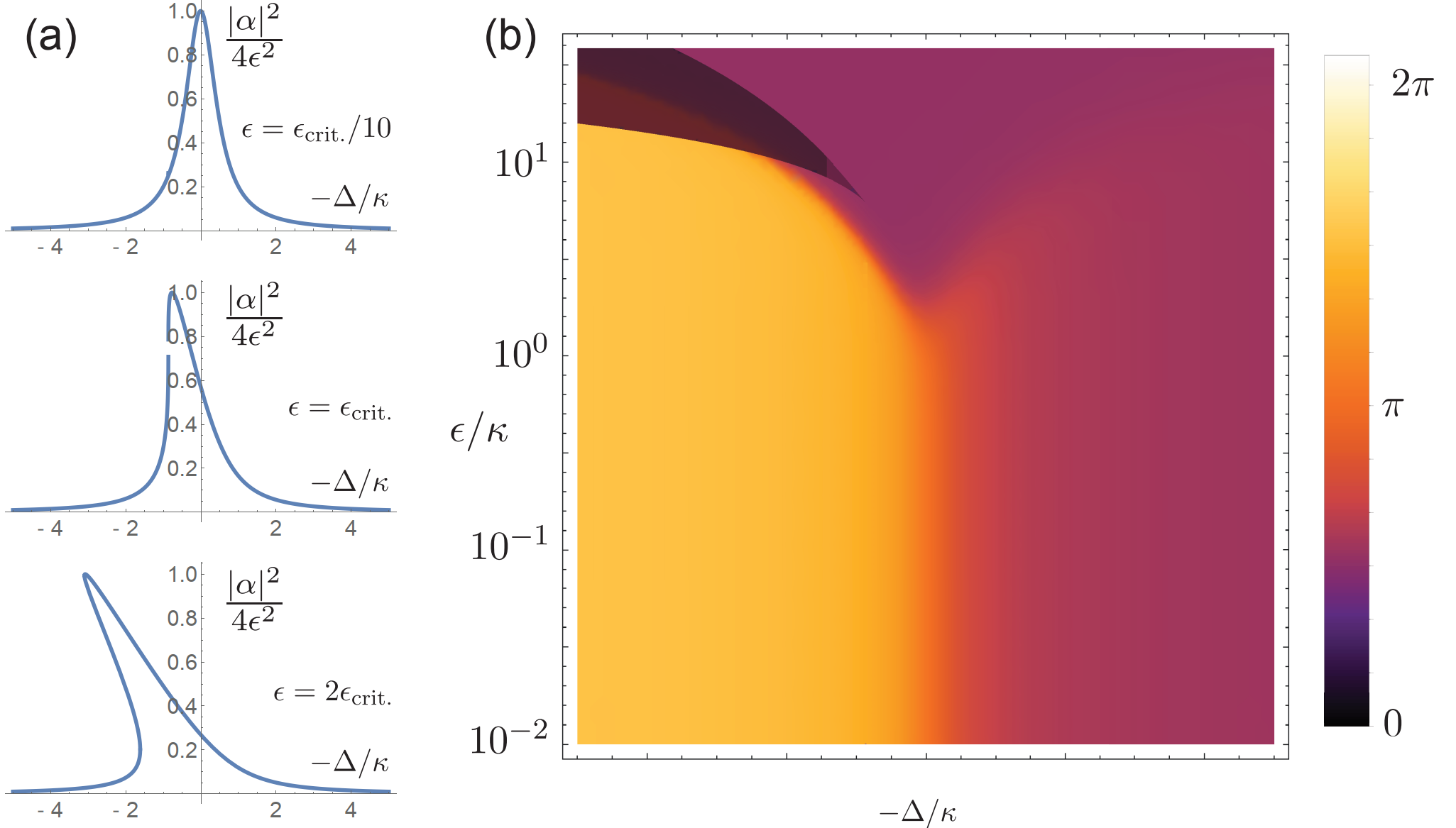}}
\caption{Behavior of a parametric amplifier when including the effects of the Kerr nonlinearity. (a) Intercavity field amplitude as a function of pump detuning for different drive powers. Multivaluedness of the bottom plot indicates bistability. (b) Phase of the internal state as a function of pump power and frequency. Region in which multiple solutions exist is shown with a darker color. 
}
\label{fig:NonIdealParamp}
\end{figure}

When there is only one real root (such as when $\zeta=0$, so that we recover the linear cavity result), then we have only one valid solution, as $|\alpha|^2$ cannot be complex. As we increase $\zeta$ or $\epsilon$, one finds a regime in which all three roots become real. To understand the physical significance of these additional roots, consider the intercavity power as a function of drive power and frequency, as plotted in Fig. \ref{fig:NonIdealParamp}a. At low powers, the Kerr nonlinearity is negligible and the intercavity field follows a Lorentzian distribution as a function of detuning. With increasing power, the resonance shifts to the left (given negative $\zeta$, as we would have for a cosine potential) and distorts slightly. At some critical power $\epsilon_\text{crit} = \sqrt{2 \kappa^3/3\sqrt{3}|\zeta|}$, the left portion of the Lorentzian becomes infinitely steep, indicating an instability. Past this power, there are drive frequencies at which all three roots are real. One of these roots corresponds to an unstable state, but the other two are stable, indicating bistability of the device. Thus before we achieve infinite gain, the system bifurcates. 

With the full solution for $|\alpha|$ on hand, we can easily compute the complex phase of $\alpha$ using \erf{eq:ParampComplexPolynomial}. In Fig. \ref{fig:NonIdealParamp}b, we plot the internal cavity phase as a function of drive amplitude and detuning. This plot is useful to measure experimentally when characterizing real devices, and may be generated using only a vector network analyzer. The dark region in the upper-left hand corner shows where bistability occurs. When tuning up a paramp in practice, it is important to work away from the bifurcation regime, and hence away from the point of peak gain. A vertical cut that maximizes gain tends to be preferable, as power sources tend to fluctuation. A dynamical analysis of the above shows that the gain profile becomes distinctly non-Lorentzian above the point of maximum gain\cite{slichter2011quantum}.

There are other nonidealities that can affect the performance of a realistic device besides the Kerr nonlinearity. Internal cavity loss essentially dropped out of the above analysis, as results only depended on the total loss rate $\kappa$. The actual output field may be recovered from the input-output relation \erf{eq:IOBInOut}, which shows that we simply lose some of the internal field. Due to interference effects, phase shifts of the output field as a function of frequency will be less than $2\pi$ in the presence of internal loss, as in Fig. \ref{fig:Ch2_InputOutputTheory}c. Another common nonideality is frequency dependence of $\kappa$, which can arise from impedance variations in the output line or external resonant modes. This can lead to different gain bandwidths as the amplifier is tuned in frequency (with a static flux bias), and also to non-Lorentzian behavior even at low powers.

The cavity-based parametric amplifier is one of many ways to amplify microwave signals. A recent development is the travelling wave parametric amplifier (TWPA), in which the nonlinear cavity from above is replaced with a nonlinear transmission line consisting of hundreds or thousands of Josephson junctions\cite{macklin2015twpa,obrien2014resonant}. The gain mechanism is the same as above, and intuitively but loosely one can think of a TWPA as a line of cavity-based amplifiers chained one after the other. An important subtlety of the device is that `sub-amplifiers' must work in concert, so that the same quadrature is amplified at each portion of the device. This requires the pump to remain in phase with the signal as the two propagate down the transmission line. Due to strong nonlinearities, the pump experiences much more dispersion than a far-detuned signal. This dispersion is cancelled out by introducing many resonators along the TWPA near the pump frequency, a technique known as phase matching. Much recent work has also focused on measuring different field operators than the field quadratures, such as photon number\cite{kono2018quantum,lescanne2019detecting}. A major contribution of this thesis is the introduction of a phase detector, which measures an observable conjugate to photon number. We describe this result in chapter \ref{ch:AdaptivePhase}.

%% file: OpenSystemsMeasurement.tex
\chapter{Open Quantum Systems and Measurement Theory}
\label{ch:OpenSystems}


The Hamiltonians of the last chapters are necessarily approximations, as they neglect interactions with the rest of the universe. In classical physics, such interactions present no great difficulty; think of including air resistance and friction when modeling the motion of a projectile. On the other hand, as quantum mechanics is fundamentally a theory of information \textbf{\cite{Nielsen1998}} (hence the mantra `information is physical'\footnote{It is tempting to elaborate on this notion further, but I couldn't do this phrase justice better than Scott Aaronson's blog post here https://www.scottaaronson.com/blog/?p=3327.}), we might expect dramatically different behavior when an external system `learns' about the state of our qubit. This behavior is the subject of this chapter, and its applications are the main subject of later chapters.

Undergraduate courses in quantum mechanics almost always omit the study of open quantum systems, which creates the impression that decoherence, weak measurement and quantum trajectories are as mysterious as the measurement problem itself. In reality however, these phenomena follow directly from the basic axioms of quantum theory, and are completely independent of the more philosophical issues regarding the interpretation of wave-function collapse. The underlying connections are quite simple to state; weak measurement of a degree of freedom (position, spin, photon number \textit{etc.}) results when one interacts that degree of freedom with another system, and then applies a standard projective measurement to that secondary system. Decoherence is just weak measurement by the environment, or to be more precise, by a system that is inaccessible to the experimenter. A quantum trajectory of a state is the state of that system as it undergoes a continuous series of infinitesimally short weak measurements. Importantly, all of these phenomena involve exchange of information. 


As long as we know how to apply the postulates of quantum mechanics, open quantum systems should present no great difficulty. However it is important to have precise statements of these postulates. Another topic that is often left somewhat nebulous is the question of \textit{when} one should apply the measurement postulate. What constitutes a measurement? We take the following operational answer: 

\begin{quote}
\definition{Projective measurement may be said to occur, and hence one may apply the Born rule, once complete information of the system's state has been irreversibly transferred to another system.}
\end{quote}

Here, irreversible could be in the thermodynamic sense, or as the result of an experimentalist directing a measurement probe to and then away from their system. We take the notion of complete information transfer to have the following definition, which although rather technical-sounding is nevertheless intuitive. There should exist a complete, orthogonal set of system states that each map the environment to one of a set of its own orthogonal states under the measurement interaction. This condition implies that there exists a hypothetical projective measurement on the environment that can perfectly distinguish the original system state. If furthermore the interaction preserves the system state, then the second part of the Born rule applies, namely that the post-measurement state is an eigenvector of the observable. In essence, we can apply the Born rule if the environment has effectively already applied a projective measurement for us. 

Notice that our answer immediately begs the question of when a process becomes irreversible.\footnote{Imagine reversing time by somehow replacing $H$ with $-H$ in the future light cone of the system. Most interpretations of quantum mechanics predict that such a process would revert the system to its original state, in contradiction with the irreversible change induced by applying the measurement postulate. However we know from statistical mechanics that such a process is extremely unlikely to occur. Regardless, the point is not that this projection has physical reality, which would be a matter of interpretation, but rather that after such an irreversible process, one is free to apply the measurement postulate.} However as long as the assumption of complete information transfer will hold at all future times, we can apply the measurement postulate whenever we like, so there is no need to ascribe a moment at which measurement occurs. Thus this ambiguity is actually a feature of the definition. It allows us skirt the ultimately philosophical question of when wave function collapse occurs and focus on the operational question of when the Born rule will agree with experimental observation.




The above discussion applies to a rather narrow range of physical processes. The Born rule as we stated it in chapter \ref{ch:IntroQuantum} pertains to projective measurements that leave the system in an eigenstate of the measurement operator. However there are plenty of processes that qualitatively resemble a measurement but satisfy neither of these properties. For example, an excited state of an atomic system that decays and emits a photon offers an obvious way to measure whether the atom is in that excited state or its ground state; wait and see if you observe a photon. This process leaves the system in its ground state regardless of the measurement outcome, in contradiction with the Born rule as typically stated. More subtly, this measurement is technically not even projective. Spontaneous emission follows an exponential decay, so formally one would have to wait an infinite amount of time to be sure that the measurement outcome is `ground state' if no photon were observed. While projective measurement can at least be recovered in the limit of infinite waiting time, our present definition refuses to acknowledge that information is acquired continuously in this process, and thus that it is a sort of measurement at shorter times too. In the following section, we will use the Born rule as a launching point to develop a more general definition of measurement that encompasses most realistic physical processes.

\section{The POVM Formalism and Weak Measurement}

Measurement and unitary evolution are usually thought of as disparate types of dynamics. However given the perspective that measurement is merely a transfer of information from one system to another, one might hope to unify them into a single equation. This is accomplished with the positive operator valued measure (POVM) formalism, which is a somewhat unfortunate choice of name given the simplicity of the underlying concept. 

The free evolution of a closed system is described by the time-independent Schr\"odinger equation
\begin{align}
\label{eq:UnitaryEv}
\rho(t) = U \rho(0) U^\dagger \\ \nonumber
U \equiv e^{-i H t}
\end{align}
%
(Throughout this chapter, we work in a system of units that has $\hbar=1$, a common convention that amounts to measuring energy in units of frequency). Projective measurement of an operator $X$ changes the state according to
\begin{align} \label{eq:ProjMeasEv}
\rho(t,i) = \frac{|i\rangle \langle i| \rho(0) | i \rangle \langle i |}{\text{Tr}[|i\rangle \langle i| \rho(0) | i \rangle \langle i |]} 
.
\end{align}
The operator $|i \rangle \langle i|$ is a projection operator into the $i$th eigenspace of $X$. We can unify equations \ref{eq:UnitaryEv} and \ref{eq:ProjMeasEv} into one expression by writing
\begin{align} \label{eq:POVM}
&\rho(t,i) = \frac{\Omega_i \rho(0) \Omega_i^\dagger}{\text{Tr}[\Omega_i \rho(0) \Omega_i^\dagger]} \\ \nonumber 
&~~P(i) = \Tr[\Omega_i \rho(0) \Omega_i^\t] \\ \nonumber
&\text{Unitary evolution: } \Omega_1 = U, ~~ \Omega_1^\dagger \Omega_1 = \hat{I} \\ \nonumber
&\text{Projective measurement: } \Omega_i = |i\rangle \langle i|, ~~ \sum_i \Omega_i = \hat{I}
\end{align}
where the latter condition is necessary to ensure that the probability of all measurement outcomes sums to 1. 
The $\Omega_i$ are known as \definition{Kraus operators}, and they are capable of modeling a much wider range of physical processes than the above two. Note that the condition
\begin{empheq}[box=\widefbox]{align} \label{eq:ValidPOVM}
    \sum_i \Omega_i^\dagger \Omega_i = I
\end{empheq}
is satisfied for both unitary evolution and projective measurement. We now set out to show that \erf{eq:ValidPOVM} in a sense \textit{defines} a valid operation on a system. The formal version of this result is known as Kraus' theorem.

If we wish to use \erf{eq:POVM} to describe the widest possible range of dynamics, we need to treat the possibility that the system interacts with an external system. For example, suppose that one introduces an auxiliary system and then applies a joint unitary on the whole. To ultimately write down an equation of motion for the system alone, we need a way to discard the auxiliary system. One way to accomplish this is with a projective measurement on the external system alone, which yields
\begin{equation} \label{eq:POVMFromUnitary}
    \rho(t,i) \otimes |i\> \< i| = \frac{(I_\text{sys} \otimes |i\> \< i|) U (\rho(0) \otimes |\psi\> \< \psi|) U^\t (I_\text{sys} \otimes |i\rangle \< i|)}{\Tr[(I_\text{sys} \otimes |i\> \< i|) U (\rho(0) \otimes |\psi\> \< \psi|) U^\t (I_\text{sys} \otimes |i\> \< i|)]}.
\end{equation}
We know the state of a system after measurement occurs, so we can simply discard it by replacing the projection operator $I_\text{sys} \otimes |i\> \< i|$ with $I_\text{sys} \otimes \langle i|$
\begin{align} \label{eq:WeakMeasPOVM}
    \rho(t,i) &= \frac{(I_\text{sys}\otimes\<i|) U (\rho(0) \otimes |\psi \>\<\psi|) U^\t (I_\text{sys} \otimes |i\>)}{\Tr[(I_\text{sys}\otimes |i\>) U (\rho(0) \otimes |\psi \>\<\psi|) U^\t (I_\text{sys}\otimes |i\>)]} = 
    \frac{\Omega_i \rho(0) \Omega_i^\t}{\Tr[\Omega_i \rho(0) \Omega_i^\t]} \\ \nonumber
    \Omega_i &= (I_\text{sys} \otimes \<i|) U (I_\text{sys} \otimes |\psi\>).
\end{align}
Importantly, $\Omega_i$ is an operator that acts on the system alone and satisfies \erf{eq:ValidPOVM}. The above $\Omega_i$ describes precisely the scenario of a weak measurement. If $U$ maps $|\psi\>$ to a different state $|i\>$ for each system subspace, then \erf{eq:WeakMeasPOVM} describes a projective measurement on the system via the auxiliary system. We will see a concrete example of such measurements in the next section.

\begin{exercise} \label{ex:GlobalMeasPOVM}
\erf{eq:WeakMeasPOVM} does not exhaust all possible processes involving interaction and measurement. Show that the following processes also yield sets of Kraus operators $\Omega_i$ that satisfy \erf{eq:ValidPOVM}: 1. Replace $U$ in \erf{eq:WeakMeasPOVM} with a global projective measurement on the whole. 2. Include in \erf{eq:WeakMeasPOVM} a unitary $U_i$ on the system that depends on the measurement outcome $i$. The latter scenario describes quantum feedback.
\end{exercise}

We have succeeded in introducing an external system $|\psi\>$ to the dynamics, but only if we make sure to measure it once we are done. The key to circumventing this constraint is to realize that if someone else were to measure it for us, we would never be able to tell by looking at the system alone; such a measurement acts as the identity on our system. Thus to develop a mathematical representation, we are free to assume that someone else performed a measurement for convenience. As we do not know the outcome of this measurement, the final state of the system is an average over all possible measurement outcomes, weighted by their probabilities
\begin{equation} \label{eq:POVMDecoherence}
    \rho(t) = \sum_i \frac{\Omega_i \rho(0) \Omega_i^\t}{\Tr[\Omega_i \rho(0) \Omega_i^\t]} P(i) = \sum_i \Omega_i \rho(0) \Omega_i^\t.
\end{equation}
The above is known as a \definition{Kraus map}, and the corresponding process is called a \definition{quantum operation}. It is notable that by averaging over measurement outcomes, we have restored linearity in $\rho$. Non-linearity only arises when we consider the state conditioned on a measurement outcome. \erf{eq:POVMDecoherence} is an example of decoherence, and shows that such processes may be modeled with a linear equation just like the Schr\"odinger equation. Kraus' theorem, and a significant related result called Choi's theorem, state that all operations that map valid density matrices to valid density matrices may be written in the form \erf{eq:POVMDecoherence}\ped{\cite{Nielsen2010}},\footnote{Such a valid map is linear and completely positive. The latter condition means that it maps positive semidefinite matrices to positive semidefinite matrices, even if one were to append an auxiliary system to $\rho$ and then apply the map to the composite system. See the discussion of completely positive maps around Eq. (8.43) of \cite{Nielsen2010}.} with one important exception discussed shortly. Thus the POVM formalism allows us to model almost all physical processes, whether they include measurement or not.

Despite Choi's and Kraus' theorems, there is still one type of process that cannot be modeled by the POVM formalism. By the way we have treated it, the transfer of information from the system is implicitly irreversible, as the measurement permanently alters the auxiliary system. Thus although \erf{eq:POVMDecoherence} can model an initial interaction with an external system, it would be incorrect to use \erf{eq:POVMDecoherence} again to model a subsequent interaction with that same system if in reality no measurement were applied in between. Such a scenario could transfer information back to the system and even reverse the decoherence induced by the first interaction. Such processes are called \definition{non-Markovian}, and arise when a system interacts with a system with which it is already entangled.\footnote{These operations also need not be completely positive.} Conversely, an interaction in which such a back-flow of information is not possible is called Markovian. The latter processes will form the basis for our model of continuous measurement, the master equation as developed in Sec. \ref{sec:MasterEquation}.

\begin{exercise}
\erf{eq:WeakMeasPOVM} assumed that the introduced auxiliary system was in a pure state. Use \erf{eq:DefRho} ($\rho \equiv \sum_i P_i |\psi_i\>\<\psi_i|$) to generalize to the mixed state case. How many Kraus operators does one need to express this process?
\end{exercise}

\begin{exercise} \label{ex:NonUniqueKraus}
Non-uniqueness of the Kraus representation: Show that two Kraus maps $\sum_i \Omega_i \rho \Omega_i^\t$ and $\sum_i \Omega_i' \rho {\Omega_i'}^\t$ are equivalent if there exists a unitary matrix $U$ such that $\Omega_i' = \sum_j U_{ij} \Omega_j$. Note that for every Kraus map, there exists an equivalent set of `canonical' Kraus operators satisfying $\Tr[\Omega_i \Omega_j^\t] = \delta_{i,j}$\cite{Nielsen2010}.
\end{exercise}

The process of discarding an external system via measurement is ubiquitous in the treatment of open quantum systems. The introduction of a hypothetical measurement is important enough to deserve its own name; the \definition{partial trace} $\Tr_B$ on a composite system $\rho_{AB}$ discards subsystem $B$, leaving us with a density matrix $\rho_A$
\begin{equation} \label{eq:PartialTrace}
    \rho_A \equiv \Tr_B[\rho_{AB}] \equiv \sum_i (I_A \otimes \<i|_B) \rho_{AB} (I_A \otimes |i\>_B)
\end{equation}
The same cancellation that occurred in \erf{eq:POVMDecoherence} has occurred in \erf{eq:PartialTrace}, again resulting in a linear equation. The partial trace generalizes the standard trace operation from linear algebra, and satisfies many corresponding identities. For example, given an operator $\mathcal{O}_B$ that only acts as the identity on subsystem $A$, we have $\Tr_B[\mathcal{O}_B \rho_{AB}] = \Tr_B[\rho_{AB} \mathcal{O}_B]$. Furthermore, for any operator that may be written as $\mathcal{O}_A \otimes \mathcal{O}_B$, we have $\Tr_B[\mathcal{O}_A \otimes \mathcal{O}_B] = \mathcal{O}_A \Tr[\mathcal{O}_B]$. The latter condition implies that if $\rho_{AB}$ is a separable \textit{i.e.} not entangled, then the partial trace does nothing but undo the tensor product operation $\rho_{AB} = \rho_A \otimes \rho_B$ and return $\rho_A$. In general if $\rho_{AB}$ is entangled between subsystems $A$ and $B$, then the purity of $\rho_A$ is lower than that of $\rho_{AB}$. This illustrates a fundamental fact about entanglement, that an entangled state looks like a classical mixture if we only have access to a subsystem.

\section{A Toy Model of Weak Measurement}
\label{sec:WeakMeas}

To make the notion of weak measurement concrete, we now turn to our first example. Although the models that we write down are initially motivated by simplicity and pedagogy, the second leads neatly to a model of continuous measurements and can be derived from realistic physical models. By the end of this section, we will have a stochastic master equation and a basic understanding of the stochastic calculus that underpins it. Continuing with the presentation style of the first two chapters, we begin by stating up front the ultimate physical models and their solutions. With intuition for the dynamics in hand, we derive them from a general model of an open quantum system in the following section, and then apply it to relevant examples in circuit QED.

Consider a pair of qubits, one playing the role of the system and the other playing the role of the auxiliary. The projective measurement limit is likely the most familiar case, so we begin with that. Let us initialize the auxiliary qubit in $|g\>$ and apply a CNOT gate with the system qubit as the control. The CNOT unitary flips the auxiliary qubit to $|e\>$ if the system qubit is in $|e\>$, so to read out the state of the system qubit, we should measure the auxiliary in the $\{|g\>, |e\>\}$ basis. To calculate the Kraus operators associated with this process, we use \erf{eq:WeakMeasPOVM} 
\begin{align}
    \Omega_g &= (I_\text{sys} \otimes \<g|) U_\text{CNOT} (I_\text{sys} \otimes |g\>) = |g\>\<g| \\ \nonumber
    \Omega_e &= (I_\text{sys} \otimes \<e|) U_\text{CNOT} (I_\text{sys} \otimes |g\>) = |e\>\<e| \\ \nonumber
    U_\text{CNOT} &= |g\>\<g| \otimes \sigma_0 + |e\>\<e| \otimes \sigma_x
\end{align}
As expected, measurement of the auxiliary qubit implements a projective measurement on the system qubit.

To apply a weak measurement, we need a unitary interaction that does less than completely transfer the system state to the auxiliary qubit. A generic Hamiltonian suited for the job resembles the Jaynes-Cummings Hamiltonian
\begin{align}
H = i \epsilon (c_\text{sys} \sigma^\t - c_\text{sys}^\t \sigma).
\end{align}
In what follows, we treat the qubit degree of freedom as the auxiliary, and let $c_\text{sys}$ be any system operator, not necessarily just a ladder operator. For our `toy model' of weak measurement $c_\text{sys}=|e\>\<e|$, we obtain $H = -\epsilon |e\>\<e| \sigma_y$, the Hamiltonian for a controlled $\sigma_y$ rotation. If we apply this Hamiltonian for a duration $\Delta t=\pi\epsilon$, then we recover the CNOT gate of our first example. If we apply $H$ for a shorter duration, then we recover a weak measurement of $|e\>\<e|$, or in general the operator $c_\text{sys}$. Another classic example to bear in mind is $c_\text{sys} = \sigma$, which turns out to model a spontaneous emission process (swapping of an excitation from system into an environment).

To see how this works, we compute the unitary generated by $H$ and plug it into \erf{eq:WeakMeasPOVM}. In the limit of a very weak interaction/measurement, we can expand $U$ to second order in the interaction time $\Delta t$
\begin{align} \label{eq:WeakSwap}
U(\Delta t) &= \exp(-i H \Delta t) \approx I + \epsilon \Delta t (c \sigma^\t - c^\t \sigma) + \frac{(\epsilon \Delta t)^2}{2}(c \sigma^\t - c^\t \sigma)^2 \\ \nonumber
&= I + \epsilon \Delta t (c \sigma^\t - c^\t \sigma) + \frac{(\epsilon \Delta t)^2}{2}(c c^\t |e\>\<e| - c^\t c |g\>\<g|)
\end{align}
where we have dropped the `sys' subscript to prevent clutter. For the moment, imagine that $\Delta t$ is small but finite. To compute a weak measurement POVM from \erf{eq:WeakSwap}, we have two choices to make: the initial state of the auxiliary qubit and the auxiliary measurement basis after interaction. For our toy model $c=|e\>\<e|$, if we align the initial state along the $\sigma_y$ axis then $U(t)$ does nothing and no measurement occurs. Conversely, starting it out in the $\sigma_x,\sigma_z$ plane maximizes the amount of information transferred to the environment, regardless of where we put it. Therefore without much loss of generality, we can simplify \erf{eq:WeakSwap} and the resulting POVM further by starting the auxiliary qubit in $|g\>$
\begin{align} \label{eq:SMEPOVM}
\Omega_i &= (I_\text{sys} \otimes \<i|) U(\Delta t) (I_\text{sys} \otimes |g\>)  \\ \nonumber
&= \<i|g\> I_\text{sys} + \epsilon \Delta t \<i|e\> c - \frac{(\epsilon \Delta t)^2}{2} \<i|g\> c^\t c.
\end{align}
If we bear in mind the example of spontantous emission, this simplification amounts to starting the environmental degree of freedon in the vacuum. Simple as it looks, the above POVM will carry us through the rest of this chapter, from the toy models below to a stochastic master equation that models countless realistic systems. 

Let us consider the dynamics under various measurement bases $\{|i\>\}_{i=1,2}$. If we measure in the $|i\> = |e/g\>$ basis, the two possible measurement outcomes produce qualitatively different results. The least likely measurement outcome $|i\>=|e\>$ causes a discrete jump in the system, just like a projective measurement
\begin{align} \label{eq:JumpPOVM}
\Omega_e = \epsilon \Delta t c.
\end{align}
In our toy model, this just collapses the system to $|e\>$. As only $|\psi_\text{sys}\> = |e\>$ leads to a rotation of the auxiliary, it makes sense that detecting a rotation of the auxiliary qubit into $|e\>$ implies with 100\% certainty that the system must be in $|e\>$ as well. However the probability for this measurement outcome is quite small. The most likely measurement outcome is $|i\> = |g\>$, as this aligns with the initial state of the auxiliary qubit. In this case, the system is only changed infinitesimally, and the Kraus operator is
\begin{align} \label{eq:NoJumpPOVM}
\Omega_g = I - \frac{(\epsilon \Delta t)^2}{2} c^\t c
\end{align}
For $c=|e\>\<e|$, $\Omega_g$ pushes the system slightly toward the ground state. Physically, this arises because this `null' measurement outcome still provides information about the system, hinting that it is in $|g\>$. 

\erf{eq:JumpPOVM} and \erf{eq:NoJumpPOVM} generally describe a \definition{quantum jump process}. If $c$ is a ladder operator for the system such that $c+c^\t$ gives the system's dipole moment, then they model detection of spontaneous emission with a photon counter, as we will see in the next section. They also provide the mathematically simplest model of an open quantum system in which we do \textit{not} measure the auxiliary, so that we must use the average evolution equation \erf{eq:POVMDecoherence}. The system evolves continuously under \erf{eq:NoJumpPOVM}, punctuated by random jumps from \erf{eq:JumpPOVM}. In some scenarios, such as in the quantum Zeno effect described in the next chapter, the jump probability is negligible and we can propagate under \erf{eq:NoJumpPOVM} alone. It is convenient to introduce the non-Hermitian
\begin{align} \label{eq:NoJumpHamiltonian}
H_\text{eff} = -i \frac{\gamma}{2} c^\t c,
\end{align}
which generates $\Omega_g$ (though with a more physical linear time dependence instead of quadratic). In the limit that the jump probability $P_e = \Tr[\Omega_e \rho(t) \Omega_e^\t]$ is small, one can add $H_\text{eff}$ to a closed system's Hamiltonian to model coupling to an environment.


Things change significantly when we measure in a basis perpendicular to the initial auxiliary qubit state. All measurement axes aligned with the equator of the Bloch sphere can be parameterized as $\sigma_\delta = \sigma_x \cos(\delta) + \sigma_y \sin(\delta$), like we defined in \erf{eq:HQubitRWA}. The eigenvectors of $\sigma_\delta$ can be parameterized as $|i\> = (\pm e^{-i \delta}, ~ 1)^\T / \sqrt{2}$, which gives the POVM
\begin{align} \label{eq:DiffusePOVM}
\Omega_\pm = \frac{1}{\sqrt{2}} \left( I_\text{sys} \pm \epsilon \Delta t e^{-i\delta} c - \frac{(\epsilon \Delta t)^2}{2} c^\t c \right)
\end{align}
Unlike quantum jumps, both measurement outcomes change the state only infinitesimally. Once again, we can visualize the action of our POVM by considering $c = |e\>\<e|$. Things will be slightly cleaner if we take $c = |e\>\<e| - I/2 = \sigma_z/2$, which maintains the interaction strength but symmetrizes the action of $H$ on the system. For $\delta = 0$, the measurement axis $\sigma_x$ lies in the plane of rotation, so our projective measurement is sensitive to the direction of rotation. Due to the presence of $I_\text{sys}/\sqrt{2}$ in both Kraus operators, both measurement outcomes occur with roughly equal probability. Depending on the system qubit state, the auxiliary qubit is rotated one way or the other, slightly biasing the measurement outcome probabilities away from 50/50. $\Omega_\pm$ kicks the qubit ever so slightly toward or away from the poles of the Bloch sphere, as we might expect from the partial acquisition of information. 

When $\delta = \pi/2$, the measurement axis aligns with the axis of rotation. This situation is peculiar. On one hand, the projective measurement gains no information about how the auxiliary qubit has been rotated, which does not look like a measurement. On the other hand, we know that some information has been transmitted from system to environment, and therefore something must happen. The resolution of this apparent paradox is that $\Omega_\pm$ is unitary in this case. Specifically, $\Omega_\pm = \exp(\pm i \epsilon \Delta t \sigma_z/2)$ to second order, indicating that the POVM generates a random rotation about the $\sigma_z$ axis. As we do not learn about the state, the qubit does not get pushed toward a measurement eigenstate, but the measurement still has a random effect on the system.

\subheading{Continuous measurement, stochastic calculus and the master equation} So far, we have worked with a somewhat unnatural model in which we interact the system and environment for a fixed amount of time, stop the interaction, and then measure the environment. As our notational use of $\epsilon$ and $\Delta t$ suggests, the above treatment is poised to provide a continuous model of measurement. It is tempting to simply take the $\Delta t \rightarrow 0$ limit, but the resulting theory runs into technical difficulties that will become clear once we move down the right path. A hint of trouble comes already if we try to compute the average evolution using \erf{eq:POVMDecoherence} in this limit, which results in $\rho(dt) = \rho(0)$ ($\dot{\rho}=0$) and hence no decoherence despite the measurement. To get a sensible theory, we need to keep terms to second order in $\epsilon \Delta t$. One ostensibly naive if unexpected way to keep second order terms in the $\Delta t \rightarrow 0$ limit is to take
\begin{align} \label{eq:DivergentCouplingForSME}
\epsilon = \sqrt{\frac{\gamma}{\Delta t}}
\end{align}
where $\gamma$ is some kind of rate. Although this seemingly pathological substitution leads to a divergent interaction Hamiltonian, it actually turns out to lead to the correct model of continuous measurement, both abstractly and phenomenologically. Physically, this divergence arises because our system interacts with an infinite number of degrees of freedom. It is essentially the same divergence that leads to the $\delta$ function commutator of the $b_\text{in/out}$ operators in input-output theory. In fact, we will essentially end up \textit{identifying} the environment operator $\sqrt{\gamma/\Delta t}\sigma$ with $b_\text{in}$ in the next section. 

It turns out that we have already derived a continuous version of the quantum jump process. As the reader may have noticed, \erf{eq:NoJumpPOVM} leads to \erf{eq:NoJumpHamiltonian} when we substitute \erf{eq:DivergentCouplingForSME}. The more subtle case and much of the interesting physics comes in when we take the continuum limit of \erf{eq:DiffusePOVM}
\begin{align} \label{eq:DiffusiveSMEPOVM}
\Omega_\pm = \frac{1}{\sqrt{2}} \left( I_\text{sys} \pm \sqrt{\gamma \Delta t} e^{-i\delta} c - \frac{\gamma \Delta t}{2} c^\t c \right).
\end{align}
If only the first and last terms were present, we could simply replace $\Delta t$ with $dt$ and write down a corresponding differential equation (which would be \erf{eq:NoJumpHamiltonian}).
Unfortunately, the second term would lead to a divergent derivative in our differential equation as we take $dt \rightarrow 0$. 

\begin{figure}
\centering
{\includegraphics[width = 160mm]{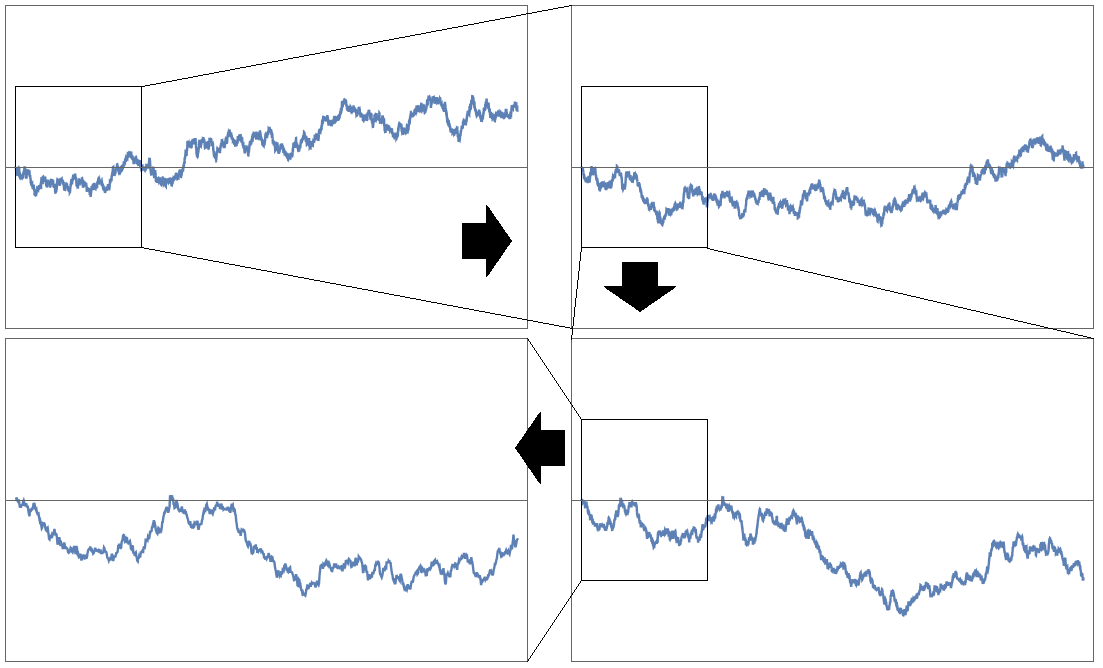}}
\caption{Sequential zoom-ins of a Wiener process. Each panel zooms in by a factor of $4$ in the horizontal axis and a factor of $\sqrt{4}=2$ in the vertical axis. Statistically, each panel is indistinguishable from the others.}
\label{fig:WienerProcess}
\end{figure}

If only one of the two possible measurement outcomes ever occurred, then we would indeed be stuck with non differentiable dynamics and no sensible continuum limit. Fortunately the randomness of the measurement outcome comes to the rescue, and the salvaged theory leads us to stochastic calculus. To see how we arrive at something sensible and finite, first consider the simplest version of such a stochastic problem, which resembles Brownian motion
\begin{align}
W(t+\Delta t) &= W(t) \pm \sqrt{\Delta t} \\ \nonumber
W(0) &= 0
\end{align}
where the $+$ and $-$ cases occur with equal probability. We can compute the variance of $W(T)$ using central limit theorem.\footnote{If $\{X_i\}$ is a set of $n$ (well-behaved) independent random variables each with variance $\sigma_i$ and mean $\mu_i$, then in the limit $n\rightarrow \infty$,  $\sum_{i=1}^n (X_i-\mu_i) \rightarrow \left(\sum_{i=1}^n \sigma_i^2 \right)^{1/2} N(0,1)$ where $N(0,1)$ is a zero-mean Gaussian distribution with unit variance.} If the interval $T$ is broken up into $n = T/\Delta t$ time steps, then the variance at time $T$ is $\sum_{i=1}^n \Delta t = T$ regardless of the value of $\Delta t$! In other words the $\Delta t^{1/2}$ dependence is just what we needed for there to be a well-defined continuum limit. As $\Delta t \rightarrow 0$, $W(t)$ is known as a \definition{Wiener process}, and exhibits a striking self-similarity as we zoom in, as depicted in Fig. \ref{fig:WienerProcess}. To think of $W$ as the position of a particle undergoing Brownian motion, the $\Delta t \rightarrow 0$ limit corresponds to simultaneously reducing the mean free path and increasing the thermal velocity so that the dynamics over a finite time interval preserve the same macroscopic diffusion process.

In the true continuum limit, the Wiener process is conventionally thought of as an integral
\begin{align}
W(t) = \int_0^t dW
\end{align}
where $dW$ is a zero-mean Gaussian distributed random variable with variance $dt$. We will not formally develop stochastic calculus here, but rather state the essential features and show how to use it in practice. For an informal explanation, see section 5B of \ped{\cite{JacobsSteck2006}}, and for a rigorous but manageable derivation, see in \cite{Oksendal2003}. For all intents and purposes, we can loosely take $dW = \pm \sqrt{dt}$. This handwaving equality suggests an odd property that turns out to be rigorously true, which is that
\begin{align} \label{eq:ItoRule}
dW^2 = dt,
\end{align}
The above relation is called the Ito rule, and it is a much stricter statement than our assertion that $\<dW^2\> = dt$. This equality holds because the fluctuations in $dW^2$ about its mean $dt$ turn out to scale as $dt^2$\ped{\cite{JacobsSteck2006}}, which equals zero exactly in the $dt \rightarrow 0$ limit. 

Using the Ito rule, we can derive or at least motivate one of the most useful equations in stochastic calculus, known as Ito's lemma. Suppose we have a twice-differentiable function $f(X)$ that we wish to evaluate $f(X+dX)$ where $X$ is a stochastic variable
\begin{align}
dX_t = \mu_t dt + \sigma_t dW.
\end{align}
By Taylor expanding $f(X+dX)$ about $X$ to first order in $dt$ and second order in $dW$, we might guess that
\begin{align} \label{eq:ItosLemma}
f(X+dX) &\equiv f(X) + df \\ \nonumber
df &= f'(X) dX_t + \frac{f''(X)}{2} dX_t^2 \\ \nonumber
&=\sigma_t f'(X) dW + \left(\mu_t f'(X) + \frac{\sigma_t^2}{2} f''(X) \right)dt
\end{align}
where we have used the Ito rule and dropped higher-order terms containing $dW dt$ or $dt^2$ in the third equality. \erf{eq:ItosLemma} is a generalization of the chain rule of ordinary calculus, and we will have many occasions to demonstrate its usefulness in what follows.

As we will have little else to say about the mathematics underlying stochastic calculus, we note that the above formalism is but one of its two forms. As we will continue to see, Ito calculus enables quick-and-dirty calculations in which we multiply differentials as if this is a sensible thing to do (soon we will make use of a generalized product rule $(f+df)(g+dg) = f g + f dg + g df + df dg$). While such manipulations would make any mathematician cringe, they are quite useful in practice, making Ito calculus surprisingly straightforward to work with. An alternative formulation is Stratonovich calculus, which is also quite useful in physics applications\ped{\cite{gardiner2004noise}}.

With a rough idea of how to take continuum limits of stochastic processes, let us return to \erf{eq:DiffusiveSMEPOVM} make use of the Wiener process. The potentially most obvious way to apply it is to simply replace $\pm \Delta t$ with $dW$. 
Anticipating a differential (or at least difference) equation, we write down the state update as
\begin{empheq}[box=\widefbox]{align} \label{eq:LinearSSE}
|\tilde{\psi}(t+dt)\> = \Omega_{dW} |\psi(t)\> = \sqrt{P(dW)}\left(I_\text{sys} + \sqrt{\gamma}e^{-i\delta} c dW - \frac{\gamma}{2} c^\t c dt \right) |\psi(t)\>.
\end{empheq}
\erf{eq:LinearSSE} is called the linearized stochastic Schr\"odinger equation, and its solution generates our first example of a quantum trajectory. We have included a tilde over $\psi$ to remind us that it is not normalized, and replaced the $1/\sqrt{2}$ normalization factor as we no longer have a 2-outcome measurement.\footnote{To justify \erf{eq:LinearSSE} more carefully, the composition $\Omega_{\pm\pm...\pm} = \prod_{i=1}^N \Omega_\pm = \frac{1}{2^{N/2}}(I + (\sum_i \pm_i\sqrt{\Delta t})M - \frac{1}{2}M^\t M \delta t)$, where we have defined $\delta t = N \Delta t$ and dropped cross terms of the form $(\pm_i 1)(\pm_j 1)M^2\Delta t$ as the variance of the sum of these terms limits to zero as $\Delta t\rightarrow 0$. We can label the possible Kraus operators as $\Omega_{\delta W}$ with $\delta W = 2n \sqrt{\Delta t}$ and $n=-N/2...N/2$ counts the number of `$+$'s and `$-$'s. For each value of $n$, there are $\begin{pmatrix} N\\N/2+n \end{pmatrix}$ identical $\Omega_{\pm\pm...\pm}$ that we can group together, so we have $\Omega_{\delta W} = \frac{1}{2^{N/2}}\frac{N!}{(N/2+n)!(N/2-n)!}(I+\delta W M-\frac{N}{2}M^\t M \delta t)$. The prefactor limits to a Gaussian as $N\rightarrow \infty$, and we recover the main result. \label{foot:SSENorm}}
$|\tilde{\psi}(t+dt)\>$ represents a physical state that could be generated by measurement. Unfortunately due to our naive substitution of $\pm \sqrt{\Delta t} = dW$, the stochastic variable does not generate $|\tilde{\psi}(t+dt)\>$ with the correct statistics; we should have $P(\tilde{\psi}(t+dt)|\psi(t)) = \<\tilde{\psi}(t+dt)|\tilde{\psi}(t+dt)\>$, but we have already fully specified $dW$ as a zero-mean Gaussian random variable and hence specified the statistics of $|\tilde{\psi}(t+dt)\>$. Nevertheless, \erf{eq:LinearSSE} is quite useful because although the dynamics that it generates do not follow the correct probability distribution implied by the Born rule, it nevertheless generates a \textit{possible} trajectory. Furthermore, it is linear, which enables closed-form solutions in many cases. Finally, we can still use the norm of $|\tilde{\psi}\>$ to compute the probability of a given trajectory occurring, as $\Omega_{dW}$ is like any other Kraus operator.

To write down an equation of motion with the correct statistics, we seek an Ito random variable with the same statistics as an actual measurement outcome. The probabilities for each of the two possible measurement outcomes from \erf{eq:DiffusiveSMEPOVM} are
\begin{align}
P(\pm) &= \<\psi(t)|\Omega_\pm ^\t \Omega_\pm |\psi(t)\> \\ \nonumber
&= \frac{1}{2} \left[1 \pm \sqrt{\gamma \Delta t} \< c e^{-i \delta} + c^\t e^{i\delta}\>\right].
\end{align}
We wish to replace the random variable $\Delta r \equiv \pm \Delta t$ appearing in \erf{eq:DiffusiveSMEPOVM} with an Ito random variable. The above probability distribution implies that
\begin{align}
&\<\Delta r\> = \sqrt{\Delta t}[P(+)-P(-)] = \sqrt{\gamma} \Delta t \< c e^{-i \delta} + c^\t e^{i\delta}\> \\ \nonumber
&\text{var}(\Delta r) = \<\Delta r^2\> - \< \Delta r\>^2 = \Delta t + \mathcal{O}(\Delta t^2).
\end{align}
Suppose we sum $\Delta r$ over many time steps, but keep $\Delta t$ small enough that $\< c e^{-i \delta} + c^\t e^{i\delta}\>$ remains constant over the longer time interval. Central limit theorem implies that the sum limits to a Gaussian distributed random variable with the above mean and variance. Thus in the $\Delta t \rightarrow 0$ limit, we can make the replacement
\begin{align}
&\Delta r = \pm \Delta t \\ \nonumber
&\rightarrow dr = \sqrt{\gamma}\< c e^{-i \delta} + c^\t e^{i\delta}\> dt + dW
\end{align}
Substituting $dr$ into \erf{eq:DiffusiveSMEPOVM} yields a stochastic Schr\"odinger equation with the right statistics
\begin{align} \label{eq:SSEBeforeNorm}
|\bar{\psi}(t+dt)\> = \left( I_\text{sys} + \sqrt{\gamma}e^{-i\delta} c dr - \frac{\gamma}{2} c^\t c dt \right) |\psi(t)\>
\end{align}
where the bar again indicates that $\psi$ is unnormalized but different from $\tilde{\psi}$.

The last step to take is to compute the normalization prefactor. As the right hand side of \erf{eq:SSEBeforeNorm} is infinitesimally closed to being normalized, we can normalize it using Ito's lemma. First, the norm of $\bar{\psi}$ is
\begin{align} \label{eq:SSENorm}
\< \bar{\psi}(t+dt) | \bar{\psi}(t+dt)\> &= 1 + \< M + M^\t \> dW + \<M + M^\t\>^2 dt \\ \nonumber
M &\equiv \sqrt{\gamma} e^{-i \delta} c
\end{align}
where we have defined $M$ for the sake of brevity. We normalize by multiplying by the inverse square root of \erf{eq:SSENorm}. It is this normalization factor that may be readily computed using Ito's lemma. Let $f(\epsilon) = 1/\sqrt{1+\epsilon}$ and take $\epsilon$ to be the infinitesimal part of \erf{eq:SSENorm}, then the normalization factor is
\begin{align}
f(\epsilon) &= 1 - \frac{\epsilon}{2} + \frac{3 \epsilon^2}{8} \\ \nonumber
&= 1 - \frac{1}{2} \<M + M^\t \> dW - \frac{1}{8} \<M + M^\t \> dt.
\end{align}
which is simply a Taylor expansion of $f$ to first order in $dt$ and second order in $dW$. The normalized state, and hence our stochastic Schr\"odinger equation is the product $f(\epsilon)|\bar{\psi}(t+dt)\>$, which when expanded out becomes
\begin{align} \label{eq:SSE}
|\psi(t&+dt)\> = |\psi(t)\> + (M - \frac{1}{2}\<M+M^\t\>) |\psi(t)\> ~ dW \\ \nonumber
&- \frac{1}{2}\left[ M^\t M - M\<M+M^\t\> + \left(\frac{\<M+M^\t\>}{2}\right)^2 \right] |\psi(t)\> ~ dt
\end{align}
The above expression is a bit messy, but it does simplify reasonably well if we take $M$ to be Hermitian, in which case 
\begin{align}
|\psi(t+dt)\> = |\psi(t)\> + (M-\<M\>)|\psi(t)\> ~ dW - \frac{1}{2}\left(M-\<M\>\right)^2 |\psi(t)\>~dt
\end{align}.

The above equations give a reasonably complete model for weak and continuous measurement. The one important aspect of open systems that we have not covered fully is decoherence. We did see how decoherences arises from a discrete weak measurement process in \erf{eq:POVMDecoherence}, but this model does not realistically capture a qubit coupled to its environment, which is almost always a continuous process. To perform the statistical average over all possible measurement outcomes as we did in \erf{eq:POVMDecoherence}, we need to work with density matrices. The stochastic master equation associated to \erf{eq:SSE} above may be computed by expanding $\rho(t+dt) = |\psi(t+dt)\>\<\psi(t+dt)|$ out to first order in $dt$ and second order in $dW$. After a bit of algebra, the result is
\begin{empheq}[box=\widefbox]{align} \label{eq:SME}
d\rho &= \mathcal{D}[M]\rho ~ dt + \mathcal{H}[M]\rho ~ dW \\ \nonumber
\mathcal{D}[M]\rho &\equiv M \rho M^\dagger - \frac{1}{2}(M^\dagger M \rho + \rho M^\dagger M) \\ \nonumber
\mathcal{H}[M]\rho &\equiv M\rho + \rho M^\dagger - \langle M+M^\dagger \rangle \rho \\ \nonumber
dr &= \<M+M^\t\> dt + dW.
\end{empheq}
We have included the expression for $dr$ to emphasize that its form is intimately tied to the form of the stochastic master equation. 

\erf{eq:SME} contains a great deal of physical intuition. Note that if the argument of $\mathcal{H}$ is antihermitian then it reduces to a commutator. We saw this feature before in \erf{eq:DiffusePOVM} when we took $\delta=\pi/2$, in which case the Kraus operator became unitary. In both cases, we identify types of measurements that look like application of a random Hamiltonian to the system, which is the model of decoherence that we used in chapter \ref{ch:IntroQuantum} when considering Ramsey measurements. As we will show below, when $M$ is Hermitian, \erf{eq:SME} describes a QND measurement, and $\mathcal{H}$ tends to push $\rho$ toward an eigenstate of $M$. In the general case, $M$ can lead to a non-QND process like spontaneous emission, which inexorably pushes $\rho$ toward a certain state or set of states. As the next exercise shows, the second term $\mathcal{D}$ leads to damping of the off-diagonal elements of $\rho$, a process that we know happens in the long-time limit when \erf{eq:SME} should reduce to a projective measurement. For QND measurements, $\mathcal{D}$ does not change the diagonal elements of the density matrix. Thus only the stochastic $\mathcal{H}$ term corresponds to acquisition of information.

\begin{exercise} \label{ex:QNDMeasurements}
Let $M$ be Hermitian, in which case it has an eigenvector decomposition $M = \sum_i m_i |i\>\<i|$. Ignoring the $\mathcal{H}$ term of \erf{eq:SME}, expand the $\mathcal{D}$ term in the eigenbasis of $M$ and show that $\dot{\rho}_{ij} = -(m_i-m_j)^2/2$. Note that this implies that the diagonal elements of $\rho$ are constant on average, the hallmark of a QND measurement.
\end{exercise}

There are many conventions for \erf{eq:SME} in the literature, but they are all equivalent as long as one modifies $d\rho$ and $dr$ together. Our convention is rather simple because we absorb all unitful quantities into $M$, so that it has units of $\sqrt{\text{frequency}}$. All other conventions may be obtained by adding a prefactor to $M$ and then pulling it out of $\mathcal{D}$ and $\mathcal{H}$. As $dr$ labels a measurement outcome, we are also free to multiply it by a non-zero constant without changing the physics. However the ratio between the $dt$ and $dW$ terms is physical, and quantifies an intrinsic signal-to-noise ratio derived from quantum fluctuations of the auxiliary qubit \textit{i.e.} the Poisson fluctuations in measurement outcomes.

To understand decoherence from \erf{eq:SME}, we can simply average over all possible measurement outcomes $dW$
\begin{align} \label{eq:AverageSME}
d\rho = \mathcal{D}[M]\rho dt.
\end{align}
Note that we recover a linear equation, just like what happened in \erf{eq:POVMDecoherence}. The more experienced reader may have recognized $\mathcal{D}$ as the Lindblad dissipation term. Indeed, Markovian open systems take the form of \erf{eq:AverageSME}, though possibly with more terms. It is an incredibly general model of open systems, and appears routinely in both artificial and natural settings.

\begin{exercise}
Show that the jump process of \erf{eq:NoJumpHamiltonian} also leads to \erf{eq:AverageSME} when we average over the jump and no-jump evolution with the correct probabilities. 
\end{exercise}

From the standpoint of quantum trajectories, \erf{eq:AverageSME} also lets us model loss of the signal, such as if a portion of the measurement signal $dr$ is washed out by classical noise or dissipated before detection. We can model these scenarios as a system undergoing two measurements, one of which we detect and the other we don't. Recall that $M$ contains a factor of $\sqrt{\gamma}$. If we break a single measurement of rate $\gamma$ into two measurements of unequal strength $M_1=\sqrt{\eta\gamma} c$ and $M_2 = \sqrt{(1-\eta)\gamma} c$, then combining \erf{eq:SME} with $M_2$ and \erf{eq:AverageSME} with $M_1$ gives
\begin{align} \label{eq:SMEEta}
d\rho &= (\mathcal{D}[\sqrt{1-\eta}M]\rho + \mathcal{D}[\sqrt{\eta}M]\rho) dt + \mathcal{H}[\sqrt{\eta}M]\rho ~ dW \\ \nonumber
&= \mathcal{D}[M]\rho ~ dt + \sqrt{\eta}\mathcal{H}[M]\rho ~ dW \\ \nonumber
dr &= \sqrt{\eta}\< M + M^\t\> + dW.
\end{align}
$\eta$ is called the \definition{quantum efficiency}, which is a crucial figure of merit in continuous measurement experiments. When $\eta=0$, \erf{eq:SMEEta} reduces to \erf{eq:AverageSME}, indicating that a zero-efficiency measurements is just a decoherence process.

\subheading{Solving the stochastic master equation} The stochastic master equations have two primary uses, one experimental and one theoretical. Theoretically, they allow us to predict possible quantum trajectories for a system and compute the average dynamics. The simplest way to compute solutions is with Monte-Carlo simulation, in which one generates $dW$ using a random number generator and then computes $\rho(t+dt) = \rho(t) + d\rho$ iteratively. There are also numerous modifications of this procedure with better numerical stability such as \cite{rouchon2015efficient} and the analytic solution obtained below. One can also calculate $P(\rho)$ directly by converting the master equation to a \definition{Fokker-Planck equation}, for which there is a standard procedure. The experimental use of the master equations is the computation of a predicted state $\rho(t,r(t))$ from a sequence of measurement outcomes $dr$. In this case, one computes $dW$ from $dr$ rather than from a random number generator.

In either case, it can be extremely useful to have an analytic solution to the master equation. \erf{eq:LinearSSE} provides an equivalent linear equation that generates the same trajectories as the full nonlinear stochastic Schr\"odinger equation, which is harder to solve directly. If $M$ is Hermitian, it is possible to solve in complete generality. Among its many uses, the solution will let us verify that these stochastic differential equations generate projective QND measurements in this case. 

Solving the stochastic Schr\"odinger equation amounts to computing a finite-time Kraus operator from a product of infinitesimal Kraus operators. This computation becomes easier if we identify an exponential that generates \erf{eq:LinearSSE}, as exponentials turn multiplication into addition. By Ito's lemma, one can verify by direct computation that the following exponential is equivalent to \erf{eq:LinearSSE}
\begin{align} \label{eq:SSEPOVMGenerator}
    \Omega_{dW} &= \sqrt{P(dW)} \exp\left(M dW - \frac{1}{2}(M^\t M dt + M^2)dt \right) \\ \nonumber
    &= \sqrt{P(dW)} \exp(M dW - M^2 dt).
\end{align}
In the second line we have used our assumption that $M=M^\t$. As $[M,M^2]=0$, the product of many copies of \erf{eq:SSEPOVMGenerator} is easily computed
\begin{align} \label{eq:SSESolution}
    \Omega_r &= \left(\frac{1}{\sqrt{2\pi T}} e^{-r^2/2T}\right)^{1/2} \exp(M r - M^2 T) \\ \nonumber
    &= \frac{1}{(2\pi T)^{1/4}} \exp\left(-\frac{(r-2M T)^2}{4T}\right)
\end{align}
where $T$ is the time interval over which we compute our solution, $r = \int_0^T dW$ is the measurement record, and we have computed the prefactor using an argument similar to that of footnote \ref{foot:SSENorm} on page \pageref{foot:SSENorm}. 
$\Omega_r$ characterizes the full solution to the master equation in the QND case. The solution demonstrates that the action of measurement over a finite time interval is only determined by the integral of the measurement record, so that $r$ uniquely determines everything.  

As we have kept track of normalization, \erf{eq:SSESolution} allows us to compute the probability for a given measurement outcome $r$ to occur.\footnote{Recall the issue with \erf{eq:LinearSSE} was that it did not reproduce the correct statistics if we took $dW$ to be a Gaussian random variable with zero mean. Here we are taking $dW$ and its integral $r$ to be unspecified, so that we may determine their probability distributions from $\Omega_r$.} Given a state $\rho$ written in the eigenbasis of $M$ as $\rho = \sum_{ij} \rho_{ij} |i\>\<j|$, we find
\begin{align} \label{eq:SSESolutionPr}
    P(r) &= \Tr[\Omega_r \rho \Omega_r^\t] \\ \nonumber
    &= \frac{1}{\sqrt{2\pi T}} \sum_i \rho_{ii} \exp\left( -\frac{(r-2m_i T)^2}{2T}\right)
\end{align}
where $m_i$ are the eigenvalues of $M$. Satisfyingly, the probability distributions are a series of Gaussians weighted by the classical probabilities $\rho_{ii}$, indicating that the measurement outcomes derive from the diagonal elements of the density matrix as expected. Note also that when $\rho$ is an eigenstate of $M$, $\rho$ and hence $\<M+M^\t\>$ are constant and the probability distribution of $\int_0^T dr$ defined in \erf{eq:SME} defined previously coincides with the above.

\begin{figure}
\centering
{\includegraphics[width = 160mm]{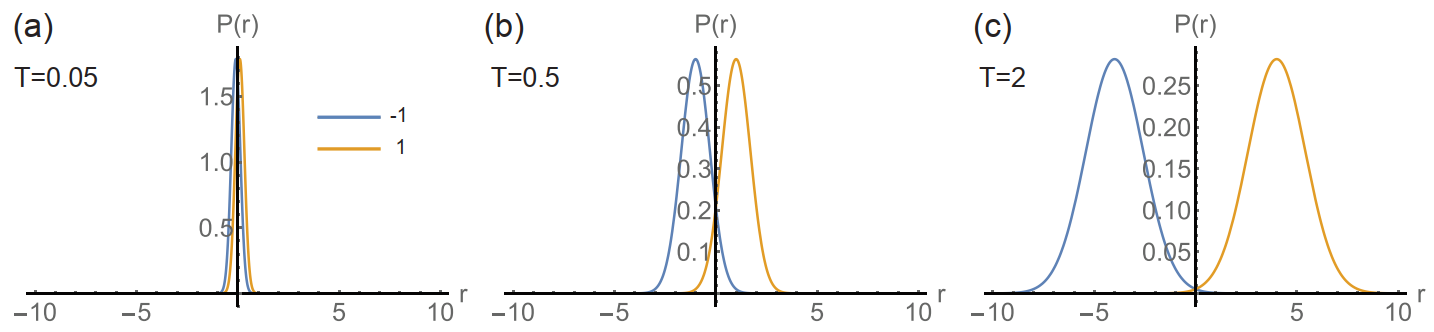}}
\caption{Probability distribution of the integrated measurement record $dr$ for three different time intervals. The signal increases faster than the noise as a function of integration interval, so that the distributions become more distinguishable.}
\label{fig:MeasOutcomes}
\end{figure}

The above solution interpolates nicely between weak and projective measurement. We plot the probabilities $P(r)$ in Fig. \ref{fig:MeasOutcomes} for $M=\sigma_z$ (so ignoring units for the moment). In the small $T$ limit, the Gaussians defined by \erf{eq:SSESolutionPr} overlap, indicating that an inconclusive measurement outcome is likely. Likewise, all diagonal elements of $\Omega_r$ are non-zero. As $T$ increases, the separation between the Gaussians scales as $T$ while the standard deviation of each Gaussian only grows as $\sqrt{T}$. In the long time limit, the overlap between the Gaussians becomes negligible, and the measurement outcome is decisive with high probability. In this same limit, when $r$ is likely near $2 m_i T$, $\Omega_r$ limits to a projection operator, recovering the projective measurement limit.

We conclude with a few general remarks on master equations. Some of the most important properties of \erf{eq:AverageSME} are mathematical and explain its generality in describing open quantum systems. Just as the Kraus representation is in a sense the most general possible quantum evolution, \erf{eq:AverageSME} is related to the most general form of Markovian continuous quantum evolution. An important generalization of \erf{eq:AverageSME} that occurs for example in systems coupled to multiple baths is
\begin{align} \label{eq:GKLSGenerator}
\dot{\rho} = -i[H,\rho] +  \sum_{i,j} \gamma_{i,j} \left(F_i \rho F_j^\t - \frac{1}{2}\left[ F_j^\t F_i \rho + \rho F_j^\t F_i\right] \right)
\end{align}
where $F_i$ form an orthonormal matrix basis ($\Tr[F_i F_j] = \delta_{ij}$) and $\gamma_{i,j}$ is called the Kossakowski matrix\ped{\cite{breuer2002theory}}\major{\cite{gorini1976completely, lindblad1976generators}}. If $\gamma_{i,j}$ is positive semidefinite, then \erf{eq:GKLSGenerator} generates positive semidefinite state evolution and hence is physically valid. Conversely, all completely positive generators can be put in this form. For finite-dimensional systems, we can diagonalize $\gamma_{i,j}$ to find
\begin{align}
\dot{\rho} &= -i[H,\rho] + \sum_i \gamma_i \left(A_i \rho A_i^\t - \frac{1}{2}\left[ A_i^\t A_i \rho + \rho A_i^\t A_i\right] \right) \\ \nonumber
&= -i[H,\rho] + \sum_i \gamma_i \mathcal{D}[A_i]\rho
\end{align}
where there are no particular restrictions on $A_i$, but $\gamma_i \geq 0$ by positivity. Thus although we derived our master equations for a specific measurement model, they are completely general, and essentially define valid Markovian dynamics. These equations, along with the Kraus representation, have the interesting property that they only make systems less distinguishable, and hence tend to erase information\cite{breuer2002theory,Nielsen2010}.





\section{Markovian Baths, Continuous Measurement and the Stochastic Master Equations} \label{sec:MasterEquation}

Now that we have a framework for treating partial measurements, we return to the physical systems considered in the first two chapters in hopes of finding a model of continuous measurement. As an excitation may leave the qubit at one time but reenter it later time, the Jaynes Cummings Hamiltonian models an example of a non-Markovian bath. This would be a bad model of a qubit measurement, as it violates our principle of irreversible information transfer. We can get around this difficulty by adding more oscillators. For a qubit coupled to two harmonic oscillators each with half the strength, it will take twice as long for a single excitation to leave and return. In the limit of an infinite number of oscillators, a lost excitation never returns and the bath becomes memoryless, or Markovian. This system yields a continuous model of measurement that is more physically natural than the discrete operations considered earlier in this chapter. 

Despite the suspicious appearance of infinities, this model underlies the entire theory of open quantum systems. It directly and to excellent approximation models countless physical systems, such as a circuit coupled to a transmission line, resistive loss, an atom coupled to the electromagnetic vacuum, a mechanical mode coupled to a bath of phonons in a solid and a nitrogen vacancy center coupled to a bath of defects or magnetic impurities.
When we assume that the bath of oscillators is unobserved, we obtain the standard form of the master equation, which is the most general equation of motion for a memoryless decoherence process. If we measure the bath, then we obtain a model for continuous measurement that yields the theory of quantum trajectories and feedback. The resulting theory and the steps used to derive it mirror our previous derivation of input-output theory in many respects. 


We begin with a generic system-bath Hamiltonian containing a continuum of harmonic oscillators
\begin{align}
H &= H_\text{sys} + H_\text{bath} + H_\text{int} \\ \nn
H_\text{bath} &= \int_0^\infty \frac{d\omega}{2\pi} \omega ~a^\dagger[\omega] a[\omega] \\ \nn
[a[\omega], a^\dagger[\omega']] &= 2\pi \delta(\omega - \omega').
\end{align}
As in chapter \ref{ch:IntrocQED}, we use the bracket notation $a[\omega]$ to emphasize that $\omega$ indexes an infinite set of operators $a$. We reserve the notation $\mathcal{O}(\omega)$ to denote the Fourier transform of a single time dependent operator $\mathcal{O}(t) = U(t)^\t \mathcal{O} U(t)$ in the Heisenberg picture. Note that as the Hamiltonian has units of $\omega$, $a[\omega]$ has units of $\sqrt{\text{time}}$. We use a more general interaction Hamiltonian that contains the multi-oscillator Jaynes-Cummings Hamiltonian as a special case
\begin{equation} \label{eq:DeriveSMEJaynesCummings}
H_\text{int} = i \int_0^\infty \frac{d\omega}{2\pi} \sqrt{\gamma(\omega)}\left(c a^\dagger[\omega] - c^\dagger a[\omega]\right)
\end{equation}
where the coupling strength parameter $\gamma$ has units of frequency and $c$ could be any system operator. Going into the interaction picture yields
\begin{align} \label{eq:DeriveSMEIntPicture}
H_\text{int}^I \equiv H_I(t) &= e^{i (H_\text{sys} + H_\text{bath})t} H_\text{int} e^{-i  (H_\text{sys}+H_\text{bath})t} \\ \nn
&= i \int_0^\infty \frac{d\omega}{2\pi} \sqrt{\gamma(\omega)}\left( c(t) a^\dagger[\omega] e^{i\omega t} - c^\dagger(t) a[\omega] e^{-i\omega t}\right)
\end{align}%
where $c(t) \equiv e^{iH_\text{sys} t} ~ c ~ e^{-i H_\text{sys} t}$. Recall that $H_\text{sys}^I = H_\text{bath}^I=0$, so Eq. \ref{eq:DeriveSMEIntPicture} fully characterizes the system. 

\begin{figure}
\centering
{\includegraphics[width = \textwidth]{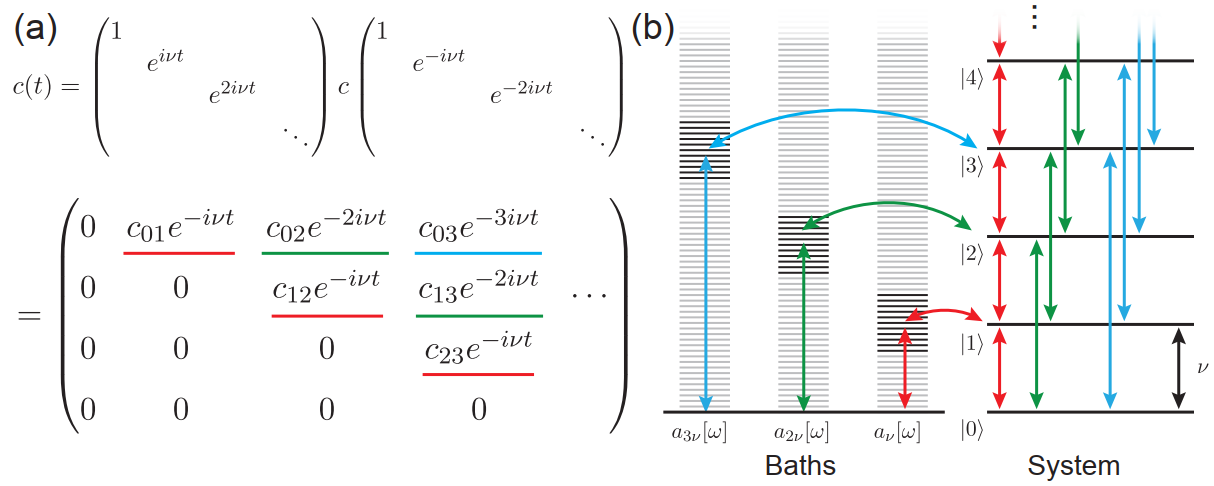}}
\caption{(a) Illustration of $c(t)$ for a harmonic oscillator with frequency $\nu$. Terms further from the diagonal drive multi-excitation transitions, cost more energy and hence couple to different components of the bath. For a non-linear system like a transmon, the $0\leftrightarrow 1$ and $1\leftrightarrow 2$ transitions could differ in frequency, and hence effectively couple to different baths as well. Elements of $c(t)$ below the diagonal have already been dropped via the rotating wave approximation. (b) Splitting the bath up into multiple independent baths, again illustrated for the harmonic case. We replace one bath coupled to all three transitions with three baths, each of which only couples to one transition. The net result is the same as one bath coupled to all three transitions.}
\label{fig:DeriveSMESHO}
\end{figure}

At this point, we have another opportunity to simplify things using the rotating wave approximation. In the eigenbasis of $H_\text{sys}$, $[c(t)]_{i,j} = e^{-i \Omega_{i,j} t} c_{i,j}$ where $\Omega_{i,j} = \nu_j - \nu_i$ is the transition frequency from eigenstate $|i\rangle$ to eigenstate $|j\rangle$ and $\nu_i$ are eigenvalues of $H_\text{sys}$ (see Fig. \ref{fig:DeriveSMESHO}a). We can divide up the problem into groups of degenerate transitions $\Omega$ by defining a set of time-independent operators $c_\Omega$ such that
\begin{equation} \label{eq:DeriveSMEcOmega}
c(t) = \sum_\Omega c_\Omega e^{-i\Omega t}.
\end{equation}
For example in the case of a harmonic oscillator, the sum would take $\Omega = \nu,~2\nu,~3\nu$ \textit{etc.} as illustrated in Fig. \ref{fig:DeriveSMESHO}a. A nonlinear system like a transmon would have its transitions subdivided further. By the rotating wave approximation, terms with $\Omega<0$ may be dropped, so that $c(t)$ and $c_\Omega$ are triangular matrices.\footnote{Whether $c$ is upper or lower triangular depends on whether the we work in a basis in which the diagonal elements of $H_\text{sys}$ are written in ascending order (as is the convention for a harmonic oscillator) or descending order (as is conventional for a qubit), respectively.} One might worry that the rotating wave approximation falls apart for $\Omega=0$, which occurs for diagonal elements of $c(t)$ and for matrix elements corresponding to degenerate transitions. Diagonal elements correspond to permanent dipole moments, which are rare.
Degenerate energy levels are often decoupled from each other and therefore have $c_\Omega=0$. For example, atomic dipole transition strength typically scale as $\Omega^3$.\footnote{To understand why, look up the Einstein A and B coefficients. There is a very general argument  for this scaling based on the thermodynamics of black body radiation.} 
Regardless, one must be careful if these special cases arise.

If we assume that the values of $\Omega_{i,j}$ are well separated relative to the coupling strengths $\gamma(\Omega) c_\Omega$,
then again by the rotating wave approximation, each set of degenerate transitions $i \leftrightarrow j$ couples to a different set of bath operators $a[\omega \approx \Omega_{i,j}]$.
Since the bath Hamiltonian is non-interacting, we can completely decouple these groups of degenerate transitions by thinking of the $a$s around each $\Omega$ as coming from an independent copy of $H_\text{bath}$ with operators $a_\Omega[\omega]$.
\begin{equation}
H_I(t) = \sum_\Omega i \int_0^\infty \frac{d\omega}{2\pi} \sqrt{\gamma(\omega)} \left( c_\Omega a_\Omega^\dagger[\omega]e^{i(\omega-\Omega)t}-c^\dagger_\Omega a_\Omega[\omega]e^{-i(\omega-\Omega)t}\right)
\end{equation}
See Fig. \ref{fig:DeriveSMESHO}. As in this equation each bath only couples to one set of system transitions, the overall dynamics will agree with those of \erf{eq:DeriveSMEIntPicture}. 

At this point, the only approximation we have made is the rotating wave approximation. To make the system tractable, we make a further series of assumptions that are essential to eliminate memory effects, just like we did in input-output theory. Firstly, we assume that $\sqrt{\gamma(\omega)}$ is constant over the relevant bandwidth (a few $\pm \gamma{\Omega}$) around each system transition $\Omega$. This is called the quasimonochromatic approximation. Secondly, we add even more harmonic oscillators to our model by extending the $\omega$ integral to $-\infty$. By the rotating wave approximation, these additional terms do not participate in the dynamics. However their inclusion helps formalize the notion that the bath is uncorrelated with itself in time, by letting us parameterize $H_I(t)$ in terms of temporal mode operators $b_\Omega[t]$
\begin{align} \label{eq:DeriveSMEMarkovH}
H_I(t) &= \sum_\Omega i\sqrt{\gamma(\Omega)} \left(c_\Omega b^\dagger_\Omega[t] - c^\dagger_\Omega b_\Omega[t]\right) \\ \nn
b_\Omega[t] &\equiv \int_{-\infty}^\infty \frac{d\omega}{2\pi} a_\Omega[\omega] e^{-i(\omega-\Omega)t} \\ \nn
[b_\Omega[t], b_\Omega^\dagger[t']] &= \delta(t-t').
\end{align}
Notice that $t$ plays a funny dual role in the above equation. It indexes the time-independent operators $b_\Omega[t]$, but also specifies the functional time dependence of $H_I(t)$.\footnote{
In reality there are two equally valid perspectives regarding time dependence of $b$. In the view we take in the main text, we change basis and think of $b[t]$ as acting on a single harmonic oscillator indexed by $t$. In the perspective that $b[t]$ is time-dependent, then it acts on the entire tensor-product Hilbert space of all $\infty$ harmonic oscillators. The point is that in the latter perspective, we work in the temporal mode basis and $b[t]$ factorizes as $I_0\otimes I_{dt} \otimes I_{2dt}...\otimes b_t \otimes....$ where $b$ acts on the single temporal mode labeled by $t$. These perspectives just boil down to where the implicit identity operators should appear, so the reader is free to take whichever perspective is most intuitive. However in contrast, $a[\omega]$ was actually time-independent, as we pulled out its time dependence $e^{-i\omega t}$ and made it explicit.} 
Physically, this tells us that at each time, the system interacts with a different temporal mode $b[t]$, just like the $b_\text{in}$ modes from input-output theory. If an excitation swaps into this mode, it never returns, so the system is memoryless. This is the essence of the Markov approximation.

By the commutation relations above, $b[t]$ behaves somewhat like a $\delta$ function, so \erf{eq:DeriveSMEMarkovH} does not make mathematical sense unless it is integrated over a finite time interval. $H_I(t)$ does not commute with itself at different times, so one cannot compute a time evolution operator using the standard relation $U(\Delta t) = \exp(-i \int_0^{\Delta t} H(t) dt)$. Fortunately this expression hold approximately if $\Delta t$ is short compared to the timescale over which the system changes, $1/\gamma$. On the other hand for the rotating wave approximation to make sense, we should have $\Delta t \gg 1/\Omega$. This establishes a hierarchy of time scales $1/\gamma \gg \Delta t \gg 1/\Omega$.
With these constraints in mind, we can write the time evolution operator over a discrete time interval $\Delta t$ as
\begin{align} \label{eq:SMEUnitaryDerived}
U(t_n, t_{n+1}) &\approx \exp\left(-i \int_{t_n}^{t_{n+1}} H(t) dt\right) = \exp\left(\sum_\Omega \sqrt{\gamma(\Omega)\Delta t}[c_\Omega b_{\Omega,n}^\dagger - c_\Omega^\dagger b_{\Omega,n}]\right) \\ \nn
b_{\Omega,n} &\equiv \frac{1}{\sqrt{\Delta t}}\int_{t_n}^{t_{n+1}} b_\Omega[t]dt \\ \nn
[b_{n,\Omega}, b_{n', \Omega'}] &= \delta_{n,n'}, \delta_{\Omega, \Omega'}
\end{align}
where we have normalized the $b_{\Omega,n}$ operators so that they satisfy the standard commutation relations for a harmonic oscillator. The appearance of $\sqrt{\Delta t}$ may look like a mathematical oddity, but it is also the crucial ingredient in modeling continuous measurements as we saw in the previous section. 

The reader may have noticed that \erf{eq:SMEUnitaryDerived} connects up directly with the previous section deriving the stochastic master equation, specifically \erf{eq:SMEPOVM} at the beginning of the derivation. The main catch is the representation of the environment as a harmonic oscillator $b_{\Omega,n}$ instead of a qubit. As long as the environment harmonic oscillator starts out in its ground state, this distinction vanishes when we consider measurement. Consider expanding out the argument of the exponential in \erf{eq:SMEUnitaryDerived}. Unlike the qubit case, we find terms like $\Delta t c_\Omega^2 b_{\Omega,n}^{\t 2}$ that do not vanish even when we put the $b_{\Omega,n}$ mode in its ground state. Via the Born rule, this leads to an $\mathcal{O}(\Delta t^2)$ probability of detecting two excitations in the $b_{\Omega,n}$ mode, which limits to exactly zero and hence may be neglected. This argument also applies to the cross terms like $c_\Omega c_{\Omega'} b_{\Omega,n}^\t b_{\Omega',n}^\t$; the probability to observe two excitations in two different modes at the exact same instant is also negligible, and only the leading-order $\sqrt{\Delta t}$ terms contribute to detection. Taking \erf{eq:SMEUnitaryDerived} and running it through the derivations of the previous section leads to multiple copies of \erf{eq:SME}, one for each mode
\begin{align} \label{eq:SMEDerived}
    d\rho &= \sum_\Omega \gamma(\Omega) \mathcal{D}[c_\Omega]\rho~dt + \sqrt{\eta\gamma(\Omega)} \mathcal{H}[c_\Omega e^{-i\phi(t)}]\rho ~dW_\Omega \\ \nonumber
    dr_\Omega &= \sqrt{\eta\gamma(\Omega)} \<c_\Omega e^{-i\phi(t)} + c_\Omega^\t e^{-i\phi(t)} \> + dW_\Omega
\end{align}
If we think of \erf{eq:SMEDerived} as coming from the measurement of a single bath, then $dr_\Omega$ is the measurement outcome on that larger bath that has been spectrally filtered around $\Omega$. If the modes are spectrally filtered before performing a measurement, then it is possible to apply a different $\phi_\Omega$ measurement on each part of the bath as well.

The above derivation connects the stochastic master equation to physical measurements on realistic systems. In our qubit model, we derived it from measurements of $\sigma e^{-i\phi} + \sigma^\t e^{i\phi}$. Replacing $\sigma$ with $b$, we recognize these measurements as quadrature detection, for example of a travelling electromagnetic wave. Thus the detection phase $\phi$ could be the phase of the pump of a parametric amplifier. An alternative method to perform such an electric field measurement is to interfere the signal with a strong coherent state on a beam splitter, resulting in heterodyne or homodyne measurement\cite{Wiseman2009book}.


We have avoided a number of important and interesting subtleties in the above derivation that are worth noting.

\begin{enumerate}
    \item We assumed that the system transitions $\Omega$ were either perfectly degenerate or well-separated in frequency. When $|\Omega_i - \Omega_j| \sim \gamma$, interference between decay pathways can lead to a beating effect, which occurs only if the final states are the same\ped{\cite{scully1999quantum}}.

    \item We have asserted a portion Markov assumption just by writing the interaction Hamiltonian as we did. Normally the coupling term would consist of an integral over the spatial extent of the system, since each part of it couples to the bath. The first Markov approximation assumes that the system is much smaller than the characteristic interaction time $\gamma$ multiplied by the time it takes for information to propagate within the bath (so $c$ for an electromagnetic field). This assumption is related to the Lamb Dicke regime from atom trapping and cooling. 
    
    
    \item Many of the assumptions we made to simplify the math, but could in principle be omitted to obtain the same result. For example, interacting baths can still be Markovian. As long as information quickly spreads in the bath, it is unlikely to return to the system and \erf{eq:SMEDerived} is likely to hold in some form. 
    
    \item We can also modify the state of the bath and still recover a Markovian master equation. There are standard master equations for illumination with thermal and even squeezed light\cite{gardiner2004noise,Wiseman2009book}. The latter case introduces pair-wise photon correlations, which in a sense reintroduces the $c^{\t 2}b^2$ terms that we have dropped.
    
    \item A good deal of physics can be extracted by going slightly beyond the Markov approximation. If the coupling strength $\gamma(\Omega)$ is relatively flat over a small bandwidth and we have the ability to tune the energy levels of our system, then we can map out $\gamma(\omega)$, the excitation spectrum of the bath and even its density of states (which we have assumed to be flat when performing our $dk$ integral with no $k$-dependent prefactor)\cite{clerk2010noise}. If the density of states goes to zero around one of the system's transition frequencies, then bath modes become spatial exponential decay instead of travelling waves, and we expect the formation of photonic bound states\cite{liu2017bandgap}.
\end{enumerate}

\begin{exercise}
As a simple application of the master equation, let $d\rho = \frac{1}{T_1}\mathcal{D}[\sigma]\rho dt$, which is the master equation for unobserved spontaneous emission from a qubit. Show that $\rho_{ee}$ decays to zero at a rate $1/T_1$ and $\rho_{eg}$ decays at half that rate. This shows that $T_2 \leq 2T_1$, a basic fact of NMR and qubit spectroscopy.
\end{exercise}

\section{Quantum Non-Demolition Measurement and Circuit QED Readout}
\label{sec:QEDReadout}

The previous section gave a general physical derivation of the master equation. However there is one important case that it does not cover. The derivation required the diagonal elements of $M$ to be zero in the system's energy eigenbasis, so that the system-environment coupling comes in at finite frequency. While this constraint is satisfied in many realistic situations, it does not hold for the dispersive Hamiltonian $H=\chi \sigma_z a^\t a$ and the resulting qubit readout mechanism described in section \ref{sec:IOTheory} of chapter \ref{ch:IntrocQED}. The reason is that a cavity is not a Markovian bath. It has is a finite correlation time set by the linewidth $\kappa$. Equivalently, the density of states is Lorentzian, whereas the master equation derived from a flat density of states.

While a cavity does not act as a Markovian bath for a qubit, the finite linewidth of a cavity does derive from a Markovian bath of some kind, such as the transmission line considered in the context of input-output theory. We can understand this system using the above formalism, which gives us a starting point to understand circuit QED readout. Starting with \erf{eq:SMEEta} with $M=a e^{-i\phi}$, the stochastic master equation for a qubit coupled to an open cavity in the rotating wave approximation is
\begin{align} \label{eq:DeriveDispersiveSME}
    d\rho &= -i [H,\rho] dt + \kappa \mathcal{D}[a]\rho~ dt + \sqrt{\eta \kappa} \mathcal{H}[a e^{-i\phi}]\rho ~dW \\ \nonumber
    H &= \omega_\text{Cav} a^\t a + \omega_q \frac{\sigma_z}{2} + g(a^\t \sigma + a \sigma^\t) + \epsilon^*(t)a 
    \\ \nonumber
    dr &= \<a e^{-i\phi} + a^\t e^{i\phi}\> dt + dW.
\end{align}
We can transform the above equation into a rotating frame using the standard techniques. As we saw in chapter \ref{ch:IntroQuantum}, the Jaynes-Cummings Hamiltonian preserves the qubit state if $\Delta \gg g$, so qubit readout works best in this regime. In this case, we can make the dispersive approximation, which replaces $H$ with $\chi a^\t a \sigma_z$. From \erf{eq:DispersiveHPert}, we saw that the Jaynes-Cummings interaction dresses qubit and cavity excitations. Exercise \ref{ex:SchriefferWolff} of that section implies that the ladder operators should therefore be dressed as well, necessitating the replacement $a \rightarrow a - [\sigma a^\t, a]g/\Delta = a + \sigma g/\Delta$. In the interaction picture, this operator becomes $\tilde{a} = a e^{-i\omega_\text{Cav} t}+\sigma e^{-i \omega_q t} \frac{g}{\Delta}$, and \erf{eq:DeriveDispersiveSME} becomes
\begin{align} \label{eq:DeriveDispersiveSMEFull}
    d\rho &= -i[H_\text{Disp}, \rho]dt + \kappa \mathcal{D}[\tilde{a}] \rho ~ dt + \sqrt{\eta \kappa} \mathcal{H}[\tilde{a}e^{-i\phi}]\rho~ dW \\ \nonumber
    &\approx -i[H_\text{Disp}, \rho]dt + \kappa \mathcal{D}[a]\rho ~ dt + \sqrt{\eta \kappa} \mathcal{H}[a e^{-i(\phi+\omega_\text{Cav}t)}]\rho ~dW \\ \nonumber
    &~~~+ \kappa\frac{g^2}{\Delta^2}\mathcal{D}[\sigma]\rho ~dt + \sqrt{\eta\kappa\frac{g^2}{\Delta^2}} \mathcal{H}[\sigma e^{-i(\phi+\omega_q t)}]\rho ~ dW \\ \nonumber
    dr &= \<a e^{-i(\phi+\omega_\text{Cav}t)} + a^\t e^{i(\phi+\omega_\text{Cav}t)}\>dt + \frac{g}{\Delta}\<\sigma e^{-i(\phi+\omega_q t)} + \sigma^\t e^{i(\phi+\omega_q t)}\>dt + dW. \\ \nonumber
    H &= \chi a^\t a \sigma_z
\end{align}
In going from the first to the second line, we have expanded the $\mathcal{D}$ term and dropped cross terms like $a\rho \sigma^\t e^{i(\omega_q-\omega_\text{Cav})t}$ via the rotating wave approximation. However we cannot drop fast-rotating terms that are proportional to $dW$. $dW$ has no correlation time, so it is equivalent to white noise when Fourier transformed. As white noise contains a non-zero spectral component at all frequencies, it leads to a resonant effect even if multiplied by a fast rotating term.

\erf{eq:DeriveDispersiveSMEFull} contains a great deal of physics and is worth taking a minute to digest. The second line contains the first-order effects evident from our starting point: cavity decay and measurement-induced back-action arising from observing the emitted field. The third line predicts qubit decay at a rate $\kappa g^2/\Delta^2$, which is known as the \definition{Purcell effect}. There is a corresponding measurement back-action term $\mathcal{H}[\sigma]$ on the qubit, which arises from observing the photons that it emits via the cavity. The last line shows that the signal emitted by the cavity contains two terms, one oscillating at the cavity frequency and one oscillating at the qubit frequency. These terms correspond to cavity and qubit photons respectively.

Purcell decay harms qubit readout and coherence, so we are better off working with a small value of $g/\Delta$. Purcell decay can also be eliminated via a Purcell filter, which decreases the density of electromagnetic states near the qubit frequency\cite{sete2014purcell}. Either way, we work in a regime in which the Purcell decay terms can be neglected. In chapter \ref{ch:AdaptivePhase}, we will see an example of quantum trajectories derived from spontaneous emission that resembles the above.

In dropping the qubit terms of \erf{eq:DeriveDispersiveSMEFull}, it seems like we are back to square one and further from our goal of describing qubit readout. To recover qubit readout, we need to solve for the qubit-conditioned dynamics of the cavity. It turns out that we can solve for these dynamics exactly if we neglect Purcell decay\cite{gambetta2008trajectory,criger2016multi}. Suppose for a moment that the qubit is in an eigenstate of $\sigma_z$. In this case, we know from our example application of input-output theory in chapter \ref{ch:IntrocQED} that the cavity remains in a coherent state at all times. 
If we add a drive term into the dispersive Hamiltonian, then the steady-state cavity displacement is given by \erf{eq:ReadoutCoherentStates}. For notational convenience, it pays to define an operator representation of this coherent displacement
\begin{align}
    \hat{\alpha} &\equiv \frac{\sqrt{\kappa}\epsilon}{i\hat{\mathcal{O}} - \kappa/2} \\ \nonumber
    \hat{\mathcal{O}} &\equiv \chi \sigma_z
\end{align}
where we have defined $\hat{\mathcal{O}}$ so that our results may be applied to systems larger than a qubit (in the general case one would take $H_\text{Disp} = \hat{\mathcal{O}} a^\t a$). If the qubit is in a superposition, then the cavity coherent displacement depends on the qubit state, and the general solution written in the eigenbasis of $\hat{\mathcal{O}}$ is
\begin{align} \label{eq:PolaronAnsatz}
    |\psi(t)\> &= \sum_i \psi_i |i\> \otimes |\alpha_i\> \\ \nonumber
    \rho(t) &= \sum_{ij} \rho_{ij} |i\>\<j| \otimes |\alpha_i\>\<\alpha_j|.
\end{align}
This solution assumes that the qubit state is not affected by decay or coherent evolution that does not commute with $\hat{\mathcal{O}}$. 

An ideal readout would effectively only depend on the qubit state and not the cavity. Now that we know the cavity dynamics, we can eliminate them with a partial trace. Starting with \erf{eq:DeriveDispersiveSMEFull} in the $g/\Delta \rightarrow 0$ limit, we can immediately eliminate the $\mathcal{D}[a]$ term
\begin{align}
    \Tr_\text{Cav}[\mathcal{D}[a]\rho] = \sum_{ij}\rho_{ij} |i\>\<j|\Tr[\mathcal{D}[a]|\alpha_i\>\<\alpha_j|] = 0.
\end{align}
That the trace is zero follows from the fact that $\mathcal{D}$ preserves the norm of the state and therefore must be traceless. The dispersive interaction term is non-zero but simplifies nicely using $\hat{\alpha}$
\begin{align}
    \Tr_\text{Cav}[&[a^\t a \hat{\mathcal{O}}, \rho]] \\ \nonumber
    &= \sum_{ij} \rho_{ij} \Tr_\text{Cav}[\hat{\mathcal{O}}a^\t a, |i\rangle\langle j| \otimes |\alpha_i\rangle\langle\alpha_j|] \\ \nonumber
    &= \sum_{ij}(\hat{\mathcal{O}}_{ii}-\hat{\mathcal{O}}_{jj})\alpha^*_j \alpha_i \langle\alpha_j|\alpha_i\rangle \rho_{ij}|i\rangle\langle j| \\ \nonumber
    &= \hat{\alpha}[\hat{\mathcal{O}}, \rho^Q]\hat{\alpha}^\t \\ \nonumber
    \rho^Q &\equiv \Tr_\text{Cav}[\rho] = \sum_{ij} \rho_{ij} \langle\alpha_j|\alpha_i\rangle |i\rangle\langle j|.
\end{align}
We have used the fact that the partial trace allows cyclic permutation on operators that act only on the degrees of freedom that we trace out. Finally, the $\mathcal{H}[ae^{-i(\phi+\omega_\text{Cav}t)}]$ term simplifies to a familiar form. If we replace $\phi$ with $\phi-\omega_\text{Cav}t$, we can cancel the explicit time dependence, which corresponds to measuring a fixed quadrature of the field in the rotating frame. Tracing out the cavity then gives
\begin{align}
    \Tr_\text{Cav}[&\mathcal{H}[a e^{-i\phi}]\rho] \\ \nonumber
    &= \sum_{ij} \rho_{ij} |i\>\<j| ~ \Tr[a e^{-i\phi} |\alpha_i\>\<\alpha_j| + |\alpha_i\>\<\alpha_j| a^\t e^{i\phi} - \<a e^{-i\phi} + a^\t e^{i\phi}\>|\alpha_i\>\<\alpha_j|] \\ \nonumber
    &= \sum_{ij} \rho_{ij}|i\>\<j| \<\alpha_j|\alpha_i\> \left[ \alpha_i e^{-i\phi} + \alpha_j^* e^{i\phi} - \left( \sum_k \rho_{kk}(\alpha_k e^{-i\phi} + \alpha_k^* e^{i\phi}) \right)\right] \\ \nonumber
    &= \mathcal{H}[\hat{\alpha}e^{-i\phi}]\rho^Q
\end{align}
Thus the stochastic master equation for the system alone is
\begin{align} \label{eq:SMETraceCav}
    d\rho^Q = -i \hat{\alpha}[\hat{\mathcal{O}}, \rho^Q]\hat{\alpha}^\t~dt + \sqrt{\eta \kappa} \mathcal{H}[\hat{\alpha}e^{-i\phi}]\rho^Q~ dW.
\end{align}

We did not derive the above equation by tracing out a Markovian bath, so in general the commutator term cannot be cast into the Lindblad form $\mathcal{D}[...]\rho$\cite{criger2016multi}. Fortunately the qubit case is especially simple. The diagonal elements of $\rho$ are not changed by the $dt$ term, and the off-diagonal terms are complex conjugates of one another since $\rho$ must be Hermitian. By computing this off-diagonal matrix element, we can match terms and derive an equivalent Lindblad form. The $\mathcal{H}$ term also simplifies nicely in the case of a qubit, leaving us with
\begin{align} \label{eq:QubitSMETraceCav}
    d\rho^Q &= - i\frac{\Delta_{AC}}{2}[\sigma_z,\rho^Q] + \frac{\Gamma_D}{2}\mathcal{D}[\sigma_z]\rho^Q ~dt + \sqrt{\eta\kappa} \mathcal{H}\left[\frac{\alpha_e-\alpha_g}{2}e^{-i\phi}\sigma_z\right]\rho^Q~dW\\ \nonumber
    \Gamma_D &= 2\chi \text{Im}[\alpha_g \alpha_e^*] \\ \nonumber
    \Delta_{AC} &= 2\chi \text{Re}[\alpha_g \alpha_e^*].
\end{align}
$\Delta_{AC}$ is an AC Stark shift resulting from the net coherent displacement of the cavity, and $\Gamma_D$ is the dephasing rate induced by measurement. 

Although our starting point assumed that the $\alpha_i$s had reached their steady-state values, our derivation follows through identically even if they change in time. \erf{eq:SMETraceCav} and \erf{eq:QubitSMETraceCav} are valid generally. If we plug in the steady state values for $\alpha_{g/e}$, we find $\Gamma_D\geq 0$. However during transients, it is entirely possible to have $\Gamma_D$ become negative, indicating non-Markovian (and non-completely positive) dynamics. This unusual situation can arise when the cavity states $\alpha_e$ and $\alpha_g$ are moving closer together, so that the joint qubit-cavity state \erf{eq:PolaronAnsatz} goes from being entangled to being separable. Negative $\Gamma_D$ increases the diagonal elements of $\rho^Q$, as we would generally expect when information flows from the environment back into a system. This classic example of non-Markovian physics is quite similar to the collapse and revival phenomena observed in the Jaynes-Cummings Hamiltonian\ped{\cite{scully1999quantum}}.

\begin{figure}
\centering
{\includegraphics[width = 0.7\textwidth]{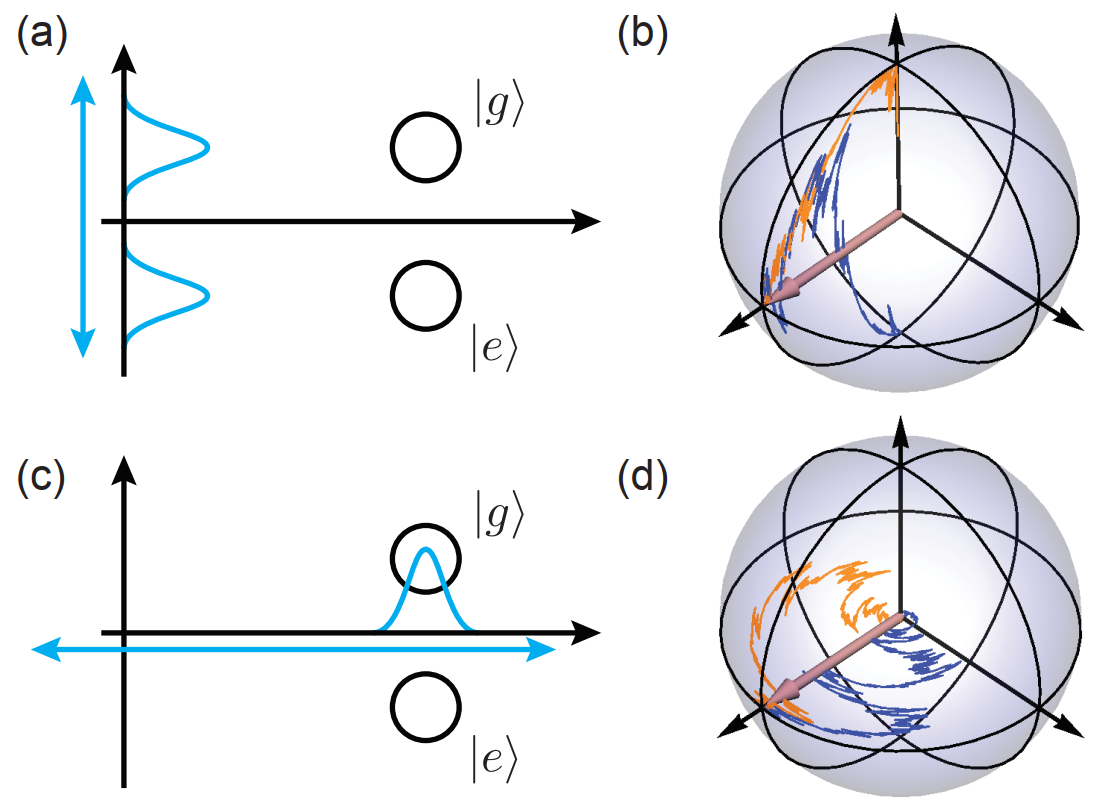}}
\caption{Simulated qubit trajectories via disperisive readout ($\eta=0.4$). (a) Measurement axis (blue) aligned with the quadrature containing qubit state information. Resulting trajectories (b) move toward the eigenstates of the measurement operator. (c) Measurement axis aligned perpendicular to the information-containing quadrature. Although qubit state information is hidden, measurement outcomes to the right end of the distribution project the cavity into a state with higher photon number. (d) Resulting quantum trajectories. Measurement-induced photon number fluctuations lead to stochastic precession around the measurement axis.}
\label{fig:QubitTrajectories}
\end{figure}

The $\mathcal{H}$ term of \erf{eq:QubitSMETraceCav} has an elegant physical interpretation. Its dependence on the difference $\alpha_e-\alpha_g$ shows that the term scales with the amount of qubit information stored in the cavity. If the argument of $\mathcal{H}$ is purely real, then $\mathcal{H}[\sigma_z]$ pushes the qubit up or down along the $\sigma_z$ axis as we would expect when we acquire qubit state information. This scenario corresponds to measuring the field quadrature that is aligned with $\alpha_e-\alpha_g$ in the complex plane, and is mathematically identical to the $\delta=0$ case of \erf{eq:DiffusePOVM}. When the argument of $\mathcal{H}$ is purely imaginary, the term reduces to a commutator $[\sigma_z, \rho^Q]dW$ \textit{i.e.} stochastic rotations about the $\sigma_z$ axis. This is exactly analogous to the $\delta=\pi/2$ case of \erf{eq:DiffusePOVM} when the Kraus operator became unitary. The unitary dynamics of these equations now has an elegant physical explanation; we are now measuring orthogonal to the quadrature that contains qubit state information, but the measurement still learns something about the intercavity amplitude.

%% file: MeasControl.tex
\chapter{Measurement as a Control Resource}
\label{ch:MeasurementControl}

Quantum trajectories provide a much more nuanced picture of decoherence and measurement. Essentially all forms of observations may be described by weak measurement when analyzed at sufficiently short time scales. Thus the dynamics considered in chapter 3 apply almost universally. However an important result that we derived is that the long-time limit reproduces standard projective measurements, and thus our improved understanding does not initially lead to many new capabilities. Quantum feedback, the process of acting on a system during the measurement process, changes this picture dramatically.

This chapter describes a series of applications of continuous measurement to quantum control. The first section describes a method for control with measurement alone using the Zeno effect. While the scheme involves applying a coherent drive during the measurement process, the way in which we apply it does not depend on the measurement outcome. The remaining sections describe actual quantum feedback, in which we drive the system in a way that depends on the sequence of measurement outcomes. Of particular significance is the PaQS feedback equations Eqs. (\ref{eq:PaQSSME},\ref{eq:PaQSCoeffs}) and their generalization Eqs. (\ref{eq:MultiPaQSSME}-\ref{eq:PaQSHessian}). These equations encapsulate and generalize many of the existing measurement-based feedback protocols in the literature and lead to some surprising new results. The final section is also of interest in light of the many developments in remote entanglement generation. The sections of this chapter depend on each other only loosely and may be read in any order.

\section{Control With Measurement Alone: Multi-Qubit Gates Using the Zeno Effect}

\subheading{The Zeno effect} As we have seen, measurement damps the diagonal elements of the density matrix when it is written in the eigenbasis of the measurement operator. This damping can actually prevent coherent transitions between measurement eigenstates. To see this effect, consider the solution to the Rabi model $H=\Omega_R \sigma_y/2$ of chapter \ref{ch:IntroQuantum}, first without any measurement
\begin{align}
    |\psi(t)\> &= \cos\left( \frac{\Omega_R t}{2}\right)|0\> + \sin\left(\frac{\Omega_R t}{2}\right) |1\> \\ \nonumber
\rho(t) &= |\psi(t)\>\<\psi(t)| = \frac{1}{2} \begin{pmatrix}
1+\cos(\Omega_R t)  & \sin(\Omega_R t) \\
\sin(\Omega_R t)    & 1-\cos(\Omega_R t)
\end{pmatrix}.
\end{align}
At $t=0$, the system starts out in $|0\>$. The off-diagonal elements of $\rho$ grow linearly in time to first order, while the population in $|1\>$ only grows quadratically. Thus off-diagonal elements build up first, and only after they are significant does population transfer between $|0\>$ and $|1\>$ begin. Intuitively, this indicates that $|\psi\>$ must pass through a state of large coherence like $|\psi\> \propto |0\>+|1\>$ in order to reach $|1\>$. If the off-diagonal elements are suppressed by a strong measurement of $\sigma_z$ with a rate $\Gamma \gg \Omega_R$, then the system will find itself frozen in the $|0\>$ state despite the coherent drive.  

\begin{exercise} \label{ex:ZenoEffect}
The Zeno effect may also be understood in the context of instantaneous projective measurements. Suppose one performs $N$ projective measurements of $\sigma_z$ while driving a qubit for a duration $T$. Show that the probability to remain in $|0\>$ is $P_\text{success} \approx \left(1-\left(\frac{\Omega_R T}{2N}\right)^2\right)^N$. Show that this limits to one as $N\rightarrow \infty$, implying again that the state becomes pinned where it starts.
\end{exercise}



\subheading{The Zeno gate} In this section, we describe a novel application of the quantum Zeno effect, aimed at demonstrating its utility for control.
In this work and throughout a large fraction of this chapter, we will rely on \definition{degenerate measurements}, which cannot distinguish between certain subspaces, and hence preserve coherence within them. In the context of the Zeno effect, a degenerate measurement allows the system to evolve within a subspace. The state is only frozen in the sense that it is trapped to lie within this subspace, which allows for much richer behavior than in standard Zeno physics. Curiously, the Zeno-restricted dynamics can show new behavior that did not appear under the unmeasured Hamiltonian. This modification of the dynamics represents the concept of Zeno dynamics, which is increasingly considered for quantum control and Hamiltonian engineering in quantum technology\cite{schafer2014experimental,raimond2012quantum}. In particular, Zeno dynamics can theoretically transform a trivial quantum system (such as an array of qubits evolving under an easily solved Hamiltonian) into a universal quantum computer\cite{burgarth2014exponential}. This result, however relies on arbitrary pulse-shaping, which makers it more similar to analog computation and is less amenable to error correction and standard compilation methods. 
In what follows, we instantiate such ideas to a more explicit model of gate-based quantum computation. We will refer to these gates as Zeno gates, which act as two- or multi-qubit Cphase gates. The gate relies only on the ability to drive a single qubit and assumes no qubit-qubit coupling. Strong measurement of the correct subspace turns this trivial, non-interacting system into one with an entangling Hamiltonian. The required measurement can be implemented in circuit QED systems using standard dispersive measurement.

The Zeno gate acting on three qubits is locally equivalent to a Toffoli gate and hence universal for quantum computation when combined with single-qubit operations\cite{shi2002both}. Due to the probabilistic nature of measurement, the Zeno effect can fail, which reduces gate fidelity as it did in the single qubit case. We show that if the measurement channel implementing the Zeno effect is monitored, the gate can be heralded, which yields a probabilistic gate of higher fidelity. Despite its finite error rate, this heralded implementation of the gate can also be used for efficient universal computation, for instance by growing cluster states and then performing adaptive measurement\cite{nielsen2002simple,barrett2005efficient}.

\begin{figure}[htp]
\centering
\includegraphics[width = \textwidth]{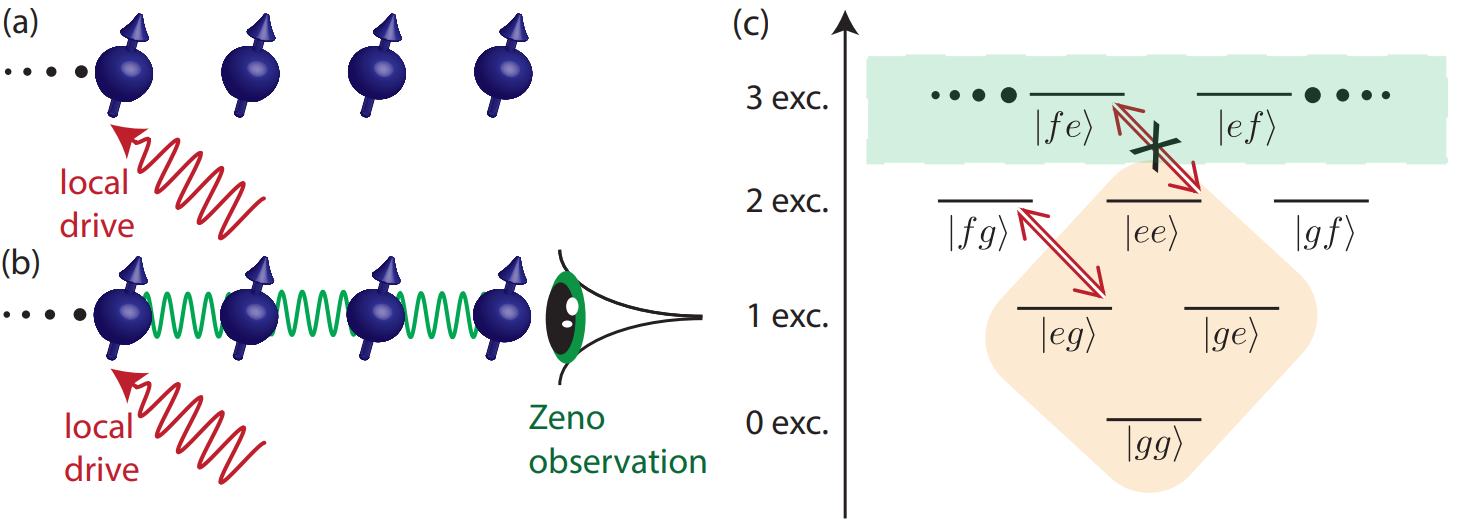}
\caption{(a) Isolated spins with local drives but no spin-spin interaction. (b) Measurement converts local drives into an entangling gate. (c) Energy level diagram for two three-level systems, explaining the basic principle of the Zeno gate. The drive Hamiltonian $H$ induces the transitions drawn in red. Measurement of $P$ blocks the $|fe\rangle$ state, as indicated in green. The combined action of $H$ and $P$ induces a drive on the first qubit conditioned on the second qubit being in $|g\rangle$. Under a full $2\pi$ rotation, a geometric phase is imparted on the $|eg\rangle$ level, which constitutes an entangling gate within the computational subspace (orange).}
\label{fig:LevelDiagram}
\end{figure}

We depict the system under consideration in Fig. \ref{fig:LevelDiagram}a. We consider N 3-level systems such as transmons, with no direct interaction. Our computational sub-space is formed by the lowest two levels of each qutrit. The energy levels are labeled $|g\rangle$, $|e\rangle$ and $|f\rangle$ in order of increasing energy as usual. For simplicity we start with only two qutrits. Our goal will be to implement an entangling gate in the qubit subspace $\{|gg\rangle,|ge\rangle,|eg\rangle,|ee\rangle\}$ by driving $|e\rangle\leftrightarrow |f\rangle$ transitions. The system will always end in the computational subspace, either because we drive full $2\pi$ rotations or because the transition is blocked by the Zeno effect.

\begin{figure}
\centering
\includegraphics[width = 0.45\textwidth]{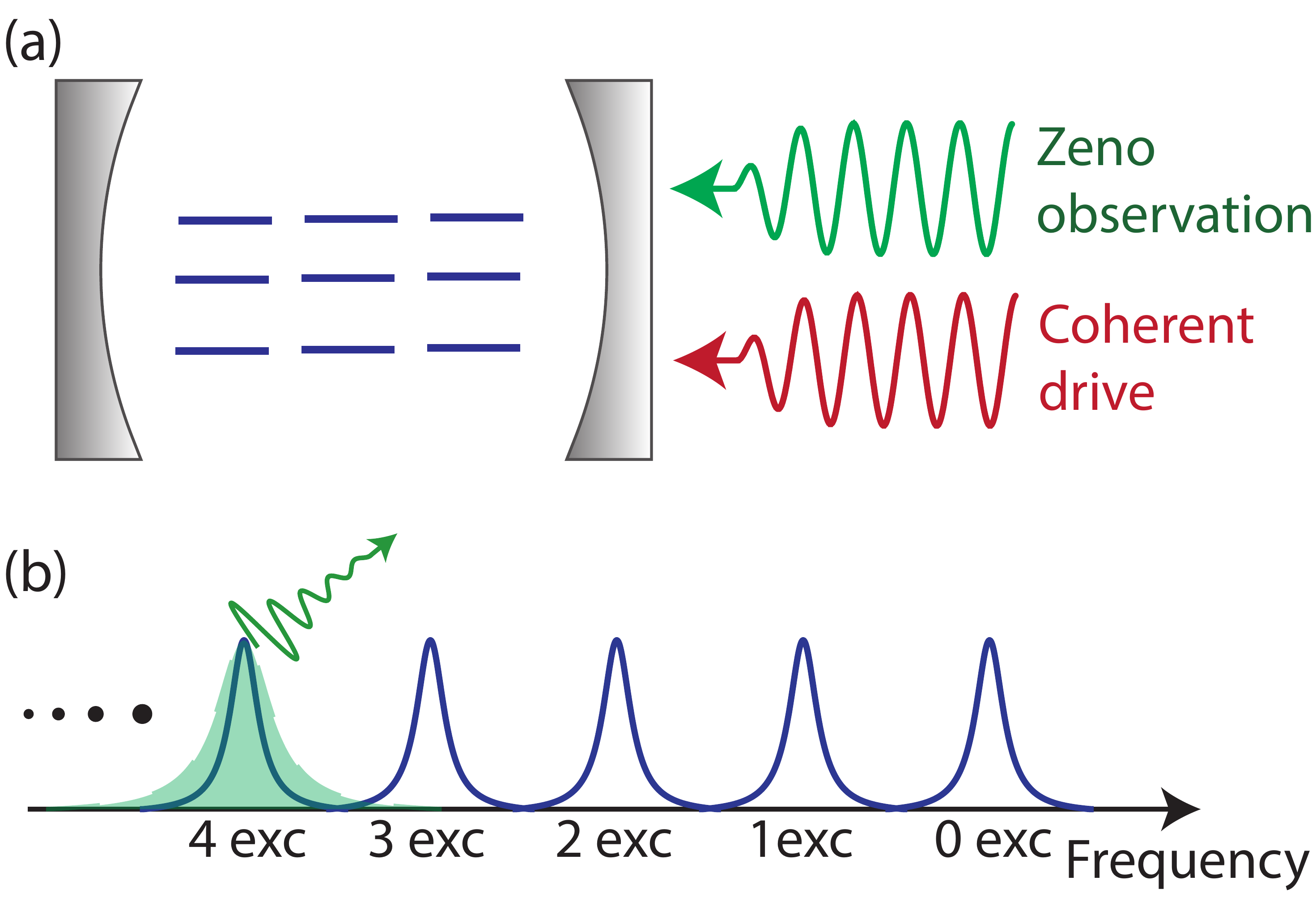}
\caption{Cavity QED implementation of the Zeno gate. (a) Several three-level atoms are coupled dispersively to a high-finesse cavity with leakage rate $\kappa$. The cavity is driven by a classical field $\omega_d$ near its resonance. (b) Cavity density of states as a function of frequency, plotted for all qubit states in the computational basis. This plot can also be interpreted as the intercavity field as a function of drive frequency. Due to the nature of the dispersive Hamiltonian, cavity resonances cluster in groups of equal qubit excitation number, facilitating measurement of $P$.}
\label{fig:cQEDImplementation}
\end{figure}

We first consider the ideal scenario when utilizing the Zeno effect, in which one repeatedly applies the following infinite strength projective measurement. The required measurement interrogates whether the system has exactly $3$ excitations, and cannot distinguish between any other subspaces
\begin{align}
\label{eq:Projector2Q}
P &= 1-|fe\rangle\langle fe| - |ef\rangle\langle ef|.
\end{align}
In the limit considered, strong measurement of $P$ prevents the system from entering the three-excitation subspace. As shown in Fig. \ref{fig:cQEDImplementation}, such a measurement can be implemented with a pair of transmon qubits dispersively coupled to a resonator. Driving the cavity resonance associated with the $|fe\>$ state implements the above measurement, assuming that it is degenerate with $|ef\>$
The gate turns out to be insensitive to the measurement's effect on the $|ef\rangle$, so this degeneracy need not be engineered in practice. During the measurement, we also drive the $|e\rangle\leftrightarrow|f\rangle$ on the first qubit alone
\begin{align}
H = i\frac{\Omega_R}{2}(|e\rangle\langle f|-|f\rangle\langle e|)\otimes I.
\end{align}

The above operations, when applied simultaneously yield a unitary map that is locally equivalent to a Cphase gate when applied for a time duration $t=2\pi/\Omega_R$. Figure \ref{fig:LevelDiagram}c depicts the basic concept. If the systems starts in the computational subspace (orange diamond), then $H$ alone drives the $|eg\rangle\leftrightarrow|fg\rangle$ and $|ee\rangle\leftrightarrow|fe\rangle$ transitions. However the latter transition is blocked by the Zeno effect under continuous measurement of $P$, so $|ee\rangle$ is left untouched. Although under a full $2\pi$ rotation, the $|eg\rangle$ component of the wave function is also mapped to itself, it picks up a geometric phase of $\pi$ in the process. This acquired phase is analogous and equivalent to the global phase acquired by a single qubit undergoing a $2\pi$ rotation. Thus the net operation applies a $\pi$ phase shift to the $|eg\rangle$ component of the state while leaving all other components unchanged. This map is equivalent to a Cphase gate when conjugated by a pair of $\pi$ pulses on the first qubit before and after the gate.

To see this effect explicitly, consider the action of $H$ projected into the 2-excitation-or-less subspace via $P$\cite{facchi2008quantum} 
\begin{align}
\label{eq:DeriveHZeno}
H&_\text{Zeno} = P H P  \nonumber \\
&= i\frac{\Omega_R}{2} P (|e\rangle\langle f|-|f\rangle\langle e|)\otimes(\underbrace{|g\rangle\langle g| + |e\rangle\langle e| + |f\rangle\langle f|}_I)P \nonumber \\
&= i\frac{\Omega_R}{2}(|eg\rangle\langle fg|-|fg\rangle\langle eg|).
\end{align}
This Hamiltonian drives Rabi oscillations on the $|e\rangle\leftrightarrow|f\rangle$ levels of the first system conditioned on the second system lying in the $|g\rangle$ state, which corresponds to the transition diagram \ref{fig:LevelDiagram}. 
%
%
As it only acts non-trivially within a 2-dimensional subspace, we can compute $U_\text{Zeno} = \text{exp}(-i t H_\text{Zeno})$ using Euler's formula
\begin{align}
\label{eq:DeriveUZeno}
U_\text{Zeno} &= (I-\Pi_{eg,fg}) + \Pi_{eg,fg}\cos(\Omega_R t/2) \\ \nonumber 
& ~~ + (|eg\rangle\langle fg| - |fg\rangle\langle eg|)\sin(\Omega_R t/2) \\ \nonumber
& = I-2 \Pi_{eg,fg}
\end{align}
where we have defined the operator $\Pi_{eg,fg} \equiv |eg\rangle\langle eg| + |fg\rangle\langle fg|$ to project into the subspace affected by $H_\text{Zeno}$ and we have chosen $t=t_\text{final} = 2\pi/\Omega_R$ for the second equality. \erf{eq:DeriveUZeno} is the unitary for a controlled phase gate that applies a $\pi$ phase conditioned on occupying the $|eg\rangle$ level, as desired. Note that we can also generate gates with any phase between $0$ and $\pi$. Any geometric phase may be attained by subtending less than a hemisphere of the Bloch sphere when driving the $|e\rangle\leftrightarrow|f\rangle$ transition. This allows one to generate infinitesimal gates, for example to generate Trotterized evolution of a desired Hamiltonian.

$U_\text{Zeno}$ generalizes straightforwardly to $N$ qubits. The relevant measurement operator and Hamiltonian are now
\begin{align}
\label{eq:ProjectorNQ}
&P_N = 1-|f e...e\rangle\langle f e...e| \\ \nonumber
&H_N = i\frac{\Omega_R}{2}(|e\rangle\langle f|-|f\rangle\langle e|)\otimes I \otimes ... \otimes I.
\end{align}
where we have only written the terms of $P_N$ that are relevant to the dynamics. As before, we compute $H_\text{Zeno}$ by conjugating $H_N$ with $P_N$. For notational compactness, we write $H_\text{Zeno}$ by specifying its action only on the relevant subspace. For all $x_i \in \{g,e\}$, we have
\begin{align}
H_{\text{Zeno},N}|g x_2...x_N\rangle &= 0 \\ \nonumber
H_{\text{Zeno},N}|e x_2...x_N\rangle &= \left\{ \begin{array}{lr}
					0 	& : x_2...x_N=e...e\\
					-i|f x_2...x_N\rangle	&  \text{otherwise}
\end{array}
\right. \\ \nonumber
H_{\text{Zeno},N}|f x_2...x_N\rangle &= \left\{ \begin{array}{lr}
					0 	& : x_2...x_N=e...e\\
					i|e x_2...x_N\rangle	&  \text{otherwise}
\end{array}
\right. \\ \nonumber
\end{align}
$H_{\text{Zeno},N}$ acts non-trivially on $2^{N-1}$ uncoupled qubit subspaces $\{|e x_2...x_N\rangle, |f x_2...x_N\rangle\}$. By the same calculation as \erf{eq:DeriveUZeno}, every component of the wave function of the form $|ex_2...x_N\rangle$ except $|ee...e\rangle$ picks up a $\pi$ phase, so that
\begin{equation}
U_{\text{Zeno},N} = \exp(-i2\pi H_{\text{Zeno},N}) = I - 2\Pi_{ex_2...x_N \neq ee...e}.
\end{equation}
Again, any phase between $0$ and $2\pi$ may be applied by subtending the corresponding solid angle on the Bloch sphere of each subspace.

$U_{\text{Zeno},N}$ is locally equivalent to an N-body Cphase gate, which in turn is locally equivalent to an N-qubit Toffoli gate. To implement a Cphase gate, simply apply another $2\pi$ rotation of $H_N$, this time without measurement. Now all kets of the form $|ex_2...x_N\rangle$ acquire a $\pi$ phase, this time including $|ee...e\rangle$. All phases imprinted by this second operation cancel with those of $U_{\text{Zeno},N}$ except that on $|ee...e\rangle$, resulting in an N-body Cphase gate. Furthermore, one may imprint an arbitrary phase on $|ee...e\rangle$ by subtending different solid angles on the first and second application of $H_N$. The full $\pi$ Cphase gate becomes a Toffoli gate when we conjugate any one qubit with a Hadamard gate before and after application of the Cphase gate. 



A significant challenge of measurement-based control is the inevitable introduction of randomness. Although in principle an infinite-strength measurement yields perfect, deterministic control, realistic implementations of the Zeno effect are of finite strength. We now analyze the effects of finite measurement rate. 
%
Continuous measurement of an operator $P$ combined with coherent evolution under Hamiltonian $H$ is described by the master equation
\begin{align} \label{eq:ZenoMasterEquation}
\frac{d\rho}{dt} &= -i[H,\rho] + \Gamma \mathcal{D}[P]\rho
\end{align}
where $\Gamma$ is the measurement rate and $\mathcal{D}[P]\rho \equiv P\rho P^\t - (P^\t P \rho + \rho P^\t P)/2$ is the standard Lindblad dissipator term that models coupling to a Markovian bath\cite{gardiner2004noise}. Note that in the conventions used, measurement damps off-diagonal elements of $\rho$ at a rate of $\Gamma/2$. \erf{eq:ZenoMasterEquation} averages over all possible measurement outcomes weighted according to their probabilities, so it takes into account the possibility that the system may enter the Zeno subspace that we are attempting to block.

When $\Gamma$ is large but finite, the primary source of infidelity is population transfer into the $|fe\rangle$ state. When this occurs, the wave function collapses entirely to $|fe\rangle$ due to measurement and the gate fidelity is zero. As the gate acts as a Rabi drive within the 2-dimensional subspace $\{|ee\rangle, |fe\rangle\}$, we can use a qubit model to estimate the probability of failure. Mapping this subspace to a single qubit $\tilde{\rho}$, the master equation analogous to \erf{eq:ZenoMasterEquation} is
%
\begin{align} \label{eq:QubitZenoME}
\frac{d\tilde{\rho}}{dt} &= -i [\tilde{H}, \tilde{\rho}] + \Gamma \mathcal{D}[|0\rangle\langle0|]\tilde{\rho} \\ \nonumber
\tilde{H} &= \Omega_R \frac{\sigma_y}{2}.
\end{align}
An exact solution is readily available. If the qubit is initialized in $|1\rangle$ and allowed to undergo a full $2\pi$ rotation ($t=2\pi/\Omega_R$), then the probability to find the qubit in $|0\rangle$ is approximately
\begin{align} \label{eq:QubitZenoSol}
\tilde{P}_0 &\approx \frac{1-e^{-4\pi \Omega_R/\Gamma}}{2}
\end{align}
where we have dropped terms of order $\Omega^2$ and terms that are exponentially small in $\Gamma$ (see reference \cite{martin2019zeno} for details). \erf{eq:QubitZenoSol} gives the probability of failure in a single qubit Zeno model. To derive the corresponding probability of failure for the full Zeno gate, we simply multiply by the probability to find the system in $|ee\rangle$, which is $\rho_{ee,ee}$. In this first-order estimate of the gate fidelity, we assume that if the Zeno projection succeeds in preventing a transition to $|ef\rangle$, then the fidelity is $1$. Putting these results together yields a first-order expression for the fidelity as a function of the initial state
\begin{align} \label{eq:FidFiniteGamma}
F_{\text{finite } \Gamma}^{(1)} &\equiv 1-\rho_{ee,ee}(0)P_0 \\ \nonumber
&\approx 1-\rho_{ee,ee}(0)\frac{1-e^{-4\pi \Omega_R/\Gamma}}{2}
\end{align}
which yields $1$ in the ideal Zeno limit.

\erf{eq:FidFiniteGamma} represents the gate fidelity as a function of the initial state. An equally useful figure of merit is gate fidelity averaged over all pure input states, denoted $\bar{F}$. By direct comparison, it can be shown that \erf{eq:FidFiniteGamma} agrees with the average fidelity if one takes the initial state to be $(|gg\rangle+|ge\rangle+|eg\rangle+|ee\rangle)/2$\cite{martin2019zeno}. The associated density matrix has $\rho_{ee,ee}(0)=1/4$, which yields
\begin{align} \label{eq:FidBarFiniteGamma}
    \bar{F}_{\text{finite }\Gamma}^{(1)} &\approx \frac{1}{8}\left(7+e^{-4\pi\Omega_R/\Gamma}\right).
\end{align}
%

\begin{figure}
\centering
\includegraphics[width = 0.45\textwidth]{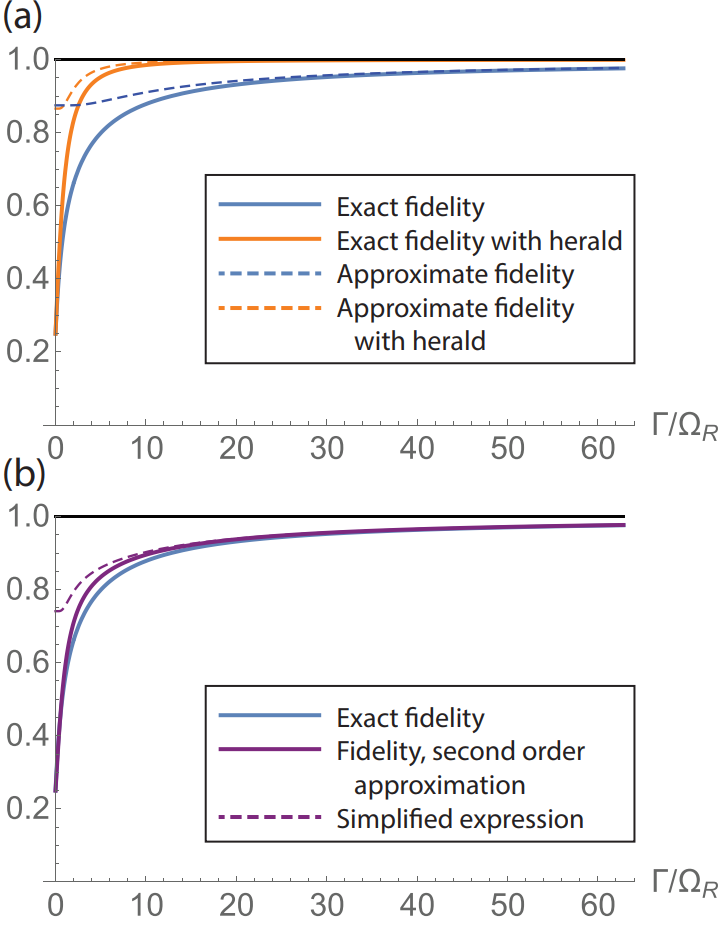}
\caption{Gate fidelity as a function of the unitless ratio $\Gamma/\Omega_R$, which must diverge to reach the ideal Zeno limit. Fidelity is calculated using the initial state $(|gg\rangle+|ge\rangle+|eg\rangle+|ee\rangle)/2$, which is known approximately or in some cases exactly to coincide with the average fidelity. (a) Fidelity for the unheralded (blue) and heralded (orange) gate. Solid blue and orange lines indicate the exact results obtained by solutions of \erf{eq:ZenoMasterEquation} and \erf{eq:HNoJump} respectively, while the blue and orange dashed lines respectively plot the approximate expressions $F_{\text{finite }\Gamma}$ (\erf{eq:FidFiniteGamma}) and $F_\text{herald}$ (\erf{eq:FHerald}). (b) Comparison of the exact fidelity (blue line, same as in (a)) with second-order approximations to it. The solid purple line plots $F_{\text{finite }\Gamma}^{(2)}$ using exact solutions inj \erf{eq:FSecondOrder}, while the dashed purple line plots \erf{eq:FSecondOrderApprox}, in which we have dropped some terms to yield a concise expression.}
\label{fig:FiniteGamma}
\end{figure}

A comparison between the exact Zeno gate fidelity on the above-mentioned initial state (which may also be solved for exactly, though the solution is complicated) and the average fidelity estimated using \erf{eq:FidFiniteGamma}
are plotted in Fig. \ref{fig:FiniteGamma}a in blue. The agreement is good for large values of $\Gamma/\Omega_R$. At smaller values, our assumption of unit gate fidelity conditioned on success of the Zeno measurement breaks down. Furthermore, even at large values of $\Gamma/\Omega_R$, convergence of fidelity to 1 is quite slow.

The above calculation assumes a unit fidelity gate if the measurement outcome indicates successful exclusion from the $|fe\rangle$ state. To go beyond this approximation, we calculate the gate fidelity post-selected on realizations in which the measurement outcome is always so. The corrected fidelity also represents the experimentally accessible scenario of post-selection. The post-selected fidelity converges to one much faster than the unconditional fidelity already calculated.

There exist many equivalent ways to unravel \erf{eq:ZenoMasterEquation}, which represent different physical implementations of the measurement\cite{Wiseman2009book}. For calculational ease, we assume a quantum jump model, in which the $|fe\rangle$ state decays to a continuum of states that we continuously monitor. If we never register population in these auxiliary states, then the system evolves under the following non-Hermitian Hamiltonian 
\begin{align} \label{eq:HNoJump}
    H_\text{no jump} &= i \frac{\Omega_R}{2}(|e\rangle\langle f|-|f\rangle\langle e|)\otimes I + i \frac{\Gamma}{2}|fe\rangle \langle fe|
\end{align}
Note that one propagates the post selected state under $H_\text{no jump}$ only using the Schr\"odinger equation, not the master equation \erf{eq:ZenoMasterEquation}. One can recover \erf{eq:ZenoMasterEquation} by evolving the state with \erf{eq:HNoJump} and randomly applying quantum jumps to $|fe\rangle$ with the right statistics.

The above Sch\"odinger equation may be exactly solved. The solution is algebraically almost identical to that of \erf{eq:QubitZenoME}, and similar simplifications may be made by dropping terms\cite{martin2019zeno}. The main difference is that we now solve for a wave function instead of a density matrix. Setting $t=2\pi/\Omega_R$, we find
\begin{align} \label{eq:HNoJumpSol}
    \psi_{gg}(t) &= \psi_{gg}(0) \\ \nonumber
    \psi_{ge}(t) &= \psi_{ge}(0) \\ \nonumber
    \psi_{eg}(t) &= -\psi_{eg}(0) \\ \nonumber
    \psi_{ee}(t) &\approx e^{-\pi \Omega_R/\Gamma}\psi_{ee}(0) 
\end{align}
where we have also assumed that $\psi_{fe}(0)=0$, which yields $\psi_{fe}(t)=0$. One also finds a non-zero solution for $\psi_{fe}(t)$. However this undesired population can be eliminated by continuing measurement for a time that is large compared to $1/\Gamma$, which exponentially damps $\psi_{fe}$. Thus we take $\psi_{fe}$ to be zero.

The heralded gate fidelity is simply the overlap squared of the above state with the initial state under application of an ideal gate as per \erf{eq:DeriveUZeno}. The only subtlety is that the post-selected wave function above must first be normalized, since evolution under a non-Hermitian Hamiltonian does not preserve the norm. The resulting heralded gate fidelity is
\begin{align} \label{eq:FHerald}
    F_\text{herald} &\approx \frac{1-|\psi_{ee}(0)|^2(1-e^{-\pi \Omega_R/\Gamma})}{\sqrt{1-|\psi_{ee}(0)|^2(1-e^{-2\pi \Omega_R/\Gamma})}}
\end{align}
where we write `$\approx$' because we have used the approximate solutions of \erf{eq:HNoJumpSol}. Due to post-selection, $F_\text{herald}$ is non-linear in $|\psi\rangle$, and thus the usual methods of calculating the average fidelity fail. $\bar{F}_\text{herald}$ is well approximated by taking $|\psi(0)\rangle = (|gg\rangle+|ge\rangle+|eg\rangle+|ee\rangle)/2$ in \erf{eq:FHerald}, just as was the case for the unheralded gate in \erf{eq:FidFiniteGamma} and \erf{eq:FidBarFiniteGamma}\cite{martin2019zeno}.

$F_\text{herald}$ plays a dual role. First interpreting it in the context of a heralded implementation of the gate, we plot the heralded fidelity in orange alongside the unheralded fidelity Fig. \ref{fig:FiniteGamma}a. The fidelity converges to unity orders of magnitude more quickly. In this setting
$\bar{F}_{\text{finite }\Gamma}^{(1)}$ may be interpreted as the success probability, or the fraction of the time in which measurement indicates to jump to $|fe\rangle$. 

Secondly, \erf{eq:FHerald} provides a second-order correction to the unheralded fidelity. As $F_\text{herald}$ explicitly removes the infidelity calculated in $F_{\text{finite }\Gamma}^{(1)}$, we can compute the total fidelity by subtracting from $1$ the infidelity associated to each effect individually. The result is
\begin{align} \label{eq:FSecondOrder}
    F_{\text{finite }\Gamma}^{(2)} &\equiv 1-(1-F_{\text{finite }\Gamma}^{(1)}) - (1-F_\text{herald}) \\ \label{eq:FSecondOrderApprox}
    &\approx \frac{1-|\psi_{ee}(0)|^2(1-e^{-\pi\Omega_R/\Gamma})}{\sqrt{1-|\psi_{ee}(0)|^2(1-e^{2\pi\Omega_R/\Gamma})}} 
    + \frac{|\psi_{ee}(0)|^2}{2}\left(1-e^{-4\pi\Omega_R/\Gamma} \right)
\end{align}
We plot this fidelity against the exact fidelity in Fig. \ref{fig:FiniteGamma}. We also include the fidelity calculated using the exact solutions to \erf{eq:ZenoMasterEquation} and \erf{eq:HNoJump}, which are omitted in the main text for brevity but appear in ref. \cite{martin2019zeno}. The three fidelity curves agree well, indicating that we have quantified the main sources of infidelity in an intuitive, closed-form expression.

The infidelity estimated in $F_{\text{finite }\Gamma}^{(1)}$ has a clear physical interpretation of leakage into the $|fe\rangle$ state. Similarly, the infidelity calculated in $F_\text{herald}$ has a simple physical explanation. Returning to \erf{eq:HNoJumpSol}, observe that the only deviation from an ideal gate is damping of the $\psi_{ee}$ component of the wave function. This damping is a result of information acquisition. Only $\psi_{ee}$ population can lead to population in $|fe\rangle$, so if we do not measure $|fe\rangle$ population, then we can infer a lower likelihood to find the system in $|ee\rangle$. This damping is analogous the case of an atom prepared in a superposition of its excited and ground states, which decays to its ground state even if it does not emit a photon.

\section{General Quantum Feedback Protocols}
\label{sec:PaQS}

In this section, we calculate the locally optimal feedback for a measurement-based control system\cite{Zhang2018Locally}. We leave the measurement and feedback operators completely general. Our optimization technique guarantees that the optimal feedback coefficients are given by ratios of two polynomials that are each linear in $\rho$. Their computation does not involve computing inverse functions, and so they may be found analytically. A key simplifying assumption is that $\rho$ (or $\bar{\rho}$, in the case of Markovian feedback) 
stays close to the locally optimal path at each time step. We check the validity of this assumption during numerical propagation of the master equation and rotate the system to the locally optimal path if necessary. With the goal of state preparation in mind, we use fidelity as our cost function throughout this section, though the calculation works with any figure of merit that is linear in the density matrix, such as the expectation value of any observable.

Suppose we have a state $\rho$ at time $t$, and that we wish to reach some target state $|\psi_T\rangle$. If we take fidelity to be our cost function, then our aim becomes finding the rotation that maximizes the fidelity after a single round measurement and feedback. We take feedback to consist of applying some Hamiltonian $H$, so that we may parameterize the feedback operation as $U(\theta) = \exp(-i H \theta)$. The fidelity after a single round of measurement and feedback is then given by
\begin{align}
\mathcal{F}_{t+dt}(\theta) = \langle \psi_T|\rho^c(\theta)|\psi_T\rangle, \\ \nonumber
\rho^c(\theta) \equiv U(\theta)(\rho+d\rho)U^\dagger(\theta).
\end{align}
Our local optimality condition is given by
\begin{align} \label{eq:G}
\mathcal{G} \equiv \frac{\partial\mathcal{F}_{t+dt}(\theta_\text{opt.})}{\partial \theta_\text{opt.}} = \Big[U'(\theta_\text{opt.})(\rho+d\rho)U^\dagger(\theta_\text{opt.}) + h.c.\Big]= -i \<\psi_T|[H, \rho^c(\theta_\text{opt.})]|\psi_T\> = 0
\end{align}
where $\theta_\text{opt.}$ is the locally optimal rotation that we wish to calculate. For systems much more complicated than a single qubit, the above equation often lacks a closed-form solution. However, by definition our protocol has applied the locally optimal rotation at the previous time step. In this case, since $d\rho$ is infinitesimal, then typically $\theta_\text{opt.}$ will be infinitesimal as well (we deal with possible exceptions below). Thus we can parameterize it as 
\begin{align}
\theta_\text{opt.} = A_1(t)dW + A_2(t)dt.
\end{align}
Inserting this parameterization into the definition of $U(\theta)$ and expanding to $2nd$ order in $dW$ according to Ito's lemma\cite{Oksendal2003} yields 
\begin{align}
U(\theta) = I-iA_1 H dW-\left(i A_2 H +\frac{1}{2}A_1^2 H^2\right)dt.
\end{align}
We now substitute $d\rho$ for the measurement stochastic master equation (\erf{eq:SMEEta}), which yields
\begin{align} \label{eq:PaQSSME}
\rho^c =&
\rho +\mathcal{D}[M]\rho dt + \sqrt{\eta}\mathcal{H}[M]\rho dW -i A_1 [H, \rho] dW + A_1^2 \mathcal{D}[H]\rho dt \\ \nonumber
&- i[H, \sqrt{\eta} A_1(M\rho+\rho M^\dagger)+A_2\rho] dt
\end{align}
The above equation is simply the Wiseman-Milburne equation\cite{wiseman1994feedback} with general coefficients. It serves as the equations of motion for a system undergoing continuous measurement, proportional feedback (via the $A_1[H,\rho]$ term) and evolution under the time-dependent Hamiltonian $A_2 H$. The $\mathcal{D}[H]$ term arises from random, noisy application of $H$ under feedback, which explains why it is present even if the measurement efficiency goes to zero.  The last term derives from correlation between the measurement back-action and the feedback operation applied as a result, so it is most readily interpreted as the feedback term.

\erf{eq:PaQSSME} is quite useful on its own and worth nothing. To find the locally optimal feedback operation, we substitute $\rho^c$ into Eq. \ref{eq:G} and solve $\mathcal{G}=0$ order by order in $dW$. Despite the multitude of terms, it is straight-forward to solve for $A_1$ and $A_2$ in complete generality. The only assumption we make is that the optimal rotation was applied at the immediately preceding time step as mentioned above, so that $\mathcal{F}_t(\theta)$ is maximized at $\theta=0$ as required for consistency with the assumption that $\theta_\text{opt.}$ is infinitesimal.
It implies that $\partial\mathcal{F}_t(\theta)/\partial \theta|_{\theta=0} = -i \langle \psi_T|[H,\rho]|\psi_T\rangle = 0$, so that the first term may be dropped. Terms proportional to $dW$ yield a linear equation in $A_1$ which is easily solved. Once $A_1$ is known, terms proportional to $dW^2=dt$ yield another linear equation, this time for $A_2$. 
The final result in full form is
\begin{align} \label{eq:PaQSCoeffs}
&A_1 = \frac{-i\sqrt{\eta}\langle \psi_T|[H, M\rho+\rho M^\dagger]|\psi_T\rangle}{\langle\psi_T|[H, [H, \rho]]|\psi_T\rangle} \\ \nonumber
&A_2 = \frac{-\langle \psi_T|[H,i\mathcal{D}[M]\rho+\sqrt{\eta}A_1[H,M\rho+\rho M^\dagger]+i A_1^2 \mathcal{D}[H]\rho]|\psi_T\rangle}{\langle \psi_T|[H, [H, \rho]]|\psi_T\rangle}
\end{align}
%
$A_1$ and $A_2$ are directly related to the proportional feedback gain and constant Hamiltonian amplitudes that would be applied experimentally. Using the expression for the measurement record $dr$, the locally optimal feedback rotation is given by
\begin{align} 
\theta_\text{opt.} = \sqrt{\eta} A_1 dr + (A_2-\sqrt{\eta}A_1 \text{Tr}[\rho(t)(M+M^\t)])dt
\end{align}
so that the first term corresponds to proportional feedback gain and the second corresponds to a constant Hamiltonian drive that does not depend on the measurement record. 

So far, we have assumed that the optimal angle is infinitesimal. However, equation \erf{eq:PaQSCoeffs} only guarantees that the solution $\theta_\text{opt.}$ is a local extremum and does not guarantee that it is necessarily a maximum. A sufficient condition for it to be a local maximum is that the second derivative of the fidelity function evaluated at $\theta_\text{opt.}$ be negative
\begin{equation}\label{eqn:2DTest}
\frac{\partial^2\mathcal{F}_{t+dt}(\theta)}{\partial \theta^2} \bigg|_{\theta=\theta_\text{opt.}}= -\<\psi_T|[H_F,[H_F,\rho^c]]|\psi_T\>\bigg|_{\theta=\theta_\text{opt.}}<0.
\end{equation}
%
Failure of this test, \textit{i.e.}, when the second derivative is positive, suggests the presence of a local minimum from the infinitesimal solution. Then we will need a large (\textit{i.e.}, non-infinitesimal) rotation, which we compute by maximizing the fidelity over the entire angular range. 

It should be noted that $A_1$ and $A_2$ can in principle become singular. However as $\rho^c_{t-dt}|_{\theta_\text{opt.}} = \rho_t$, the denominator diverges only when the second derivative test failed at the previous time step (compare \erf{eqn:2DTest}) to the denominator of \erf{eq:PaQSCoeffs}). Thus this divergence is typically prevented by the global search described above. A special case is $[H_F, \rho_t]=0$, in which case feedback has no effect on the state, so that we may simply set $\theta_\text{opt.}=0$.

To simulate this form of feedback in practice, we assume that the controller chooses the rotation angle $\theta_\text{opt.}$ that ensures a global maximum of $\mathcal{F}_{t=0}$ at the initial time step. During evolution of the state, the above protocol typically continues to pick $\theta_\text{opt.}$ as the global maximum of $\mathcal{F}_t$ and thus maintains the system on a locally (time-)optimal trajectory. However even if \erf{eqn:2DTest} remains negative, it is possible that the nearest local maximum of $\mathcal{F}_t(\theta)$ can fail to be the global maximum. The only way to catch such instances is to occasionally undertake a brute-force maximization of $\mathcal{F}$ and to thereby check whether the local maximum identified by equation \erf{eq:PaQSCoeffs} is also a global maximum. In practice, such global maximization procedures are often unnecessary; table \ref{tab:PaQSTable} shows many combinations of $M$ and $H$ that allow \erf{eq:PaQSCoeffs} and \erf{eqn:2DTest} to reproduce the indicated feedback protocols established in the control literature. Global searches are only required when indicated by equation \erf{eqn:2DTest}. We explore some of these applications further in the next section.

\begin{table} \label{tab:PaQSTable}
\begin{center}
\begin{tabular}{|c|c|c|c|}
\hline
\textbf{Feedback} & \textbf{Measurement} & \textbf{Feedback} & \textbf{Target state} \\
\textbf{protocol} & \textbf{operator} ($M$) & \textbf{Hamiltonian} ($H$) & ($|\psi_T\>$) \\
\hline
Adaptive phase & $\sigma$ & $\sigma_z$ (Heisenberg & $(|0\rangle + i|1\rangle)/\sqrt{2}$ \\
measurement\cite{wiseman1995adaptive}$^*$ &  &  picture) &\\
\hline
Rapid qubit & $\sigma_z$ & $\sigma_y$ & $(|g\rangle + |e\rangle)/\sqrt{2}$ \\
purification\cite{Jacobs2003}$^*$ & & &\\
\hline
Half-parity Bell & $\sigma_{z,1}+\sigma_{z,2}$ & $\sigma_{y,1} + \sigma_{y,2}$ & $(|eg\> + |ge\>)/\sqrt{2}$ \\
state preparation\cite{HPFPRA}$^*$ & & &\\
\hline
Full-parity Bell state & $\sigma_{z,1}\sigma_{z,2}$ & $\sigma_{x,1}$ & $(|gg\> + i|eg\>$ \\
 preparation \cite{Hill2008entanglement, martin2017optimal}$^*$& & & $+i|eg\>+|ee\>)/2$\\
\hline
N-qubit Dicke states& $\sum_i \sigma_{z,i}$ & $\sum_i \sigma_{y,i}$ & N-qubit Dicke state\\
 \cite{Thomsen2002,stockton2004dicke,Wei2015,Zhang2018Locally} & & & with n excitations \\
\hline
N-qubit GHZ & $\sigma_{z,i}-\sigma_{z,j}$ & $\sigma_{y,i}-\sigma_{y,j}$ & $N$-qubit GHZ state \\
states$^*$ & for all $i,j$ & for all $i,j$ & \\ \hline
Hong-Ou-Mandel & $i(\sigma_1+\sigma_2)$,  & $\sigma_{x,1}+\sigma_{x,2}$ & $(|gg\>+|ee\>)/\sqrt{2}$ \\ 
Bell state$^*$ & $\sigma_1-\sigma_2$ & $\sigma_{y,1}-\sigma_{y,2}$ & \\ \hline
\end{tabular}
\end{center}
\caption{Summary of some measurement-based feedback protocols that may be derived from \erf{eq:PaQSCoeffs} or its generalization in the following section. Protocols in which perfect noise cancellation (cancellation of $dW$ terms) can occur are marked with an asterisk. Note that for $N$-qubit GHZ states, noise cancellation only occurs for $N=3$.}
\end{table}

\begin{figure}
\centering
\includegraphics[width = \textwidth]{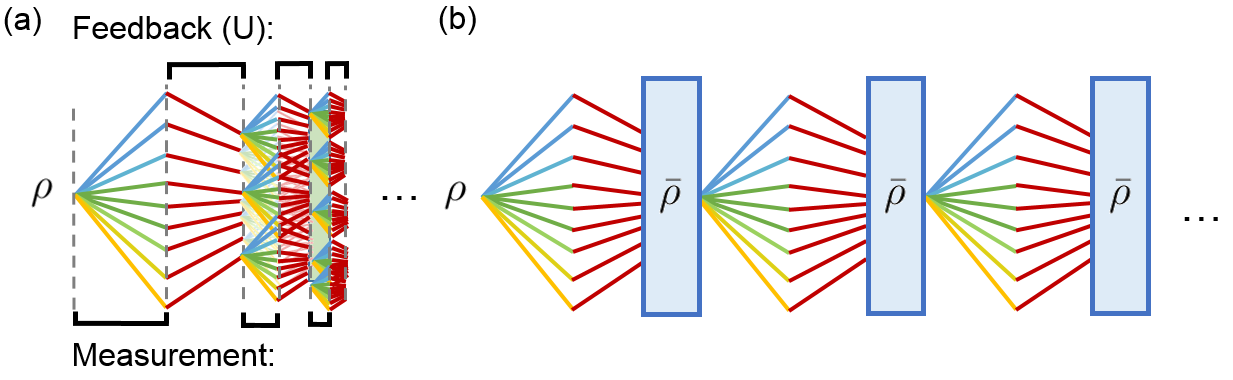}
\caption{(a) A cartoon illustration of a non-Markovian feedback protocol. The number of possible states, and hence the number of potentially distinct feedback operations needed in response scale exponentially in time. (b) The behavior of a forgetful feedback controller, which only applies controls based on the evolution of the average state and the most recent measurement outcome.}
\label{fig:ASLO}
\end{figure}

The functions $A_1$ and $A_2$ are dependent on both the initial and target states, as well as on the state at time $t$. Dependence on the current state implies implicit dependence on the full measurement record, yielding a potentially non-Markovian feedback protocol in general. Non-Markovian protocols are more difficult to implement in practice. The feedback controller must either calculate $\rho(t)$ in real time or perform an exponential amount of precomputation to determine the optimal action for all possible measurement records, as illustrated in Fig. \ref{fig:ASLO}a. However we can also use the feedback master equation to simulate a memoryless controller, which is only capable of implementing Markovian feedback. We illustrate the basic concept in Fig. \ref{fig:ASLO}b. At the initial time step, many possible measurement outcomes $dW$ are possible, leading to many possible conditioned states (red lines). A memoryless controller knows that it has applied feedback at the previous time step, but does not recall the specific measurement outcome. It therefore only has access to the average state $\bar{\rho}$, which may be computed by averaging over \textit{i.e.} dropping the $dW$ term in \erf{eq:PaQSSME}. At the next time step, the feedback controller computes $\theta_\text{opt.}$ based on $\bar{\rho}$ instead of $\rho$, which amounts to replacing $\rho$ with $\bar{\rho}$ in \erf{eq:PaQSCoeffs}. Thus we may generate a Markovian feedback protocol simply by dropping the $dW$ terms in \erf{eq:PaQSSME} and propagating the state. We call such a protocol average sense locally optimal, or ASLO for short. 

In essence, ASLO protocols exploit the subjectivity of $\rho$, so that we can use different calculations to simulate different states of knowledge of our feedback controller. An ASLO feedback controller is memoryless in the same way as a general Markovian open system. The ASLO concept emphasizes that $dW$ is also a subjective quantity. The measurement outcome $dr$ is objective, but the right hand side $\sqrt{\eta}\<M + M^\t\>dt + dW$ contains an expectation value that differs depending on if we use $\rho$ or $\bar{\rho}$. To reproduce the same observable $dr$ no matter the state of knowledge of the observer, $dW$ must also differ based on which we use.

\section{Bi- and Multipartite Entanglement Generation}

Quantum feedback has many potential application, particularly as it can be implemented in almost any measurement process, at least in principle. A natural starting point to apply it is in situations where measurement is already known to be a useful control resource and to see if feedback offers any fundamentally new capabilities.

\begin{figure}
\centering
\includegraphics[width = \textwidth]{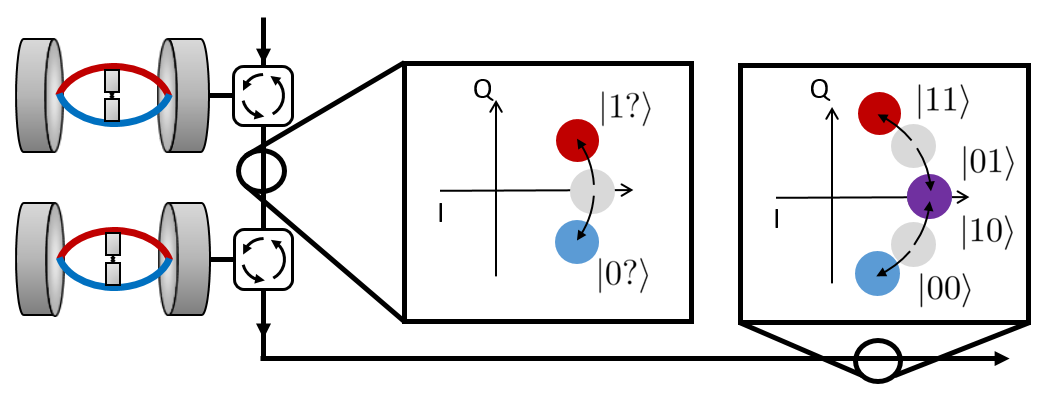}
\caption{A graphical illustration of the bounce-bounce remote entanglement experiment performed in \cite{Roch2014}. A coherent measurement tone interacts sequentially with two cavities, each housing a superconducting transmon qubit. The final coherent state picks up two conditional phase shifts, and cannot distinguish between the $|01\>$ and $|10\>$ states.}
\label{fig:BounceBounce}
\end{figure}

One of the most important applications of quantum measurement from a control standpoint is remote entanglement generation. Entanglement occurs naturally and even automatically in interacting systems. Two atoms in close proximity will hybridize, and two dipole-coupled transmon qubits form Bell states when resonantly exchanging energy. A much more difficult task is to create entanglement between non-interacting or even remote systems. The standard method involves interference between signals emitted by the two systems. If the signals are indistinguishable, so that we cannot tell from which system it came, then we can implement a \textit{degenerate measurements}. An example of such an experiment performed in superconducting circuits is shown in Fig. \ref{fig:BounceBounce}\cite{Roch2014}. In the scheme, a measurement signal interacts sequentially with two different cavities separated by over a meter of cable. The cavities are dispersively coupled to superconducting qubits, so the coherent tone picks up a phase shift conditioned on the qubit states. If the phase shifts are equal in magnitude, then the final output signal cannot distinguish the $|01\>$ and $|10\>$ states. Performing a homodyne measurement of this output field implements measurement of the operator\cite{Motzoi2015}
\begin{align}
M = \sqrt{\frac{\Gamma}{2}}\frac{\sigma_{z,1} + \sigma_{z,2}}{2}.
\end{align}
The above operator is called a half-parity measurement, and it is degenerate in the $|01\>$, $|10\>$ subspace. This can be used to probabilistically generate entanglement by first preparing the separable uniform superposition state $|\psi_0\> = \frac{1}{2}(|00\>+|01\>+|10\>+|11\>)$ and then projectively measuring $M$. Half of the time, the measurement outcomes $\pm1$ occur (dropping unitful quantities) and we collapse into the separable states $|00\>$ or $|11\>$. The other half of the time, the system collapses into the $|01\>$, $|10\>$ subspace while retaining its coherence within that subspace. The result is a Bell state $|\psi^+\> \equiv \frac{1}{\sqrt{2}} (|01\>+|10\>)$. The above process, and many like it, are intrinsically probabilistic. Even if implemented ideally, the success rate cannot exceed 50\%, which means that the unheralded state $\bar{\rho} = \frac{1}{4}|00\>\<00| + \frac{1}{4}|11\>\<11| + \frac{1}{2}|\psi^+\>\<\psi^+|$ has no entanglement. 

The above procedure can be made to produce a Bell state deterministically if we add in continuous feedback during the measurement process. Although not originally derived in this way, we can use \erf{eq:PaQSCoeffs} to derive the globally optimal protocol for this task. All we have to do is choose a suitable feedback Hamiltonian. First and foremost, we must restrict our feedback Hamiltonian to be local, as nonlocal Hamiltonians are simply not available when there is no interaction. Feedback should correct for measurement perturbations toward the undesired states $|00\>$ and $|11\>$. We can rotate between these states and $|\psi_0\>$ with a $\pi/2$ rotation of $H = (\sigma_{y,1}+\sigma_{y,2})/2$, and from $|\psi_0\>$ we can probabilistically prepare the target state. It seems reasonable to guess that infinitesimal rotations of $H$ could steer the measurement outcome toward the desired subspace of $M$ continuously, so we take it as our feedback Hamiltonian. Initializing the system in $|\psi_0\>$ and taking $\eta=1$ for now, the state evolves as
\begin{align} \label{eq:HPFSolution}
    |\psi(t)\> = \frac{1}{\sqrt{2}}\left(e^{-\Gamma t/4}|\phi^+\> + \sqrt{2-e^{-\Gamma t/2}}|\psi^+\>\right)
\end{align}
where $|\phi^+\> \equiv \frac{1}{\sqrt{2}} (|00\>+|11\>)$. We have $|\psi(0)\> = |\psi_0\>$ and $|\psi(t\gg 1/\Gamma) = |\psi^+\>$, indicating that we can deterministically prepare an entangled state from a separable state using feedback. One can derive \erf{eq:HPFSolution} by taking general coefficients and substituting a state of this form into the feedback master equation. This ansatz amounts to assuming that feedback maintains the $0 \leftrightarrow 1$ symmetry of the initial state. A detailed derivation is provided in \cite{HPFPRA}, and also in chapter \ref{ch:OptimalControl} where we prove global optimality of the resulting protocol. For now, we note a striking feature of the result, which is that $|\psi(t)\>$ only depends on time, and not on any function of the measurement record. This occurs because the stochastic terms of the master equation, one coming from measurement back action and one from feedback, cancel exactly. This cancellation is somewhat surprising, given that feedback is nonlocal while measurement is not. We remark on the generality of this phenomenon at the end of this section.

\begin{figure}[htp]
\centering
\includegraphics[width = 0.7\textwidth]{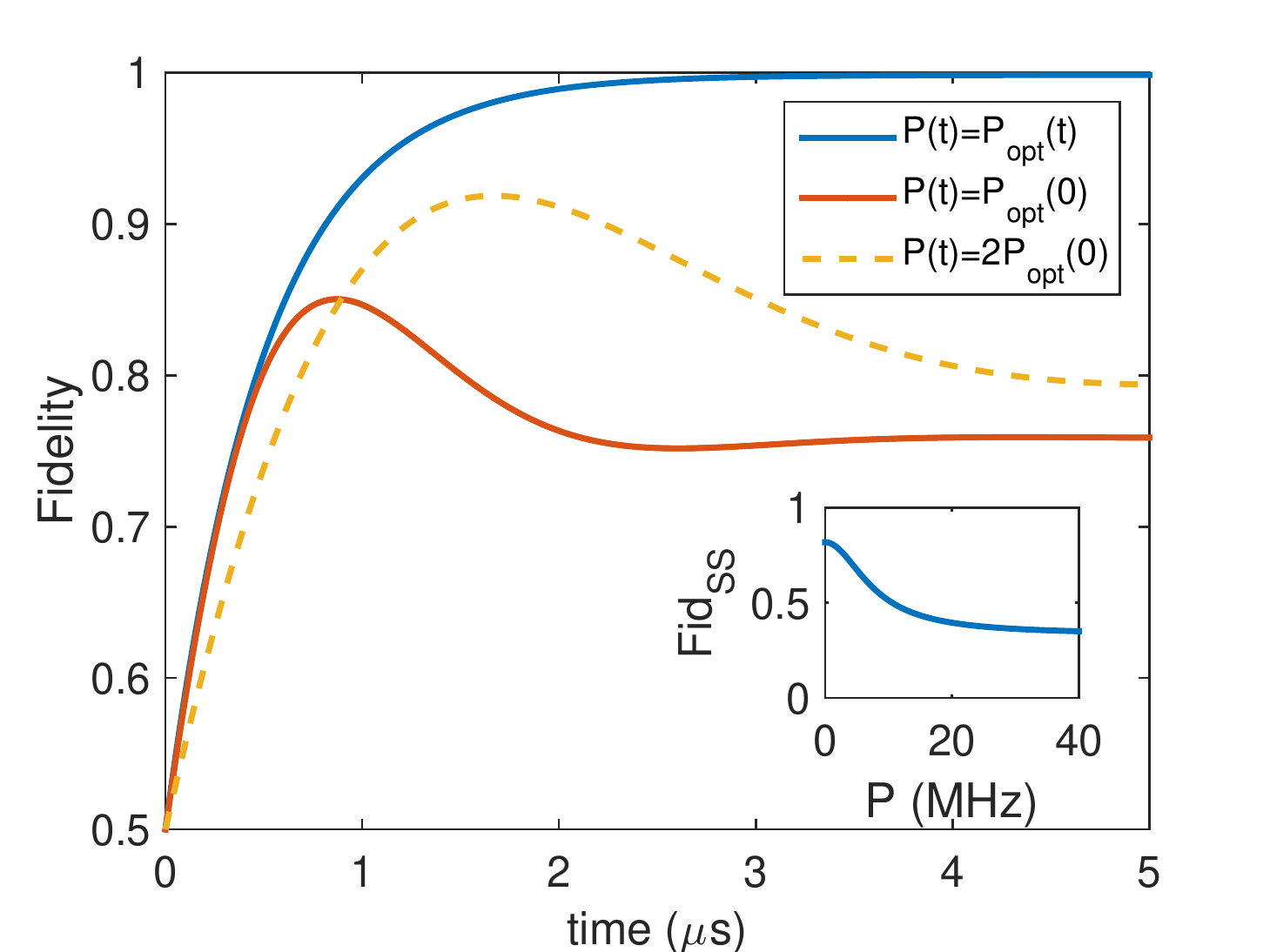}
\caption{Performance of the half-parity measurement under locally optimal feedback (blue curve) and suboptimal protocols that enforce $P$ to be fixed as a function of time. Fidelity is with respect to the target state $|\psi^+\> = (|01\>+|10\>)/\sqrt{2}$. Inset shows the steady state fidelity for fixed $P$. The highest achievable value is $9/11$.}
\label{fig:HPFFidelity}
\end{figure}

In Fig. \ref{fig:HPFFidelity} we plot the fidelity as a function of time under the optimal feedback protocol. Feedback has $A_2=0$ and maintains $\<M+M^\t\>=0$, so it requires only proportional feedback. We label the proportional feedback coefficient $P$, as is standard in control theory. Fig. \ref{fig:HPFFidelity} also plots the fidelity when $P$ is held constant at a few fixed values. Constant feedback is a common strategy in quantum control, as it is often argued that the steady state solution is the most important in application. However in this case, the fidelity does not reach 1 for any fixed value of $P$, indicating that a time-dependent strategy is essential for satisfactory performance. Performance in the presence of loss is given in \cite{HPFPRA}, which confirms that the protocol could be successfully applied with the experimental parameters already achieved in \cite{Roch2014}.

\subheading{Dicke state generation} An obvious extension of the above result is to more than two qubits. There are two particularly simple generalizations of the Bell state to N qubits: Dicke states and GHZ states. We begin with the Dicke states, which turn out to be easier to prepare given the measurement operators that tend to be accessible in nonlocal or weakly interacting systems. An $n$-excitation Dicke state of $N$ qubits is defined to be
\begin{align}
    |N,n\> = \frac{1}{\sqrt{\binom{N}{n}}}\Sigma_{P\in S_N} P(|0\>^{\otimes (N-n)}\otimes |1\>^{\otimes n})
\end{align}
where $P$ is an operator belonging to the permutation group $S_N$ on $N$ qubits. $|N,n\>$ is a uniform superposition over all states with the same number of excitations, and is also known as a spin squeezed state. They have proposed applications in a wide range of sensing protocols, including very long baseline interferometry\cite{gottesman2012longer} and Heisenberg-limited measurement sensitivity. When $n=1$, we have the usual W state.
\begin{align}
    |W\>=\frac{1}{\sqrt{N}}(|10\cdots 0\>_N+|01\cdots 0\>_N+\cdots +|00\cdots 1\>_N)
\end{align}

Several previous works have demonstrated deterministic Dicke state preparation with feedback, but these protocols were state-based and hence non-Markovian\cite{Thomsen2002,stockton2004dicke,Wei2015}. It is interesting to ask how well a Markovian protocol performs at the same task, particularly since dynamical state estimation becomes exponentially more challenging for larger systems. The most straightforward generalization of our two-qubit protocol is to simply add more operators to $M$ and $H$
\begin{align}
    M_N &= \sigma_{z,1}+\sigma_{z,2}+\cdots + \sigma_{z,N} \\ \nonumber
    H_N &= \sigma_{y,1}+\sigma_{y,2}+\cdots + \sigma_{y,N}.
\end{align}
We have dropped units, which amounts to working in a unit system in which $\Gamma=2$. $H_N$ is still local as required. $M_N$ can be measured by appending more qubits in cavities to the scheme outlined in Fig. \ref{fig:BounceBounce}, or by coupling many qubits dispersively to the same cavity. The latter method (without continuous feedback) has been applied to generate spin squeezing in cold neutral atoms coupled to an optical cavity\cite{Cox2016}. 

\begin{figure}[htp]
\centering
\includegraphics[width = 0.7\textwidth]{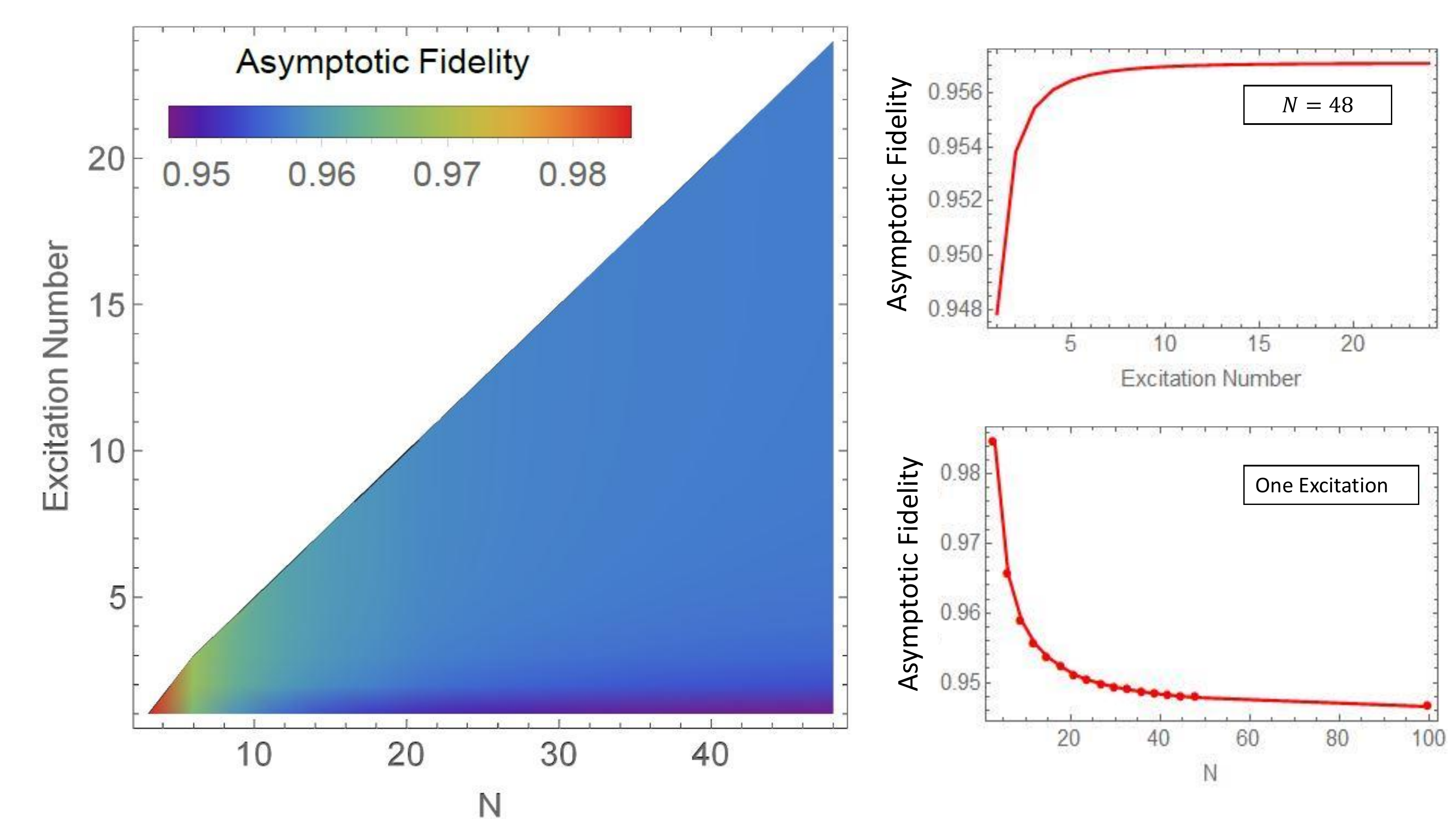}
\caption{Fidelity with respect to the $n$th Dicke state consisting of $N$ ($N$ takes values from $3$ to $48$ in increment of $3$ qubits under the ASLO (Markovian) protocol.
Insets show cutouts along $N=48$ (top) and $n=1$ (bottom), where the latter includes also the N=100 case.}
\label{fig:DickeFidelity}
\end{figure}

Once again, direct application of the feedback equations of section \ref{sec:PaQS} produces excellent results. We plot the asymptotic fidelity in Fig. \ref{fig:DickeFidelity} for many values of $N$ and $n$. To simulate up to 100 qubits, we work in a symmetry-reduced subspace spanned by $|N,n\>$, which shrinks the Hilbert space exponentially. Unlike in the two-qubit case, the average state $\bar{\rho}$ does not remain pure under ASLO feedback. This indicates that the stochastic terms do not cancel in \erf{eq:PaQSSME}, and therefore the unaveraged dynamics (\textit{i.e.} conditioned on the entire measurement record) depend on the entire measurement record. Curiously, the final fidelity remains above 94\% despite the fact that the system does not take a predictable path through Hilbert space, and hence an ASLO feedback controller does not know the true state. The efficacy appears to be due in part to the highly symmetric nature of the problem. The symmetry reduction to $|N,n\>$ has removed all degeneracy from the measurement operator $M_N$, so that the measurement outcome now uniquely determines the state.

From a practical standpoint, it should be noted that although the ASLO protocol does not achieve unit fidelity beyond $N=2$, one can still produce unit-fidelity states by adding a final projective measurement. Although the success probability under the ASLO protocol is less than one, it has been significantly enhanced by feedback. 

\subheading{GHZ states and multiple measurements} Generation of GHZ states using the above method poses new challenges and leads to an interesting generalization of the results of section \ref{sec:PaQS}. The GHZ state is defined as 
\begin{align}
    |GHZ\> \equiv \frac{1}{\sqrt{2}}(|000\> + |111\>).
\end{align}
with the obvious generalization to more than 3 qubits. As we have argued above, measurement operators that are linear combinations of single-body observables like $\sigma_z$ are easiest to implement in remote and weakly interacting settings. The GHZ state is an eigenvector of operators of the form
\begin{align}
    M = 2\sigma_{z,1} - s\sigma_{z,2} - (1-s)\sigma_{z,3}.
\end{align}
However the above operator does not satisfy the permutation symmetry of the target state, leaving us with no obvious symmetry reduction and no way to break the degeneracy of $M$. In practice, this choice of measurement operator does not seem amenable to GHZ state generation with continuous feedback.\footnote{This includes attempts that involve beyond-locally optimal strategies and strategies based on entanglement monotones such as the tangle. See \cite{Zhang2018Locally} for details.}

One way to reintroduce permutation symmetry to the problem is to allow for multiple measurement operators. The GHZ state is an eigenstate of the following operators
\begin{align}
    M_{(i,j)} = \frac{\sigma_{z,i} - \sigma_{z,j}}{2}, ~ i\neq j
\end{align}
which are just like the half-parity measurement operators but are instead degenerate in the $|00\>$, $|11\>$ subspace of each $i,j$ pair of qubits. 

To find the locally optimal protocol, we need to generalize the expression for $\theta_\text{opt.}$ to handle multiple simultaneous measurements $M_i$ and feedback operations $H_j$. In general there is no preferred pairing between them, so that the $i$th measurement outcome may affect how we apply the $j$th feedback Hamiltonian. This forces us to rederive the feedback master equation with a more general feedback unitary
\begin{align}
	U &= \exp \left( -i \sum_{ij} A_{ij}(\rho)H_j dW_i - i \sum_i B_i(\rho) H_i dt \right) \\ \nonumber
    &= I - i\sum_{ij} A_{ij} H_j dW_i - \left[ i\sum_i B_i H_i + \frac{1}{2} \sum_{ijk} A_{ij} A_{ik} H_j H_k \right]dt
\end{align}
where $A_{ij}$ and $B_i$ are analogous to $A_1$ and $A_2$ respectively, and our goal is to find their locally optimal values. The presence of cross terms in the above modifies the resulting feedback master equation
\begin{align} \label{eq:MultiPaQSSME}
  \rho(t+dt) &= \rho + \sum_i \Big[ \mathcal{D}[M_i]\rho ~ dt + \sqrt{\eta_i}\mathcal{H}[M_i]\rho~ dW_i - i \sum_{j} A_{ij} [H_j, \rho] dW_i  -i B_i [H_i,\rho] dt\\ \nonumber
  &-i \sum_j \sqrt{\eta_i} A_{ij}[H_j, M_i\rho + \rho M_i^\t]dt + \sum_{jk} A_{ij} A_{ik} [H_k\rho H_j - \frac{1}{2}(H_j H_k \rho + \rho H_j H_k)]dt \Big]
\end{align}
In contrast to \erf{eq:PaQSSME}, the last term is not in Lindblad form due to the presence of cross terms, though it may be cast into Lindblad form by defining the modified feedback operators $\tilde{H}_i = \sum_j H_j$. For simplicity, we assume that the control Hamiltonians commute pairwise, though the end result is essentially unmodified if one relaxes this assumption\cite{martin2019single}.\footnote{In general, one must have that the $H_i$ form a Lie algebra \textit{i.e.}, that the vector space formed by $H_i$ is closed under commutation.} The locally optimal feedback coefficients must satisfy $\<\psi_T|[H_\alpha, d\rho]|\psi_T\>=0$, this time for all $\alpha$. Dealing first with the $dW$ terms, we find
\begin{align}
\sum_i \left[ \sum_j -i A_{ij} \underbrace{\<\psi_T|[H_\alpha,[H_j,\rho]]|\psi_T\>}_{c_{j\alpha}} + \underbrace{\sqrt{\eta_i}\<\psi_T|[H_\alpha,M_i\rho + \rho M_i^\t]|\psi_T\>}_{a_{i\alpha}} \right] dW_i = 0
\end{align}
%
The solution is evident if we rewrite the expression in matrix form with the help of $a$ and $c$, which are state-dependent
\begin{align} \label{eq:MultiPaQSA}
-i A c + a = 0 ~~ \implies A = -i a c^{-1}
\end{align}
In general, one should use the Moore-Penrose pseudoinverse to handle the case in which $c$ is not invertible\cite{martin2019single}. With a solution for $A$ in hand, we can solve for the $B$ coefficients by collecting $\mathcal{O}(dt)$ terms from $\<\psi_T|[H_\alpha, d\rho]|\psi_T\>=0$ 
\begin{align} \label{eq:MultiPaQSB}
&b_\alpha \equiv \<\psi_T|[H_\alpha,\sum_i \Bigg[ \mathcal{D}[M_i]\rho -i \sum_j \sqrt{\eta_i}A_{ij}[H_j,M_i\rho+\rho M_i^\t] \\ \nonumber
&~~~~~~~~~~~~~~~~~~+ \sum_{jk}A_{ij}A_{ik} [H_k\rho H_j - \frac{1}{2}(H_j H_k \rho + \rho H_j H_k)]\Bigg]]|\psi_T\> \\ \nonumber
&-i\sum_i B_i c_{i\alpha} + b_\alpha = 0 ~~\implies \vec{B} = -i\vec{b}c^{-1}
\end{align}
treating $\vec{b}$ and $\vec{B}$ as row vectors. \erf{eq:MultiPaQSA} and \erf{eq:MultiPaQSB} generalize \erf{eq:PaQSCoeffs}. They even allow for the unusual situation in which we apply simultaneous non-commuting measurements, a situation that we consider experimentally in the following chapter.

A significant advantage of this generalization is the extension to multiple feedback Hamiltonians, even if there is only one measurement operator (note that we did not have to assume that the number of measurement and feedback operators need not coincide). In the previous formulation, we had to guess the best feedback Hamiltonian based on intuition, whereas now we can let $H_i$ be a Hamiltonian basis for the system and let the equations of motion determine the locally optimal Hamiltonian for us. 
As before, we should discard Hamiltonians that commute with the state. The condition to have found a local maximum as opposed to a minimum or saddle point is that the Hessian matrix
\begin{align} \label{eq:PaQSHessian}
    -\<\psi_T|[H_\alpha,[H_j,\rho^c]]|\psi_T\>
\end{align}
should be negative definite.

To generate GHZ states, we need to choose a set of feedback operators. For simplicity, we limit ourselves to local $\sigma_y$ rotations. $c$ can fail to be invertible if there are to many $H_i$s, and we have learned the importance of symmetry, so some care is required in choosing them. One obvious choice of Hamiltonian basis is $H_i = \sigma_{y,i}$, one for each qubit. However, this basis does not enforce the $|11...1\> \leftrightarrow |00...0\>$ symmetry of the GHZ state and the $M_{(i,j)}$s. A better choice would be a set orthogonal to $\sum_i \sigma_{y,i}$, such as
\begin{align} \label{eq:GHZHFs1}
    H_i = \frac{\sigma_{y,i}-\sigma_{y,i+1}}{2}, ~~~
    i=1...N-1
\end{align}
where $N$ is the number of qubits. Although this basis appears to break permutation symmetry, any Hamiltonian of the form 
\begin{align} \label{eq:GHZHFs2}
H_{(i,j)} = \frac{\sigma_{y,i}-\sigma_{y,j}}{2}
\end{align}
can be written as a linear combination of them. Thus the locally optimal solution can retain permutation symmetry (and should, unless the protocol explicitly breaks symmetry, which is possible and would happen in any case).

\begin{figure}[htp]
\centering
\includegraphics[width = 1\textwidth]{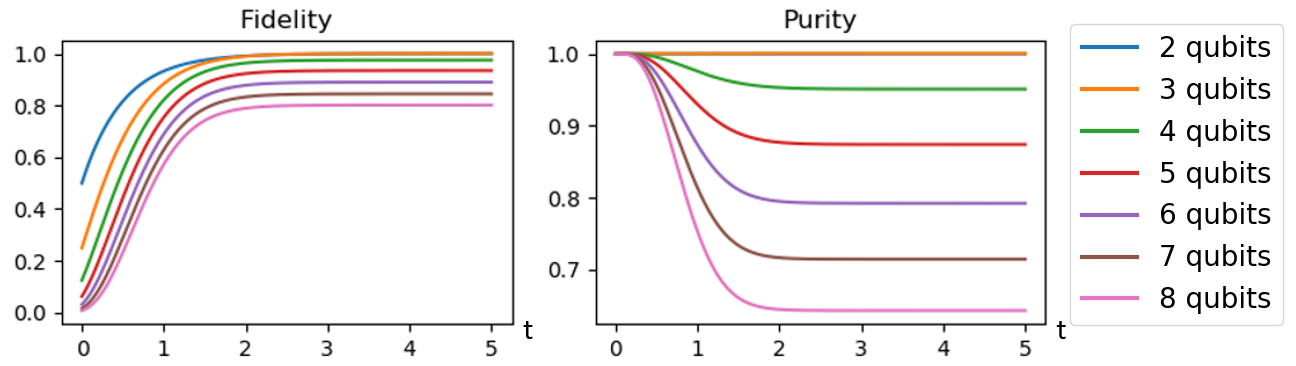}
\caption{Fidelity and purity of the average state under the feedback protocol described in the text. Purity remains exactly $1$ for $N=2$ and $N=3$.}
\label{fig:GHZFidelity}
\end{figure}

GHZ generation under ASLO feedback is plotted in Fig. \ref{fig:GHZFidelity}. The 2-qubit case reproduces the half-parity measurement under the transformation $\sigma_{z,2} \leftrightarrow -\sigma_{z,2}$, while the remaining results $N=3$ through $N=8$ are novel. Although we use \erf{eq:GHZHFs1} as our feedback Hamiltonians, the locally optimal protocol takes linear combinations of the form \erf{eq:GHZHFs2}, so that each $H_{(i,j)}$ is paired with its corresponding $M_{(i,j)}$. We speed up the numerics by taking this observation into account, so that many terms in Eqs. (\ref{eq:MultiPaQSSME}-\ref{eq:PaQSHessian}) may be set to zero. Perhaps most interestingly, the $N=3$ case reaches exactly unit fidelity, and the purity of the average state remains 1 throughout the feedback process, as in the example at the beginning of this section. One can derive an analytic solution in this case. We suspect that the protocol is globally optimal, but have not yet taken the time to prove this potential result.

While we have focused on application to entanglement generalization, the full scope of Eqs. (\ref{eq:MultiPaQSSME}-\ref{eq:PaQSHessian}) extends to virtually any application of measurement-based feedback. The ability of these equations to reproduce all existing feedback protocols that we have tried them on and the ease with which they generate new protocols suggests that there is much more to be gained from them. An interesting future direction would be to characterize more broadly the types of states that may be generated efficiently using these results. Entangled state generation in the context of Markovian dissipation has yielded interesting general results in other contexts\cite{johnson2017exact,ticozzi2018alternating}, and would be natural to apply and extend here. An important direction for future applied work would be to classify what states can be prepared deterministically (under a suitable definition) using local feedback and measurement operators that are linear combinations of local observables.

\subheading{Deterministic evolution} We have seen a number of intriguing examples in which feedback perfectly eliminates the randomness of the measurement process. These cases are highlighted with asterisks in table \ref{tab:PaQSTable}. All of these protocols are globally optimal for the task at hand, except for the 3-qubit GHZ protocol that we nevertheless conjecture to be optimal. They also yield analytic solutions in all cases, further indicating that they are a natural class of protocols in continuous measurement.

It is straightforward to introduce a strict necessary condition for whether or not deterministic evolution is possible given fixed measurement and feedback operators. This condition also provides a simple tool for finding solutions to the Wiseman-Milburne feedback master equation. Suppose that we apply proportional feedback of the form $U=\exp(-i H(P dr+c dt))$ for time-dependent constants $c$ and $P$. Applying $U$ to the stochastic Schr\"odinger equation yields a pure-state equation of motion for $|\psi\>$ valid for $\eta=1$
\begin{align}
    |\psi(t+dt)\> = \Big[I&+(M-m-iPH)dW - \frac{1}{2}(M^\t M-2M m+m^2+P^2 H^2)dt \\ \nonumber
    &-i H(P(M+m)+c)dt \Big] |\psi(t)\>
\end{align}
where $m\equiv \<M+M^\t\>/2$. We should allow for the solution to be deterministic up to a global phase, so we make the substitution
\begin{align}
    H \rightarrow H+V_0.
\end{align}
For a deterministic solution, the $dW$ terms of the stochastic Schr\"odinger equation must cancel \textit{i.e.}
\begin{align} \label{eq:DetdWCondition}
    [M-m-iPH-iPV_0]|\psi\> = 0.
\end{align}
This is a nonlinear constraint on $|\psi\>$, as $m$ depends on the state. It can only be satisfied if $|\psi\>$ is a right eigenvector of $M-i P H$. To see that this necessary condition is also sufficient, compute the corresponding eigenvalue $\lambda$ by taking an expectation value
\begin{align}
    \<\psi|&[M-i P H] |\psi\> = \lambda = \<M\>-i P\<H\>
\end{align}
and then compute the left-hand side of \erf{eq:DetdWCondition}
\begin{align}
[M-m-iPH-iPV_0]|\psi\> &= \Big[\<M\>-iP\<H\>-\frac{\<M+M^\t\>}{2}-iP V_0\Big]|\psi\> \\ \nonumber
&=\Big[\frac{\<M^\t-M\>}{2}-i P \<H\> - i P V_0\Big]|\psi\>
\end{align}
As $\<M^\t-M\>$ is purely imaginary, we can always find a $V_0$ such that the above expression is zero. Thus feedback can exactly cancel the measurement-induced noise on $|\psi\>$ if and only if $|\psi(t)\>$ is a right eigenvector of $M-i P H$. As $P$ is a free parameter, we can check if this condition is possible for any value of it.

With the stochastic terms cancelled, the next step is to see if the deterministic part of the equations of motion can maintain the above condition for a finite time interval. Given the above result, we can search for such solutions within each eigenspace of $M-i P H$ separately (the eigenspaces will typically evolve continuously as a function of $P$).  Let $\{|v_i(P(t))\>\}$ be a set of orthonormal eigenvectors associated with a chosen eigenspace. To obtain a deterministic solution, we must have
\begin{align}
    |\psi(t)\> = \sum_i c_i(t) |v_i\>.
\end{align}
Now consider the deterministic part of the equation of motion. Using the above expansion, we have
\begin{align}
    \frac{d|\psi\>}{dt} &= \sum_i \dot{c}_i(t)|v_i\>+c_i(t) \frac{dP}{dt}\frac{d|v_i\>}{dP} \\ \nonumber
    &= \left[\frac{1}{2}(M^\t M-2M m+m^2+P^2 H^2)dt - i H(P(M+m)+c)dt\right]|\psi(t)\>.
\end{align}
As $m=\text{Re}(\lambda)$ and $\lambda(t)$ is fixed by working within a fixed eigenspace, the above equation of motion is linear in $|\psi\>$, which significantly simplifies the finding of a solution. One can use the above procedure to identify candidate solutions, which appear to be plentiful and relatively unexplored. For example, the 3-qubit GHZ state protocol yields an analytic solution under the above procedure.

%% file: NonCommutingObservables.tex
\chapter{Simultaneous Measurement of Non-Commuting Observables}
\label{ch:XZ}

In the previous chapters, we discussed a significant generalization of the more basic projective, or Von Neumann description of measurement. Continuous measurement and POVMs lead to a great deal of novel physics that would not even be well defined in the standard formalism. We now turn to our first experimental application of this body of research. We revisit one of the most elementary and fundamental predictions of quantum mechanics, the Heisenberg uncertainty principle. 

The Heisenberg microscope thought experiment and the resulting uncertainty principle were among the earliest realizations of the importance of non-commuting observables in quantum mechanics. As it became clear that the basic properties of a system cannot be simultaneously determinate, doubt arose as to whether or not it was even possible to simultaneously measure them. Indeed, in the Von Neumann interpretation of measurement as instantaneous collapse of the system into an observable's eigenstate, the notion of simultaneous non-commuting measurements is ill-defined, because a system cannot simultaneously exist in the eigenstates of two non-commuting observables even mathematically.

A more accurate and subtle understanding arose in the following decades. Researchers came to realize that physical measurements occur over a finite timescale, and that interactions with multiple baths can indeed lead to non-trivial combinations of measurements. In fact, because most systems couple to their environments in multiple different ways, they are consequently measured in non-commuting bases in their natural settings. For example, charge fluctuations, phonons and spin baths all potentially form Markovian reservoirs, and can couple via different system operators. A ubiquitous example is the duo of dephasing ($T_2$) and relaxation ($T_1$), induced by the non-commuting operators $\sigma_z$ and $\sigma$ respectively. However, despite the fact that this process is happening in all realistic systems, researchers have not resolved multiple non-commuting dissipation channels as the measurements that they are until the following work.

In the experiment of this chapter, we begin with a transmon qubit dispersively coupled to two modes of a 3D cavity. These two modes act as two separate baths for our system. 
However because our experiment relies on the dispersive coupling to two cavity modes, we natively only couple to $\sigma_z$. Thus having two modes yields simultaneous measurement of two commuting observables. In order to access non-commuting observables, we devised a scheme that lets us change the measurement axis, which we call a single-quadrature, or stroboscopic measurement. We implement this scheme by adapting a technique from optomechanics known as a back-action evading measurement\cite{clerk2010noise}. We also describe a natural generalization to arbitrary quantum systems, which we term a multi-quadrature measurement. 

In section \ref{sec:XZDynamics}, we apply the above technique to measure two non-commuting observables of a qubit. We derive a novel connection to the uncertainty principle that governs the dynamics of the state by enforcing a lower bound on the measurement-induced disturbance. Consequently, as a function of the measurement non-commutativity, the dynamics transition smoothly from standard wave function collapse to isotropic, persistent diffusion. Although evolution of the state now differs drastically from that of a conventional measurement, information about both non-commuting observables is extracted by keeping track of the time ordering of the measurement record. We use maximum likelihood techniques to perform a novel form of quantum state tomography that does not involve alternating measurements in different bases.

Before going into experimental details specific to this work, we begin in section \ref{sec:HighQE}, outlining the crucial steps and techniques for implementing high-quantum efficiency measurements in circuit QED. This section is important for any quantum trajectory, quantum feedback or adaptive measurement experiment, including the experiment of chapter \ref{ch:AdaptivePhase}. The following sections describe work first presented in \cite{hacohen2016noncommuting}, with an elaborated description of single-quadrature measurements and the introduction of multi-quadrature measurements (not described in the original work). In section \ref{sec:XZDynamics}, we present the bulk of the experimental work, which includes a novel connection between measurement back-action and the uncertainty principle. A more lengthy and technical calculation, the polaron transformation deriving our effective stochastic master equation, is relegated to the last section, section \ref{sec:XZPolaron}.

\section{Achieving high quantum efficiency measurements}
\label{sec:HighQE}

In principle, continuous measurement are implemented in all experiments at some level. While they may be difficult to resolve temporally in some physical systems, the bandwidths of microwave and optical detectors easily reach the GHz range, and time resolutions for some photodetectors are measured in picoseconds. Loss presents a much more serious problem. The difficulty of a realistic quantum trajectories experiment is that by the time an experimentalist's detector has acquired a finite amount of information about the system in question, the environment at large has likely already learned far more. Once complete information has been transferred to the environment, then the residual coherence that makes quantum trajectories and feedback interesting has been destroyed. Thus the primary experimental parameter limiting the observation of quantum trajectories is quantum efficiency. Only if a large fraction of the total amount of information content emitted by the measured system is in our hands can we experimentally observe quantum trajectories and their consequences.

\begin{figure}
\centering
{\includegraphics[width=0.8\textwidth]{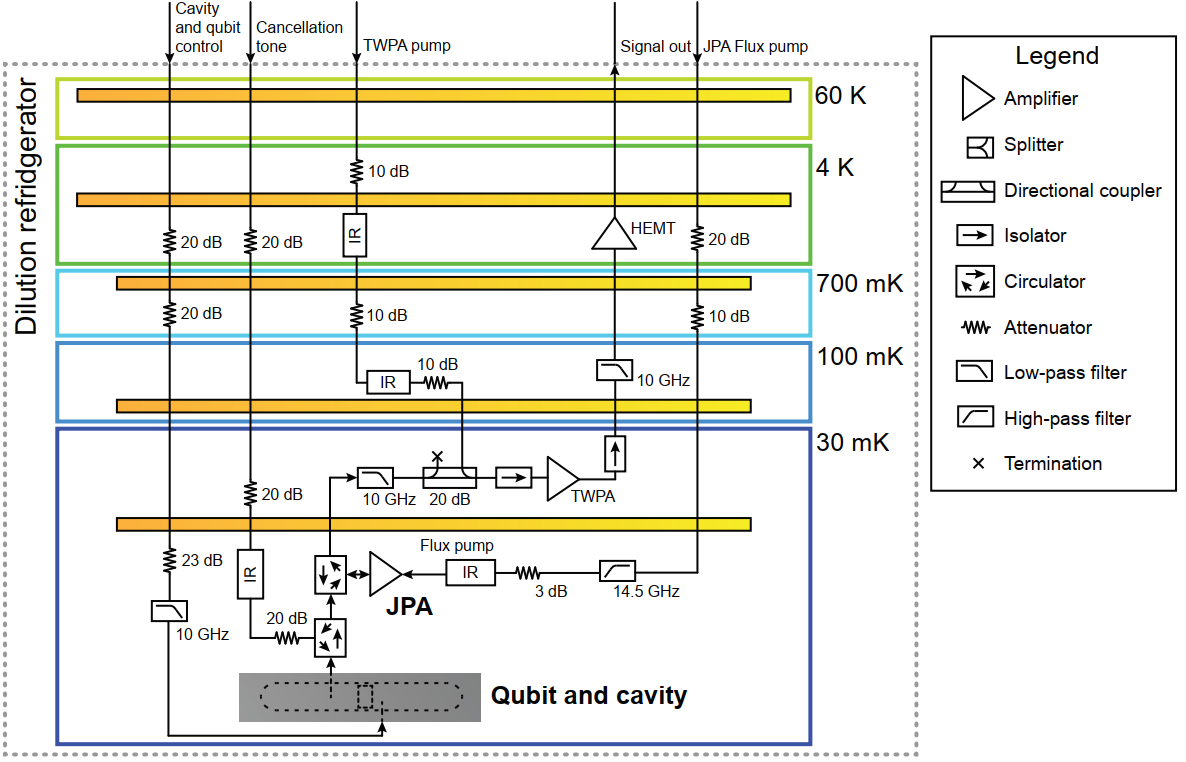}}
\caption{Wiring diagram for the experiment outlined in chapter \ref{ch:AdaptivePhase}, which captures the essential features of a high quantum efficiency setup. The JPA amplifies in reflection and is pumped and feedback-controlled via the flux pump line. IR filters are Eccorsorb or similar.}
\label{fig:HighEfficiencyFridge}
\end{figure}

Fig. \ref{fig:HighEfficiencyFridge} shows the low-temperature wiring diagram for a quantum trajectories experiment. The diagram is actually for the experiment of the following chapter, which is less cluttered because it contains only one measurement chain instead of two. The first critical aspect of the setup is the assortment of input line attenuators distributed throughout the refrigerator. To maintain low temperatures at the qubit and parametric amplifier, we must filter out blackbody radiation from higher temperature stages. At microwave frequencies, we eliminate power with standard microwave attenuators. To eliminate infrared power, light-tight sealing and Eccosorb filters are essential\cite{birenbaum2014cshunt}. 

To determine the required level of attenuation, we can use equipartition theorem to estimate the amount of thermal power travelling through the lines. Over a fixed bandwidth at high temperature, $E \propto k_B T$, so it is sensible to ignore prefactors between power and temperature and measure of blackbody radiation power in units of Kelvin.\footnote{If absolute or total integrated powers are needed, it is important to note that the $T^4$ power dependence of the Stefan Boltzmann law is modified when we consider a 1-dimensional system like a microwave transmission line.} At low temperatures (relative to the photon quanta energy), we will tend to overestimate the total power, which is better for making conservative estimates. The power emitted by an attenuator consists of the attenuated incident thermal power plus the blackbody radiation emitted by the attenuator itself. Taking the top right attenuator of Fig. \ref{fig:HighEfficiencyFridge} as an example, it is held at a temperature of $4$ K and receives $300$ K noise at its input. $20$ dB $=$ $\times 100$, so the emitted power is $T_\text{out} = T_\text{in}/100 + 99 T_\text{atten.}/100\approx 7$ K. Note that this only applies if the attenuator is thermally anchored to the $4$ K stage, which requires a metallic clamp around the portion of the attenuator that dissipates power. Noise power at the base-temperature circuits should be comparable to the base plate temperature of $30$ mK. The energy scale of a $1$ GHz photon is 48 mK, so GHz-scale circuits can be quite sensitive to thermal power. For example, thermal population of a readout resonator is often a dominant source of dephasing\cite{wang2019cavity}.

Control line attenuation can be large, as we may compensate for it by increasing the control power during operation. We also naturally gain the potentially weak coupling of the control line to the device as an additional source of attenuation, as well as the spectral filtering of a cavity if one is present. Achieving proper thermal isolation for the output line is more challenging. Attenuation is not an option if we wish to attain high quantum efficiency, and we also require strong coupling of the output line to the device itself (relative to the control line coupling strength, so that information does not leak out of the input ports). The output line in Fig. \ref{fig:HighEfficiencyFridge} is filtered with circulators, wide-band isolators and low-pass filters (such as K\&L brand). The presence of amplifiers further restricts what components must be used. For instance, ideally a single circulator would suffice between the qubit and the JPA, but two often appear to be necessary in practice. The amplifier emits a substantial amount of power over its bandwidth, and reflections can interfere with ideal operation. The HEMT (high electron mobility transistor amplifier) dissipates enough power that it must remain at the $4$ K stage cooled by the pulse tube. Aluminum does not superconduct at this temperature, so unlike other lines, one must use another material such as niobium for the connection.

\begin{figure}
\centering
{\includegraphics[width=\textwidth]{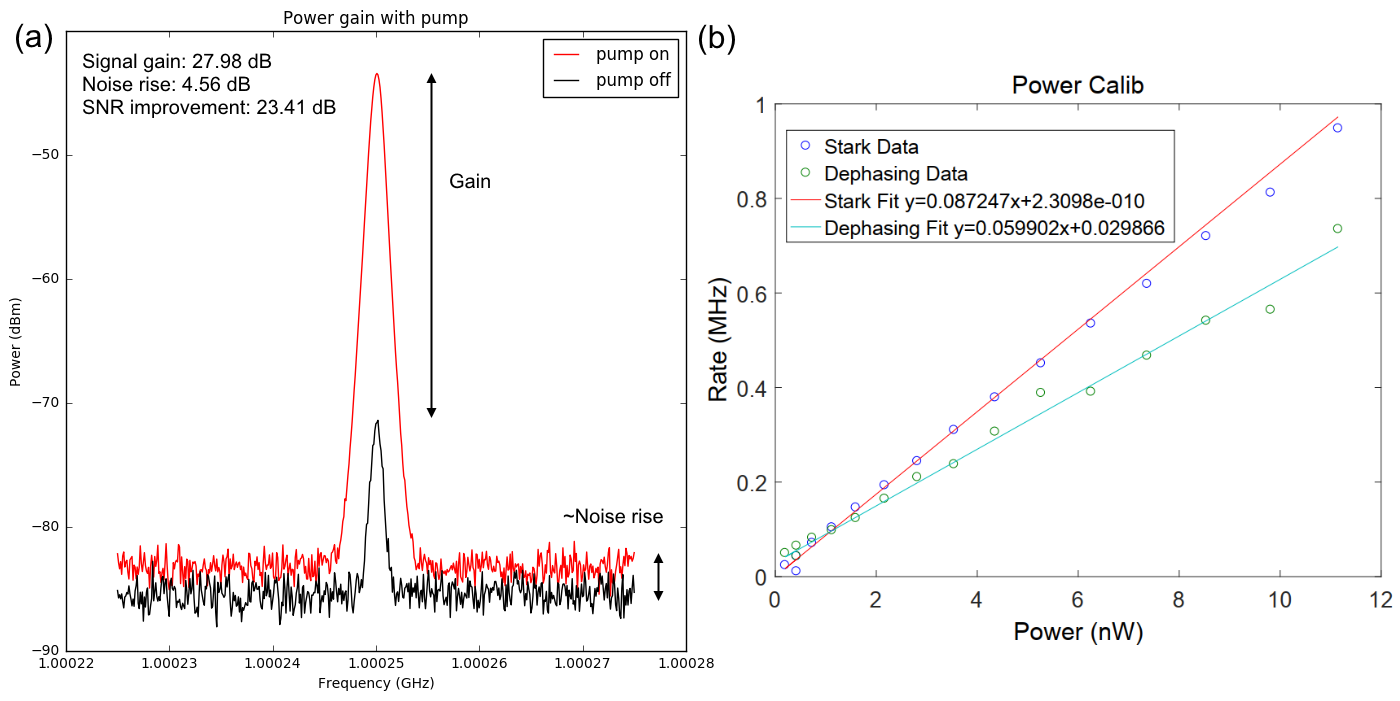}}
\caption{(a) Spectrum analyzer data from an SNR improvement measurement of a single-pumped JPA. Axes plot power, and hence are not accurate for noise power (which would have units of mW/Hz, not mW. A weak signal is applied at 1.00025 GHz, which is by necessity detuned from the pump frequency of 1 GHz. The SNR improvement is unusually high, indicating loss or a high noise temperature at the HEMT. Data courtesy of Akel Hashim. (b) $\chi$ and photon number calibration data. Stark shift and dephasing rate fits taken from Ramsey measurements are fit to a calibrated power axis. The slopes of the Stark shift and dephasing rate curves are $2\chi/A$ and $8\chi^2/\kappa A$ respectively (assuming $\Delta_\text{AC}$ is measured as an angular frequency. If not, we must include a factor of $2\pi$ in \erf{eq:StandardRamseyFit}).}
\label{fig:NoiseRise}
\end{figure}

In moving a microwave signal from $30$ mK to room temperature, amplifiers are absolutely necessary. We can model a linear amplifier as an ideal amplifier preceded by a noise source. Each must amplify the signal above the noise level of the subsequent device, so that we can eventually digitize the signal with a room temperature analog-to-digital converter.  A standard way to characterize the performance of an amplifier is a signal-to-noise (SNR) improvement measurement using a spectrum analyzer. Data from a low-frequency JPA are shown in Fig. \ref{fig:NoiseRise}a. To measure the SNR improvement, we apply a weak coherent tone to the amplifier with it powered off and record the signal power and noise power. We do not require a calibrated power measurement here, though it is important to use a noise marker to measure the noise power quantitatively.\footnote{Spectrum analyzers work by filtering the input signal over a narrow bandwidth and then plotting the transmitted power as a function of the filter center frequency. They are calibrated to report the absolute power assuming that the input field is infinitely narrow in frequency. As noise is actually broadband, the detected signal power depends on the transmission profile of the filter. A noise marker calibrates out the filter profile and reports the noise level in units of power per unit bandwidth $=$ energy, which is the appropriate set of units for noise.} We then measure the signal and noise powers again with the amplifier on. Noise rise is defined to be the ratio of noise powers, and gain is of course the ratio of signal powers. SNR improvement is the ratio of these ratios. At low gain, the SNR increases with increasing gain, but the measured noise is dominated by devices after the amplifier and remains constant. At high gain, the dominant noise source becomes noise amplified by the amplifier itself, so gain and noise rise increase in tandem. To achieve high quantum efficiency, we should work in the latter regime.

SNR improvement measurements provide a way to check the sufficiency of an amplifier, but do not provide absolute metrics on device performance. A key figure of merit is noise temperature, which is the effective temperature at the input, assuming that all output noise arises from ideally amplified input noise from a hypothetical source. For this we need some kind of calibrated power source. A simple method is to source the amplifier with an impedance matched termination at a known temperature. In practice however, we would like to measure the amplifier performance while it is connected to something more interesting. Fortunately, if the measurement chain is connected to a qubit in a cavity, then the latter system provides a convenient calibrated power source via the dispersive Jaynes Cummings Hamiltonian. The essential ingredients for this process are that $H=\chi \sigma_z a^\t a$ provides conversion between photonic energy and qubit frequency, and the cavity linewidth $\kappa$ provides a conversion between internal cavity energy and output power. Once we know the output power from the cavity, we can convert the measured power at a spectrum analyzer to absolute power units referenced to the amplifier input. For example, if we read a power of -40 dBm at the spectrum analyzer but know that the true input power is $\kappa \<a^\t a\>  = -128$ dBm, then this provides a conversion factor to infer the effective noise power at the amplifier input from a room temperature noise power measurement. In reality, some noise is due to loss and subsequent amplifiers, but ideally noise from the amplifier closest to the source dominates. Due to the uncertainty principle, the lowest possible noise power at a given frequency is $\omega \times \hbar/2$, the power per unit bandwidth of vacuum fluctuations.

$\kappa$ is easy to measure, as it is the full-width half-maximum of the cavity transmission profile (multiplied by $2\pi$), in accordance with input-output theory. The challenge of the above measurement is the precise determination of $\chi$ and $\<a^\t a\>$.  If $\chi > \kappa$, then the qubit resonance splits into many resonances each separated by $2\chi$, one for each value of the intercavity photon number $a^\t a$ with non-negligible population. By driving the cavity with a probe tone matched to the cavity resonance (conditioned on the qubit being in its ground state) and then sweeping another tone over the qubit resonance, we can observe these peaks. When the qubit tone hits a qubit resonance, the cavity frequency is shifted, which in turn shifts the phase of the cavity tone. We can detect the resulting phase shift in the cavity tone if we apply it using a vector network analyzer (VNA). The distribution of peaks and their relative heights is determined by the coherent amplitude of the cavity tone via \erf{eq:CoherentState}, and $\<a^\t a\> = |\alpha|^2$. 

The procedure is somewhat more involved when $\chi < \kappa$, as the large cavity bandwidth masks the discrete photon-number peaks\cite{Vijay:2012ua}. To measure $\chi$, we linearly sweep the power of a cavity-resonant tone and monitor the resulting qubit Stark shift and dephasing rate with a Ramsey measurement. Using \erf{eq:QubitSMETraceCav} and the standard input-output relation for a cavity ($\dot{\alpha} = -i \Delta \alpha - \kappa \alpha/2 + \epsilon$), one can derive the following expressions for the Stark shift and dephasing rates in the limit $\chi \ll \kappa$
\begin{align} \label{eq:SmallChiKappaRates}
\Delta_\text{AC} &= 2\chi \bar{n} \\ \nonumber
\Gamma_D &= \frac{8\chi^2 \bar{n}}{\kappa}
\end{align}
where $\bar{n} = |\alpha_{g/e}|^2$ is the intercavity photon number, which takes the same value regardless of the qubit state. The general expressions for any $\chi$, $\kappa$ and cavity-drive detuning $\Delta$ are straightforward to derive as well. Experimentally, we can measure $\Delta_\text{AC}$ and $\Gamma_D$ for any $\bar{n}$ by fitting the Ramsey data to
\begin{align} \label{eq:StandardRamseyFit}
\<\sigma_z\> = e^{-\Gamma_D \tau}\cos(\Delta_\text{AC}\tau + \delta)
\end{align}
where $\tau$ is the delay between $\pi/2$ pulses in the Ramsey measurement. By fitting the linear relations \erf{eq:SmallChiKappaRates} together, we can extract $\chi$ and a calibration factor $A$ between actual power $P$ applied from a room temperature source and the intercavity photon number $\bar{n}$ (\textit{i.e.} $P = A \bar{n}$). $A$ provides the calibration factor needed to produce a known power at the input of an amplifier coupled to the cavity.

\subheading{Quantum efficiency} The above procedures offer a starting point for debugging unexpected features in a high quantum efficiency measurement chain. Quantum efficiency itself is the ratio of the measurement-induced dephasing rate to the rate at which information is acquired. Formally, if we write the master equation for a qubit with $M=\sqrt{\Gamma_D/2}\sigma_z$ as
\begin{align}
d\rho &= \mathcal{D}[M]\rho ~ dt + \sqrt{\eta} \mathcal{H}[M]\rho ~ dW \\ \nonumber
&= \frac{\Gamma_D}{2}\mathcal{D}[\sigma_z]\rho ~ dt + \sqrt{\frac{\Gamma_M}{2}} \mathcal{H}[\sigma_z]\rho ~ dW
\end{align}
then the quantum efficiency is typically defined as
\begin{align} \label{eq:EtaDefinition}
\eta \equiv \frac{\Gamma_M}{\Gamma_D}.
\end{align}
Notice that $\eta$ is not quite a signal-to-noise ratio, but rather a ratio of rates.

$\Gamma_D$ measures the total rate at which information is extracted from the system by everything else in the universe, and we already know how to estimate it using a Ramsey measurement. Given the convention above, it is also the rate at which the off-diagonal elements of the density matrix $\rho_{e,g}$ and $\rho_{g,e}$ decay, which can be measured by fitting a Ramsey trace to \erf{eq:StandardRamseyFit}. Measured in this way, $\Gamma_D$ also includes damping due to the intrinsic qubit dephasing time $T_2^*$. One can compute the true dephasing rate as $\Gamma_D = \Gamma_{D,\text{fit}} - 1/T_2^*$, or measure $\Gamma_D$ at multiple readout powers ($\bar{n}$) to measure and subtract off the zero-intercept.

We can infer $\Gamma_M$ from the signal-to-noise ratio of the integrated measurement record
\begin{align}
dr &= \sqrt{\eta}\<M + M^\t\>dt + dW \\ \nonumber
&= \sqrt{2\Gamma_M} \< \sigma_z \>dt + dW
\end{align}
If we integrate the signal over a fixed time interval $\tau$, we obtain two Gaussians with variances $\sigma^2 = \tau$ and means $\mu_{g/e} = \pm \sqrt{2 \Gamma_M}\tau$ conditioned on the prepared qubit state. The measurement can then be inferred with
\begin{align}
    \Gamma_M = \frac{(\mu_e-\mu_g)^2}{8\tau \sigma^2}.
\end{align}
This formula is a ratio of signal amplitudes, and so any arbitrary multiplicative factor between the numerical value of the signal and $dr$ cancels. In practice, linear fits in $\tau$ are advisable, as finite measurement bandwidth introduces offsets.\footnote{On timescales short compared to the measurement signal correlation time, $dW$ no longer approximates its statistics well, and the variance scales as $\tau^2$ instead of $\tau$.}

\section{Single and multi-quadrature measurements}

As described in chapter \ref{ch:IntrocQED}, dispersive readout is the standard method of readout in circuit QED. Here we present an alternative scheme that provides several novel and useful capabilities. Firstly, it enables dynamical changing of the measurement axis even during measurement\cite{hacohen2018incoherent}, a useful capability for adaptive measurement and quantum state tomography. Secondly, it provides a method to implement longitudinal ($a+a^\t$) as opposed to transverse ($a^\t a$) coupling to a resonator without modification of the underlying circuit. This enables enhancement of qubit readout with squeezed light \cite{eddins2018stroboscopic}, and stabilizer measurements for the Gottesman-Kitaev-Preskill qubit encoded in a cavity\cite{eickbusch2019grid, touzard2019grid}. Finally, we show how our scheme has a built-in method for rapidly and deterministically depopulating the cavity after readout, enabling complete measurement in a duration shorter that the inverse cavity bandwidth. In this chapter, we will be primarily be interested in the ability to change the measurement axis in-situ.

Conveniently, the following measurement scheme derives from the dispersive Jaynes Cummings Hamiltonian, meaning that we can implement it in the standard circuit QED setup. The basic concept involves Rabi driving the qubit fast enough that the dispersive coupling term averages out and effectively decouples. We `reactivate' it by applying sidebands to the cavity, which restore a resonant coupling in a different basis. For a qubit, it allows us to measure any operator orthogonal to the Rabi drive basis. For a multi-level system like a transmon or a set of qubits, it enables measurement of system operators which are \textit{hollow} in the basis of the drive Hamiltonian, meaning that the diagonal elements of the matrix are zero. By Schur-Horn theorem, any traceless matrix can be cast into hollow form by change of basis. As adding a multiple of the identity to a measurement operator does not change its eigenbasis, it follows that our scheme can measure any system operator. 

An important advantage of multi-quadrature measurements is that they can also be applied concurrently, as in the present experiment. The only restriction on the set of possible simultaneous measurements is the requirement that all measured operators must be hollow in the same basis. We note that an extension of the scheme presented in reference \cite{Vool2016} can be applied to measure in the eigenbasis of the drive Hamiltonian, thus providing access to an operator with a non-zero diagonal. Combining these schemes allows for a form of continuous tomography in which the full set of tomographic measurements can be applied simultaneousely. Even without this scheme however, the hollowness constraint is not severe. Many important sets of measurement operators satisfy this property, including the parity measurements $\{XX,YY\}$, the repetition code stabilizers $\{XXI,IXX\}$ and the five-qubit code stabilizers $\{XZZXI, IXZZX, XIXZZ, ZXIXZ\}$ capable of protecting against single bit- and phase-flip errors on the encoded logical qubit.

\begin{figure}
\centering
{\includegraphics[width = 0.7\textwidth]{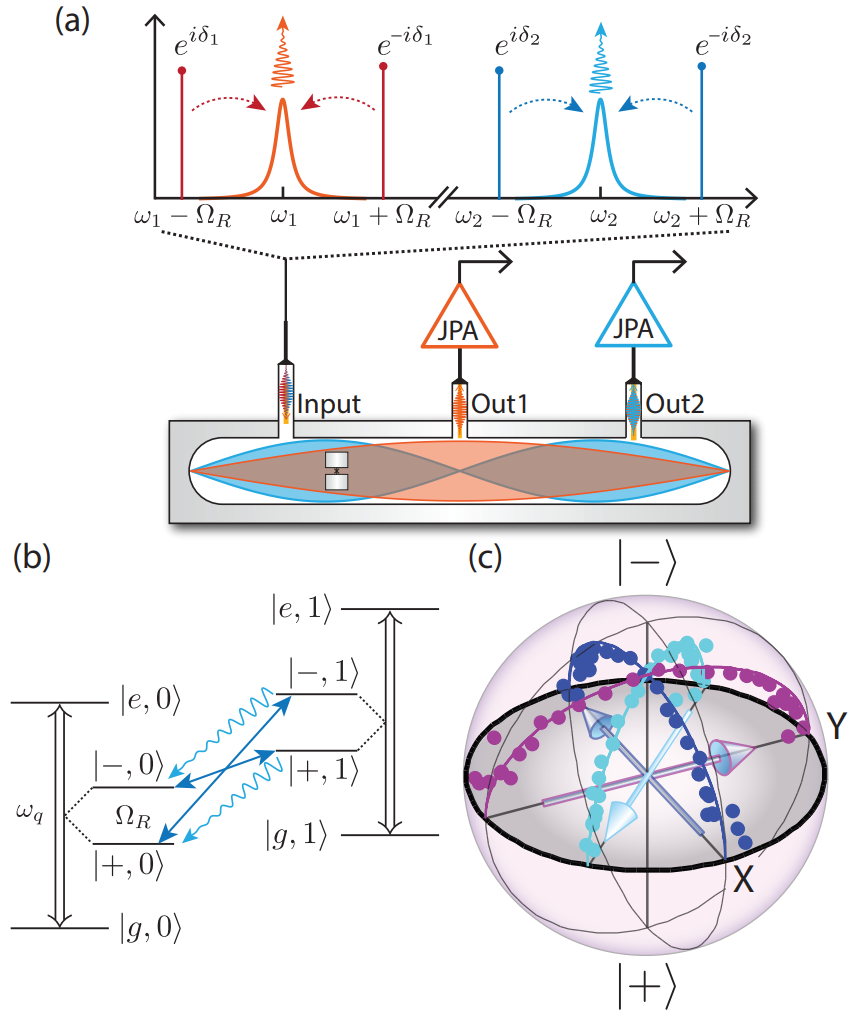}}
\caption{\textbf{Experimental implementation of a multimode single quadrature measurement (SQM)} (a) A transmon qubit in an aluminum cavity. Via the input port, we drive Rabi oscillations at frequency $\Omega_R$ and apply $\pm\Omega_R$ sideband tones to each of the two lowest cavity modes. Each output is monitored using a lumped-element Josephson parametric amplifier (LJPA)\cite{castellanos2008amplification}.
Coupling is designed so that $\sim$90\% of each signal is routed to its corresponding LJPA. This system yields two separate measurement channels, each with a controllable measurement basis. (b) Dressed state picture of the SQM scheme for one of the cavity modes. Sideband tones drive transitions indicated by the solid lines, and undulating lines represent cavity decay, which we detect. (c) Tomographic reconstruction after a $1~\mu \mathrm{s}$ SQM applied only to the lower mode, showing collapse along three separately chosen measurement axes $\sigma_{\delta_1}$, $\delta_1 = \{ 0,\pi/4,\pi/2 \} $. Circles plot tomographic data and lines give theory based on a dephasing rate of $\Gamma_{1}/2 \pi = 122 ~\mathrm{kHz}$ and quantum efficiency of $\eta_1 = 0.41$.}
\label{fig:XZSetup}
\end{figure}

\subheading{Single-quadrature measurements} We begin with a detailed derivation of a single-quadrature measurement (abbreviated SQM), and then outline the multi-quadrature (MQM) generalization. The system described in this chapter involves two single-quadrature measurements applied to the same qubit. The experimental setup, shown in Fig. \ref{fig:XZSetup}a consists of a Rabi-driven 3D transmon qubit coupled to the two lowest frequency modes of a 3D superconducting cavity. The key aspect of an SQM are the cavity sidebands depicted above the cavity. Altogether, the system is described by two of the standard dispersive Hamiltonian with time-dependent drives
\begin{align}
\label{eq:HTot}
& H = H_q + H_\text{drive} + \sum_{i=\{1,2\}} (\omega_{c,i} + \chi_i\sigma_z)a_i^\dagger a_i \\ \nonumber
& H_q = \frac{\omega_q}{2} \sigma_z + \Omega_R(\sigma + \sigma^\dagger)\cos(\omega_d t) \\ \nonumber
H_\text{drive} &= \sum_{i=\{1,2\}} \epsilon_i(t) a_i^\dagger + \epsilon_i(t)^* a_i
\end{align}
%
where $\epsilon_i(t) \in \mathbb{C}$ is the coherent drive that will represent the SQM sideband drives. 
Details of physical parameters are given in \cite{hacohen2016noncommuting}. For now we note that $\Omega_R \gg \kappa$ is an important physical constraint. It prevents sidebands from populating the cavity significantly, and ensures that an application of the rotating wave approximation holds in what follows.

An intuitive description of the SQM is depicted in Fig. \ref{fig:XZSetup}b. Suppose we wish to create an effective longitudinal interaction like $\sigma_x (a+a^\t) = (\sigma+\sigma^\t)(a+a^\t)$. Such a Hamiltonian is typically not relevant in nature because of the rotating wave approximation, which averages out the $\sigma a$ and $\sigma^\t a^\t$ terms. However it is possible to generate a resonant interaction if both the qubit and cavity operators have commensurate time dependence. To create this situation, we drive the qubit at a frequency $\Omega_R$. In the rotating frame of the cavity, the sideband tones have a time dependence given by the sideband detuning. If we set this detuning to be $\pm\Omega_R$, then we can expect a resonant interaction. As visualized in Fig. \ref{fig:XZSetup}b, the drive Hamiltonian leads to dressed states with an effective energy splitting of $\Omega_R$ in the rotating frame of the qubit. The lower ($\omega_c - \Omega_R$) simultaneously excites the cavity and removes energy from the qubit, taking it from the $|+\>$ to the $|-\>$ dressed states. This process sounds just like the Jaynes-Cumming interaction $\sigma a^\t + \sigma^\t a$, so we are back to where we started. However we can also add an upper sideband ($\omega_c + \Omega_R$) that simultaneously creates or destroys qubit and cavity excitations $\sigma^\t a^\t + \sigma a$. Thus the combination of an upper and a lower sideband gives us the the longitudinal coupling that we wanted. Moreover the relative phase between these two terms can be controlled by varying the relative phase of the sidebands, which yields a controllable measurement axis. In what follows, we justify the above intuition mathematically.

If the cavity decay channels $\kappa_i$ are monitored with parametric amplifiers with quantum efficiencies $\eta_i$, the qubit+cavity dynamics are modelled by the following stochastic master equation
\begin{align}
\label{eq:CavSME}
d\rho = -i [H, \rho] dt 
+ \sum_{i=\{1,2\}} \kappa_i \mathcal{D}[a_i] \rho ~dt + \sqrt{\kappa_i \eta_i} \mathcal{H}[a_i e^{i \phi_i}] \rho ~dW_i
\end{align}
where $\phi_i$ is the amplification axis of the phase sensitive amplifier used to monitor mode $i$. 
For the derivation, it suffices to consider just one of the two cavity modes. To measure a single quadrature of the qubit, we drive the cavity with a pair of sidebands detuned by $\pm \Omega_R$ from the cavity resonance, and with phases 
$\pm (\delta-\text{atan}(\kappa/2\Omega_R))$. The second term in the phase cancels the lag between the external drive and the internal cavity field. As $\Omega_R \gg \kappa$, it can be ignored in our experiment. We choose this drive $\epsilon(t) = -i \bar{a}_0 \sqrt{\Omega_R^2 + \kappa^2/4}\sin(\Omega_R t + \delta - \text{atan}(\kappa/2\Omega_R))e^{-i \omega_c t}$ so that the cavity internal displacement due to the sidebands takes the relatively simple form
%
\begin{equation}
\label{eq:SidebandDrive}
\bar{a}(t) = \bar{a}_0 \cos(\Omega_R t + \delta) e^{-i \omega_c t} ~ \rightarrow ~ \bar{a}_0 \cos(\Omega_R t+\delta).
\end{equation}
The right arrow indicates that we have transformed into the interaction picture with respect to the cavity. We include the complicated prefactor in $\epsilon(t)$ to compensate for the cavity-drive detuning, and soon leads to simplifying cancellations that justify \erf{eq:SidebandDrive}. We take $\bar{a}_0$ to be real for simplicity. 

To understand the effects of the cavity drive, we transform into a displaced frame in which the cavity is in the vacuum (neglecting for a moment the cavity-qubit interaction). This time dependent unitary transformation is $U(t) = D[-\bar{a}(t)] = \exp(\bar{a}^* a - \bar{a} a^\t)$. As we noted in chapter \ref{ch:IntroQuantum} around \erf{eq:DisplacementOperator}, this transformation maps $a \rightarrow a+\bar{a} = D[-\alpha] a D[-\bar{a}]^\t$, which greatly simplifies calculations. We first transform the cavity dissipation terms, in direct correspondence with \erf{eq:RotatingFrame}
\begin{align}
\kappa \mathcal{D}[a]\rho &\rightarrow U(t) 
\kappa \mathcal{D}[a + \bar{a}]\rho = \kappa \mathcal{D}[a]\rho - i[H_\text{dis}, \rho] \\ \nonumber
H_\text{dis} &\equiv -\frac{i \kappa}{2}(\bar{a} a^\dagger - \bar{a}^* a) \\ \nonumber
\sqrt{\kappa \eta} \mathcal{H}[a e^{i \phi}] \rho &\rightarrow 
\sqrt{\kappa \eta} \mathcal{H}[a e^{i \phi}]\rho.
\end{align}
%
\erf{eq:RotatingFrame} also tells us that we must add a term $H' = -i U(t)\dot{U}^\t(t) =  i(\dot{\bar{a}} a - \dot{\bar{a}} a^\dagger)$ to the Hamiltonian. A bit of algebra shows that $H_\text{dis}$ and $H'$ cancel the drive term $H_\text{drive}$ from \erf{eq:HTot}, which completely eliminates cavity drive terms from the Hamiltonian. In this sense, $\bar{a}$ is a steady state, which confirms that \erf{eq:SidebandDrive} gives the internal cavity displacment. Transforming the remaining terms of $H$ for one cavity mode into the interaction picture (\textit{i.e.} transforming the Hamiltonian by $\text{exp}[i \omega_c a^\dagger a t + i \omega_d \sigma_z t/2 ]$), applying the rotating wave approximation, and then applying the displacement transformation above yields
\begin{align}
H_\text{int} &= \chi a^\dagger a \sigma_z + H_{q, \text{int}}\\ \nonumber
&= \chi[(a^\dagger + \bar{a}^*)(a+\bar{a})]\sigma_z  + H_{q, \text{int}} \\ \nonumber
&= \chi \Big{[}\bar{a}_0 \cos(\Omega_R t + \delta)(a^\dagger + a) + \frac{\bar{n}_0}{2} + \bar{n}_0\frac{\cos(2\Omega_R t + 2\delta)}{2} + a^\dagger a \Big{]} \sigma_z  + H_{q, \text{int}} \\ \nonumber
H_{q, \text{int}} &= \frac{1}{2}[(\omega_q-\omega_d)\sigma_z + \Omega_R \sigma_x]
\end{align}
where we have defined $\bar{n}_0 = \bar{a}_0^2$. Choosing the Rabi drive to be resonant (\textit{i.e.} $\omega_q -\omega_d = -\chi \bar{n}_0$, which accounts for the Stark shift term), we diagonalize the qubit drive term of the Hamiltonian by going into the Hadamard frame ($\sigma_z \leftrightarrow \sigma_x$), and then going into a frame rotating at the Rabi frequency ($\text{exp}[i\Omega_R \sigma_z t/2]$). These transformations eliminate $H_{q, \text{int}}$ and map $\sigma_z$ to $\sigma e^{-i \Omega_R t} + \sigma^\dagger e^{i\Omega_R t}$, so the Hamiltonian becomes
\begin{align}
H_\text{q-frame} &= \frac{\chi \bar{a}_0}{2}(e^{i\Omega_R t + i \delta} + e^{-i \Omega_R t - i\delta}) (a^\dagger + a)(\sigma e^{-i \Omega_R t} + \sigma^\dagger e^{i \Omega_R t})  \\ \nonumber
& + \chi \Big{[} \bar{n}_0 \frac{\cos(2\Omega_R t + 2\delta)}{2} + a^\dagger d \Big{]}(\sigma e^{-i \Omega t} + \sigma^\dagger e^{i\Omega_R t}).
\end{align}
Dropping terms which rotate at $\Omega_R$ or $2\Omega_R$, we are left with
\begin{align}
\label{eq:HEffSuppMeth}
H_\text{q-frame} &= \frac{\chi \bar{a}_0}{2}(a^\dagger + a)(\sigma e^{i\delta} + \sigma^\dagger e^{-i \delta}) = \tilde{g}\sigma_\delta(a^\dagger + a) \nonumber \\
\sigma_{\delta} &\equiv \cos(\delta) \sigma_x + \sin(\delta) \sigma_y \nonumber \\
\tilde{g} &\equiv \frac{\chi \bar{a}_0}{2}
\end{align}
which is a qubit-state-dependent cavity drive Hamiltonian. In the same way that a qubit-state-dependent phase shifts induced by $\sigma_z a^\t a$ yields a $\sigma_z$ measurement of the qubit, \erf{eq:HEffSuppMeth} yields a measurement of $\sigma_\delta$. We defer the remainder of this calculation to section \ref{sec:XZPolaron}, though note that the methods of section \ref{sec:QEDReadout} would also yield a valid stochastic master equation. The most important detail of the result is the measurement rate
\begin{align}
\Gamma = 2\chi^2 \bar{a}_0^2 / \kappa.
\end{align}

\erf{eq:HEffSuppMeth} also yields a simple method to turn off the measurement quickly. In dispersive readout, one typically waits a duration longer than $1/\kappa$ after turning off the coherent cavity drive. There are methods to depopulate the cavity coherently after readout, but the simplest scheme suffers from the fact that the optimal depopulating displacement depends on the measurement outcome. In an SQM, the $\bar{a}_0$ prefactor gives us a way to invert the sign of $H$ and hence undo its action. A large pulse can depopulate the cavity relatively quickly. This scheme is particularly convenient for the MQM scheme, which benefits from a smaller $\kappa$.

One minor experimental imperfection arises which is worthwhile considering theoreticall. In practice, we generate the cavity sideband tones by amplitude-modulating a cavity-resonant tone that we call the local oscillator at a frequency $\Omega_R$ using an arbitrary waveform generator and a diode-based microwave mixer. In principle, amplitude modulation creates two sidebands at $\omega_c \pm \Omega_R$ and eliminates the local oscillator completely. In practice, diode mixers must be nulled with a DC bias. This nulling is rarely perfect, which leads to a small coherent tone resonant with the cavity in addition to $\epsilon(t)$.  Adding this LO leakage to Eq. \ref{eq:SidebandDrive} gives $\bar{\alpha}_0 \cos(\Omega_R t+\delta) + \bar{a}_\text{LO}$, and the effective Hamiltonian would instead be
\begin{align}
H_\text{q-frame} = \tilde{g}\sigma_\delta(a^\dagger + a + 2\text{Re}[\bar{a}_\text{LO}]).
\end{align}
Thus LO leakage induces unwanted rotations about the $\sigma_\delta$ axis. 

To demonstrate the above scheme, and in particular control of the measurement axis, we perform a single SQM using the lower cavity mode. Further details of the experimental setup are given in \cite{hacohen2016noncommuting}. We prepare the $|-\rangle$ state and read out for $1~\mu\text{s}$, which is comparable to the inverse measurement time of $1.3 \mu$s. Because we stop the on the timescale of the rate of collapse, the system exhibits residual coherence. We observe this residual coherence by including one of six possible pulses and a projective measurement using a standard dispersive readout of $\sigma_z$ in the lab frame at the end of each experimental sequence. These pulses, called quantum state tomography, allow us to effectively measure $\sigma_x$ and $\sigma_y$ in addition to $\sigma_z$, which enables reconstruction of the density matrix. Fig. \ref{fig:XZSetup}c shows the results of these tomographic measurements for three different measurement axis orientations. We compute the data points by using the stochastic master equation to predict $\rho$ for each experimental run, clustering these datasets into groups with similar final states, and then plotting the Bloch vector for each subset. The final states lie along 3 arcs for the three measurement axes chosen. The final state is somewhat impure because the measurement efficiency is less than 1. Note that the three measurements applied in Fig. \ref{fig:XZSetup}c come from three separate experiments, so we have not yet applied simultaneous non-commuting measurements.

\begin{figure}
\centering
{\includegraphics[width = 0.8\textwidth]{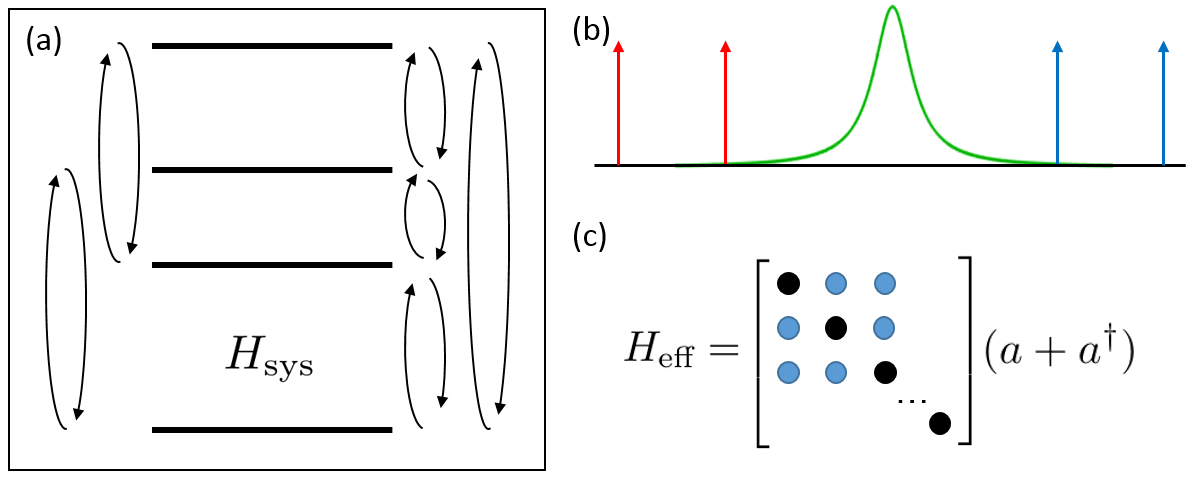}}
\caption{(a) Possible transitions that may be driven and stroboscopically read out via a multi-quadrature measurement in a 3-level system. (b) Red- and blue-detuned sidebands applied to a cavity. The sideband detuning is greater than the cavity linewidth. (c) A schematic illustration of the multi-quadrature measurement Hamiltonian. Only off-diagonal elements of the coupling operator are time-dependent and therefore may be addressed by a sideband.}
\label{fig:MultiQuadratureMeas}
\end{figure}

\subheading{Multi-quadrature measurements} Before describing the dynamics of simultaneous non-commuting measurements, we turn to the concept of a multi-quadrature measurement. We called the above measurement single-quadrature because it singles out one quadrature $\sigma_\delta$ of the qubit Rabi oscillations. This arose from driving the sideband transitions depicted in Fig. \ref{fig:XZSetup}. For an $N$-level system, there are up to $N(N+1)/2$ distinct transitions, each of which can be driven and coupled to the cavity via sidebands (see Fig. \ref{fig:MultiQuadratureMeas}a and b). The picture is somewhat complicated by the fact that these transitions interact with one another, but we can nevertheless handle this system analytically in the general case.

Consider a completely general $N$-level system dispersively coupled to a cavity, written in the cavity frame and the qubit drive frame
\begin{align} \label{eq:MQMInitialH}
H &= H_\chi a^\dagger a + H_d 
\end{align}
where $H_d$ is an external drive Hamiltonian on the $N$-level system and $H_\chi$ is a system operator representing the dispersive coupling to the cavity. In writing \erf{eq:MQMInitialH} to be time-independent, we have implicitly taken $H_\chi$ to be diagonal in the system energy eigenbasis, as is usually the case in practice (for example $H_\chi=\chi \sigma_z$ for a qubit). Note that the coherent system drive term $H_d$ could include off-resonant drive terms, so it need not be hollow (see for example \erf{eq:HQubitRWADetunedDrive} with $\Delta \neq 0$). 
We now transform $H$ into the eigenbasis of $H_d$, so that $V H_d V^\dagger = \Lambda$ is diagonal. In this basis, it is simple to go into the frame of the drive Hamiltonian, as $U(t)=\exp[i \Lambda t]$ is also diagonal. In this frame, $H$ becomes
\begin{align}
H = U(t)V H_\chi V^\dagger U^\dagger(t) a^\dagger a. 
\end{align}
As $U$ is time-dependent, it will couple to sideband tones applied on the cavity. Much of the first derivation of this section amounted to showing that we can put a resonator in a specified time-dependent coherent state using coherent drives. We take this fact for granted and begin by assuming a specific time-dependence of the internal cavity state $\bar{a}(t)$. We now transform into the displaced frame
\begin{align}
H &= U(t)V H_\chi V^\dagger U^\dagger(t) (a+\bar{a}(t))^\dagger (a+\bar{a}(t)) \\ \nonumber
&= Q(t) (a+\bar{a}(t))^\dagger (a+\bar{a}(t)) \\ \nonumber
Q_{ij}(t) &\equiv \sum_{ij} (V H_\chi V^\t)_{ij} e^{-i(\Lambda_{jj} - \Lambda_{ii})t}
\end{align}
We suppose that the dressed state transition frequencies $\Lambda_{jj} - \Lambda_{ii}$ are non-zero and nondegenerate, as would typically be the case for a generic $H_d$. In this case, each pair of off-diagonal elements of $Q_{ij}$ rotate at a different frequency, and thus may be addressed individually. Under a few assumptions on $H_d$, we can measure the hollow, Hermitian operator $X$ using the following choice of sideband tones
\begin{align} \label{eq:MQMSidebands}
\bar{a}(t) = \sum_{i \neq j} \left| \frac{X_{ij}}{Q_{ij}} \right| \cos((\Lambda_{jj} - \Lambda_{ii})t + \delta_{ij})
\end{align}
where $\delta_{ij} = \text{arg}(X_{ij})$. If $Q_{ij}=0$, then the singularity in the above expression means that $X_{ij}=0$. \erf{eq:MQMSidebands} leads to the following effective Hamiltonian under the rotating wave approximation
\begin{align} \label{eq:MQMHEff}
H_\text{eff} = X (a+a^\t)
\end{align}
The primary assumption going into \erf{eq:MQMHEff} is that $V H_\chi V^\t$ is hollow, so that $Q(t)$ has no time-independent matrix elements and we can therefore drop $Q(t) |\bar{a}(t)|^2$ and $Q(t) a^\t a$. We also assume that the sideband frequencies work out such that the crossterms within $|\bar{a}(t)|^2$ do not lead to additional resonances with $Q(t)$.

To check hollowness experimentally, one can apply $H_d$ without sideband tones, apply a standard dispersive readout and then look for state-dependent cavity frequency shifts via the term $Q(t)a^\t a$. We can also derive an appropriate $H_d$ with the following necessary and sufficient condition for $H_\chi$ to be hollow in the eigenbasis of $H_d$
\begin{align} \label{eq:FindHollowOp}
\Tr[H_d^n H_\chi] &= 0\text{~~~ for $n=0,...N-1$} \\ \nonumber
\Lambda_{ii} &\neq 0 \\ \nonumber
\Lambda_{ii} &\neq \Lambda_{jj}.
\end{align}
The second constraint can be satisfied by adding an inconsequential multiple of the identity to $H_d$, and we already assumed the third constraint in order to ensure that all matrix elements of $Q$ can be addressed with a sideband of non-zero frequency. This formula lets us solve a linear system of homogeneous polynomial equations to find an appropriate $H_d$ given a fixed $H_\chi$. In practice, one can generate many solutions to \erf{eq:FindHollowOp} based on any additional experimental constraints, and then pick a solution that also satisfies the second and third constraints. 

To derive \erf{eq:FindHollowOp}, note that if we write it in the eigenbasis of $H_d$, then $\Tr[\Lambda^n V^\t H_\chi V] = \Tr[\Lambda^n Q(0)] = 0$ is trivially satisfied if $Q$ is hollow. To prove the reverse implication, rewrite it as
\begin{align}
\Tr[\Lambda^n Q(0)] = \sum_i \Lambda_{ii}^n Q_{ii}(0) = 0
\end{align}
This equation implies that $Q_{ii}(0)=0$ if
\begin{align}
\text{det}\begin{pmatrix}
1	 & 1 	& \hdots & 1 \\
\Lambda_1 & \Lambda_2 & \hdots & \Lambda_N \\
\vdots 		& & & \\
\Lambda_1^{N-1} & \Lambda_2^{N-1} & \hdots & \Lambda_N^{N-1}
\end{pmatrix} = \prod_{1 \leq i < j \leq N-1} (\Lambda_j - \Lambda_i) \neq 0.
\end{align}
The above matrix is called a Vandermonde matrix, and it is clearly invertible if (and only if) the $\Lambda_i$s are distinct.

For the purposes of measuring non-commuting observables, \erf{eq:FindHollowOp} also provides a way to check if a set of operators $\{X_m\}$ can be made hollow in the same basis, which determines if we measure them simultaneously. One searches for a Hermitian matrix $H_d$ satisfying \erf{eq:FindHollowOp} simultaneously for each $X_m$ in place of $H_\chi$. We are unaware of any simpler condition that can perform the same test. This is in contrast to the question of whether a set of operators can be made diagonal in the same basis, which is possible if and only if they commute. Nevertheless \erf{eq:FindHollowOp} is usable for small systems to which we might end up applying an MQM.

\section{Dynamics under incompatible measurements and a connection to the uncertainty principle}
\label{sec:XZDynamics}

\begin{figure}
\centering
{\includegraphics{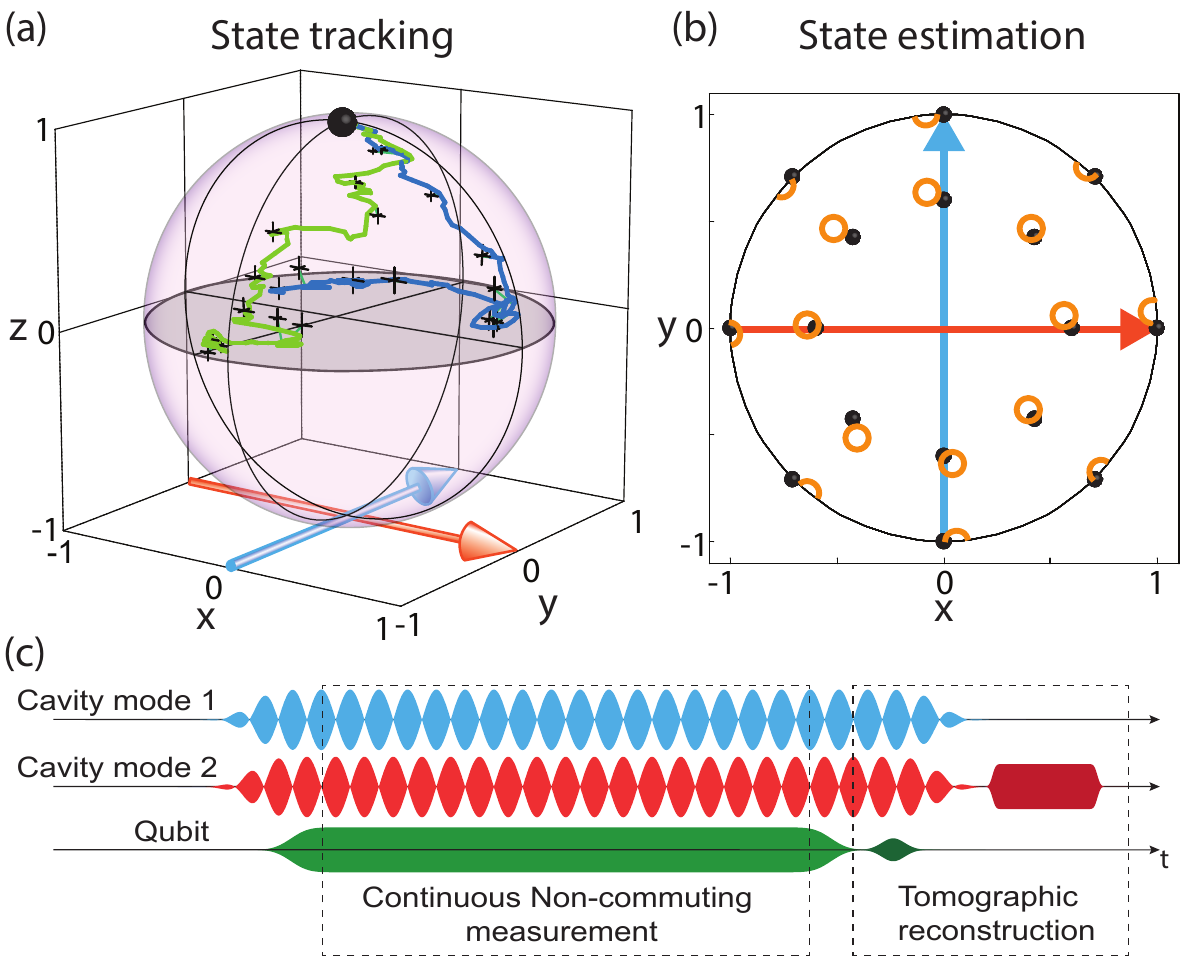}}
\caption{\textbf{Validation of simultaneous non-commuting measurements} (a) Reconstruction of two quantum trajectories initialized at state $|-\>$ (black sphere) and their associated tomographic validation. Arrows indicate axis of each measurement. Tomography points are taken every $200~\text{ns}$, and are linked to the corresponding trajectory time point with a green line. All trajectories ending within $\pm 0.11$ around a given point along the plotted trajectory are used for the corresponding tomography reconstruction. Error bars are generated from statistical uncertainty arising in qubit readout. (b) Estimation of 16 initially unknown state preparations (black spheres) using non-commuting measurements. Orange circles mark 95\% confidence intervals, generated by applying maximum likelihood estimation to $\sim$10,000 trajectories each. (c) Pulse sequence, showing cavity sideband tones and Rabi drive. Tomographic reconstruction consists of a tomography qubit pulse and projective readout.}
\label{fig:XZTomography}
\end{figure}

To simultaneously measure two non-commuting observables, we apply the SQM to the two lowest modes of the cavity housing our transmon qubit. Each pair of sidebands sets a measurement axis, and thus we are able to control them independently. Because non-commutativity leads to dependence on the time ordering of measurement outcomes, the integral of the output signal is no longer sufficient to uniquely determine the qubit state as it was in the previous section. Nevertheless, the combined measurement records encode complete information, and we are able to track quantum trajectories of an initially known state. We calculate the trajectories not using the master equation, but instead using the single-measurement solution to the master equation over a short (16 ns) time interval and alternatingly applying them iteratively. Both of these objects are treated in the next section, but are faimiliar to the reader from chapter \ref{ch:OpenSystems}. This procedure is more numerically stable and guarantees positivity of the density matrix. We validate this theory using quantum state tomography and the post-selection process described in the previous section. Figure \ref{fig:XZTomography}a shows two such traces with their tomographic validation from data in which the measurement axes are set to be orthogonal, $\sigma_{\delta_1}=\sigma_x$ and $\sigma_{\delta_2}=\sigma_y$. 

Due to the disturbances introduced by measurement incompatibility, extraction of an initially unknown quantity, such as a Hamiltonian parameter or system observable, requires use of the \textit{combined} measurement records and their full time orderings.
In particular, estimation methods must rely not only on the statistics of the measurement records, but also on some estimate of this disturbance. We encode this information into a single Kraus operator $\Omega_{r_1(0;T), r_2(0;T)}$ associated with each measurement record. To compute it, we use a Kraus operator $\Omega_{r_1(t),r_2(t)}$ that generates the master equation (\erf{eq:XYMeasOpp} derived in the following section) and build up the full Kraus operator from the measurement outcomes integrated over $\Delta t =16$ ns intervals
\begin{align} \label{eq:XZREcordPOVM}
\Omega_{r_1(0;T),r_2(0;T)} = \Omega_{r_1(T), r_2(T)} \Omega_{r_1(T-\Delta t), r_2(T-\Delta t)}... \Omega_{r_1(0), r_2(0)}
\end{align}
Using \erf{eq:XZREcordPOVM}, we can compute the probability for an observed record $\{r_i(0;T)\}$ to occur as $P(r_1, r_2) = \Tr[\Omega_{r_1,r_2} \rho \Omega_{r_1,r_2}^\t]$. By maximizing this probability for all measurement records over all possible initial states, we perform maximum likelihood estimation of the initial state. Figure \ref{fig:XZTomography}b shows reconstruction of sixteen state preparations again in the case of orthogonal measurement axes. Agreement within the confidence interval demonstrates that despite the complicated dynamics induced by competing observables, our scheme performs as a measurement, and that it extracts information about $\sigma_x$ and $\sigma_y$ simultaneously. 

\begin{figure}
\centering
{\includegraphics[width=\textwidth]{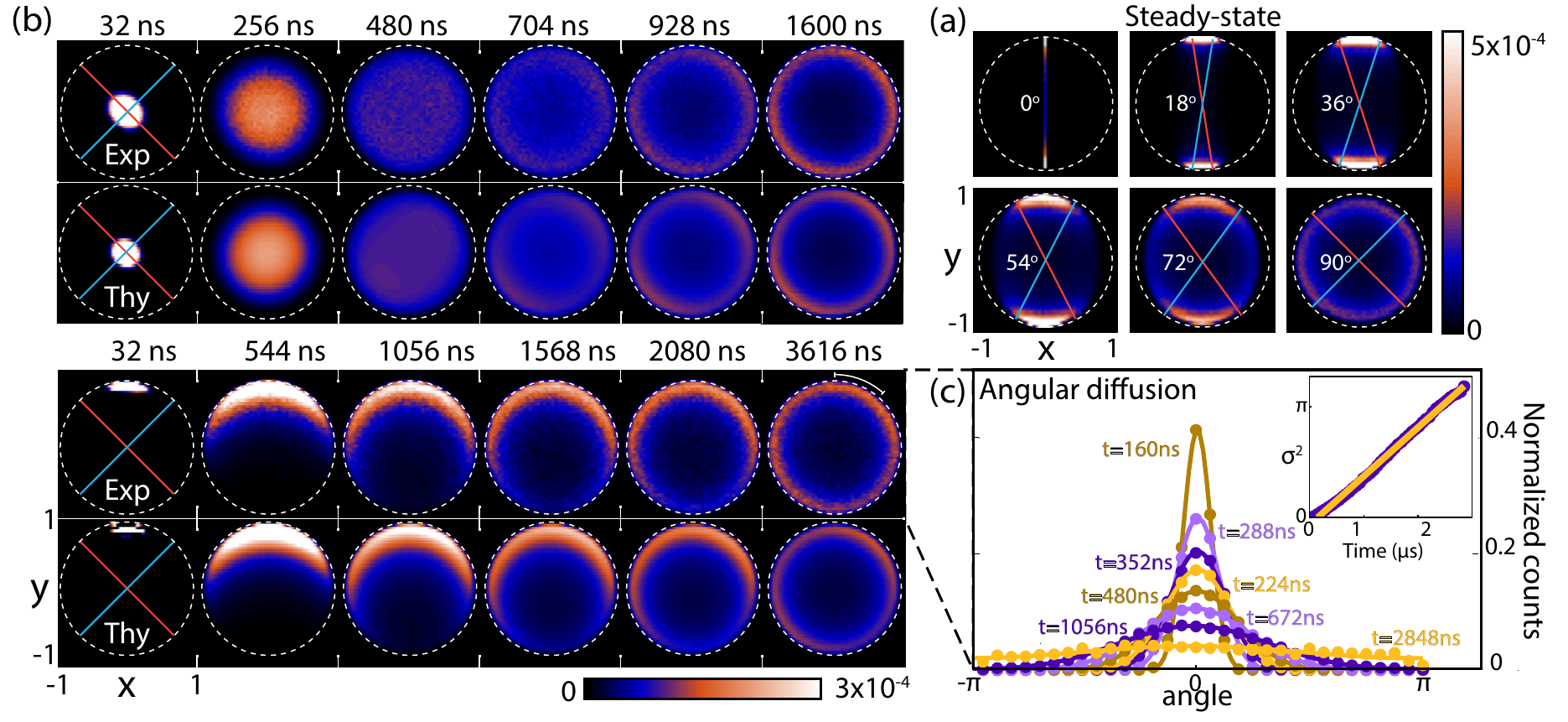}}
\caption{\textbf{Probability distribution of the density matrix} (a) in the steady-state as function of angle, between the axes, demonstrating loss of collapse for non-commuting observables. (b) as function of time for perpendicular measurement axes. We prepare a mixed state for observing the radial dynamics (top), and a state with purity of $P = 0.89$ for observing the azimuthal dynamics (bottom). Upper row of each is experimental data, and bottom row is theoretical comparison derived from the Fokker-Planck equation (see methods). (c)
Angular probability distributions for states within a ring of inner radius 0.86 and outer radius 0.92, showing that dynamics match those of a random walk. Points are normalized counts and lines are fits to normal distributions convolved with a $2 \pi$ periodic Dirac comb. Inset -- variance as a function of time (violet) and linear fit (yellow) yielding a slope of 1.4 $\mu \mathrm{s^{-1}}$. The expected slope for a perfect random walk in our system is 1.5 $\mu \mathrm{s^{-1}}$ with the main source of uncertainty due to measurement rate uncertainty of about 10\%.
}
\label{fig:XZFokPlanck}
\end{figure}

A state undergoing a non-commuting measurement exhibits dynamics beyond those of usual wavefunction collapse. We directly observe this evolution by measuring the probability distributions of the resulting density matrices. Figure \ref{fig:XZFokPlanck}a shows the steady state probability distributions. When the measurement axes align, the system collapses to one of two points at the poles of the Bloch sphere as expected for commuting measurement operators. When the axes are separated by a finite angle less than 90 degrees, the state does not collapse to any point, but rather 
localizes to regions of finite size defined by the measurement axes.
A salient feature is that when the axes are orthogonal and hence maximally incompatible, the location of the measurement axes no longer leaves any imprint on the state evolution. The distinct regions merge and we lose all notion of collapse.

Figure \ref{fig:XZFokPlanck}b shows probability distributions as a function of time for this canonical case. Starting in a mixed state, we see that the state purifies isotropically to a mean steady-state radius given by $r = \sqrt{\eta}$. 
High quantum efficiency in our system allows us to observe the azimuthal dynamics by preparing a state of purity $P = 0.89$, the most likely steady-state purity. As predicted by Ruskov \textit{et al.} \cite{Ruskov2010} in the case of unit quantum efficiency, competition between the maximally incompatible observables leads to diffusive motion given by a uniform random walk. Figure \ref{fig:XZFokPlanck}c shows the angular distribution as a function of time, which agrees quantitatively with this prediction.

\begin{figure}
\centering
{\includegraphics{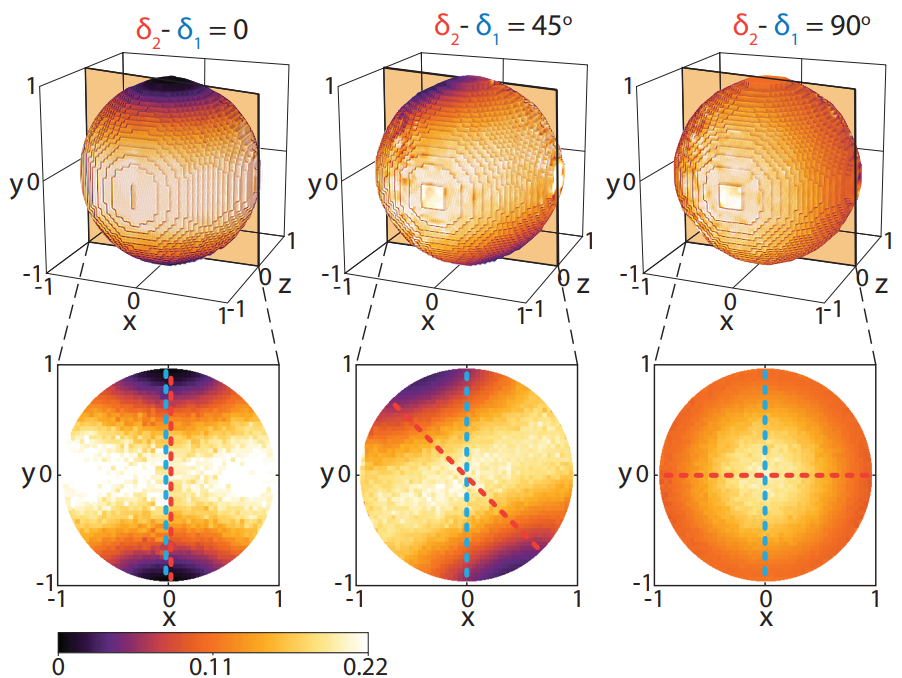}}
\caption{\textbf{Magnitude of the quantum back-action.} The amplitude of disturbance induced by measurement ($\text{Tr}[d\rho^\dagger d\rho]$) at each point on the Bloch sphere for three relative measurement angles $\delta_2-\delta_1$. (top) Disturbance at the surface of the Bloch sphere, which is bounded from below by the uncertainty principle. (bottom) A slice through the $z=0$ plane. Dotted lines mark measurement axes. For non-commuting observables ($\delta_2-\delta_1 \neq 0$), no point of zero disturbance exists within the entire volume of the Bloch sphere, indicating that the state must diffuse indefinitely.}
\label{fig:XZDistMap}
\end{figure}

The diffusive behavior seen in Fig. \ref{fig:XZFokPlanck}c suggests that even once the probability distributions have reached steady state, the system continues to evolve. To quantify this measurement-induced disturbance, we plot the norm of the state change $d\rho$ over an interval of $64~\mathrm{ns}$ versus position on the Bloch sphere in Fig. \ref{fig:XZDistMap}. When the measurements are compatible, two points on the Bloch sphere exhibit zero disturbance, which indicate mutual eigenstates of the measurements and hence points of collapse. No such points exist when the measurements are incompatible, implying that the state diffuses indefinitely. If we calculate this disturbance directly from the stochastic master equation for the system, we find the following relation valid for all initially pure states
\begin{align}
\label{eq:XZDistMap}
\text{Tr}[d\rho^\dagger d\rho] = (\Delta \sigma_{\delta,1}^2 \Gamma_{1}\eta_1 + \Delta \sigma_{\delta,2}^2 \Gamma_{2}\eta_2) dt \\ \nonumber
\geq |\langle[\sigma_{\delta_1}, \sigma_{\delta_2}]\rangle|\sqrt{\eta_1 \eta_2 \Gamma_{1}  \Gamma_{2}} dt
\end{align}
which holds for any Hermitian operators of any system. The right hand side of the equality closely resembles the original Heisenberg uncertainty relation, but contains the sum of the variances instead of the product. Unlike the latter, the sum can be bounded by a stronger inequality which is never trivial for non-commuting measurements\cite{MacCone2014Stronger}. 
This shows that the persistent diffusion observed in Fig. \ref{fig:XZFokPlanck}c is a universal consequence of the uncertainty principle and can be quantitatively derived from it.

There are several interesting applications of our work that we have not yet explored. First, with a modest improvement in quantum efficiency, one could purify a quantum state faster with non-commuting observables than with a single measurement\cite{Ruskov2012, Combes2010Replacing}. The ability to purify a state faster using quantum trajectories is a fundamental prediction of the field that has yet to be realized\cite{Jacobs2003}. This enhancement scales favorably with system size, suggesting an application for MQMs as well. 

Second, existing work on quantum foundations has focused on testing the validity of essential features of quantum mechanics, such as contextuality and the various bounds on error and disturbance. Our work presents the possibility of exploring the consequences such concepts in realistic systems\cite{Nishizawa2015ContinuousErrorDisturbance, Dressel2016LG}, which tend to interact continuously with their environments via multiple non-commuting decoherence channels. Error-disturbance bounds would provide a natural starting point, as we have separately extracted state information in Fig. \ref{fig:XZTomography}b and quantified state disturbance in Fig. \ref{fig:XZDistMap}, but have not yet investigated the intrinsic trade-off between these concepts.

\section{The Polaron Transformation}
\label{sec:XZPolaron}

\erf{eq:HEffSuppMeth} largely justifies thinking of SQMs as measurements. It couples the qubit to the cavity via a Hermitian operator, and we saw a very similar Hamiltonian lead to a QND measurement in the dispersive case. To go the rest of the way, we need to treat the cavity as an open quantum system and derive the stochastic master equation that we relied upon in the previous section. The master equation for a cavity which is damped at rate $\kappa$ and monitored with quantum efficiency $\eta$ is
\begin{align}
\label{eq:XZCavSME}
d\rho = -i [H_\text{q-frame}, \rho] dt 
+ \kappa \mathcal{D}[a] \rho ~dt + \sqrt{\kappa \eta} \mathcal{H}[a e^{i \phi}] \rho ~dW.
\end{align}
As before, we treat the cavity modes one at a time.
If the cavity starts in the ground state when this Hamiltonian is turned on, then it remains in a coherent state at all times. To derive an effective master equation for the qubit, we must eliminate the cavity entirely from \erf{eq:XZCavSME}. This may be accomplished by first applying a transformation which displaces the cavity to its ground state, which is known as the polaron transformation\cite{gambetta2008trajectory}
\begin{align}
\label{eq:XZEOM}
U &= |e_\delta\rangle\langle e_\delta| D[\alpha_{e}] + |g_\delta\rangle\langle g_\delta| D[\alpha_{g}] \\ \nonumber
\dot{\alpha}_{e/g} &= -i \epsilon_{e/g} - \frac{\kappa}{2} \alpha_{e/g} ~~~~~ \epsilon_{e/g} = \pm\tilde{g}
\end{align}
where $D$ is the cavity displacement operator \erf{eq:DisplacementOp}, $|e_\delta/g_\delta \rangle$ are the eigenstates of $\sigma_\delta$ and $\alpha_{e/g}$ is the cavity displacement conditional on the qubit state. Because the second line is the classical equation of motion for a resonantly driven cavity with drive rate $\epsilon$ and damping $\kappa$, this transformation maps the cavity to its ground state at all times, allowing us to trace it out. Transforming back from the polaron frame to the qubit frame, we find that the effective stochastic master equation for the qubit is
\begin{align}
\label{eq:XZRTerms}
d\rho &= \mathcal{L} \rho ~dt + \frac{\sqrt{\Gamma_{M}}}{2} \mathcal{H}[\sigma_\delta]\rho ~dW - i \frac{\sqrt{\Gamma_{M}'}}{2}[\sigma_\delta, \rho] dW \\ \nonumber
\mathcal{L}\rho &= -i\frac{B}{2}[\sigma_\delta, \rho] + \frac{\Gamma}{2} \mathcal{D}[\sigma_\delta] \rho \\ \nonumber
B &= 2 \tilde{g} \text{Re}(\alpha_e+\alpha_g) ~~~~~ \Gamma = -2 \tilde{g} \text{Im}(\alpha_e - \alpha_g) \\ \nonumber
\sqrt{\Gamma_{M}} &= \sqrt{\kappa \eta} |\beta| \cos(\phi - \theta_\beta) ~~~~~ \sqrt{\Gamma_{M}'} = \sqrt{\kappa \eta} |\beta| \sin(\phi-\theta_\beta) ~~~~~ \\ \nonumber &\beta = \alpha_e-\alpha_g
\end{align}
where $\theta_\beta = \text{arg}(\beta)$ is the angle of the cavity displacement axis $\beta$ in the IQ plane and $\Gamma$ is the dephasing rate. Substituting the equations of motion for the cavity \erf{eq:XZEOM} into the remaining expressions of \erf{eq:XZRTerms}, we find
\begin{align}
\label{eq:XZSME}
d\rho &= \frac{4 \tilde{g}^2}{\kappa}(1-e^{-\kappa t/2}) \mathcal{D}[\sigma_\delta]\rho ~dt + \sqrt{\frac{4 \tilde{g}^2}{\kappa} \eta}(1-e^{-\kappa t/2}) \mathcal{H}[e^{i \phi} \sigma_\delta]\rho ~dW \\ \nonumber
 &\implies \Gamma = \frac{8 \tilde{g}^2}{\kappa} = \frac{2\chi^2 \bar{n}_0}{\kappa}.
\end{align}
We align the LJPA amplification axis to the axis of displacement arising in \erf{eq:HEffSuppMeth}, so that $\phi = \theta_\beta$. Application of sidebands to another mode of the cavity does not change the above derivation except to add an additional measurement also modeled by \erf{eq:XZSME}. Neglecting time-scales of order $1/\kappa$, measurement of two observables $\sigma_{\delta_1}$ and $\sigma_{\delta_2}$ is modeled by
\begin{align}
\label{eq:XZSMEFinal}
d\rho = \sum_{i=\{1,2\}} \frac{\Gamma_i}{2} \mathcal{D}[\sigma_{\delta_i}]\rho ~dt + \sqrt{\frac{\Gamma_i \eta_i}{2}}\mathcal{H}[\sigma_{\delta_i}]\rho~dW \\ \nonumber
r_i dt = \langle \sigma_{\delta_i} \rangle dt + \frac{dW_i}{\sqrt{2 \eta_i \Gamma_i}}
\end{align}
where $r_i$ is the measurement signal at time t, normalized appropriately. Equation (\ref{eq:XZSMEFinal}) can also be converted to a Fokker Planck equation, which propagates probability distributions of $\rho$. We use the latter to generate theory plots for Fig. \ref{fig:XZFokPlanck}. This stochastic master equation is generated by the following measurement operator
\begin{align}
\label{eq:XYMeasOpp}
\Omega_{r_1(t),r_2(t)} &= \exp \left[\sum_{i={1,2}} -\frac{\Gamma_{i} \eta_i}{2} \left(r_i(t)-\sigma_{\delta,i}\right)^2 dt \right] \\ \nonumber
\rho(t+dt) &= \mathcal{E}_{1-\eta_i}\frac{\Omega \rho(t) \Omega^\dagger}{\text{Tr}[\Omega \rho(t) \Omega^\dagger]} \\ \nonumber
\end{align}
where $\mathcal{E}_{1-\eta_i}$ is a superoperator which models dephasing due to finite quantum efficiency. To assure positivity of the state when $dt$ is taken to be finite, we use \erf{eq:XYMeasOpp} to numerically propagate quantum trajectories and also to calculate probabilities for maximum likelihood reconstruction. 

It is natural to ask if the above calculation generalizes to multi-quadrature measurements. It turns out that for systems larger than a qubit, the polaron transformation does not necessarily lead to a Markovian master equation\cite{criger2016multi}. Nevertheless, the methods of chapter \ref{ch:OpenSystems}, section \ref{sec:QEDReadout} can be used to show that the effective dynamics still damp off-diagonal elements and project into an eigenstate of $X$. Thus we can still use MQM measurements for quantum trajectories and feedback.

%% file: AdaptivePhase.tex
\chapter{Adaptive Measurements and the Canonical Phase Measurement}
\label{ch:AdaptivePhase}

Much of modern metrology and communication technology encodes information in electromagnetic waves, typically as an amplitude or phase. While current hardware can perform near-ideal measurements of photon number or field amplitude, to date no device exists that can even in principle perform an ideal phase measurement, as we noted in chapter \ref{ch:IntrocQED}. However phase is no less fundamental than amplitude and power, and underlies a huge range of technological applications, from FM radio and optical communication to metrological interferometry such as gravitational wave detection and even long-baseline telescopes.

In this chapter, we describe the implementation of a single-shot canonical phase measurement on a one-photon wave packet. As defined below, the canonical phase measurement defines the quantum-mechanically ideal POVM for measurement of phase. The measurement surpasses the standard method of heterodyne detection and is optimal for single-shot phase estimation. Our system adaptively changes the measurement basis of a parametric amplifier (operated in phase-sensitive mode) during photon arrival. Changing the measurement basis during the measurement process constitutes a form of quantum feedback, and provides much more control to optimize sensitivity to phase. We use a superconducting qubit to emit our photon signal, which allows us to validate the detector's performance by tracking the quantum state of the source. This gives us an opportunity to put to practical use the physics of quantum trajectories.

These results are part of a broader effort to understand and demonstrate the full capabilities of adaptive measurement. In chapter \ref{ch:MeasurementControl}, we saw many examples in which quantum feedback offered fundamentally enhanced control over state preparation as compared to what can be achieved with measurement alone. In these cases, we applied feedback operations to the device that we wished to control, such as a register of qubits. The canonical phase measurement gives an example in which quantum feedback fundamentally enhances the capabilities of a measurement device. In particular, our use of feedback converts a standard JPA (with a TWPA as backup) into a theoretically ideal phase detector. In contrast to control applications, the protocol works by changing the measurement basis (and hence POVM) in response to the incoming signal. In the broadest sense, we can think of adaptive measurement as a kind of `POVM-generating machine,' which widens the range of observables that may be accessed by a detector.\footnote{Thanks to John Preskill for coining this term.} The following work demonstrates that quantum feedback can both enhance the precision of a detector and enable it to measure new classes of physical observables. Excitingly, the full theoretical capabilities of adaptive measurement are not yet known.

Even the definition of phase as an observable in quantum mechanics is surprisingly subtle, and has lead to a protracted debate that took decades to resolve\cite{London1926, London1927, susskind1964phase, nieto1993quantum}. In chapter \ref{ch:IntrocQED}, we saw the definition of a somewhat suspicious phase operator
\begin{align} \label{eq:PhaseOpNoBound}
    \hat{\varphi} = \int_0^{2\pi} \varphi |\varphi\>\<\varphi| \\ \nonumber 
    |\varphi\> = \frac{1}{\sqrt{2\pi}} \sum_{n=-\infty}^{\infty} e^{i\varphi n} |n\>.
\end{align}
As $\varphi$ is $2\pi$ periodic, the limits of the integration could just as well have been chose as $-\pi$ and $\pi$ or any other $2\pi$ interval, and therefore $\hat{\varphi}$ has no unique definition. Fortunately, this ambiguity posed no problem for the physically relevant operators of that section, such as $\cos(\hat{\varphi})$, but it does raise some subtleties in the discussion of phase measurements.

Things get even more complicated when one considers a truncated version of the above Hilbert space, such as a harmonic oscillator. This case is perhaps the most important from an applied standpoint, as it models the phase of an electromagnetic wave. As the lower bound on $n$ is now zero instead of $-\infty$, $|\varphi\>$ no longer forms an orthonormal basis. As the eigenbasis of a Hermitian operator is necessarily orthonormal, this implies that there is no Hermitian operator corresponding to the phase basis $|\varphi\>$, and thus the standard measurement formalism cannot apply here.

Fortunately, the POVM formalism introduced in chapter \ref{ch:OpenSystems} provides the generalization of measurement that naturally works with phase. The set of Kraus operators
\begin{align}
\Omega_\phi &= |\phi\>\<\phi| \\ \nonumber
|\phi\> &\equiv \frac{1}{\sqrt{2\pi}} \sum_{n=0}^\infty e^{i\phi n} |n\>
\end{align}
satisfy the required normalization condition
\begin{equation}
    \int_0^{2\pi} \Omega^\t \Omega d\phi = \<\phi|\phi\> \int_0^{2\pi} |\phi\>\<\phi| d\phi = 2\pi \<\phi|\phi\> I.
\end{equation}
This relation also holds if we further truncate the Hilbert space down to a qudit, meaning that it defines a phase measurement on any finite system as well. 

There are several appealing features of the canonical phase basis. Firstly, if a signal is encoded to a state by applying phase shifts, the canonical phase measurement is optimal when no prior information on the phase is available, as is the case in communication\cite{hall1991quantum}. Secondly, although for many reasons we cannot write an expression like $[\hat{\phi},\hat{n}] = i$, there is a clear though unique sense in which $\phi$ is canonically conjugate to $\hat{n}$. Phase is a $2\pi$ periodic variable, so the standard definition of variance does not make sense if we wish to write down a Heisenberg uncertainty relation like $\Delta \phi^2 \Delta n^2 \geq 1/4$. However there is a generalization of variance that is compatible with periodic variables, called the Holevo variance
\begin{align} \label{eq:HolevoVariance}
\Delta \phi^2 \equiv |\<e^{i\phi}\>|^{-2} - 1.
\end{align}
For a uniform distribution over $\phi$, $\Delta \phi^2=\infty$. Furthermore in the limit of a narrowly peaked distribution, \erf{eq:HolevoVariance} reproduces the standard definition of variance. Most importantly, if we use this definition of variance, then the above uncertainty is principle exact\cite{berry2002adaptive}. Although $\phi$ is not an operator, $\phi$ labels the measurement outcome $\Omega_\phi$, and thus the number-phase uncertainty relation is exactly analogous to the Heisenberg uncertainty relation. Intuitively, this implies a bizarre property of the canonical phase measurement, which is that it is completely insensitive to the incident power. An ideal phase detector would be fundamentally incapable of distinguishing the vacuum from a bright, phase-incoherent light, but it could optimally detect the coherent phase of a laser beam.

Canonical phase measurements fit into the broader context of quantum phase estimation\ped{\cite{pezze2014quantum}}. Here we focus on performing an optimal \textit{single-shot} phase measurement, meaning that given a single copy of a state to which we have applied some phase shift $e^{i\phi a^\t a}$, we wish to optimally estimate $\phi$. This formulation underlies more intricate versions of the phase estimation problem, such as estimating $\phi$ given multiple copies of the same state, or given the ability to choose the input state to the phase shift operation. The latter enables Heisenberg-limited scaling of the measurement uncertainty, for example by applying $e^{i\phi a^\t a}$ to multiple subsystems of an entangled state\cite{giovannetti2006quantum}.

In the absence of an instrument capable of implementing a canonical phase measurement, heterodyne detection, in which one measures a rapidly varying quadrature of the input, serves as the standard technique for estimating the phase of an unknown signal. Several schemes have surpassed heterodyne detection 
\cite{yonezawa2012phasetracking,wheatley2010adaptive,armen2002adaptive,iwasawa2013mirrormotion}, 
 however these protocols also acquire undesired photon number information, and thus cannot reach the quantum limit or implement a canonical phase measurement.
In this work, we implement a feedback-controlled quantum-limited amplifier which dynamically updates its amplitude measurement in response to the incident field. When the system continuously optimizes this measurement basis for phase sensitivity, it implements a canonical phase measurement on an incoming single-microwave-photon state\cite{wiseman1995adaptive}. We verify implementation of a canonical phase measurement using the entanglement between the emitted photon and its source, which allows us to confirm that acquisition of photon number information is suppressed. The system surpasses heterodyne detection by $15\pm 2$\%. 
The first section of this chapter present the experimental work, and the second section derives the theory behind the experiment.

These results are part of a broader effort to understand quantum feedback applied to adaptive measurement, which is still an exploratory direction of research even theoretically. Given a set of POVMs that may be implemented by a given detector, it is not known how to construct the set of POVMs that may be implemented by adaptive measurement. In the following work, we start with the set of POVMs generated by infinitely short homodyne measurements in an arbitrary basis and generate the canonical phase measurement. The result is one of many examples given in this thesis in which quantum feedback enhances experimental capabilities in the presence of linearity constraints. In deterministic remote entanglement generation, we required quantum feedback to circumvent the impossibility of a linear optics Bell measurement\cite{Calsamiglia2001maximum}. We speculate that there are many further examples in which quantum feedback can improve a linear optics experiment or other system with severe constraints on the set of POVMs that may be implemented. Conversely, when strong nonlinearities are already available, the potential gains of adaptive measurement narrow substantially. A formal piece of evidence for this suggestion is Neumark's dilation theorem, which shows that \textit{any} POVM may be implemented by applying a unitary operation on the system plus some auxiliary system and then applying a projective measurement on the latter, as in \erf{eq:POVMFromUnitary}. This implies that there exists a digital quantum circuit that can implement any POVM, circumventing any possible need for adaptive measurement. This result certainly does not obviate the need for adaptive measurement. Quantum computers are a long way off, and it will not necessarily be possible to load the quantum state of the system that we wish to measure into it.

\section{Implementation and Verification of a Canonical Phase Measurement}

\begin{figure}
\centering
{\includegraphics[width = 0.8\textwidth]{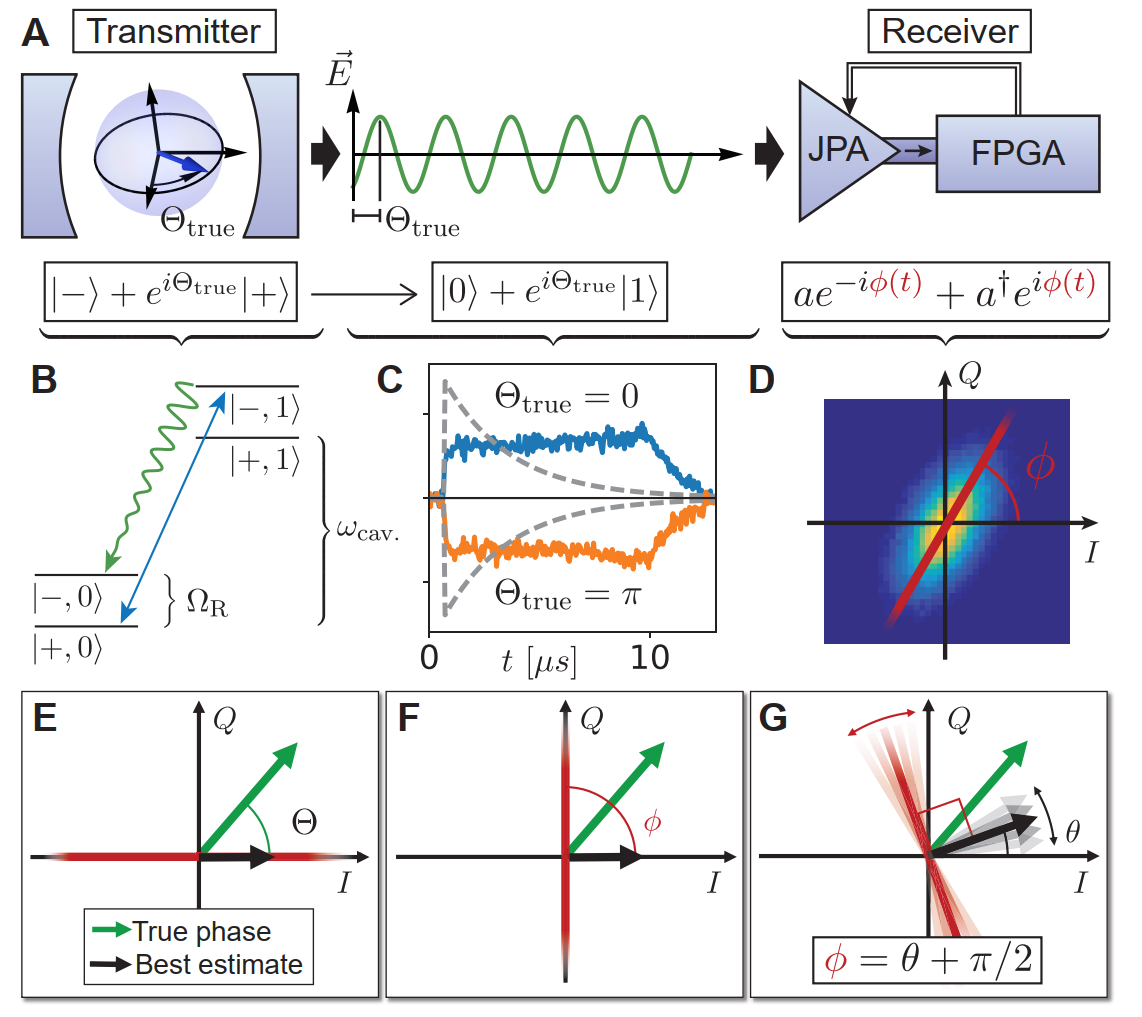}} 
\caption{
Experimental implementation.
(a) Atom in a cavity, with phase $\Theta_\mathrm{true}$ encoded into its dipole moment. The atom decays and emits a photon into a 1D waveguide with phase encoded into the electric field as shown. The JPA receives the photon and measures an amplitude quadrature selected by the FGPA.
(b) Sideband cooling scheme to emit photon. Sideband converts a qubit excitation to a cavity excitation, which is then emitted as a single photon at the cavity frequency.
(c) Measured mode shape (E-field envelope) of emitted photon. Dashed line shows mode shape if constant cooling rate were used instead.
(d) Output of JPA. Signal is amplified along measurement axis $\phi$ and squeezed along the other. 
(e-g) Estimating and tracking state by changing measurement basis. Receiver attempts to maintain the phase measurement condition $\phi=\theta+\pi/2$. See text for details.
}
\label{fig:ExperimentalSetup}
\end{figure}

As shown in Fig. \ref{fig:ExperimentalSetup}a, our system consists of a transmitter, which encodes a variable $\Theta_\mathrm{true}$ into the phase of a single-photon electromagnetic signal, and a receiver, which uses a continuous feedback protocol to guess this phase in a single shot using an adaptive feedback protocol. A superconducting transmon qubit\cite{koch2007transmon} embedded in a 3D aluminum cavity acts as the transmitter. We use coherent bath engineering \cite{murch2012cavity} of this artificial atom to generate our photonic state, which yields more process control than direct spontaneous decay. To implement this scheme, we Rabi drive our qubit at $\Omega_\mathrm{R}/2\pi = 20$ MHz, which creates an effective low-frequency qubit. Simultaneously, we apply a cavity sideband at $\omega_\mathrm{cav.}+\Omega_\mathrm{R}$, where $\omega_\mathrm{cav.}$ is the cavity resonance frequency. As shown in Fig. \ref{fig:ExperimentalSetup}b, the sideband drives a transition from the $|+,0\rangle$ state to $|-,1\rangle$ state, where $|\pm\rangle \equiv (|e\rangle \pm i|g\rangle)/\sqrt{2}$ are the dressed states of qubit under driving and $0,1$ count the number of photons in the cavity. The cavity then decays, emitting a photon and leaving the system in the $|-,0\rangle$ state, which is not affected by the sideband. We ensure that the cavity decay rate is fast compared to the sideband-induced coupling, so that the qubit's effective decay rate from $|+\rangle$ to $|-\rangle$ is limited by the sideband amplitude. By modulating the sideband amplitude during photon emission, we generate a photon with a flat modeshape (Fig. \ref{fig:ExperimentalSetup}c), which greatly ameliorates the detrimental effects of feedback delay at the receiver\cite{pozza2015deterministic}. The details of mode shape control are derived in the following section. To encode the phase $\Theta_\mathrm{true}$, we prepare the qubit in a superposition state of the form $(|-\rangle+e^{i\Theta_\mathrm{true}}|+\rangle)/\sqrt{2}$, which decays by emitting the photonic state $(|0\rangle + e^{i\Theta_\mathrm{true}}|1\rangle)/\sqrt{2}$.

Our receiver consists of a Josephson parametric amplifier (JPA) (like that of chapter \ref{ch:IntrocQED} pumped at twice its resonance frequency by a field-programmable gate array (FPGA), which serves as a classical feedback controller (Fig \ref{fig:ExperimentalSetup}a). To maintain high measurement bandwidth for quantum feedback, we operate the JPA at a relatively low gain of $6$ dB, which yields a gain bandwidth of $45$ MHz, and follow it with a traveling wave parametric amplifier\cite{macklin2015twpa} (not shown) to boost the signal strength and maintain a quantum efficiency of $\eta = 0.4$. The JPA measures field amplitude via the quantum mechanical quadrature operator $a e^{-i\phi(t)} + a^\dagger e^{i\phi(t)}$, where $a$ is the quantum mechanical annihilation operator of the incident field and $\phi(t)$ is the instantaneous phase of the parametric pump. 

To perform a canonical phase measurement on the incident field, the feedback controller continuously adapts the measurement axis $\phi(t)$ as the photon arrives at the receiver\cite{wiseman1995adaptive}. The measurement axis is chosen to maximize the acquisition of phase information as follows. Before the photon reaches the JPA, the receiver has no information and therefore chooses $\phi$ arbitrarily. Upon arrival of a portion of the photon, the JPA detects a small positive (or negative) fluctuation, which then informs the system that the true phase is likely oriented along (or opposite) the measurement axis (Fig. \ref{fig:ExperimentalSetup}e). At this point, any further measurement in this basis interrogates the amplitude of the incident field and thus yields undesired photon number information. Ideally, the system would then rotate the measurement axis by 90 degrees (Fig. \ref{fig:ExperimentalSetup}f), so that a small deviation between the current best estimate of the phase $\theta(t)$ and the true phase $\Theta_\mathrm{true}$ would be detectable as a positive or negative fluctuation in the signal. As the photon continues to arrive, the feedback controller gains more information and updates the phase $\phi(t)$ to maximize sensitivity to phase (Fig. \ref{fig:ExperimentalSetup}g). If the phase measurement condition $\phi(t) = \theta(t) + \pi/2$ is maintained at all times, then the system acquires no photon number information and implements a canonical phase measurement.

\begin{figure}
\centering
{\includegraphics[width = 100mm]{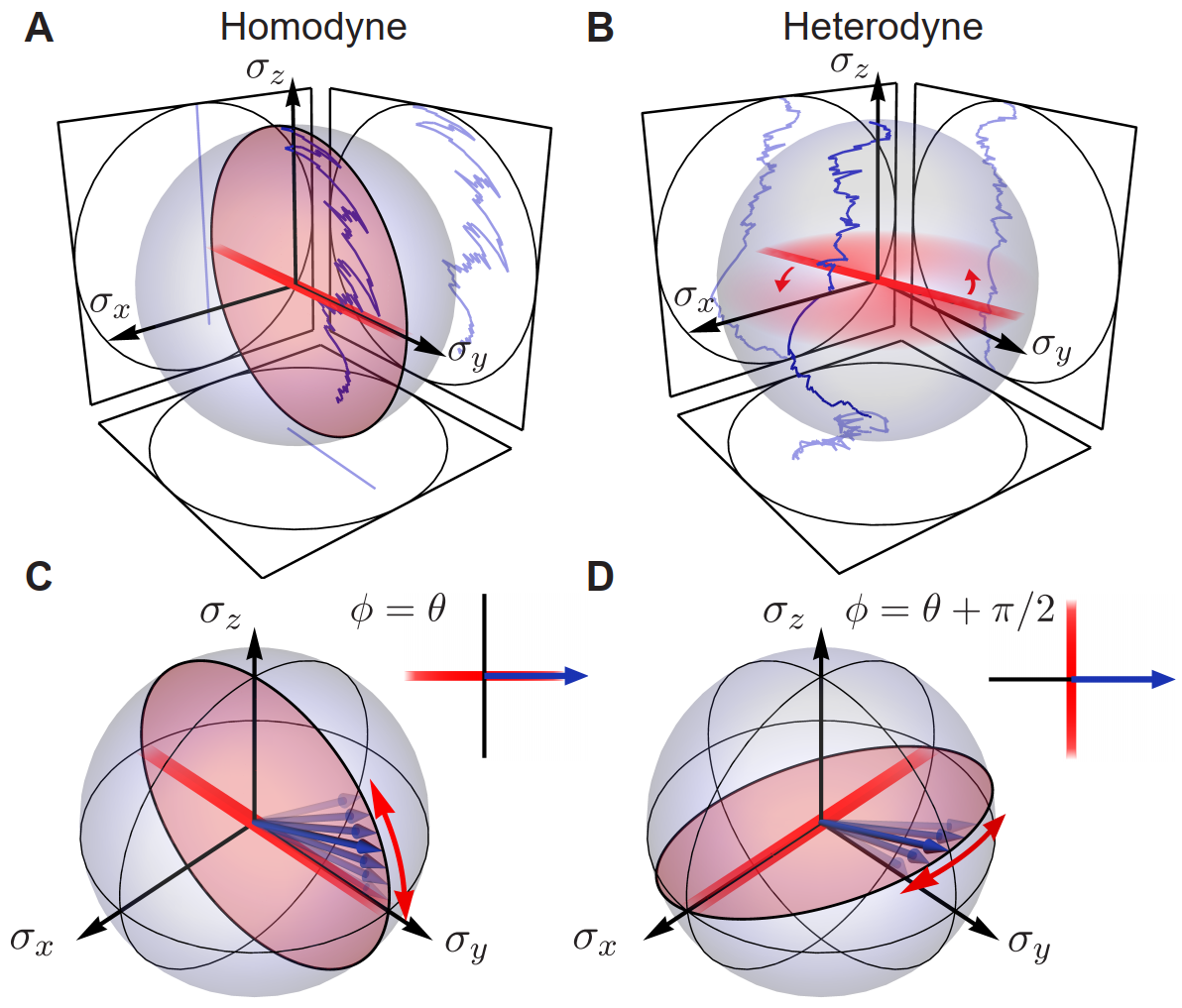}}
\caption{
Measurement back-action and quantum trajectories. Coordinate axes are chosen so that the atom decays to $\sigma_z=-1$. (a) A single homodyne quantum trajectory ($\phi(t) = 0$). State only propagates in the plane of the measurement axis. (b) A single heterodyne trajectory ($\phi(t) = \omega_\mathrm{het.} t$). The qubit is initialized in $|+\rangle$ for both trajectories. (c) Amplitude back-action, which occurs when the measurement axis (red line) is aligned to the best estimate of the state (blue arrow). (d) Phase back-action, which occurs when the phase measurement condition is satisfied. 
}
\label{fig:BackAction}
\end{figure}

To track the best estimate of the phase, the feedback controller must continuously update its best guess of the atom's state based on the measurement signal starting with no prior information \textit{i.e.} it should track the quantum trajectories of the system\cite{Murch:2013ur,campagne2016observing} given an initially maximally mixed state. 
We begin by observing and verifying quantum trajectories for homodyne ($\phi(t)=0$) and heterodyne ($\phi(t) = \omega_\mathrm{het.} t$, $\omega_\mathrm{het.}/2\pi =0.5$ MHz) detection\cite{campagne2016observing}. Example trajectories are plotted in Fig. \ref{fig:BackAction}a,b and tomographically validated in \cite{martin2019adaptive}. These data allow us to characterize measurement back-action and check consistency with theory. The stochastic component of the back-action always lies in the plane of the instantaneous measurement basis, as is clear from the homodyne data.

The presence of back-action not only governs how to adapt the measurement axis $\phi(t)$, but also offers a method to independently validate the receiver's implementation of a canonical phase measurement. 
Because an ideal phase measurement acquires maximal phase information and no photon-number information, it maximally disturbs the atomic dipole phase while minimally disturbing the atomic excitation probability. This effect is directly visible in the quantum trajectories, as illustrated conceptually in Fig. \ref{fig:BackAction}c and d. When the measurement axis is aligned with the best estimate of the phase ($\phi = \theta$), the resulting acquisition of amplitude information manifests as a random disturbance of the qubit state along the axis of decay (Fig. \ref{fig:BackAction}c). Conversely, when the phase measurement condition is satisfied ($\phi = \theta+\pi/2$), then only the phase of the qubit state is subject to noise (Fig. \ref{fig:BackAction}d). In this way, we can verify the performance of our receiver by characterizing the dynamics of the transmitter. This capability is uniquely quantum, and arises from entanglement between the atom and its emitted photon.

\begin{figure}
\centering
{\includegraphics[width = 0.7\textwidth]{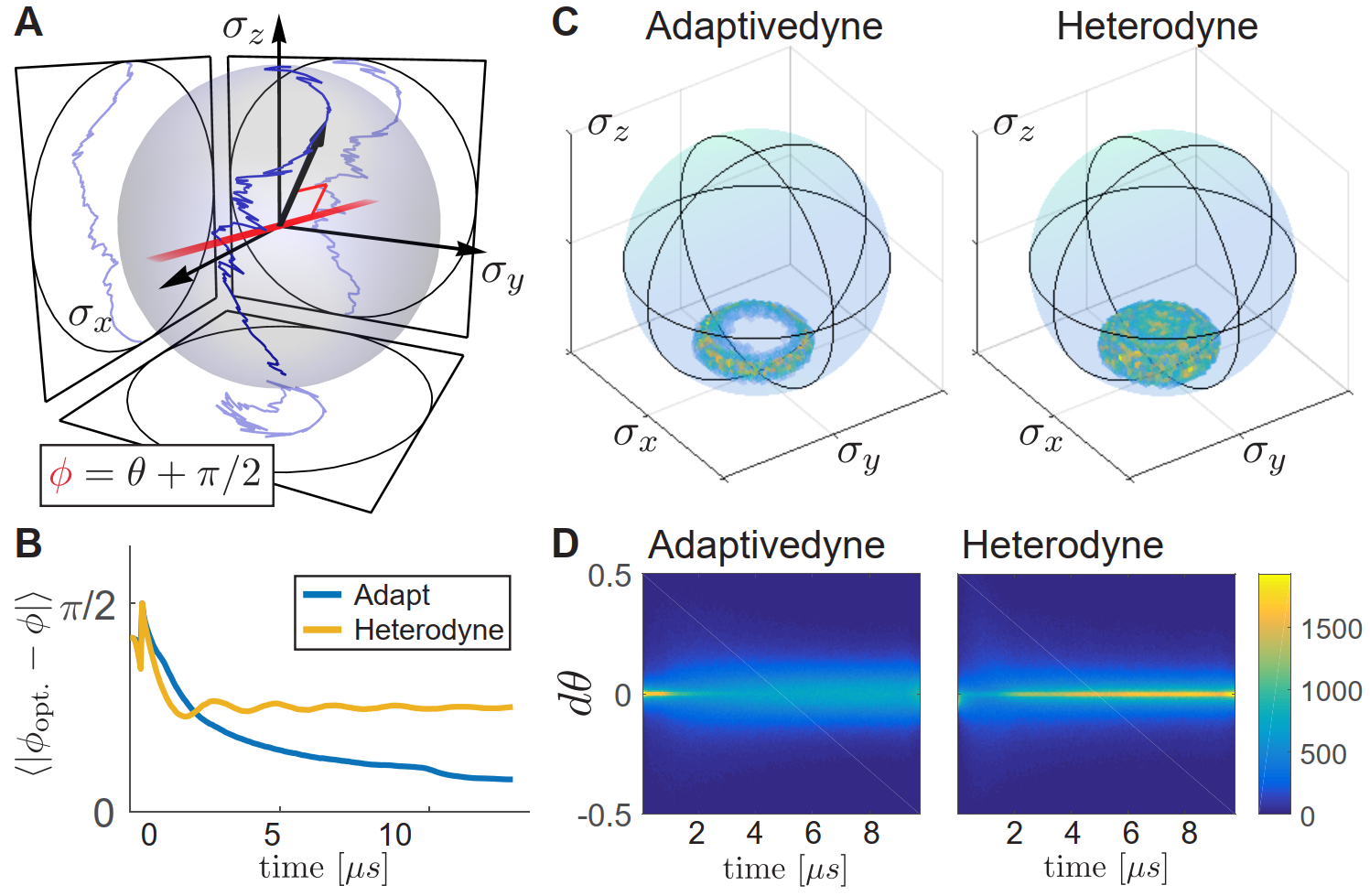}}
\caption{
Back-action and measurement validation. (a) A single adaptive-dyne quantum trajectory. The red right-angle bracket emphasizes orthogonality between the measurement axis and the state. (b) Quality of tracking for heterodyne and adaptivedyne, where $\phi_\mathrm{opt.} = \theta(t)+\pi/2$. Adaptivedyne significantly outperforms the heterodyne and comes close to the ideal phase by $T=13 \mu s$. The difference $\phi_\mathrm{opt.}-\phi$ is cut to lie on the interval $[-\pi/2,\pi/2]$. (c) Distribution of trajectories at $t=10 \mu s$. Due to suppression of photon-number back-action, adaptivedyne trajectories cluster in a ring at late times.
(d) Statistics of the phase back-action $d\theta$ for adaptivedyne and heterodyne. 
On average, the phase back-action is significantly larger for adaptivedyne. 
}
\label{fig:AdaptiveBackaction}
\end{figure}

We show the results of this verification scheme in Fig. \ref{fig:AdaptiveBackaction}. Fig. \ref{fig:AdaptiveBackaction}a shows a single quantum trajectory under adaptivedyne detection, in which $\phi(t)$ is continuously adapted by the feedback controller. Fig. \ref{fig:AdaptiveBackaction}b shows the difference between the ideal quadrature phase and the measured phase, which shows that the feedback controller approximately maintains the phase measurement condition. To interpret the dynamics, we plot the ensemble statistics of the phase back-action as a function of time in Fig. \ref{fig:AdaptiveBackaction}d, with the heterodyne detection case included for comparison. It can be seen that the phase back-action $d\theta$ is significantly larger for adaptivedyne detection. Fig. \ref{fig:AdaptiveBackaction}c shows the ensemble statistics of the state at $t=10 ~\mu$s. As observed in \cite{campagne2016observing}, the quantum trajectories of a decaying atom evolve on a spherical shell that shrinks deterministically to the south pole of the Bloch sphere. Due to the suppression of back-action along the decay axis, adaptivedyne trajectories are further confined, exhibiting something closer to a ring-like structure. This feature presents an unambiguous signal that our system approximately implements a canonical phase measurement.









\begin{figure}
\centering
{\includegraphics[width = 120mm]{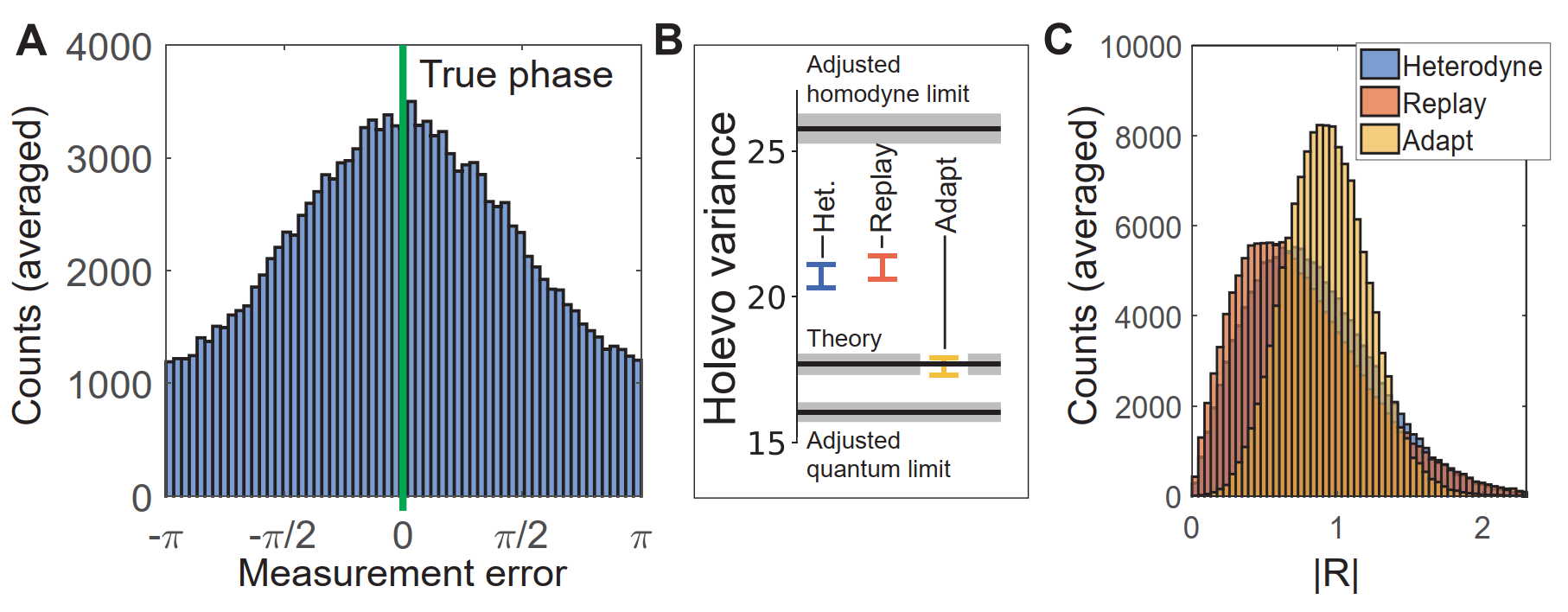}}
\centering
\caption{
Phase-estimation performance. (a) Histogram of the difference between the measurement outcome and the true phase \textit{i.e.} $\theta(T)-\Theta_\mathrm{true}$ (b) Performance is evaluated by computing the Holevo variance of this distribution. Quantum limit (bottom black line) homodyne limit (top black line) and absolute theory prediction based on feedback delay are inferred from the performance of heterodyne, with corresponding error bars shown as gray rectangles. (c) Distribution of the amplitude information. The distribution is significantly narrower for adaptivedyne, indicating suppression of this information channel. %
}
\label{fig:PhaseEstimation}
\end{figure}

A canonical phase measurement should outperform heterodyne detection in estimating the phase $\Theta_\mathrm{true}$. To verify superior performance, we prepare our qubit in one of 8 equally spaced points along the equator of the Bloch sphere. From each shot, the receiver optimally\cite{pozza2015deterministic} estimates the phase of the photon by computing the following quantity
\begin{equation} \label{eq:R}
R = \int_0^T e^{i \phi(t)}\sqrt{u(t)}r(t) dt
\end{equation}
where $u(t)$ is the photon mode shape, and $T$ is the duration of each experimental run and $r(t)$ is the measurement signal read out from the JPA normalized such that its variance is $dt$. The best estimate of the photon's phase in a single shot is given by the complex argument $\theta(T) = \arg(R)$. Fig. \ref{fig:PhaseEstimation}a plots a histograms of this best estimate for adaptivedyne detection, which exhibits the $\cos(\theta-\Theta_\mathrm{true})$ dependence expected theoretically\cite{pozza2015deterministic}.

We compare the performance of adaptivedyne and heterodyne detection by plotting the Holevo variance of each underlying distribution in Fig. \ref{fig:PhaseEstimation}b. We also include data for what we term replay detection, in which $\phi(t)$ from an adaptivedyne shot of the experiment is replayed instead of feeding back based on the current signal. In this way, we can confirm that it is the correlations between $\phi(t)$ and the state that yield enhanced performance, rather than the independent statistics of $\phi(t)$. For additional confirmation, we independently measure the signal-to-noise ratio of our amplifier chain for heterodyne and adaptivedyne detection and verify that it remains the same to well within 1\%\cite{martin2019adaptive}. Heterodyne and replay perform equally well, and are both significantly surpassed by a canonical phase measurement implemented via adaptivedyne detection. Adaptivedyne does not reach the quantum limited Holevo variance of 3 due to a combination of loss, qubit decoherence and feedback delay. However from our heterodyne data we infer an adjusted quantum limit given our quantum efficiency and purity of the emitted photon, as well as the hypothetical homodyne limit. The canonical phase measurement comes significantly closer to this adjusted quantum limit than any other scheme, limited almost entirely by feedback delay.

We infer the sensitivity of each scheme to photon-number information from the distributions of $|R|$, which are shown in Fig. \ref{fig:PhaseEstimation}. The distributions for heterodyne and replay are almost identical, while the adaptivedyne histogram is substantially narrower, indicating that the latter is less sensitive to this undesired information\cite{pozza2015deterministic}.

The above experiment gives an important example of the power of quantum feedback applied to adaptive measurement. There are a number of tangible, immediate applications to quantum information. Firstly, the implementation of quantum feedback on a detector is known to allow enhanced readout of superconducting circuits\cite{sarovar2007adaptive}. Furthermore, the ability to perform a canonical phase measurement enables linear-optics preparation of the $|0\rangle+|1\rangle$ photonic state, which is a major experimental challenge of single-rail linear optics quantum computing\cite{ralph2005adaptive}. More broadly, it is known that adaptive measurements are universal\cite{oreshkov2005weak}, meaning that many relatively simple measurement devices augmented with quantum feedback can perform any measurement allowed by quantum mechanics. Thus our extension of a standard amplitude measurement device to a theoretically ideal phase measurement represents one of potentially many possible directions for future work in this field. 

\section{Adaptive Measurements and Quantum Feedback}

We now derive the feedback protocol that enables the implementation of a canonical phase measurement in the $|0\>$, $|1\>$ manifold by solving the stochastic master equation for the source. Before diving into the math, we outline where we are going with this calculation and what we hope to extract from it. A general photonic state with a well-defined phase in the subspace considered can be written as
\begin{align}
|\psi\> = \frac{1}{\sqrt{2}} \left[ |\vec{0}\> + e^{i\Theta} \int_{-\infty}^t \sqrt{u(s)}  a^\t[s] |\vec{0}\>\right]
\end{align}
where $|\vec{0}\>$ denotes the electromagnetic vacuum and $u(t)$ is a mode shape that square-integrates to 1. In solving the stochastic master equation for the artificial atom with a time-dependent cooling rate $\gamma(t)$, we will derive the atomic excitation probability as a function of time, which fixes $u(t)$ by conservation of energy. We can also use a solution to the master equation to determine the POVM implemented by the experiment. Given a measurement record $r(t)$ occurring experimentally, the unnormalized stochastic Schr\"odinger equation produces an unnormalized final state $|\tilde{\psi}(\infty)\>$. We can use this to compute probability for $r(t)$ to occur given any arbitrary initial state, which in turn lets us determine the measurement operator $E[r]\equiv \Omega^\t_r \Omega$ associated to $r(t)$
\begin{align} \label{eq:POVMFromTrajSol}
P[r(t)|\psi(0)] = \<\tilde{\psi}(\infty)|\tilde{\psi}(\infty)\> = \Tr(E[r(t)] |\psi(0)\>\<\psi(0)|]
\end{align}
By matching terms for different initial states $|\psi(0)\>$, we can compute $E[r(t)]$ and confirm that adaptive measurement leads to the canonical phase measurement. In principle we can also use the solution to verify quantum trajectories, though in practice we use numerics to take decoherence and finite measurement efficiency into account.

To solve for the dynamics under adaptive measurement, we use the full master equation to derive the required $\gamma(t)$, which also lets us calculate the optimal feedback strategy for a given photon mode shape. Many of these results are also derived in \cite{pozza2015deterministic}, but we include a variant of that derivation here for completeness. As photon loss does not affect the decay dynamics or the best estimate of the phase, we assume $\eta=1$ for this analysis. We also neglect other forms of decoherence, which have a negligible effect on the decay dynamics. This allows us to perform our computations with a pure state, so we begin with the unnormalized stochastic Schr\"odinger equation for an atom observed via homodyne detection\cite{Wiseman2009book}, which provides a state update from the acquisition of an infinitesimal amount of information via homodyne detection of the atom's spontaneous emission
\begin{align} \label{eq:SSEAdaptivePhase}
\frac{d}{dt}|\tilde{\psi}\rangle = \left[-\frac{1}{2} \gamma(t) \sigma^\dagger \sigma + \sqrt{\gamma(t)}e^{-i \phi(t)}\sigma r(t) \right]|\tilde{\psi}\rangle.
\end{align}
%
$r(t)$ is the measurement record, $\sigma=|-\rangle\langle+|$ and $|\tilde{\psi}\rangle$ is the unnormalized pure state describing the state of the atom\footnote{It is a slight abuse of notation to write \erf{eq:SSEAdaptivePhase} as a differential equation, as $r(t)$ is actually an unbounded stochastic quantity, but this convention is nevertheless common in the literature.}. As in most references, we have dropped the normalization prefactor $\sqrt{P(dW)}$ that we included in \erf{eq:LinearSSE}. This normalization factor turns out to be unimportant for reasons that will be clarified later. 
If we write $|\tilde{\psi}\rangle$ as $|\tilde{\psi}\rangle=c_-|-\rangle+c_+|+\rangle$, then the equation of motion for $c_+$ is
\begin{align} \label{eq:EOMcplus}
\frac{dc_+}{dt} = -\frac{1}{2}\gamma(t)c_+ ~~\rightarrow~~ c_+(t) = c_+(0)e^{-\frac{1}{2}\int_0^t\gamma(s)ds}
\end{align}
where we have assumed that $\gamma(t<0)=0$. Recall that our system decays from $|+\rangle$ to $|-\rangle$. Although in general \erf{eq:SSEAdaptivePhase} does not preserve the norm of $|\tilde{\psi}\rangle$, one nevertheless derives the correct equation of motion for the average population from the above in the absence of measurement. 
The result is $d|c_+|^2/dt = \gamma(t)|c_+|^2$, which coincides with the expectation based on a standard rate equation for decay of the excited state population. We identify the mode shape with the instantaneous emitted intensity, assuming for a moment that the atom was initialized with $c_+=1$, $c_-=0$
\begin{align}
u(t) \equiv \gamma(t)|c_+|^2 = \gamma(t) e^{-\int_0^t\gamma(s)ds}.
\end{align}
Notice that $u(t)$ integrates to 1 for any $\gamma(t)$. If we demand a flat mode shape so that $u(t)$ is constant, then $\gamma(t) = 1/(\tau-t)$, where $\tau=10\mu s$ parameterizes the photon's duration. As $\gamma(t)$ diverges at $t=\tau$, we set a maximum cooling rate of $1.4$ MHz and cool at this maximum rate for several microseconds longer than $\tau$, such that more than 99\% of the excited state population has decayed by $T=13\mu s$. 
The constant $\gamma$ portion when $\gamma/2\pi=1.4$ MHz coincides with the portion of the photon that decays exponentially, as can be seen in \ref{fig:ExperimentalSetup}C.

Now that we have developed the necessary tools for emitting a flat photon, we derive the optimal feedback protocol given our photon. The equations of motion for $c_-$ determine the best estimate of the phase
\begin{align} \label{eq:EOMcminus}
\frac{dc_-}{dt} = c_+ \sqrt{\gamma(t)}e^{-i\phi(t)}r(t) ~~ \rightarrow ~~ c_-(t) = c_-(0) + c_+(0)\int_0^t e^{-i\phi(s)} \sqrt{u(s)} r(s) ds.
\end{align}
Notice the similarity between \erf{eq:EOMcminus} and \erf{eq:R} of the previous section.
For feedback, we wish to compute the best estimate of the atomic dipole phase at time $t$ assuming that the controller initially has no information about the phase. This best estimate coincides with the best estimate for the phase of the emitted photon after that time. To compute it, we note that the dynamics are trivial if the system is initialized in $|-\rangle$, so that the dipole phase evolution of the zero-knowledge mixed state $\rho_0 = (|-\rangle\langle-| + |+\rangle\langle+|)/2$ is entirely determined by the dynamics of the second term. Again taking $c_+=1$, $c_-=0$, the dipole phase is given by the relative complex phase between $c_+$ and $c_-^*$. The complex phase of $c_+$ remains constant, so the dipole moment phase is simply
\begin{align} \label{eq:RIntegralSM}
\theta(t) = \arg(R), ~~ R(t) \equiv \int_0^t e^{i\phi(s)} \sqrt{u(s)} r(s) ds
\end{align}
in agreement with \erf{eq:R}. In principle, \erf{eq:R} 
and the phase measurement condition $\phi(t) = \theta(t)+\pi/2$ define the optimal protocol. For ease of implementation, this protocol may be further simplified by solving for the absolute value and complex argument of $R$ individually as follows. If the controller maintains the phase measurement condition, then we have $\exp(i\phi(s)) = i R/|R|$. Making this substitution and differentiating with respect to $t$ yields
\begin{align} \label{eq:dR}
dR = i \frac{R}{|R|}\sqrt{u(t)} r(t)dt.
\end{align}
To compute a differential equation for $|R|$, one must be aware that $r(t)$ is a random variable. $r(t)dt$ is unbounded, and the standard chain rule of differential calculus must be replaced with Ito's lemma, which looks like the chain rule but expanded to higher order like a Taylor series. As $r(t)$ is normalized to have a variance $dt$, $(r(t)dt)^2 = dt$ and we have
\begin{align} \label{eq:RSol}
d|R|^2 = u(t)dt ~~ \rightarrow ~~ |R(t)|^2 = \int_0^t u(s)ds.
\end{align}
Thus the time evolution of $|R|$ is deterministic. Substituting this solution into \erf{eq:dR} yields
\begin{align} 
dR = i R P(t) r(t)dt, ~~ P(t) \equiv \sqrt{\frac{u(t)}{\int_0^t u(s)ds}}.
\end{align}
Finally, we use Ito's lemma one more time to compute the differential increment of $\theta = \arg(R) = \mathrm{Im}[\log(R)]$
\begin{align} \label{eq:PofT}
d\theta = \mathrm{Im}(d\log(R)) = \mathrm{Im}\left[i P(t) r(t)dt + \frac{P(t)^2}{2}dt\right] = P(t) r(t) dt.
\end{align}
As $d\phi(t) = d\theta(t)$, \erf{eq:PofT} states that the instantaneous angular frequency of the measurement axis is proportional to the measurement outcome. Thus in the limit that the feedback delay is small, the process of computing a quantum trajectory and then calculating the optimal phase may be reduced to applying proportional feedback. We implement this feedback law in the FPGA, as described in \cite{martin2019adaptive}.

In the above, we have maximized phase sensitivity using an intuitive argument to justify $\phi(t) = \theta(t)+\pi/2$. To show that this indeed leads to a canonical phase measurement, we return to \erf{eq:POVMFromTrajSol}, substituting our solution \erf{eq:EOMcplus} and \erf{eq:EOMcminus}
\begin{align}
P[r] &\propto \<\psi(\infty)|\psi(\infty)\> = |c_-(0) + c_+(0) R(\infty)^*|^2 \\ \nonumber
&= |c_-(0)|^2 + c_-(0)c_+^* R(\infty) + c_-(0)^* c_+(0)R^*(\infty) + |c_+(0)|^2|R(\infty)|^2
\end{align}
using $c_+(\infty) = 0$. We have written $\propto$ instead of $=$ because we dropped a normalization prefactor in \erf{eq:SSEAdaptivePhase}. We can then compute the measurement operator by matching terms
\begin{align}
P[r] &= \Tr[E[r]|\psi(0)\>\<\psi(0)] \propto \Tr\left[ 
\begin{pmatrix}
E_{--} & E_{-+} \\
E_{+-} & E_{++}
\end{pmatrix}
\begin{pmatrix}
|c_-(0)|^2 & c_-(0) c_+^*(0) \\
c_-^*(0) c_+(0) & |c_+(0)|^2
\end{pmatrix}
\right] \\ \nonumber
&= E_{--} |c_-(0)|^2 + E_{-+} c_-^*(0) c_+(0) + E_{+-} c_-(0) c_+^*(0) + E_{++} |c_+(0)|^2 \\ \nonumber
& \implies E(r) \propto \begin{pmatrix}
1 & R^*(\infty) \\
R(\infty) & |R(\infty)|^2
\end{pmatrix}.
\end{align}
We have $|R(\infty)|^2=1$ by \erf{eq:RSol}, so
\begin{align} \label{eq:WeDidCanonicalPhase}
P[r] \propto 1+ \cos(\arg(c_-^*(0)c_+(0)) - \theta(\infty))
\end{align}
where $\theta(\infty) = \arg(R(\infty))$ is by definition the measurement outcome. This expression coincides with the probability distribution for a canonical phase measurement. On might worry that we have dropped something in the proportionality factor $P(dW_1)P(dW_2)...$ that biases the measurement result away from \erf{eq:WeDidCanonicalPhase} on average. However it is easily shown that this prefactor (often called $P_\text{ost}$ for the ostensible statistics of the measurement) leads to a uniform distribution in $\arg(R^*)$, so the results are not affected.

%% file: OptimalMeas.tex

\chapter{Optimal measurement-based control}
\label{ch:OptimalControl}

In previous chapters, we have developed several methods for finding measurement-based feedback protocols, and used them to derive explicit control strategies. We have mostly concerned ourselves with finding protocols that achieve a certain task, and in some cases applied substantial constraints on the system in order to narrow the search space to something tractable. While the ASLO and symmetry restrictions lead to tractable calculations, we have no reason to expect that these simple solutions couldn't be improved upon by a less restrictive approach. 

The following sections address this concern by proving that some of the presented protocols are theoretically optimal. In proving these results, we derive a powerful method for deriving measurement-based feedback protocols on bipartite systems, which was the technique originally used to derive the Hong-Ou-Mandel based feedback protocol mentioned in chapter \ref{ch:MeasurementControl}.\footnote{We have also generalized this method to tripartite systems based on the associated Schmidt decomposition\cite{acin2000generalized}. However there are now five entanglement invariants instead of one, which appears to limit the usefulness of this technique. The resulting coupled, nonlinear stochastic differential equations did not provide any novel protocols, though they may provide a way to prove optimality of the GHZ protocol presented in chapter \ref{ch:MeasurementControl}.}

To date only a limited number of measurement-based quantum feedback protocols have been established to be globally optimal, owing in part to the non-linear nature of the problems. A feedback scheme for qubit purification was discovered by Jacobs \cite{Jacobs2003} and its optimality properties were studied in a number of subsequent works \cite{Wiseman2006reconsidering, Wiseman2008, Teo2014, Li2013}. A protocol for rapid purification of a qudit was shown to exist in \cite{Combes:2006hx}, and constructed explicitly in \cite{Shabani2008locally}. Upper bounds on qudit and multiqubit purification speedups are known \cite{Combes2010rapid}, but proving global optimality for all situations remains an open challenge except for the qutrit case, which was solved in \cite{Shabani2008locally}. A globally optimal protocol for rotating a monitored qubit to a desired state was given in \cite{Sridharan2012optimal}, but the precise way in which it is optimal is somewhat unnatural and several basic questions remain open.

Regarding feedback for entanglement generation, some optimality results are known for linear quadratic Gaussian systems \cite{Serafini2010determination, Mancini2007optimal, Genoni2013optimal}. Feedback protocols for enhancing the rate of entanglement generation using half-parity and full-parity measurements were given in \cite{HPFPRA} and \cite{Hill2008entanglement} respectively, but without proof of optimality. Both protocols generate entanglement faster than is possible with measurement alone.
They are also of interest because they drive the system along a deterministic path in Hilbert space; feedback is chosen to exactly cancel the randomness introduced by measurement, allowing for deterministic entanglement generation. This is particularly significant for the half-parity measurement, which has only a 50\% success rate in the absence of feedback.

\section{Optimal Entanglement Generation}

In this section, we prove that the half-parity measurement protocol given in \cite{HPFPRA} and chapter \ref{ch:MeasurementControl} is globally optimal for several practically relevant tasks listed in the following section. We also consider entanglement generation under a continuous full-parity measurement $X_F = \sigma_{z1}\sigma_{z2}$, which behaves like the half-parity measurement except that it is also degenerate in the $\{|00\>,|11\>\}$ subspace. For the full-parity measurement, we derive a protocol that is related to that given in \cite{Hill2008entanglement} but which is designed for initially pure states. Elaborating on the connection between rapid entanglement generation and rapid purification first discussed in \cite{Hill2008entanglement}, we derive a natural mapping between this two-qubit system and measurement of an effective qubit, in which the concurrence \cite{Wootters1998} exactly maps to this qubit's Bloch vector length. We then use this mapping to prove several optimality results for the proposed full-parity protocol.
We also show that our full-parity protocol 
is optimal among all measurement-based entanglement generation schemes that acquire single qubit information at a fixed rate. This result sets a useful standard by which the effectiveness of an entanglement generation protocol may be assessed.

We assume a controller has the ability to apply feedback only in the form of local unitary (LU) rotations on the measured qubits, so that measurement is the only source for entanglement generation. This restriction substantially simplifies experimental implementation and is essential when the qubits are remotely separated. 
The protocol for the full and half-parity measurements $X_F$ and $X_H$ are given below, and the half-parity protocol is also given in chapter \ref{ch:MeasurementControl}. We name these $P_F$ and $P_H$ respectively. Each protocol involves applying proportional feedback, \textit{i.e.}, local unitary rotations with a rotation angle that is proportional to the most recent measurement outcome. In this section, we prove the following global optimality properties:\\

\textit{Half-parity measurement} $X_H = (\sigma_{z1} + \sigma_{z2})/2$:

$P_H$ is globally optimal for the following tasks.
\begin{enumerate}
\item Max. concurrence goal: $P_H$ maximizes the expectation value of the entanglement (quantified using the concurrence $\mathcal{C}$ \cite{Wootters1998}) reached at a chosen stopping time $T$.
\item Max. fidelity goal: Same as above but for fidelity $\mathcal{F}$ \cite{Jozsa1994fidelity} with respect to any target state which is pure and maximally entangled.
\item Min. time goal for concurrence: $P_H$ minimizes the expected time to reach a desired concurrence $\mathcal{C}_\text{Threshold}$, so long as $\mathcal{C}_\text{Threshold}\leq 1/\sqrt{2}$.
\item Min. time goal for fidelity: Same as above but for a desired fidelity $\mathcal{F}_{\text{Threshold}}$, so long as $\mathcal{F}_{\text{Threshold}} \leq (1 + \sqrt{2})/\sqrt{8}$
\end{enumerate}

\textit{Full-parity measurement} $X_F = \sigma_{z1}\sigma_{z2}/2$:
\begin{enumerate}
\item Max. concurrence goal: $P_F$ is globally optimal.
\item Max. Fidelity goal: $P_F$ is globally optimal
\item Min. time goal ($\mathcal{C}$ or $\mathcal{F}$): Measuring without feedback is globally optimal.
\end{enumerate}

The min. time results for concurrence also apply to any monotonic function of $\mathcal{C}$, such as entanglement of formation. Finally, we show that $P_F$ is optimal for the max. concurrence goal among all protocols that acquire single qubit information at a fixed rate. A precise statement of this result is given with its proof. 

In what follows, we make the simplifying assumptions that $\eta=1$ and that the initial state is pure. In the absence of additional decoherence, the state remains pure at all times, which allows us to use a stochastic Sch\"odinger equation instead of the corresponding master equation
\begin{equation}
\label{eq:PsiMCont}
\psi(t+dt) = \Big[ -\frac{1}{2}(X-\langle X \rangle)^2 dt + (X-\langle X \rangle) dW(t) \Big]|\psi(t)\rangle,
\end{equation}
where we have set the measurement rate $\Gamma=2$ for simplicity, a convention that we will retain in what follows.


Our objective is to quantify how entanglement changes under arbitrary feedback protocols for binary systems, so that we can identify which are optimal. We characterize entanglement using the concurrence, which for pure states is defined as \cite{Wootters1998} 
\begin{equation}
\label{eq:Concurrence}
\mathcal{C} \equiv |\langle \psi^*| \sigma_y \otimes \sigma_y |\psi \rangle|
\end{equation}
with $\langle\psi^*|$ the complex conjugate of $\langle\psi|$, or equivalently the transpose of $|\psi\rangle$. Like all valid entanglement measures, $\mathcal{C}$ is invariant under local unitary rotations, so our allowed feedback operations leave it unchanged. For the time being, we take $\mathcal{C}$ to be our figure of merit. 

Since all bipartite pure states with the same concurrence are equivalent up to LU operations \cite{Nielsen2010}, it is possible to parameterize any pure state in terms of $\mathcal{C}$ and single qubit rotations. Such a parameterization is useful in this context because 
feedback can directly control the latter quantities. Therefore we can model feedback to directly set these qubit rotation parameters to desired values without specifying the Hamiltonian necessary to prepare the resulting state.
The Schmidt decomposition provides an explicit example of such a parameterization \cite{Nielsen2010}. By expressing the Schmidt coefficients in terms of the concurrence, we can write a general two-qubit state as
\begin{align}
\label{eq:StateParam}
\psi &(\mathcal{C}, \theta_1, \theta_2, \phi_1, \phi_2, \gamma_1, \gamma_2)  \\ \nn
&= U_1 \otimes U_2 \Bigg{[} \sqrt{\frac{1+\sqrt{1-\mathcal{C}^2}}{2}} |00\rangle - \sqrt{\frac{1-\sqrt{1-\mathcal{C}^2}}{2}} |11\rangle \Bigg{]} \\ \nn
&U_i \equiv \exp(-i \gamma_i \sigma_z/2) \exp(-i \sigma_y \theta_i/2) \exp(-i \phi_i \sigma_z/2), \\ \nn
\end{align}
where we have written $U_i$ in terms of the Euler angles $\{\phi_i, \theta_i, \gamma_i\}$. For convenience in subsequent calculations, we break the local unitaries into symmetric and antisymmetric rotations by defining $\theta \equiv (\theta_1 + \theta_2)/2$, $\Delta \theta \equiv (\theta_1 - \theta_2)/2$, and likewise for $\phi$ and $\gamma$. The final expression does not depend on $\Delta \phi$ because the state in brackets is invariant under antisymmetric rotations about $\sigma_z$. This separation of variables into an LU-invariant quantity depending on $\mathcal{C}$ and LU rotations depending on $\{\theta_i,\phi_i,\gamma_i\}$ motivates an analogy with parameterization of a qubit in terms of a Bloch vector $\vec{r}$ in spherical coordinates; $\mathcal{C}$ is analogous to $|\vec{r}|$, which is invariant under unitary operations, while $\{\theta_i,\phi_i,\gamma_i\}$ can be set arbitrarily using Hamiltonian feedback like $\theta$ and $\phi$. When considering the full-parity measurement, 
we will find that this analogy admits an explicit mapping in which $\mathcal{C}$ and $r$ obey the same equations of motion under measurement.

To study how entanglement changes under measurement, we substitute \erf{eq:StateParam} into \erf{eq:PsiMCont} and compute the concurrence of the resulting state $\psi(t+dt)$. It is not necessary to compute how the five angles parameterizing $\psi$ evolve, since we will model the controller to set them according to some feedback protocol. The computation is further simplified by the fact that $\sigma_z$ rotations commute with $X_F$ and $X_H$, so that the resulting equations of motion do not depend on $\gamma$ or $\Delta \gamma$. We first focus on the half-parity measurement $X_H$. As is detailed in the appendix of \cite{martin2017optimal}, 
application of Ito's lemma yields
\renewcommand{\arraystretch}{1.5}
\begin{align}
\label{eq:dCHPF2}
d\mathcal{C} = \left\{ \begin{array}{ll}
					2\mathcal{C}\sqrt{1-\mathcal{C}^2} u v~dW \\
					~~+ [(v^2-u^2)w-\mathcal{C}(v^2+u^2)]dt 	~~ &|~\mathcal{C}>0 \\
					|v^2-u^2|~dt	&|~\mathcal{C}=0
\end{array}
\right. 
\end{align}
where we have defined the control variables $u = \cos(\theta)$, $v = \cos(\Delta \theta)$ and $w=\cos(2\phi)$. 

In what follows, we implicitly assume that at any time, the controller can instantaneously set the angles $\theta, \Delta \theta$ and $\phi$ to any desired value. This is equivalent to assuming that the controller can implement any local unitary on both qubits with no time delay. For many systems such as superconducting qubits, single qubit rotations can be performed much faster than the measurement time scales, which makes this approximation appropriate. The assumption of zero time delay can be satisfied as long as the propagation delay between the qubits is small relative to the inverse measurement rate. However, this restriction can be relaxed for the full-parity protocol, as 
it will turn out that the protocol can be implemented by applying feedback on only one of the two qubits.
Thus, \erf{eq:dCHPF2} is an equation of motion for the concurrence of the state under arbitrary feedback protocols that may be specified by choosing a particular, set $u(t,\mathcal{C})$, $v(t,\mathcal{C})$ and $w(t,\mathcal{C})$. 

Although there is no reason \textit{a priori} that a locally optimal strategy is also globally optimal, it often turns out to be so in practice. This will be the case in the results that we prove here. The locally optimal protocol maximizes the expectation value of the concurrence at time $t+dt$. To find it, one simply chooses the values of $u$, $v$ and $w$ to maximize the $dt$ term of \erf{eq:dCHPF2}. This occurs for $\{u=0,v=1,w=1\}$ with all values of $\gamma$ and $\Delta \gamma$ allowed.\footnote{$\{u=1,v=0,w=-1\}$ is also a solution,  but since it is actually equivalent to the first if one makes the transformation $\gamma \rightarrow \gamma-\pi/2$, $\Delta\gamma \rightarrow \Delta \gamma + \pi/2$, we ignore it. Solutions corresponding to $u=-1$ or $v=-1$ are similarly redundant.} We henceforth refer to this protocol as $P_H$. The resulting equations of motion under this set of control parameters may be easily solved:
\begin{equation}
\label{eq:PH}
d\mathcal{C} = (1-\mathcal{C})dt ~~\implies \mathcal{C}(t) = 1-(1-\mathcal{C}(0))e^{-t}.
\end{equation}
The evolution of the state under this feedback protocol may be computed by substituting this solution and the above control values into \erf{eq:StateParam}. Setting $\mathcal{C}(0) = 0$ for simplicity, the state evolution under feedback is thus
\begin{align}
\label{eq:HPFPsiSol}
\psi(t) = \frac{1}{2}
 \begin{bmatrix}
e^{-i \gamma} \sqrt{e^{-t}}	\\
e^{-i \Delta\gamma} \sqrt{2-e^{-t}}	\\
e^{i \Delta \gamma} \sqrt{2-e^{-t}}	\\
e^{i \gamma} \sqrt{e^{-t}} 	\end{bmatrix}.
\end{align}
For $\gamma=\Delta \gamma = 0$, this state evolution exactly matches the solution to the feedback protocol given in \cite{HPFPRA} and chapter \ref{ch:MeasurementControl}, indicating that they are the same feedback protocol. 

The two methods that we have used to obtain this result differ substantially. The method of chapter \ref{ch:MeasurementControl} parameterizes feedback operations according to what an experimental controller would actually implement. Being a proportional feedback strategy, \erf{eq:PaQSSME} calculates $d\rho$ by computing state updates due to infinitesimal measurement and feedback. In contrast, the above method applies infinitesimal measurement to determine the change in entanglement, but otherwise discards all information about the state at each time step. In a sense, \erf{eq:dCHPF2} is an equation of motion for an equivalence class of states, and the state is parameterized not by the components of $\psi$, but rather by the unitary operations that one would perform to map $\alpha|00\>-\beta|11\>$ (the representative element singled out in \erf{eq:StateParam}) to the physical state. Each method has strengths and weaknesses. PaQS allows for treatment of mixed states, average evolution and an arbitrary Hilbert space, while the present method allows us to more easily maximize over all possible protocols, including those with discontinuous evolution. The latter feature is necessary for optimality proofs, where we essentially need to maximize over all possible protocols.

Equation (\ref{eq:HPFPsiSol}) can be more easily understood by writing down the states for $t=0$ and $t \rightarrow \infty$. Up to a global phase, these states are
\begin{align}
\psi(0) &= \frac{1}{2}\big(|0\rangle + e^{i\gamma_1} |1\rangle \big) \otimes \big( |0\rangle + e^{i \gamma_2}|1\rangle \big) \nn \\
\psi(\infty) &= \frac{1}{\sqrt{2}}\big( |01\rangle + e^{2i \Delta \gamma} |10\rangle \big).
\end{align}
Thus this protocol involves preparing both qubits in a separable state polarized along some axis of the equator of their respective Bloch spheres. Feedback deterministically projects the qubits into a Bell pair with an arbitrary relative phase. 

Before proving global optimality results for the above protocol, we first repeat the above analysis for the full-parity measurement. Following similar steps to those above but now for $X_F$, the equations of motion for the concurrence can be shown to be
\begin{align}
\label{eq:EOMsF}
d\mathcal{C}  = \left\{ \begin{array}{ll}
	 (1-\mathcal{C}^2)\big[(u^2-v^2)w~\mathcal{C}~dW \\ 
	~~~~ + (u^2-v^2)^2(1-w^2)dt/2\mathcal{C}\big]~~~	& |~\mathcal{C}\neq0 \\
					(u^2-v^2) dW	&  |~\mathcal{C}=0,
\end{array}
\right. 
\end{align}
A minor technicality in the derivation detailed in the appendix of \cite{martin2017optimal} has forced us to allow $\mathcal{C}$ to take on negative values in general. However since $\psi(\mathcal{C})$ only depends on $\mathcal{C}^2$, this fact presents no further difficulty. We shall simply interpret $|\mathcal{C}|$ as the concurrence instead of $\mathcal{C}$.

For the full-parity measurement, two distinct sets of locally optimal parameters emerge: $\{u=0,v=1,w = 0\}$ and $\{u=1,v=0,w=0\}$. The resulting state evolutions are equivalent up to a complex conjugation of $\psi$, so the underlying dynamics are essentially identical. We may therefore only consider the protocol $P_F:~\{u=0,v=1,w = 0\}$ and ignore the other solution.
As for the half-parity case, the equations of motion for $\mathcal{C}(t)$ are deterministic and again yield an analytic expression with easy solution:
\begin{equation}
\label{eq:PF}
d\mathcal{C} =  \frac{1-\mathcal{C}^2}{2\mathcal{C}}dt ~~\implies \mathcal{C}(t) = \pm\sqrt{1-[1-\mathcal{C}(0)^2]e^{-t}}.
\end{equation}
Validity of this solution at $\mathcal{C}(0)=0$ is most easily established by deriving equations of motion for 
$\mathcal{C}(t)^2$ 
at $\mathcal{C}=0$ and showing that the solutions coincide. State evolution under this protocol is given by
\begin{equation}
\label{eq:FPFPsiSol}
\psi(t) = \frac{1}{\sqrt{8}}
 \begin{bmatrix}
e^{-i (\gamma+\pi/4)} \big(\sqrt{1-e^{-t/2}} - i \sqrt{1+e^{-t/2}} \big) \\
e^{-i (\Delta\gamma+\pi/4)} \big(\sqrt{1-e^{-t/2}} + i\sqrt{1+e^{-t/2}} \big) \\
e^{i (\Delta \gamma-\pi/4)} \big(\sqrt{1-e^{-t/2}} + i\sqrt{1+e^{-t/2}} \big)	\\
e^{i (\gamma-\pi/4)} \big( \sqrt{1-e^{-t/2}} - i\sqrt{1+e^{-t/2}} \big)	\end{bmatrix} .
\end{equation}
Again, taking early- and late-time limits of the solution gives some insight into the induced dynamics. Here
\begin{align}
\psi(0) &= \frac{1}{2}\big(|0\rangle + e^{i\gamma_1} |1\rangle \big) \otimes \big( |0\rangle - i e^{i\gamma_2}|1\rangle \big) \\
\psi(\infty) &= \frac{1}{2}\big( e^{-i\gamma} |00\rangle + e^{-i \Delta \gamma}|01\rangle + i e^{i\Delta \gamma} |10\rangle - i e^{i \gamma} |11\rangle \big), \nn
\end{align}
where as before we have dropped a global phase from the states. The optimal initial state is again to prepare both qubits in an equal superposition of $|0\rangle$ and $|1\rangle$. Somewhat counterintuitively, the final state produced by feedback is not an eigenstate of the measurement operator. This is somewhat analogous to Jacobs' purification speedup protocol, in which the state is maintained to be in an equal superposition of the measurement eigenstates \cite{Jacobs2003}. Application of Jacobs' protocol to an encoded qubit led the authors of \cite{Hill2008entanglement} to a method for converting a classically correlated mixed state into a maximally entangled state. The protocol presented here performs the analogous task for an initially pure state. $\mathcal{C}(t)$ under application of $P_F$ coincides with that given in \cite{Hill2008entanglement}, as shown in \cite{martin2017optimal}. Below we establish an even more precise connection between $P_F$ and Jacobs' rapid qubit purification protocol. 
\begin{figure}
\centering
\includegraphics[width = 0.5\textwidth]{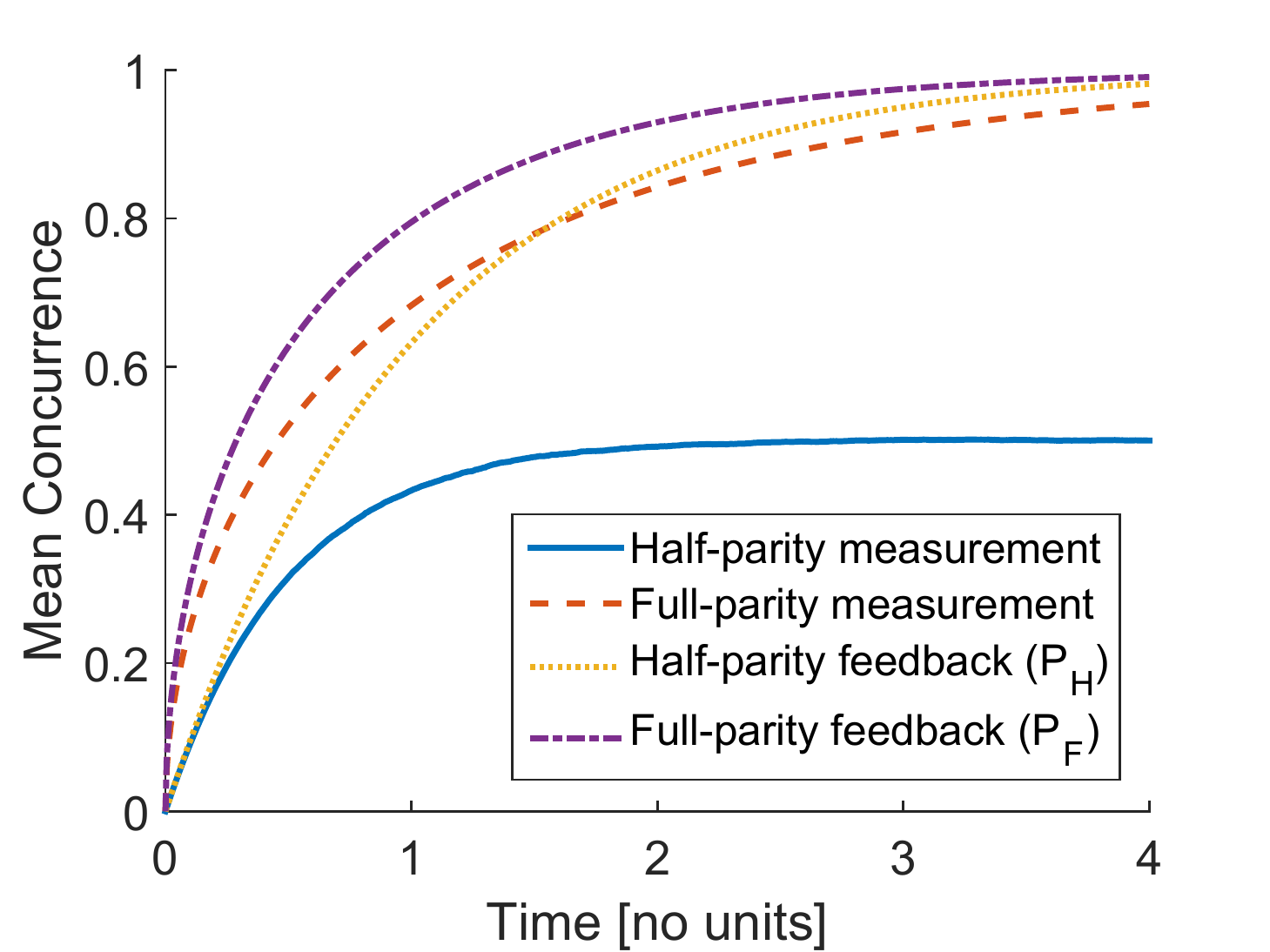}
\caption{Expectation value of the concurrence while applying feedback or by applying measurement alone. Using the method of linear quantum trajectories \cite{Jacobs1998linear}, concurrence under full-parity measurement without feedback can be shown to take the form $\mathcal{C}(t) = \text{erf}(\sqrt{t/2})$. Concurrence under half-parity measurement without feedback is calculated numerically by averaging $10000$ trajectories.}
\label{fig:CompareFPF}
\end{figure}

The performance of the half-parity and full-parity feedback protocols are plotted in Fig. \ref{fig:CompareFPF}. We also plot the average concurrence without feedback for both measurement operators. We have normalized the measurement operators $X_F$ and $X_H$ so that the dephasing rate on one of the two qubits (tracing out the other qubit and assuming no feedback) are the same under measurement, allowing quantitative comparison the half- and full-parity performance. It is evident that the full-parity measurement with feedback strictly outperforms all other protocols.
  

We now set out to prove the optimality results stated above. 
For the half-parity feedback protocol, we apply the standard verification theorems often used in control theory to check whether a protocol is globally optimal \cite{Shabani2008, Wiseman2008}. While one might expect that proving such a result would require knowledge about all allowed strategies, it turns out that knowing the performance of the trial protocol is sufficient, so long as one knows how it behaves for any initial state. 

Before using the verification theorems, we first provide a brief summary following Ref. \cite{Shabani2008}. A common feedback goal is to minimize the expectation value of some cost function after applying feedback for a fixed time interval $T$. The cost function can depend on the final state of the system as well as the resources used to apply feedback. From the cost function, one defines the cost-to-go $c(P, \mathbf{x}, t)$ where the vector $\mathbf{x}$ represents the system state at time $t$. Here $c$ is defined to be the expectation value of the cost function at time $T$, assuming that the system was in the state $\mathbf{x}$ at time $t$ and that the controller used feedback protocol $P$ from $t$ to $T$. To check optimality, one first generically writes the equations of motion for $\mathbf{x}$ in terms of their deterministic and stochastic parts as
\begin{align}
\label{eq:dx}
d\mathbf{x} = \mathbf{A}(t, \mathbf{v}_P(t), \mathbf{x}(t))~dt +\mathbf{B}(t, \mathbf{v}_P(t), \mathbf{x}(t))~dW
\end{align}
where $\mathbf{v}_P$ are the feedback control settings and parameters specified by protocol $P$. In the context of quantum feedback, $\mathbf{v}$ can represent Hamiltonian, measurement and dissipation parameters. The procedure can easily be generalized to include multiple noise processes if needed. If $\partial c(P)/\partial t$ and $\partial^2 c(P)/\partial \mathbf{x}^2$ are continuous, then global optimality of $P$ follows from the following two conditions. Firstly, assuming the cost function only depends on the final state, the Hamilton-Jacobi-Bellman equation for the cost-to-go $c(P,\mathbf{x},t)$, given by
\begin{align}
\label{eq:HJB_G}
&G(t, \mathbf{u}, \mathbf{x}, P) \equiv \\ \nn
& -\frac{1}{2} \mathbf{B}(t,\mathbf{u},\mathbf{x})^\top \frac{\partial^2 c(P)}{\partial \mathbf{x}^2} \mathbf{B}(t,\mathbf{u},\mathbf{x}) - \mathbf{A}(t,\mathbf{u},\mathbf{x})^\top \frac{\partial c(P)}{\partial \mathbf{x}}
\end{align}
and
\begin{align}
\label{eq:HJB}
\frac{\partial c(P)}{\partial t} = \text{max}_{\mathbf{u}} [G(t,\mathbf{u}, \mathbf{x}, P)].
\end{align}
must be satisfied for all time $0$ to $T$ and for all states $\mathbf{x}$. Secondly, $\mathbf{u} = \mathbf{v}_P$ must maximize $G$ for all time and states.

In order to apply the verification theorems, we parameterize the state $|\psi\rangle$ as a vector $\mathbf{x} = \{\mathcal{C}, u, v, w \}$ using \erf{eq:StateParam} and the definitions of $u$, $v$ and $w$ given below \erf{eq:dCHPF2}. Applying a specific local unitary feedback protocol is equivalent to setting $\mathbf{v} = \{u, v, w\}$ equal to specific functions of $t$ and $\mathcal{C}$. 
  
  \textbf{Half-parity measurement: }  For the max. concurrence goal (maximizing $\langle \mathcal{C}(T)\rangle$), we define our cost function to be $1-\mathcal{C}(t)$.
Choosing the locally optimal protocol for the half-parity measurement $P_H: \{u = 0, v = 1, w = 1\}$, we can use the analytic solution \erf{eq:PH} to find the cost-to-go when evolving according to $P_H$ as
\begin{align}
c(P_H, \mathbf{x}, t) = (1-\mathcal{C}(t))e^{t-T}.
\end{align}
The Hamilton-Jacobi-Bellman equation uses the following derivatives of $c$
\begin{align}
\label{eq:HPFPartials}
\frac{\partial c}{\partial t} = (1-\mathcal{C}(t))e^{t-T}, ~~ \frac{\partial c}{\partial\mathbf{x}} = \{-e^{t-T}, 0, 0, 0\}, ~~ \frac{\partial^2 c}{\partial \mathbf{x}^2} = 0
\end{align}
which satisfy the continuity conditions required by the verification theorem. We divide the equations of motion under feedback and half-parity measurement \erf{eq:dCHPF2} into their deterministic and stochastic parts as
\begin{align}
\label{eq:HPFAB}
d\mathcal{C} &= A_\mathcal{C}~dt + B_\mathcal{C}~dW \nn \\
A_\mathcal{C} &= \left\{ \begin{array}{ll}
		[(v^2-u^2)w-\mathcal{C}(v^2+u^2)]dt 	~~ &|~\mathcal{C}>0 \\
		|v^2-u^2|~dt	&|~\mathcal{C}=0
\end{array}
\right. \\ \nn
B_\mathcal{C} &= \left\{ \begin{array}{ll}
		2\mathcal{C}\sqrt{1-\mathcal{C}^2} u v~dW 	~~ &|~\mathcal{C}>0 \\
		0	&|~\mathcal{C}=0
\end{array}
\right. \\ \nn
\end{align}
where we have only calculated the $\mathcal{C}$ components of $\mathbf{A}$ and $\mathbf{B}$. Due to the form of $G$ in \erf{eq:HJB_G}, the other components are unnecessary, since $\partial c/\partial \mathbf{v} = 0$, \textit{i.e.}, our cost function is invariant under the feedback control settings. Substituting Eqs. (\ref{eq:HPFPartials}) and (\ref{eq:HPFAB}) into Eqs. (\ref{eq:HJB_G}) and (\ref{eq:HJB}), we find the condition for $P_H$ to be globally optimal is that it satisfies the maximization condition
\begin{align}
(1-&\mathcal{C}(t))e^{t-T}  \\ \nn
 &= \text{max}_{\{u, v, w\}} [(v^2-u^2)w - \mathcal{C}(t)(v^2+u^2)]e^{t-T}.
\end{align}
The maximum occurs for the values of $u$, $v$ and $w$ specified by $P_H$, and the equation then is satisfied. This proves that the half-parity protocol $P_H$ is globally optimal for maximizing the concurrence at fixed time $T$. Because $c(P_H,\mathbf{x},t)$ is linear in $\mathcal{C}$, one can also show that global optimality follows directly from local optimality in this case \cite{Teo2014}, but this proof method is not applicable in general.

A similar calculation can be performed in which we test for global optimality with respect to the min. time goal \cite{Shabani2008} (minimizing the expected time at which $\mathcal{C}$ reaches some desired value $\mathcal{C}_\text{Threshold}$, called the expected hitting time). One finds that $P_H$ maximizes G only when $\mathcal{C} \leq 1/\sqrt{2}$, which implies that $P_H$ is not globally optimal in general. If one chooses $\mathcal{C}_\text{Threshold}\leq 1/\sqrt{2}$ however, then only the dynamics of the system when $\mathcal{C}\leq 1/\sqrt{2}$ are relevant for determining the expected hitting time. Thus we can instead ask if $P_H$ is globally optimal within the constraint that $\mathcal{C}\leq 1/\sqrt{2}$. Restricted to this parameter space, the Hamilton-Jacobi-Bellman equation Eqs. \ref{eq:HJB_G}-\ref{eq:HJB} is satisfied for all allowed values of $\mathbf{x}$, and $P_H$ maximizes $G$. This proves that $P_H$ is globally optimal for the min. time goal when $\mathcal{C}_\text{Threshold}\leq 1/\sqrt{2}$. As the expected hitting time is the same whether one considers $\mathcal{C}$ or some arbitrary monotonic function $f(\mathcal{C})$ and the corresponding threshold $f(\mathcal{C}_\text{Threshold})$, the min. time proof applies to all monotonic functions of $\mathcal{C}$, such as entanglement of formation.

Concurrence has the favorable property of being invariant under the control parameters, which makes it amenable to methods for proving global optimality. For many tasks however, a specific target state is desired, in which case fidelity is a more relevant figure of merit. We now show that the global optimality proofs given above extend to the corresponding fidelity goals. 

We take any maximally entangled state $|\Psi\rangle$ to be the target state and define $P_H^\Psi$ to be a variant of $P_H$ which rotates $\psi$ to have the maximal fidelity with respect to $\Psi$ at the final time. For the max. fidelity goal, the final time is simply $T$. For the fidelity min. time goal, the final time is the earliest time at which $\mathcal{F}_\text{Threshold}$ can be reached. For both of these goals, we need to know the fidelity of a given state $\psi$ maximized over all local unitaries. For pure states, this maximal fidelity is uniquely determined by the concurrence of $\psi$, and is given by \cite{Verstraete2002fidelity}
\begin{align}
\label{eq:FidConcRelation}
\mathcal{F}_\text{max} = \frac{\mathcal{C}+1}{2}.
\end{align}
In the context of entanglement distillation, $\mathcal{F}_\text{max}$ is often called the singlet fraction, and measures the usefulness of a general quantum state for the task.

To prove that $P_H$ is globally optimal for the fidelity min. time goal, we note that any optimal protocol must have $\mathcal{F} = \mathcal{F}_\text{max}(\mathcal{C})$ at the hitting time. If this were not the case, the protocol could have applied local unitaries to $\psi$ at an earlier time that increase $\mathcal{F}(\psi)$ to $\mathcal{F}_\text{max}(\mathcal{C}(\psi))$, and hence achieve an earlier hitting time. As $\mathcal{F}_\text{max}$ is a monotonic function of $\mathcal{C}$, this relation suffices to prove optimality of $P_H$ with respect to the fidelity min. time goal when $\mathcal{F}_\text{Threshold} \leq \mathcal{F}_\text{max}(1/\sqrt{2}) = (1+\sqrt{2})/\sqrt{8}$. 
  
To prove global optimality of the max. fidelity goal, we assume the evolution of the system is well-approximated by a discrete protocol in which one measures for a small but finite duration before applying feedback \cite{Teo2014}. As realistic implementations of feedback inevitably suffer from feedback delay and finite bandwidth effects at sufficiently short timescales, the continuum limit may not be a good physical model, and we shall therefore not focus on this here. In the discrete approximation, we can express the expectation value of the fidelity at the stopping time $T$ as an integral over all possible measurement outcomes $\{\mathbf{V}\}$. Suppose that some hypothetical feedback protocol $P'$ is globally optimal for the max. fidelity goal. We write
\begin{align}
\label{eq:FidProof1}
\langle \mathcal{F}(P', T)\rangle &= \int \mathcal{F}_\mathbf{V}(P', T) p(\mathbf{V}) d\mathbf{V} \nn \\
 &= \int \mathcal{F}_\text{max}(\mathcal{C}_\mathbf{V}(P', T)) p(\mathbf{V}) d\mathbf{V}
\end{align}
where $\mathcal{F}_\mathbf{V}(P', T)$ and $\mathcal{C}_\mathbf{V}(P', T)$ are respectively the fidelity and concurrence at time $T$, assuming protocol $P'$ was applied and measurement outcome $\mathbf{V}$ occurred. $p(\mathbf{V})$ is the probability of $\mathbf{V}$ occurring (note that $p$ implicitly depends on $P'$ and the state evolution, but this is of no consequence for the proof). The last equality follows from the fact that the globally optimal protocol could perform at least as well if it maximizes the fidelity of each possible final state. If it does not do so on a set of non-zero measure, then a better protocol exists which does. We continue by relating the performance of $P'$ to that of $P_H^\Psi$:
\begin{align}
\label{eq:FidProof2}
 &= \int \frac{\mathcal{C}_\mathbf{V}(P', T)+1}{2} p(\mathbf{V}) d\mathbf{V} \nn \\
 &\leq \int \frac{\mathcal{C}_\mathbf{V}(P_H^\Psi, T)+1}{2} p(\mathbf{V}) d\mathbf{V} \nn \\
 &= \int \mathcal{F}_\text{max}(\mathcal{C}_\mathbf{V}(P_H^\Psi, T)) p(\mathbf{V}) d\mathbf{V} = \int \mathcal{F}_\mathbf{V}(P_H^\Psi, T) p(\mathbf{V}) d\mathbf{V} \nn \\
 &= \langle \mathcal{F}(P_H^\Psi, T)\rangle.
\end{align}
The inequality follows from global optimality of $P_H$ with respect to the max. concurrence goal. Note that relating the performance of $P'$ to that of $P_H$ relies on the fact that $\mathcal{F}_\text{max}$ is linear in $\mathcal{C}$. The second-to-last equality follows from the fact that at time $T$, $P_H^\Psi$ rotates $\psi$ to have maximum fidelity with respect to the target state. Together Eqs. (\ref{eq:FidProof1}) and (\ref{eq:FidProof2}) imply that $P_H^\Psi$ performs at least as well as any potential protocol $P'$, and therefore that $P_H^\Psi$ is globally optimal for the task. 

\textbf{Full-parity measurement: } A connection between Jacobs' rapid purification protocol, and rapid entanglement using full-parity measurement was first established in Ref. \cite{Hill2008entanglement}. To prove optimality of the full-parity protocol $P_F$, we observe that the dynamics of the concurrence of a two-qubit pure state under full-parity measurement and local feedback are precisely those of the Bloch vector length of a single continuously monitored qubit with feedback. We then use existing optimality results regarding rapid qubit purification \cite{Wiseman2008} to prove the analogous two-qubit results. 

To see the correspondence between 2-qubit concurrence $\mathcal{C}$ and the Bloch vector length of a single qubit, consider a qubit undergoing a continuous measurement of $\tilde{\sigma}_z$ at a rate $\tilde{\Gamma}$. We assume that some feedback controller can instantly apply any unitary operation at any time, as we have assumed for the two-qubit case. Parameterizing the qubit as a Bloch vector in spherical coordinates $\{\tilde{r}, \tilde{\theta}, \tilde{\phi}\}$, one can use the stochastic master equation to derive an equation of motion for the Bloch vector length $\tilde{r}$ as a function of $\tilde{\theta}$ and $\tilde{\phi}$ \cite{Li2013}
\begin{align}
\label{eq:OneQubit}
d\tilde{r} = (1-\tilde{r}^2) \Big[ \frac{\tilde{\Gamma}}{4\tilde{r}}(1-\tilde{u}^2)dt + \sqrt{\frac{\tilde{\Gamma}}{2}} \tilde{u} ~dW \Big]
\end{align}
where $\tilde{u} = \cos(\tilde{\theta})$ and $\tilde{\theta}$ is the angle the Bloch vector makes with the measurement axis $\tilde{\sigma_z}$. We do not derive equations of motion for $\tilde{\theta}$ and $\tilde{\phi}$ because we assume feedback can set them to their desired values at any time. Making the following identifications
\begin{align}
\label{eq:FPFMapping}
\mathcal{C} &\rightarrow \tilde{r} \nn \\
w &\rightarrow \tilde{u} \nn \\
2(u^2-v^2)^2 &\rightarrow \tilde{\Gamma},
\end{align}
the equation of motion for concurrence under full-parity measurement and feedback, \erf{eq:EOMsF} becomes \erf{eq:OneQubit}. This mapping reveals several interesting features of the dynamics.
As observed in Ref. \cite{Hill2008entanglement}, entanglement generation can be turned on and off by rotating the system into one of the decoherence free subspaces of the measurement operator $X_F$. This is evident from the dependence of $\tilde{\Gamma}$ on $u$ and $v$, as $u=v=1$ and $u=v=0$ correspond to states fully localized to the even and odd parity subspaces, respectively. Of particular interest is the direct mapping between $\mathcal{C}$ and $\tilde{r}$; the concurrence coincides exactly with Bloch vector length of the effective qubit. We provide more details on this effective qubit in the appendix.

Although the effective qubit we consider here is different from that discussed in Ref. \cite{Hill2008entanglement}, rapid entanglement generation corresponds to applying Jacobs' protocol to the effective qubit in both mappings. $\tilde{u}=0$ is globally optimal for maximizing the linear entropy $\langle \tilde{r}(T) \rangle$ as shown in \cite{Wiseman2008}. Although that work did not consider allowing the measurement rate $\tilde{\Gamma} \leq 2$ to vary, it is straightforward to extend the proofs of \cite{Wiseman2008} to this case by repeating their calculation with $\tilde{\Gamma}$ as a control variable bounded from $0$ to $2$. 
Thus the mapping between equations of motion for $\tilde{r}$ and $\mathcal{C}$ implies that $P_F$ is globally optimal for the max. concurrence goal.

It is also the case that not applying feedback to the effective qubit yields the same equations of motion as not applying feedback to the two-qubits. Since it was proved in \cite{Wiseman2008} that not applying feedback is globally optimal for the min. time goal for linear entropy, the analogous result applies to the full-parity measurement of two qubits for the concurrence min. time goal. The arguments used to extend the max. concurrence and the concurrence min. time results of $P_H$ to the corresponding fidelity results (\textit{i.e.} Eqs. \ref{eq:FidConcRelation}-\ref{eq:FidProof2}) also apply to $P_F$ without modification. This completes the proof of the enumerated results given in at the beginning of this section.

\begin{figure}
\centering
\includegraphics[width = 0.7\textwidth]{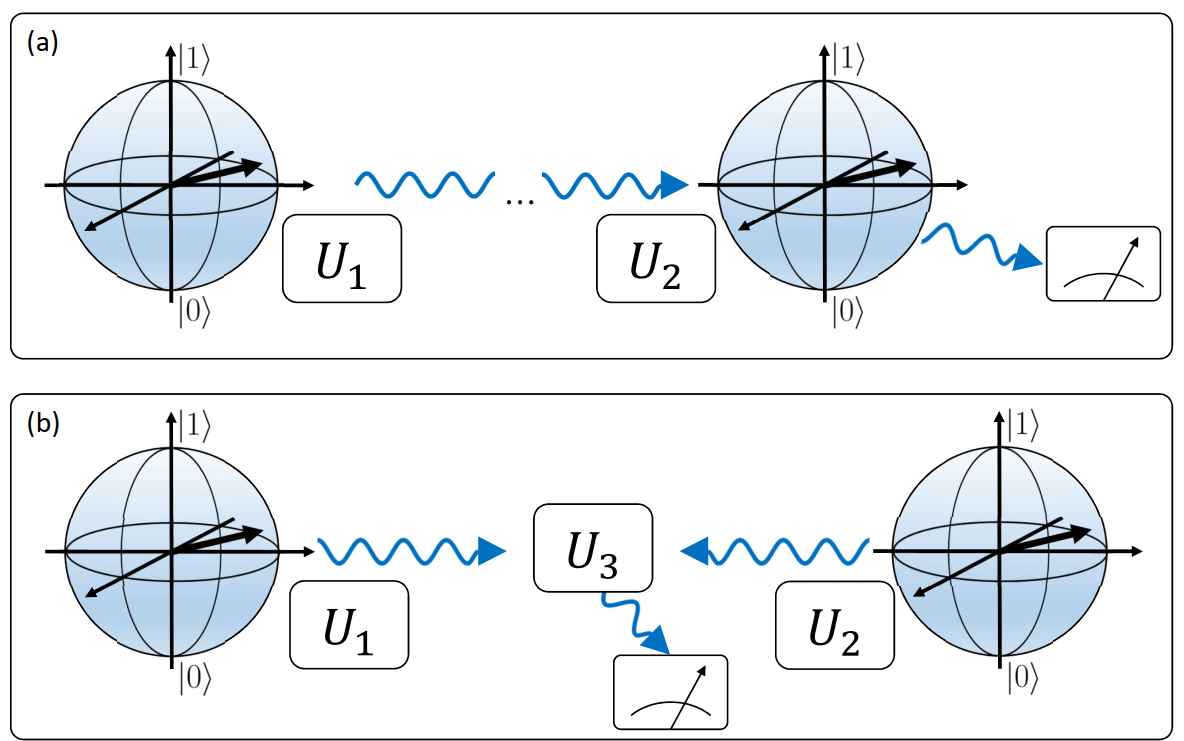}
\caption{Remote 2-qubit measurement implementations. (a) Remote 2-qubit measurement in which the signal degrees of freedom (wavey line) propagate from qubit 1 to qubit 2. $U_1$ and $U_2$ represent the interactions between the signal and each qubit, while the dial represents projective measurement of the signal after qubit interactions. These three operations determine the effective 2-qubit measurement operator. (b) Remote measurement based on entanglement swapping. $U_3$ is the interaction between the incoming signals (such as a beam splitter) that erases which-path information. Note that in both schemes, the rate of entanglement generation is bounded by $U_1$.}
\label{fig:RemoteEntanglement}
\end{figure}

\textbf{Upper-bound on measurement-based protocols:} So far, we have focused on optimality given a fixed measurement operator. However, one may ask whether a different measurement operator could offer superior performance using similar resources. In the context of remote entanglement generation, when entanglement is created using some signal degree of freedom as an intermediary (see Fig. \ref{fig:RemoteEntanglement}), one could ask whether entanglement is transferred with unit efficiency.
Motivated by these questions, we now prove a more general result which sets an upper limit on the entanglement entropy achievable generally under a much larger class of measurement-based protocols. 
We will find that the bound is only saturated by the full-parity feedback protocol $P_F$. 

We consider any system in which the action of the measurement on qubit 1 is of the form
\begin{align}
\label{eq:MERho1}
d\rho_1 = (\Gamma_\text{deph.}/2) \mathcal{D}[\vec{\sigma}\cdot \hat{n}]\rho_1~dt
\end{align}
where $\rho_1$ is the state of qubit 1 unconditioned on the measurement outcome and tracing out qubit 2. Physically, the dephasing rate $\Gamma_\text{deph.}$ sets an upper bound on the amount of information that can be extracted from the measurement \cite{clerk2010noise}, so this restriction fixes the rate at which information about qubit 1 is transferred to the rest of the system. Thus $\Gamma_\text{deph.}$ defines a physically meaningful reference that lets us compare the performance of $P_H$ and $P_F$ to more general protocols. By tracing out qubit 2 in the stochastic master equation, one arrives at \erf{eq:MERho1} with $\Gamma_\text{deph.} = 1/2$ for measurement of both $X_H$ and $X_F$. 

The entanglement entropy of a state is given by $E_1 = \text{Tr}[\rho_1 \log_2(\rho_1)]$, the Von Neumann entropy of qubit 1. Note that if the joint state is pure, then this expression also equals the entanglement of formation. In order for the entanglement to increase, the entropy of the subsystem must increase, as it does under the action of \erf{eq:MERho1}. As measurement constitutes the only available non-local interaction, \erf{eq:MERho1} fully determines how entanglement may change. At time $t$, the increase in entropy of $\rho_1$ is maximized if the state is unbiased with respect to the measurement axis $\vec{\sigma}\cdot \hat{n}$. For example, if $\vec{\sigma}\cdot \hat{n}=\sigma_z$, then optimal states would be of the form $\rho_1 = (x \sigma_x + y \sigma_y + \sigma_0)/2$ where $\sigma_0$ is the identity matrix. Assuming this condition is satisfied at all times, the entropy as a function of time may be derived by solving \erf{eq:MERho1} with the initial condition $\rho_1 = (|0\rangle + |1\rangle)(\langle 0| + \langle 1|)/2$. Taking $\Gamma_\text{deph.} = 1/2$, we find
\begin{align}
\label{eq:E}
E_1(t) =& -\frac{1-\langle \sigma_{x1} \rangle}{2}\log_2\Big[\frac{1-\langle \sigma_{x1} \rangle}{2} \Big]  \\ &-\frac{1+\langle \sigma_{x1} \rangle}{2}\log_2\Big[\frac{1+\langle \sigma_{x1} \rangle}{2}\Big] \nn \\
\langle \sigma_{x1} \rangle &= e^{-t/2}.
\end{align}
This Von Neumann entropy sets an upper bound on the entanglement entropy that can be achieved by the action of a given measurement operator (as well as on the entanglement of formation if the joint evolution is pure). 

The main result of this section is that by tracing out qubit 2 from 
\erf{eq:StateParam} and substituting \erf{eq:PF},
one can show that the entanglement entropy of qubit 1 is exactly \erf{eq:E}. For pure states, a similar result can be derived for concurrence, which has a one-to-one relation with entanglement entropy in this case \cite{Wootters1998}.
Thus $P_F$ saturates the  bound \erf{eq:E} given by the dephasing rate of $X_F$, and hence no measurement-based protocol of the form \erf{eq:MERho1} can generate entanglement faster than $P_F$. It can also be shown that $P_H$ does not saturate this bound, nor does measurement of $X_F$ or $X_H$ without feedback. 

This bound may be intuitively be understood by considering a remote implementation of the measurement, as depicted in Fig. \ref{fig:RemoteEntanglement}.
\erf{eq:MERho1} then governs how much entanglement is generated between qubit 1 and the auxiliary qubit. An ideal entangling protocol would transfer all of this entanglement to qubit 2, so that measurement of the auxiliary does not decrease the entanglement entropy. Evidently only $P_F$ fully transfers the entanglement from the measurement signal to qubit 2. For the other protocols, the remaining entanglement is destroyed when the signal is measured. 

Note that if the effect of measurement on qubit 1 yields a dephasing operator that is not normal (\textit{i.e.}, $[X,X^\dagger]\neq0$), then it cannot be put in the form of \erf{eq:MERho1} and our derivation does not apply. 
Thus our bound does not apply for quantum-demolition measurements such as a spontaneous emission process on the first qubit \textit{i.e.}, $d\rho_1 = \mathcal{D}[\sigma]\rho_1~dt$, though a similar bound could be derived for such cases. An example of the latter is entanglement generation via Hong-Ou-Mandel interference. 

\section{Optimal Hamiltonian-Free Control}

\begin{figure}
    \centering
    \includegraphics[width = 0.7\textwidth]{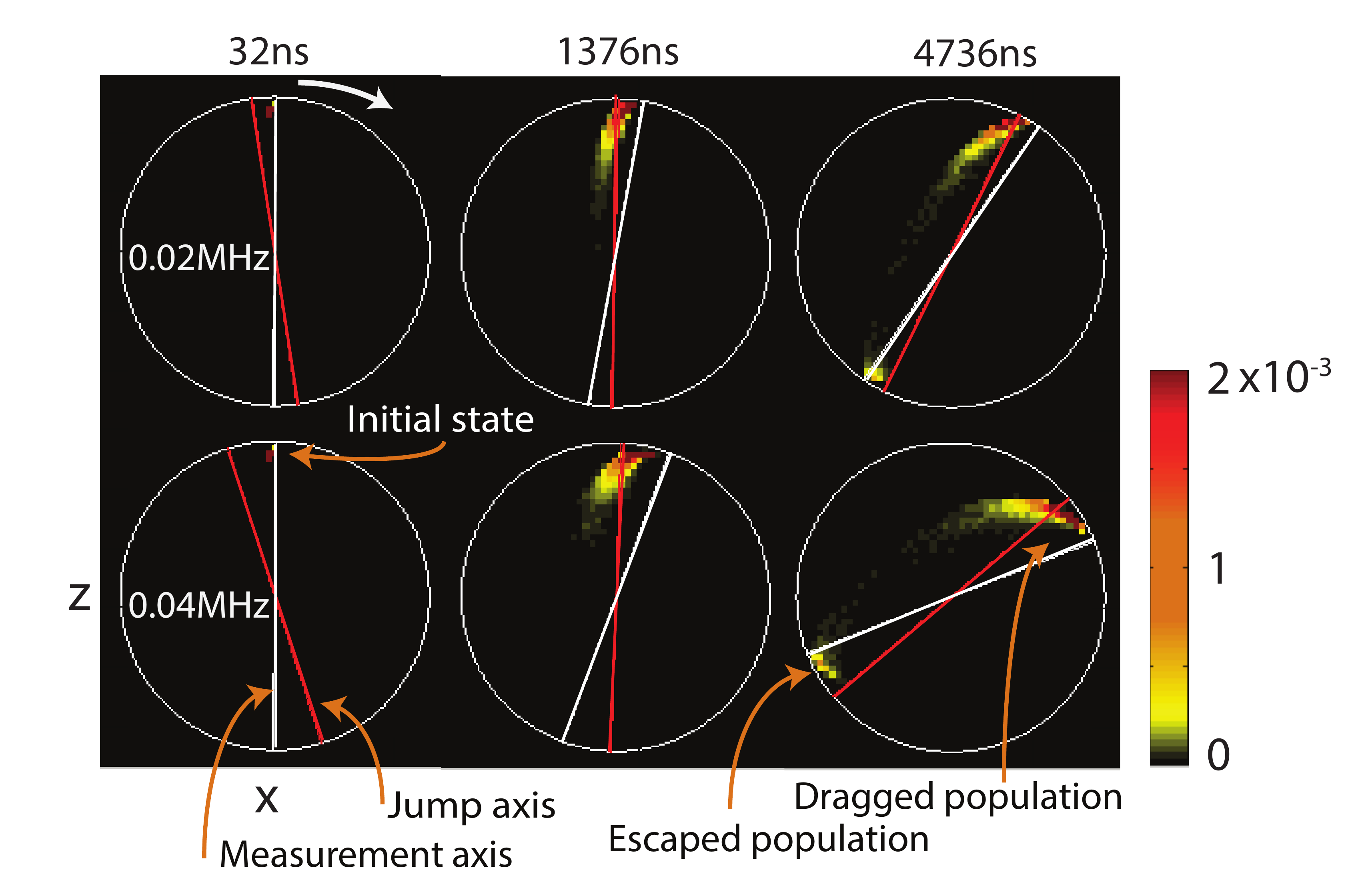}
    \caption{Histograms in the XZ plane of the Bloch vector at different time points in the Zeno dragging experiment. State is initialized at the North pole of the Bloch sphere at $t=0$, and the measurement axis is aligned with the state. Measurement axis is gradually rotated with the angular frequency shown in each pane, which drags the state along with it via the Zeno effect. Red lines show a theoretically calculated `jump axis' discussed in ref. \cite{hacohen2018incoherent}.}
    \label{fig:ZenoDraggingHists}
\end{figure}

Control with measurement alone can serve as a proxy to understand the ultimate advantages of measurement-based feedback and control. By understanding what is possible using arbitrary, potentially adaptive measurements, we gain insight into the limits of quantum feedback and derive a kind of `quantum speed limit' for measurement-based control. In what follows, we develop a feedback protocol based on some of our experimental work\cite{hacohen2018incoherent} extending the results of chapter \ref{ch:XZ}. The experiment involves using the single-quadrature measurement to control a quantum state. As the axis of a single-quadrature measurement can be changed during the measurement process, we can use the Zeno effect to `drag' the Bloch vector of a qubit along the measurement axis. Experimental data showing this procedure are plotted in Fig. \ref{fig:ZenoDraggingHists}. In the limit of infinite measurement strength, the Bloch vector remains aligned to the measurement axis, enabling arbitrary state rotations simply by continuously changing the measurement axis. When considering measurements of finite strength, the state has a finite probability to jump to the other pole of the measurement axis, as is visually apparent from the histograms.

Given an upper bound on the measurement strength, there should exist an optimal strategy for minimizing quantum jumps and maximizing the fidelity to the target state. In general, this strategy could involve adaptive measurement, in which we choose the measurement axis based on past measurement outcomes. In the following, we imagine that the feedback controller maintains the measurement axis half way between the current and target states within a fixed plane of the Bloch sphere. For the time being, we restrict the Bloch vector and measurement axis to lie in the XZ plane. A variant of this protocol was given in ref. \cite{balouchi2014optimal}, though the work only considered the case when the current state is infinitesimally close to the target state. In what follows, we calculate the dynamics under this protocol and show that it is globally optimal for control with measurement alone. 

We parameterize the state and measurement axis with the angular variables $\chi$ and $\delta$ respectively as
\begin{align} \label{eq:ZenoDragParameterization}
    |\psi\> &= \begin{pmatrix}
    \psi_e \\
    \psi_g
    \end{pmatrix} = 
    \begin{pmatrix}
    \cos(\chi/2) \\
    \sin(\chi/2)
    \end{pmatrix} \\ \nonumber
    M &= \cos(\delta) \sigma_z + \sin(\delta) \sigma_x.
\end{align}
Given our usual convention, the effective measurement rate is implicitly $\Gamma=2$. If the initial state is pure and $\eta=1$, then we can simplify calculations by working with a stochastic Schr\"odinger equation
\begin{align}
    d|\psi\> = \left[-\frac{1}{2} (M-\<M\>)^2 dt + (M-\<M\>) dW\right]|\psi\>.
\end{align}

We do not expect an exact solution to exist for an arbitrary feedback strategy. However things simplify nicely if we maintain the measurement axis halfway between the current and target states. If the target state is $|e\>$ (corresponding to $\chi=0$), then the corresponding feedback strategy is $\delta = \chi/2$. The stochastic equation for $\psi_e$ becomes
\begin{align}
    d\psi_e = (1-\psi_e^2)\left(\frac{\psi_e}{2} dt + dW\right).
\end{align}
$\psi_g$ is uniquely determined by $\psi_e$ given our restriction to the XZ plane. Using Ito's lemma, we can derive an equation of motion for $\mathcal{F}_e = \psi_e^2$, the fidelity with respect to the target state
\begin{align}
    d\mathcal{F}_e = (1-\mathcal{F}_e)dt + 2\sqrt{\mathcal{F}_e}(1-\mathcal{F}_e) dW.
\end{align}
As the drift ($dt$) term is linear in $\mathcal{F}_e$, averaging over $dW$ on both sides yields a very simple equation of motion for $\<\mathcal{F}_e\>$
\begin{align}
    d\<\mathcal{F}_e\> &= (1-\<\mathcal{F}_e\>)dt + 2\<\sqrt{\mathcal{F}_e}(1-\mathcal{F}_e)\>\<dW\> = (1-\<\mathcal{F}_e\>)dt \\ \nonumber
    &\implies \<\mathcal{F}_e(t)\> = 1-(1-\<\mathcal{F}_e(0)\>)e^{-t}.
\end{align}
Not only does linearity enable a simple solution to the equations of motion, but it also greatly simplifies the proof of optimality. The cost-to-go for the above protocol is
\begin{align}
    c(t) = (1-\<\mathcal{F}_e(t)\>)e^{t-T}
\end{align}
where $T$ is the stopping time as usual. The corresponding Hamilton-Jacobi-Bellman equation is
\begin{align} \label{eq:QubitHJB}
    \frac{\partial{c}}{\partial t} = \max_v\left[ -\frac{\partial c_v}{\partial t} A_{e,v} - \frac{1}{2}\frac{\partial^2 c_v}{\partial t^2} B_{e,v}^2 \right]
\end{align}
The second term drops due to linearity. As $A_{e,v}$ is positive, the above maximization over $c_v$ is then equivalent to the condition that $u$ is locally optimal. It can be shown algebraically that the above protocol is locally optimal, even if we maximize over all possible orientations of the measurement axis. By \erf{eq:QubitHJB}, it is also globally optimal.

We can also generalize the above feedback protocol to control an arbitrary quantum system. Suppose the current state is $|\psi\>$ and we wish to take it to $|\psi_T\>$. One can always find a basis in which $|\psi\> = \cos(\chi/2) |\psi_T\> + \sin(\chi/2) |\psi_T^\perp\>$, where $|\psi_T^\perp\>$ is orthogonal to $|\psi_T\>$. In the multidimensional case, imposing a maximum measurement rate amounts to upper-bounding the difference between the largest and smallest eigenvalues of $M$. Without loss of generality, let the maximal and minimal eigenvalues be $\pm 1$. Let us orient the measurement operator $M$ such that
\begin{align}
    &M = (|\psi_T\>\<\psi_T| - |\psi_T^\perp\>\<\psi_T^\perp|)\cos(\delta) + (|\psi_T^\perp\>\<\psi_T| + |\psi_T^\perp\>\<\psi_T|)\sin(\delta) + \mathcal{O} \\ \nonumber
    &\<\psi_T|\mathcal{O}|\psi_T\> = \<\psi_T^\perp|\mathcal{O}|\psi_T^\perp\> = \<\psi_T^\perp|\mathcal{O}|\psi_T\> = \<\psi_T|\mathcal{O}|\psi_T^\perp\> = 0.
\end{align}
In general, there will exist infinitely many unitary changes of basis that put $M$ in this form. The first two terms of $M$ exactly coincide with $\sigma_z$ and $\sigma_x$ in \erf{eq:ZenoDragParameterization}, and the eigenvectors associated to this subspace are $\pm 1$. As $|\psi\>$ has no overlap with $\mathcal{O}$, none of the measurement outcomes associated to the eigenvectors of $\mathcal{O}$ can occur, and the problem reduces to the qubit case considered above. In essence, we have rotated $M$ so that the maximally distinguished eigenvectors are `in-plane' with the current and target states, effectively maximizing the measurement strength. If this choice of basis (with $\delta = \chi/2$ as before) is locally optimal, then it follows that this protocol is also globally optimal by the same argument as for a single qubit. This result would imply a measurement-based quantum speed limit for an arbitrary quantum system. We have numerical evidence of local optimality, but have not proven the result in full generality.



%% file: Outlook.tex
\chapter{Outlook}
\label{sec:Outlook}

What new capabilities arise from the ability to resolve quantum measurements as the continuous processes that they are? We have identified and expanded on three potential areas, adaptive measurement, quantum foundations and entanglement generation. In what follows, we give a perspective on what may come next, and introduce one relatively new direction for future research.

A common theme in the work of this thesis is quantum feedback's ability to circumvent the limitations of linear optics. In chapter \ref{ch:MeasurementControl}, we saw numerous examples in which feedback enhanced the success rate of entanglement generation in remote or weakly interacting systems. The reason that these experimental systems cannot create entanglement deterministically to begin with goes back to a fundamental result in linear optics that prevents deterministic implementation of a Bell measurement\cite{Calsamiglia2001maximum}. This theorem forces some fraction of the measurement outcomes to project into separable states. Feedback provides an effective nonlinearity, which allows us to steer the system toward entangled outcomes with up to 100\% efficacy. These results provide evidence that there is something fundamental to be gained from applying quantum feedback to these types of problems. We have many examples in which these schemes can be applied in realistic experimental systems. We believe that most of the protocols developed in this thesis could be readily applied in superconducting circuit systems with parameters that have already been demonstrated. Expanding beyond superconducting circuits, we have developed an improved version of the standard Barrett-Kok scheme for remote entanglement generation in atomic and other solid state systems. This improved protocol retains the standard method's robustness to all common experimental imperfections, and will be the subject of a future manuscript. 

The present work focused primarily on the preparation of relatively simple entangled states such as Bell, Dicke and GHZ states. Beyond these states, there is ample opportunity to apply the formalism developed in chapter \ref{ch:MeasurementControl} to more advanced systems. In a way, generation of these highly symmetric states states serves mostly as a proof of usefulness of the feedback equations derived here, and we expect them to be capable of far more. Interesting applications could be toward cluster states and tensor network states, where there is a locality structure to consider and exploit in the context of remote entanglement and quantum networks. Furthermore, just as we see feedback benefiting systems constrained by linearity, we expect the converse to also hold to some extent; systems in which strong nonlinearities are available will likely benefit far less from quantum feedback. We have given some justification for this assertion in previous chapters, for instance by noting that any POVM can be synthesized by a quantum circuit.

Our implementation of a canonical phase measurement provides another example in which feedback fundamentally enhances the capabilities of a linear optics system. Adaptive measurement enables exactly reaching the canonical phase measurement in our experiment, and more generally can enhance the precision of a detector\cite{armen2002adaptive, wheatley2010adaptive}. Much progress has been made in theoretically mapping out the POVMs that may be implemented with continuous measurement\cite{oreshkov2005weak}. However the question of what POVMs can be implemented given a measurement device and the ability to adapt is still largely an open theoretical problem. This question ties into the more specific question of determining the minimal resources to implement a general canonical phase measurement. Both of these questions would be interesting to address with stochastic control theory methods like the one developed in this thesis. Just as we addressed optimal control for state generation, it should be possible to apply these same techniques to optimal generation of a POVM.

There are a number of potential and known experimental applications of adaptive measurement worth noting. Firstly, the universality of weak measurements for generating arbitrary POVMs can be demonstrated elegantly in superconducting circuits with existing technology\cite{dressel2014implementing}. There are also protocols for enhanced state purification\cite{Jacobs2003} and faster readout of qubit registers\cite{Combes2008rapid} with forms of adaptive measurement. More broadly, implementation of general quantum computation with linear optics has been a theoretically rich direction for quantum information\cite{knill2001scheme} and a still nascent but promising experimental platform\cite{rudolph2017optimistic}. The canonical phase measurement fills a need in photonic quantum information processing by enabling the generation of arbitrary single-rail photonic qubits\cite{pozza2015deterministic}. It will be interesting to see if adaptive measurement offers further novel capabilities for the technology moving forward, particularly for state generation.

The third major topic addressed in this thesis is the simultaneous measurement of non-commuting observables. While this novel capability may have applications for metrology and quantum state tomography, we feel that the most exciting future direction is in quantum foundations. When viewed in a fixed basis, quantum mechanics largely reduces to classical mechanics, and interesting phenomena like contextuality and nonlocality do not manifest themselves. Typical experiments probing these properties involve sequential or discrete measurements, but these do not mimic how the phenomena would manifest in a natural setting. In contrast, most physical systems are continuously measured by their environments. Thus the experimental direction here provides an opportunity to merge the study of quantum foundations and the study of decoherence in natural systems. Several works have begun investigating this line of research\cite{Dressel2016LG, Nishizawa2015ContinuousErrorDisturbance}. As we noted in chapter \ref{ch:XZ}, `Error-disturbance bounds would provide a natural starting point, as we have separately extracted state information in Fig. \ref{fig:XZTomography}b and quantified state disturbance in Fig. \ref{fig:XZDistMap}, but have not yet investigated the intrinsic trade-off between these concepts.'


All existing quantum trajectory experiments have been performed with one or two qubits. Is it possible to scale them to many-body systems? At first glance, the problem does not look scalable. To validate quantum trajectories experimentally, we repeated each experiment hundreds of thousands of times, acquiring statistics for all possible final states and then post-selecting on realizations that lead to the same final state. Quantum state tomography scales exponentially in the number of qubits, but trajectory validation scales exponentially in the number of orthogonal system states, which itself scales exponentially in the number of qubits! Fortunately, there are far less costly ways to test a trajectory model experimentally. One can post-select on states that lead to the same value of some particular operator, and then measure that operator repeatedly, for example. This method does away with all exponential scaling, except in theoretical simulation of the system itself. One can also use feedback to limit the space of possible final states. In a feedback-based control problem, the final number of states is ideally just the target state.

\subheading{Quantum trajectories in AMO and many-body physics} While our focus has been on elucidating new capabilities, quantum trajectories also offer a tool to more deeply understand existing, or even `canonical' systems. One existing example is the ability to use weak measurements to measure correlators in many body systems\cite{halpern2018quasiprobability}. More generally, quantum trajectories should offer both an experimental tool and a unique theoretical perspective. As only a handful of quantum trajectory experiments have been performed, many of the most basic physics effects at the foundation of AMO and many-body physics have only been observed at the ensemble level. Unravelling the decoherence channels in these systems as diffusing quantum trajectories offers a new perspective on even the most basic phenomena.

There are many standard experiments that one could imagine unravelling. The Mollow triplet has been observed in many atomic and artificial atomic systems, but quantum trajectories would allow direct observation of the transitions between dressed states that lead to this characteristic spectrum. Photon absorption and more generally atom-field dynamics in the presence of non-classical light could yield interesting experiments in which one monitors the trajectories of the joint atom-field state. Arguably, the recent observation of quantum jumps as a continuous process is a step in this direction\cite{minev2018catch}. A particularly interesting potential avenue would be the unravelling of superradiance and superradiant lasing. In the many-body limit, this phenomena may be understood as a nonlinear synchronization process between atoms, which is mathematically analogous to the synchronization of fireflies, for example\cite{xu2016theory}. At the level of a few-atom system, entanglement between atoms should begin to play a role. Observing entanglement generation and spontaneous synchronization with trajectories would be interesting not only from a fundamental physics perspective, but could also provide new directions for the development of superradiant lasers. Such devices would have unprecedented linewidths, and a trajectories approach could offer practical benefits via feedback.

One broader generalization of quantum trajectories experiments would be to study truly many-body physics. Phase transitions induced by dissipation, such as the superradiant phase transition would be natural candidates. However in principle virtually any phase transition could be continuously monitored by observing the bath with which it exchanges heat, volume, particles and entropy. While observing the microscopic dynamics in a truly macroscopic material is daunting, much could be learned in cold-atom systems such as quantum gas microscopes, or in systems simulated via superconducting circuits.